\documentclass[prb,twocolumn,showpacs]{revtex4-1}

\usepackage{graphicx} \usepackage{dcolumn} \usepackage{bm}

\begin{document}

\title{Fractional quantum Hall effects in  bilayers in the presence of inter-layer tunneling and charge imbalance}

\author{ Michael R. Peterson$^1$, Z. Papi\'{c}$^{2,3,4}$ and S. Das Sarma$^1$} 
\affiliation{$^1$Condensed Matter Theory Center, Department of
Physics, University of Maryland, College Park, MD 20742 USA}
\affiliation{$^2$Institute of Physics, University of Belgrade, Pregrevica 118,
11 000 Belgrade, Serbia} 
\affiliation{$^3$Laboratoire Pierre Aigrain, Ecole Normale Sup\'{e}rieure,
CNRS, 24 rue Lhomond, F-75005 Paris, France}
\affiliation{$^4$Laboratoire de Physique des Solides, Univ. Paris-Sud, CNRS,
UMR 8502, F-91405 Orsay Cedex, France}

\date{\today}

\begin{abstract}
Two-component fractional quantum Hall systems are providing a 
major motivation for a large section of the physics community.  Here we 
study two-component fractional quantum Hall systems in
spin-polarized half-filled lowest Landau level (filling factor 1/2) and second Landau level 
(filling factor 5/2) with exact
diagonalization utilizing both the spherical 
and torus geometries.  The two distinct two-component systems we consider are the
true bilayer and effective bilayers (wide-quantum-well).  In each model (bilayer and
wide-quantum-well) we completely take into account inter-layer tunneling 
and charge imbalancing terms.  We find that in the half-filled lowest Landau 
level, the FQHE is described by the two-component Abelian Halperin 331 
state which is remarkably robust to charge imbalancing.  In the half-filled second 
Landau, we find that the FQHE is likely described by the non-Abelian 
Moore-Read Pfaffian state which is also quite robust to charge imbalancing.  Furthermore, 
we suggest the possibility of experimentally tuning from an Abelian to non-Abelian 
FQHE state in the second Landau level, and comment on recent experimental studies of FQHE in wide quantum well structures.
\end{abstract}

\pacs{73.43.-f, 71.10.Pm}

\maketitle

We theoretically investigate two-component spin-polarized fractional quantum 
Hall states in the bilayer half-filled
first and second (first-excited) Landau levels considering the effects of both inter-layer 
 tunneling and charge imbalance tunneling 
using exact diagonalization.  The 
aspect that is different in this work compared to previous 
studies~\cite{he-1,he-2,nomura,papic,mrp-sds-bilayer} is 
that we consider inter-layer tunneling 
along with charge imbalance tunneling.  
The reason we consider this \emph{extra} tunneling term (charge imbalance), along with 
the an inter-layer tunneling term, 
is because recent experimental efforts~\cite{shabani-1,shabani-2} have 
achieved bilayer FQHE systems 
where both the inter-layer and 
charge imbalance tunneling terms can be controlled by varying different system parameters 
(such as gate voltages) 
and a theoretical investigation of the physics of this system is both 
timely and important.  Furthermore, the basic effects 
produced by a charge imbalance term in a theoretical exact diagonalization context 
is currently lacking.  Before we 
delve into our results we provide an introduction, motivation, 
and historical perspective of this broad subject.

\section{Fractional quantum Hall effect in the one- and 
two-component variety}

The fractional quantum Hall effect~\cite{tsui-stormer-gossard,laughlin} (FQHE) 
has proved to be the paradigm for emergent quantum physics for the 
nearly 30 years of its existence.  It occurs when electrons are confined to a 
(quasi-)two-dimensional plane (such as in 
semiconductor structures, e.g., GaAs/AlGaAs, with electron densities 
on the order of $10^{10}$-$10^{11}$ cm$^{-2}$) and a strong perpendicular 
magnetic field is applied (usually on the order of tens of Tesla--sometimes up to 
$\sim40$ T).  
Phenomenologically, the FQHE manifests as 
a quantized plateau in the Hall resistance $R_{xy}$ (quantized to parts per billion) and 
a concomitant vanishing (or deep minimum that displays activated behavior) 
of the longitudinal resistance $R_{xx}$.  The FQHE is said to occur at rational 
fractional filling factor $\nu$ when 
the quantized value of the Hall resistance 
is $R_{xy}=h/(\nu e^2)$, where $\nu=\rho/\phi_0$ (here $\rho$ is the electron 
density and $\phi_0=hc/eB$ is the magnetic 
flux quantum and $B$ is the magnetic field strength, 
hence, $\nu$ is the number of electrons per magnetic flux quanta).  The perpendicular magnetic 
field quantizes the two-dimensional kinetic energy into 
Landau levels 
separated in energy by $\hbar\omega_c=\hbar eB/mc$ and when the filling 
factor $\nu$  is 
made to be fractional (like it is for the FQHE), 
by either adjusting the electron density and/or magnetic field 
strength, the  kinetic energy is a constant and completely 
flat bands obtain.  In the limit that $\hbar\omega_c\rightarrow\infty$ (or
 the extreme quantum limit) the electron-electron 
Coulomb interaction is the dominant term in the Hamiltonian for electrons 
fractionally filling a Landau level (LL).  

The one-component FQHE in the 
lowest orbital electronic Landau level is the most often discussed since it has been 
observed  in the form of approximately 80 odd-denominator FQHE 
states and is  well 
understood~\cite{perspectives,heinonen,cf-book,laughlin,jain-prl}.
Essentially, the FQHE occurs due to the non-perturbative and 
electron-electron interaction driven formation of an emergent topologically ordered~\cite{wen} 
incompressible 
quantum fluid at certain filling factors $\nu$ with non-zero energy gaps.  This 
(bulk) energy gap, along with some sample disorder, explains the FQHE~\cite{perspectives}.

Recently, the FQHE 
in half-filled Landau levels has reinvigorated the community due to its possible 
connection to topological quantum computing, non-Abelian quasiparticles, and  the 
requisite cutting edge material science advances that have produced much of this 
new physics and continue to coax nature into revealing her secrets.  In particular, the 
FQHE at filling factor 5/2~\cite{willett}, which corresponds to filling factor 1/2 in the second 
orbital electronic Landau level, has arguably produced most of the excitement.

\begin{figure}[]
\begin{center}
\includegraphics[width=8.cm,angle=0]{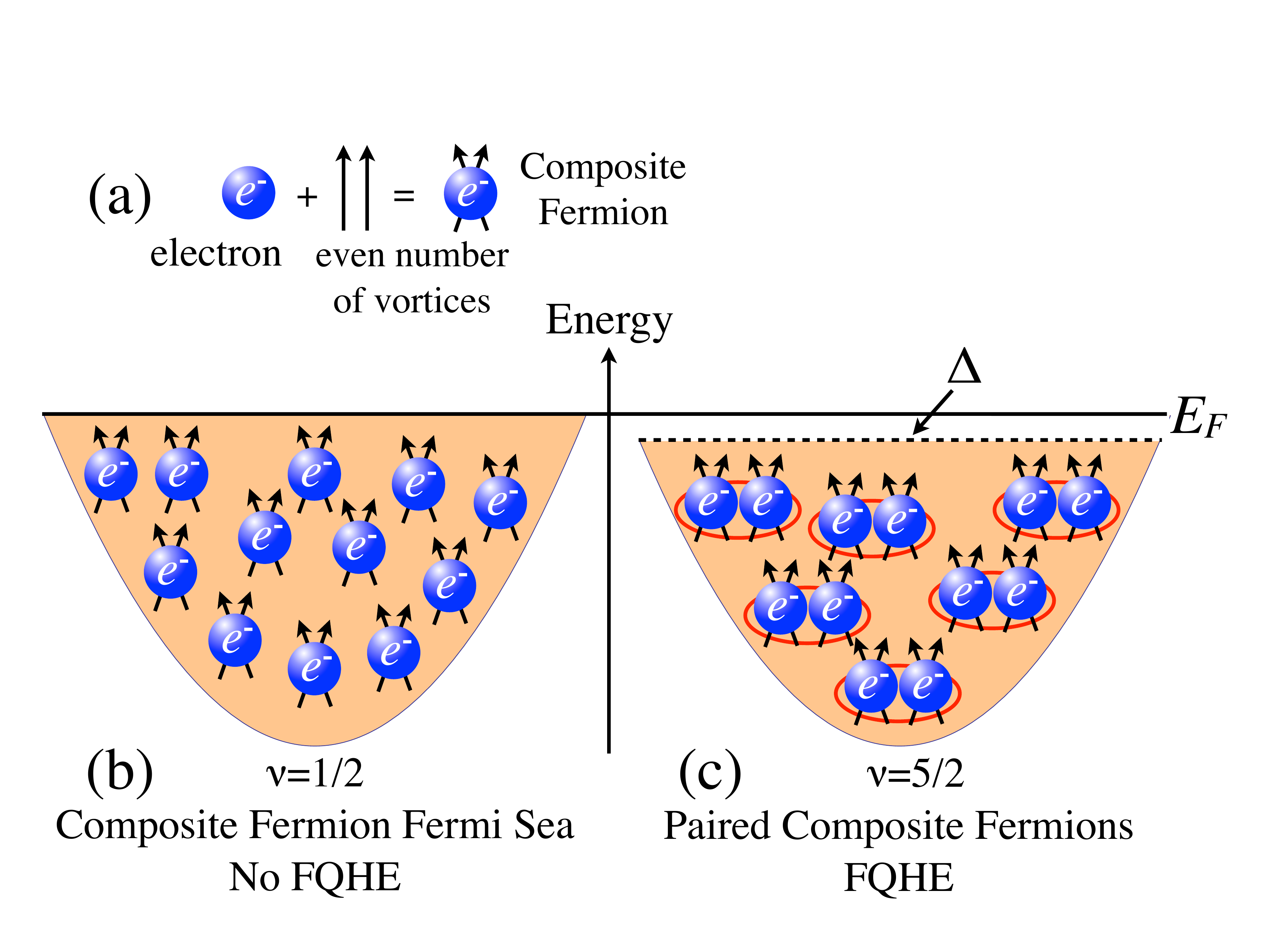}
\caption{(Color online) (a) A Composite Fermion is an electron 
bound (or attached) to an even number  of quantum mechanical vortices of the 
many-body wavefunction, see Ref.~\onlinecite{jain-prl,cf-book}.  (b) A compressible 
Fermi sea of Composite Fermions forms for a one-component system at filling factor 
$\nu=1/2$ in the lowest Landau level and produces \emph{no FQHE} 
since a (Composite Fermion) Fermi sea 
has no energy gap.  (c) In the half-filled 
second Landau level at $\nu=5/2$ the electron-electron interaction is modified 
compared to the electron-electron interaction in the lowest Landau level (see text) causing the 
weakly interacting Composite Fermions to pair into a spin-polarized $p$-wave 
BCS state described by the Moore-Read Pfaffian wavefunction.  This state, due to the 
quasiparticle pairing, has an energy gap and thus exhibits the FQHE. }
\label{fig1}
\end{center}
\end{figure}

The denominator of $\nu$ for the ``standard" fractional quantum Hall 
states is always 
odd and reflects the fermionic nature of the quasiparticles.   At $\nu$=1/(even), naive, 
zeroth-order, 
theory asserts that no FQHE would be expected.  This 
is because, at $\nu$=1/(even), the 
weakly interacting quasiparticles of the 
FQHE (Composite Fermions~\cite{jain-prl,cf-book}) experience 
a zero effective magnetic 
field and form a \emph{gapless} exotic Fermi sea~\cite{hlr,kalmeyer,rezayi-read}. 
In fact, no FQHE is experimentally observed
in one-component systems at $\nu=1/2$ instead showing signatures of the 
exotic Fermi sea~\cite{willett-1,du,kang-1} 
(see Fig.~\ref{fig1}).  

The FQHE at $\nu=2+1/2=5/2$ (the 2 comes from completely filling the spin-up and spin-down 
Landau levels) is particularly interesting and the only known and 
well established violator of the ``odd-denominator"  rule for the one-component 
FQHE.     A one-component FQH state 
with an even-denominator 
suggests some kind of 
interaction driven pairing among the weakly interacting emergent (quasi-)fermions and does not find a 
description within the standard FQHE theory.  All FQHE states observed 
in the second Landau level (SLL), such as 5/2, 7/3=2+1/3,  8/3=2+2/3, 
12/5=2+2/5, etc. (there are about 8 
FQHE states observed in the SLL in all) are fragile when compared to the FQHE 
states in the lowest Landau level.  The 5/2 FQHE is one of the, if not the, strongest 
FQHE states in the SLL and yet has a measured activation gap only about 0.5 K 
or less.  This is despite being measured in the world's cleanest two-dimensional 
electron gas samples with mobilities over $30\times10^6$ cm$^2$/Vs at temperatures 
of less than 100 $m$K.  Hence, there is a strong experimental drive to produce 
cleaner samples and apparatus capable of doing 
measurements at exceedingly low temperatures.  
Most early experimental 
observations of the FQHE at 5/2 indicated the state to be one-component in 
nature~\cite{eisenstein-52-1,gammel,pan-prl,eisenstein-52-2,xia,csathy,choi,nuebler}. However, 
care needs to be taken when making broad 
absolute statements and there is some 
experimental evidence~\cite{eisenstein-52-1,csathy,dean-2} that puts the one-component 
interpretation of the 5/2 FQHE into question, but, 
the interpretation of these results is notoriously 
difficult~\cite{sds}.  

Recently, the FQHE was observed~\cite{dean} at $\nu=5/2$ 
in a low magnetic field where the ratio of the cyclotron 
energy $\hbar\omega_c$ to the Coulomb 
energy $e^2/\epsilon l$ (where $e$ is the electron charge, $\epsilon$ is the dielectric constant of 
the host semiconductor and $l=\sqrt{c\hbar/eB}$ is the characteristic length scale called the 
magnetic length) is near unity or less.  In that situation, it is not clear how the system would be 
completely spin-polarized and/or would not experience significant Landau level 
mixing.  
Recent experiments~\cite{dean-3,rhone,stern} have begun to seriously 
tackle these two issues.   Not surprisingly,  
theoretical groups have 
enthusiastically taken up the (considerable) challenge of understanding the role of 
Landau level mixing~\cite{bishara-nayak,simon-rezayi,wojs-toke-jain,wang-sheng-haldane}.  That 
being said, the FQHE at 5/2 has also been seen~\cite{pan-ssc,zhang-prl} at magnetic 
field strengths of more than 10 T.  A FQH state at fields that high is likely to be 
spin-polarized and, hence, one-component.   Adding 
to the confusion is the fact that \emph{all} theoretical 
work~\cite{morf,feiguin} to date have established that, 
within a wide range of parameter space, the electrons are 
completely spin-polarized at $\nu=5/2$.  
It is extremely difficult theoretically (computationally) to consider fully spin un-polarized FQHE 
and until there is definitive experimental evidence indicating the 
real system to be spin un-polarized 
we will theoretically consider the electron spin to be absolutely polarized throughout 
this work.

Arguably, the reason the FQHE at 5/2 is so fascinating is its connection to topological 
quantum computing~\cite{kitaev,tqc-rmp}.  An intriguing ansatz called the 
Moore-Read Pfaffian~\cite{mr-pf} is thought to describe the FQHE at 5/2 and this ansatz has non-Abelian quasiparticle 
and quasihole excitations.  It is proposed~\cite{sds-freedman-nayak} that the 
world-lines of these non-Abelian 
excitations can be braided around each other, thus changing the ground state 
in the degenerate manifold of ground states, 
and certain quantum computing gates can be achieved.  This degenerate 
manifold of states is separated from the continuum by an energy gap that is 
topological in nature, since it is a FQHE state, thus, any 
sort of local disturbance to the system, like those caused from 
typical noise encountered in an experiment, will not be able to destroy the state due to its 
topological origin.  Of course, if the disturbances are of an energy that is larger than the protective 
gap then the above does not hold.  Using the non-Abelian quasiparticles and quasiholes of the 
Moore-Read (MR) Pfaffian (Pf) description of the 5/2 state of the FQHE to  
achieve quantum computing gates that are topologically protected  (called 
fault-tolerant topological quantum computation) is a major 
research goal~\cite{tqc-rmp}.  
(Note that the non-Abelian quasiparticles that might exist for the FQHE 
at 5/2 are so-called Ising anyons and are unable to be used for \emph
{universal}  quantum computation--see Ref.~\onlinecite{tqc-rmp} for more details on this point and 
how Ising anyons can be augmented to achieve a version of universal 
fault-tolerant topological quantum computation.)

Intuitively, the existence of the Moore-Read Pfaffian can be understood (see 
Fig.~\ref{fig1}).  Any time a Landau 
level is half-filled (not just at $\nu=1/2$ as mentioned above but also at $\nu=5/2=2+1/2$) 
naive zeroth-order 
theory tells us that a (Composite Fermion) Fermi sea~\cite{hlr,kalmeyer,rezayi-read} can form.  
(Note that in this discussion we assume that the two completely filled 
Landau levels are inert which is the same as assuming that $\hbar\omega_c\rightarrow\infty$.)  As 
is well known, any system of weakly interacting fermionic quasiparticles is unstable to pairing via 
the BCS~\cite{bcs,bcs-2} pairing mechanism and, if the quasiparticles 
are spin-polarized, the simplest 
pairing  is chiral $p$-wave symmetry ($p_x+ip_y$) 
(see the work by Read and Green 
in Ref.~\onlinecite{read-green} for a discussion of the BCS mean field 
description of the FQHE in a half-filled Landau level.)   The resulting 
paired state has an energy gap and the FQHE will occur as long as the system is clean 
enough and cold enough such and the Fermi energy lies within the FQHE gap.  The 
Moore-Read Pfaffian wavefunction encapsulates 
this physics.

We briefly note that recently it has been pointed 
out~\cite{apf-1,apf-2} that a competing state for the FQHE at $\nu=5/2$ is the so-called 
anti-Pfaffian which is the particle-hole conjugate of the Moore-Read Pfaffian.  The 
anti-Pfaffian is topologically distinct from the Pfaffian due to the fact that the two 
have different edge state behavior which, in principle, makes it possible to 
tell the two apart experimentally.  We emphasize that both the Moore-Read Pfaffian 
and anti-Pfaffian \emph{both} support non-Abelian quasihole and quasiparticle 
excitations so both could perform as platforms for fault-tolerate topological 
quantum computation.  Whichever one happens to be responsible 
(if either) for the FQHE at $\nu=5/2$ depends on how, and if, the particle-hole symmetry 
is broken~\cite{mrp-park-sds} as well as experimental and material details 
such as disorder and Landau level mixing~\cite{bishara-nayak,simon-rezayi,wojs-toke-jain}.

(All results in this work concerning the Pfaffian apply equivalently to the anti-Pfaffian within the 
constraints of our approximation scheme since the model we use does not distinguish between 
Pfaffian and anti-Pfaffian. The reason for this is that our Hamiltonian is particle-hole symmetric, 
since we do not consider any particle-hole breaking terms such as those that might arise due to 
Landau level mixing. Since the Pfaffian and anti-Pfaffian are particle- hole conjugates of one 
another, their (bulk) physics, the subject of our current work, is identical.)

A natural question is why the FQHE is observed at $\nu=5/2=(2+1/2)$ and not, 
so far, at $\nu=1/2$ for one-component systems.  Theoretically, this is due to 
the physics of the lowest Landau level being different compared 
to the physics of the second Laudau level~\cite{abrumenil,macdonald,scarola-park-jain,toke-1}, 
i.e., details matter.  For the FQHE at filling factor 5/2, the inert 
electrons in the lowest spin-up and spin-down Landau levels partially screen the electron-electron 
Coulomb interaction between the electrons half-filling the second Landau level--in fact, the 
electrons in the lower Landau level \emph{over screen} 
the Coulomb interaction~\cite{scarola-park-jain} which leads to a slight attraction among 
the quasiparticles 
 causing them to form a BCS state--the Moore-Read Pfaffian state.  Furthermore, the 
 experimental details of a real quantum well--the single-electron wavefunction in the 
 direction perpendicular to the two-dimensional 
surface has a finite extent (the so-called finite thickness of the quasi-two-dimensional 
electron system where typical experimental 
systems have widths ranging from approximately $\sim20$ nm for the thinnest 
samples to $\sim60$ nm for the wide-quantum-well samples) which produces subtle effects that 
make the effective interaction felt by the electrons in the half-filled 
second Landau level even more amenable to forming a non-Abelian Moore-Read Pfaffian 
FQHE at $\nu=5/2$.  Recently, it has been theoretically shown that (within 
certain approximate models)
the effects of Landau level mixing~\cite{bishara-nayak,simon-rezayi,wojs-toke-jain} 
might produce non-trivial and subtle effects that drive the system 
to the Pfaffian~\cite{simon-rezayi} or the anti-Pfaffian~\cite{wojs-toke-jain} state.

We reiterate that experimentally the FQHE at $\nu=1/2$ has not been observed 
in \emph{one-component systems} and  
theory~\cite{storni,papic} suggests that it will not occur for a pure Coulomb interaction.  However, 
one cannot rule out a Moore-Read Pfaffian FQHE at $\nu=1/2$ if the 
system parameters are tuned in the perfect way~\cite{mrp-ft-prl,mrp-ft-prb,papic-2} or, 
perhaps, at extremely low temperatures not currently experimentally accessible.

However, about the same time the experimental 
observation of the 5/2 FQHE was realized, the FQHE was 
also observed by Suen \emph{et al.}~\cite{suen-1,suen-2,suen-3} 
and Eisenstein \emph{et al.}~\cite{eisenstein-bilayer} 
in the lowest Landau level at filling factor $\nu=1/2$ in systems that were later 
determined by He \emph{et al.}~\cite{he-1,he-2,nomura} 
to be spin-polarized two-component systems.
It turns out that the FQHE at $\nu=1/2$ in two-component systems is described by the 
(Abelian) Halperin 331 two-component wavefunction~\cite{halperin-331}.
The system of Eisenstein~\cite{eisenstein-bilayer} 
was an actual bilayer (two quasi-two-dimensional systems separated from 
one another by a tunneling barrier--typical parameters 
are quasi-two-dimensional electron systems of width $\sim20$ nm separated from 
one another by a tunneling barrier of thickness $\sim3-10$ nm with 
electron densities on the order of $10^{11}$ cm$^{-2}$), thus, the two-components are the two 
layers.  The FQHE at $\nu=1/2$ in bilayers occurs by a two-part process 
where Laughlin~\cite{laughlin} states, at filling factor 1/3, are
formed in each layer and the fermions between layers form pairs.  In the 
experiments by Suen \emph{et al.}~\cite{suen-1,suen-2,suen-3} the electrons 
in the wide-quantum-well (a width of order $\sim$70 nm and electron densities  
of $\sim10^{11}$ cm$^{-2}$) minimize their single-particle energy
by reorganizing into effectively 
two ``layers" and, again, the FQHE at $\nu=1/2$ occurs as described above.
Recent observations~\cite{luhman,shabani-1,shabani-2} 
of $\nu=1/2$ and 1/4 in wide-quantum-wells 
have rekindled interest in this rich system.

The discovery and identification of the FQHE at $\nu=1/2$ was an exciting advance 
in the FQHE because it opened up the possibility of a new fractional quantum Hall 
state outside those given in the ``standard" theory~\cite{perspectives,laughlin,jain-prl,cf-book}.  
However, the Halperin 331 state is closely 
related to the standard theory being that it is essentially a 
pairing of Laughlin states.   Scarola and 
Jain~\cite{scarola-jain} 
further generalized the Halperin 331 state to pairings of FQHE states belonging 
to the Jain sequence producing states for bilayer systems at filling 
factors other than $\nu=1/2$--they also produced \emph{partially} 
pseudo-spin polarized states at $\nu=1/2$ that are distinct from the Halperin 
331 state. 

Along with the recent observations~\cite{luhman,shabani-1,shabani-2} 
of $\nu=1/2$ and 1/4 in wide-quantum-wells 
are observations~\cite{shabani-1,shabani-2} 
of bilayer FQHE at filling factor 1/2 in the lowest Landau 
level with \emph{asymmetric} charge distributions--so-called ``tilted samples".  
A tilted sample is created by an additional charge gate that 
produces a charge asymmetry by ``pushing" electrons from one side of the wide-quantum-well 
to the other.  In the simplest approximation, this asymmetry produces 
a charge imbalance between one layer and the other while keeping the total 
filling factor $\nu$ fixed.  That is, the individual filing factors in 
each layer, $\nu_1$ and $\nu_2$, respectively, (keeping $\nu_1+\nu_2=\nu$ fixed) 
can be varied between $(\nu_1,\nu_2)=(\nu,0)$, to $(\nu_1,\nu_2)=(\nu/2,\nu/2)$, 
to $(\nu_1,\nu_2)=(0,\nu)$.

Of course, care should be taken when considering the effects of a charge gate that 
presumably causes charge asymmetry in a wide-quantum-well.  Recently, 
Scarola \emph{et al.}~\cite{vito} have 
used the local-density-approximation (LDA) to get a more realistic handle on exactly 
what tilting does to the charge distribution and relative single-electron 
energy levels in the quantum well.  They went on 
to analyze the system with the use of variational wavefunctions 
finding that the recent experimental observation of the FQHE at $\nu=1/2$ by 
Shabani \emph{et al}.~\cite{shabani-2} at intermediate  
charge imbalancing is likely described by a partially psuedo-spin polarized  
bilayer state~\cite{vito,scarola-jain}.  

The purpose of the present 
work  is to understand the general effects of charge imbalancing in the most 
minimal model one can consider to describe the bilayer FQHE in an 
exact context--that is, using exact diagonalization.  
Our work compliments recent work~\cite{vito} since 
we are solving the Hamiltonian exactly instead of using a 
combination of LDA and variational wavefunctions.  As such, we are restricted to only 
certain ansatz, namely the Moore-Read Pfaffian and the Halperin 
331 state.  We emphasize that our solution is general and exact.

In Sec.~\ref{sec-models} we present our theoretical model in the form of 
a Hamiltonian that can describe either a true 
bilayer and effective (wide-quantum-well) bilayer where we consider 
both inter-layer and charge imbalance tunneling terms.  Four (two plus two) variational 
states are discussed in Sec.~\ref{sec-ansatz} that are thought to describe our 
bilayer and effective bilayer Hamiltonian.   Secs.~\ref{sec-LLL} and ~\ref{sec-SLL} 
present our results for the bilayer and wide-quantum-well model in both the 
lowest Landau level (in both the spherical and torus 
geometry) and second Landau level (in the spherical geometry).  
Finally, in Sec.~\ref{sec-conc}, we present our conclusions.

\section{Theoretical models: true bilayer and effective bilayer}
\label{sec-models}

\begin{figure}[]
\begin{center}
\includegraphics[width=4.cm,angle=0]{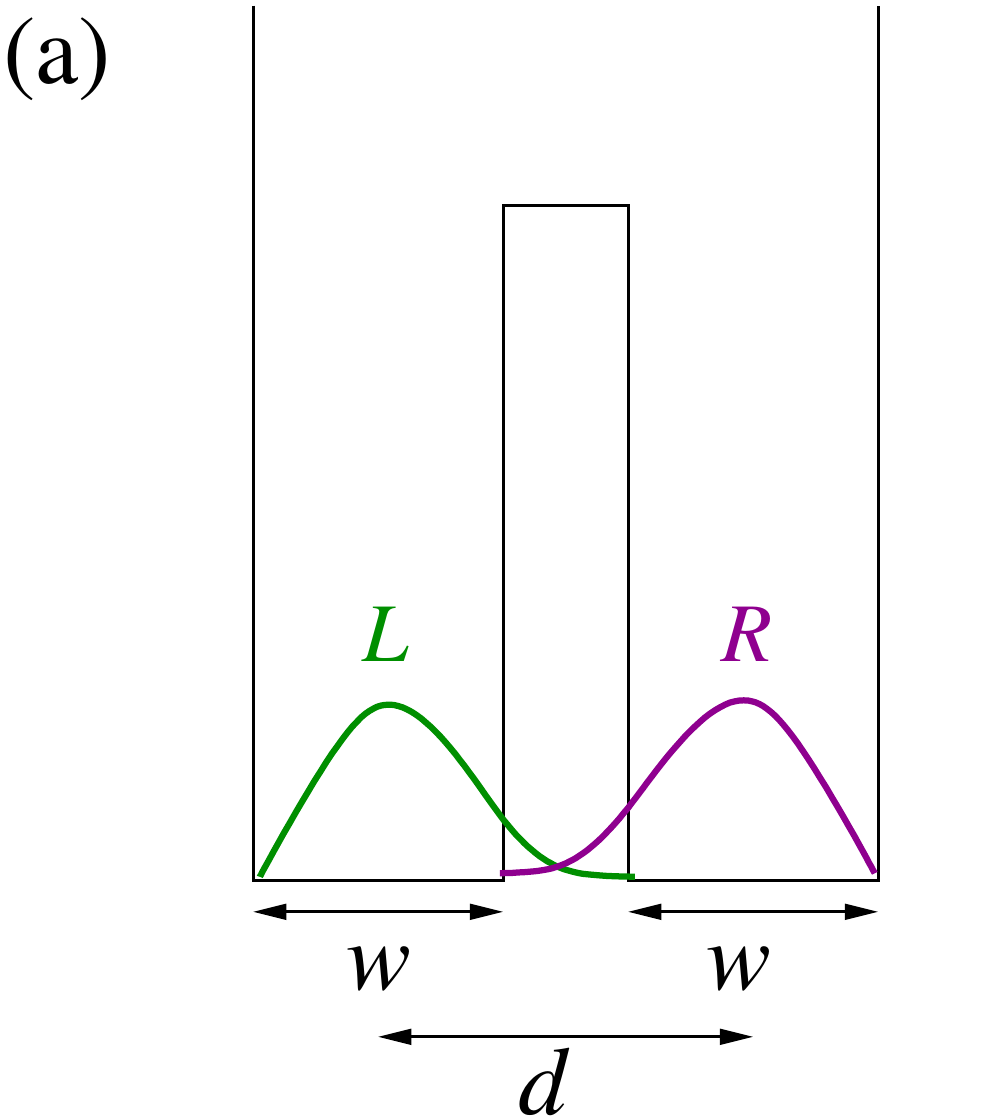}
\includegraphics[width=4.cm,angle=0]{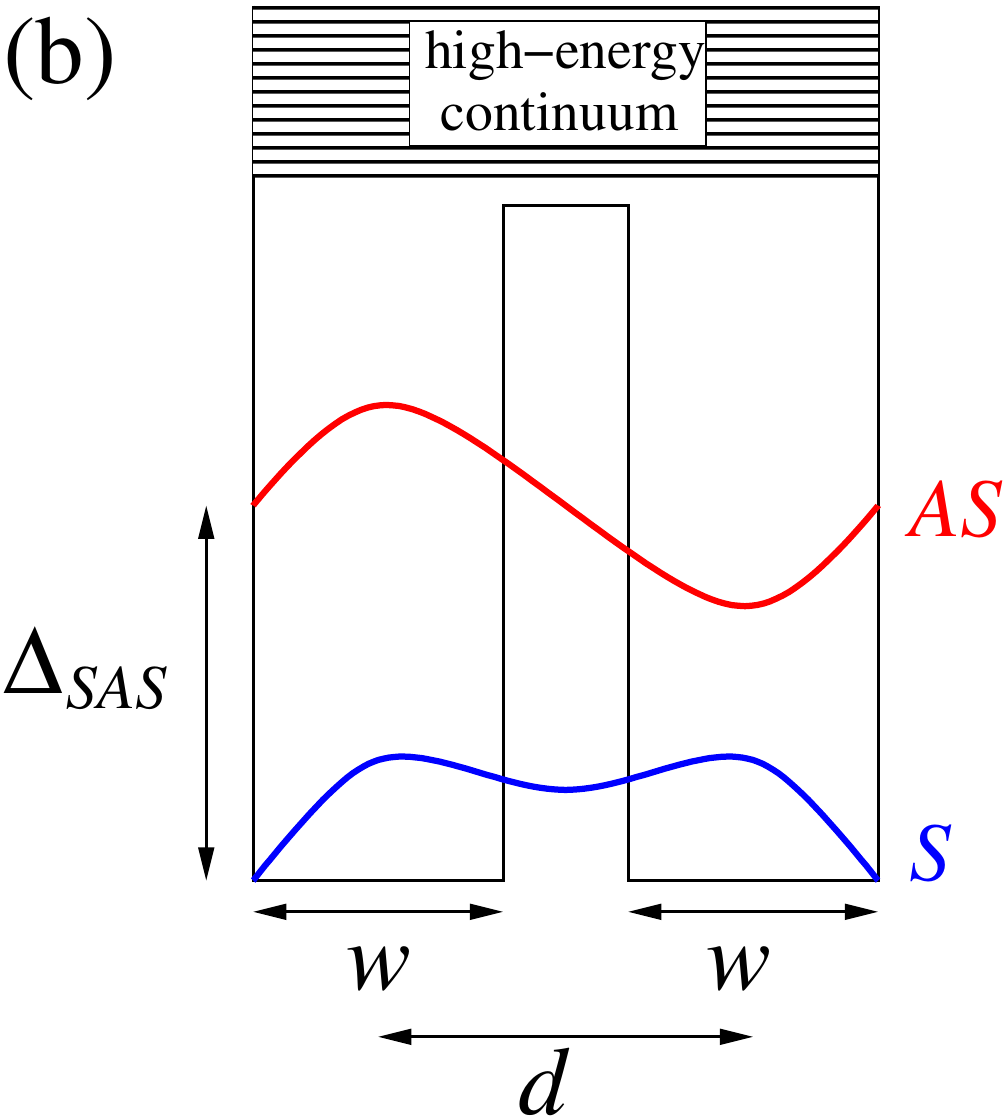}
\caption{(Color online) A bilayer system consisting of two 
(quasi-)two-dimensional quantum wells of width $w$ separated from one 
another by $d$ ($d\geq w/2$).  (a) Typical density profile of 
an electron localized in either the right (R) or left (L) layer, respectively.  (b) 
The ground state single particle wavefunction (the symmetric ($S$))
state and the next higher energy state (the anti-symmetric ($AS$)) 
state separated in energy from one another by the so-called 
symmetric-anti-symmetric energy gap $\Delta_{SAS}$. }
\label{fig-bilayer}
\end{center}
\end{figure}

There are two experimental systems in which two-component FQHE systems can be produced.  
One of them is to manufacture a true bilayer system which consists of two 
parallel (quasi-)two-dimensional electron systems 
of width $w$ separated from one another by a tunneling barrier of thickness $d$.  This 
system is used extensively by the Eisenstein group at CalTech~\cite{eisenstein-bilayer} and 
is what one generally thinks about when contemplating a bilayer system.  The 
height of the barrier can be adjusted such that electrons are either localized in separate 
layers or delocalized between the two layers, i.e., large energetic barrier or small energetic 
barrier, respectively.  In other words, the symmetric-anti-symmetric energy gap, 
$\Delta_{SAS}$, can be controlled independently of the individual well width $w$ and 
the layer separation $d$.  In this system, an electron can be in the right (R) or left (L) state, 
written as $|R\rangle$ or $|L\rangle$, respectively.  Fig.~\ref{fig-bilayer}
shows a schematic diagram of a bilayer system consisting of two parallel two-dimensional 
electron systems where the electrons in each layer are confined to two-dimensions 
by a quantum well of width $w$ (both layers have the same width) and 
the quantum-well-center to quantum-well-center separation is  $d$ ($d\geq w/2$).  A typical 
density profile of electrons in the right or left layer ($\langle R|R\rangle$ or 
$\langle L|L\rangle$) is also shown in Fig.~\ref{fig-bilayer}a.  The ground 
state of the single-electron bilayer Hamiltonian is the symmetric state $S$ given by
\begin{eqnarray}
|S\rangle=\frac{|R\rangle+|L\rangle}{\sqrt{2}}
\label{s-state}
\end{eqnarray}
 and the next higher energy state 
is the anti-symmetric state given by  
\begin{eqnarray}
|AS\rangle=\frac{|R\rangle-|L\rangle}{\sqrt{2}}
\label{as-state}
\end{eqnarray}
(assuming no charge imbalance term in the Hamiltonian).  
The two states ($S$ and $AS$) are separated by 
an energy difference $\Delta_{SAS}$ and their respective (typical) density 
profiles ($\langle S|S\rangle$ and $\langle AS|AS\rangle$) are shown in Fig.~\ref{fig-bilayer}b.  

\begin{figure*}[]
\begin{center}
\includegraphics[width=4.cm,angle=0]{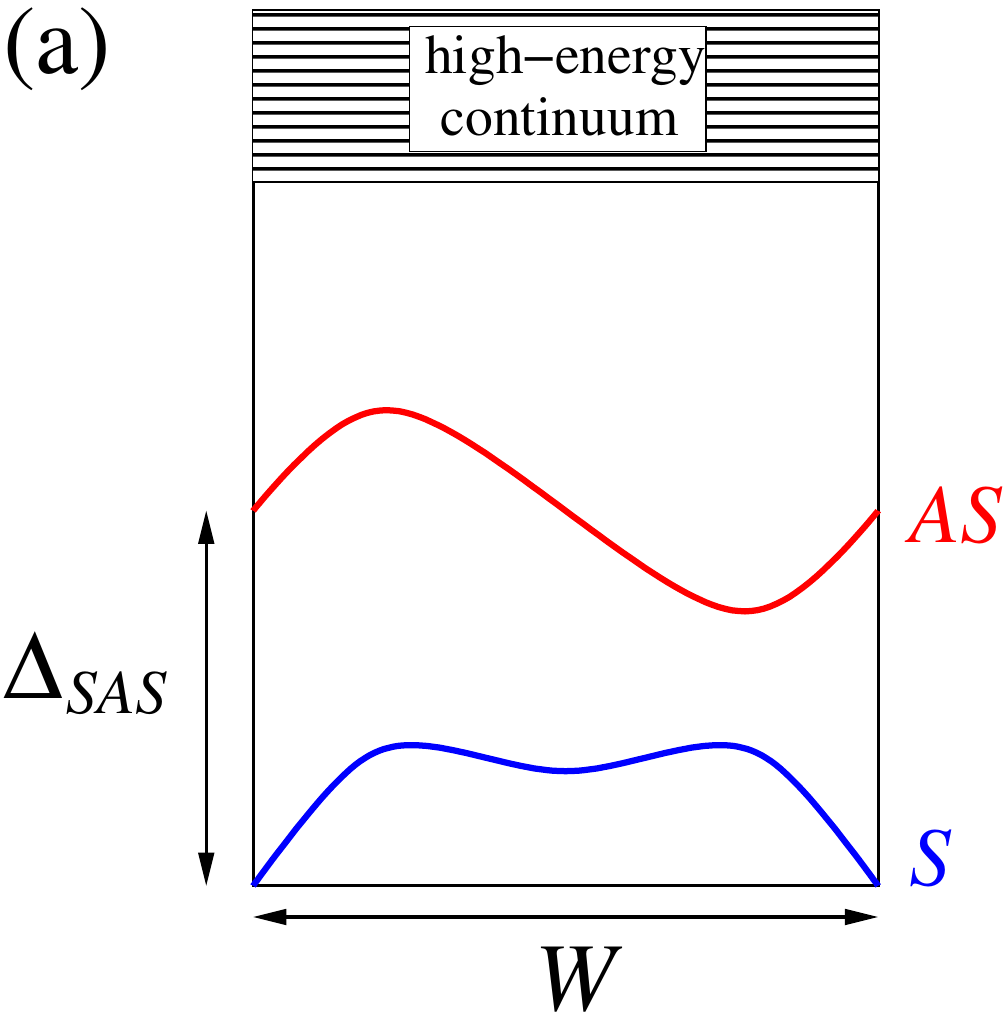}
\includegraphics[width=4.cm,angle=0]{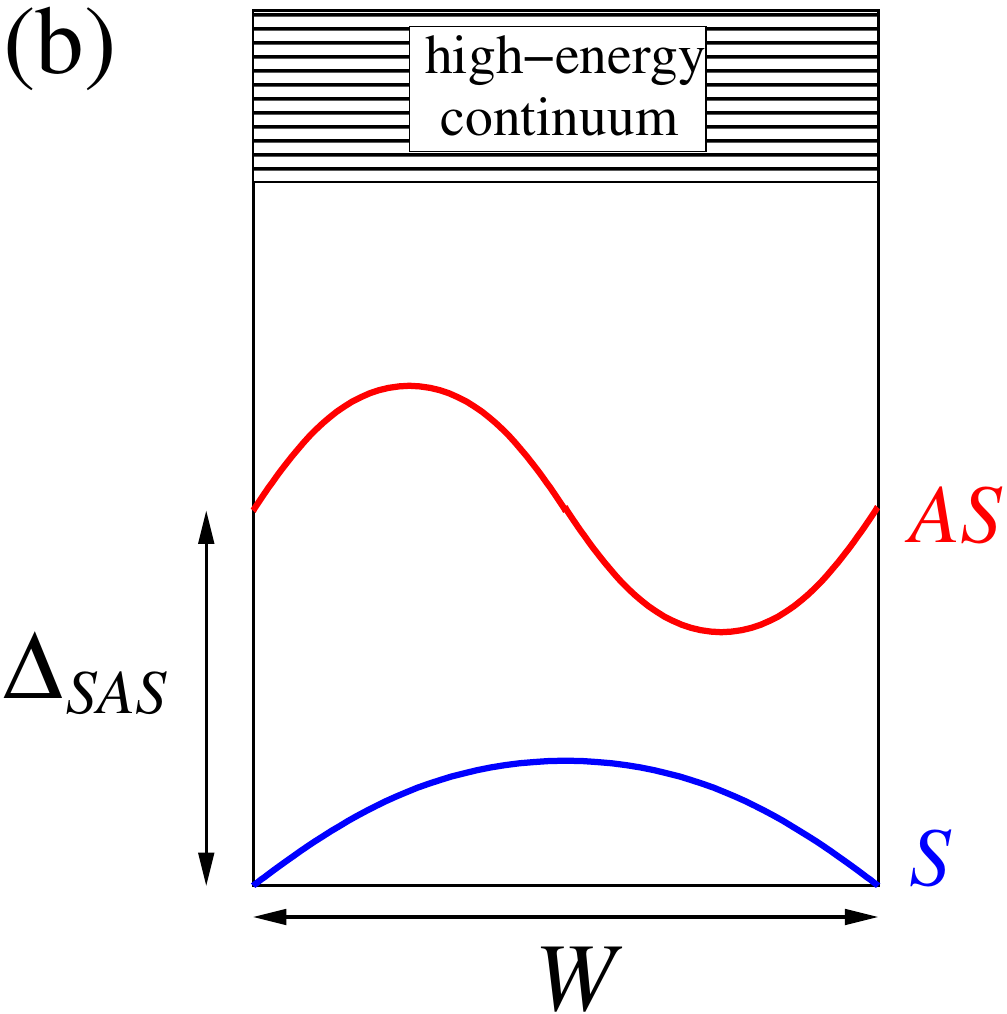}
\includegraphics[width=4.cm,angle=0]{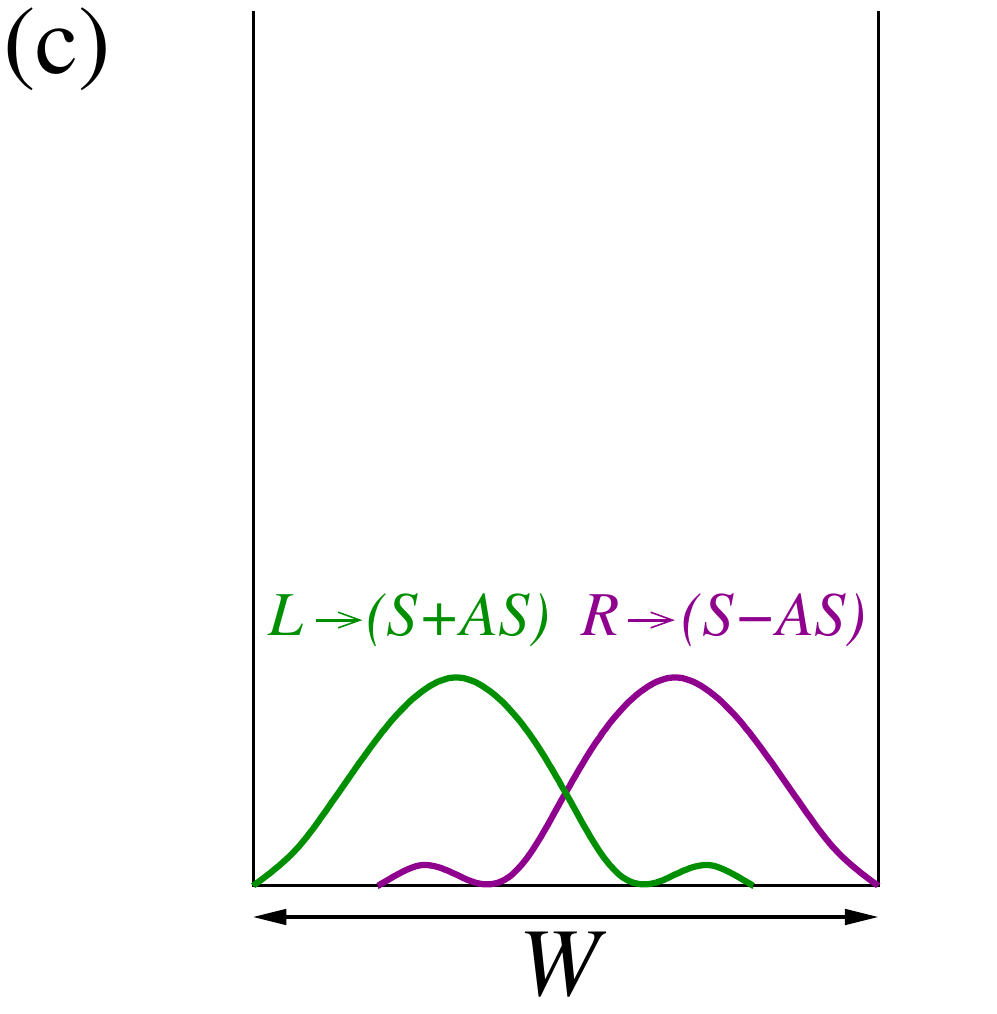}
\caption{(Color online) (a) A two-component system created 
by a wide-quantum-well (WQW).  The electrostatic potential generated by the 
electrons self-consistently creates an effective bilayer system by pushing the 
electrons towards the walls of the WQW.  Note that (a) is \emph{not} identical 
to Fig.~\ref{fig-bilayer}b even though they are very similar--specifically, $\Delta_{SAS}$ is a 
strong function of $W$ and the electron density in WQW systems as opposed to 
bilayers.  (b) The model of  Papi\'{c} \textit{et al}.~\cite{papic} which 
can be transformed into the layer-basis shown in (c).}
\label{fig-wqw}
\end{center}
\end{figure*}

The other way to experimentally create a two-component FQHE system 
is to make a single (quasi-)two-dimensional electron gas of large 
width, a so-called wide-quantum-well (WQW), as done by the Shayegan group at 
Princeton~\cite{suen-1,suen-2,suen-3,shabani-1,shabani-2}.  In this case, the 
electrostatic interaction between the electrons (at the Hartree-Fock level) compels the 
system to behave as an
 effective bilayer.  The electron density for the ground state is maximum  (and symmetric)
near the edges of the wide-quantum-well  and depleted in the middle.  The 
energy level diagram and typical density profiles are similar to those shown in 
Fig.~\ref{fig-wqw}a.  Note that in this system, $\Delta_{SAS}$ is a strong function 
of the width $W$ (and electron density) of the wide-quantum-well.  This contrasts the 
bilayer system described above where, in principle, $\Delta_{SAS}$, the individual 
well width $w$, and the layer separation $d$ can all be independently 
adjusted.

A simplified model for the WQW is due to Papi\'{c} \textit{et al}.~\cite{papic} 
where the ground state of 
a wide-quantum-well of width $W$ is taken to be a symmetric state (for example, 
$\langle z|S\rangle\sim\sin(\pi z/W)$) and the next excited state an anti-symmetric state 
(for example, $\langle z|AS\rangle\sim\sin(2\pi z/W)$)--both the $S$ and $AS$ states 
are written on the interval $z\in[0,W]$.  (The coordinates of the (quasi-)two-dimensional 
plane will always be denoted by $x$ and $y$ and 
the coordinates of the wide-quantum-well width, bilayer 
layer separation, or individual quantum-well thickness will be denoted by $z$.)  
The density profiles for the WQW model 
are shown in Fig.~\ref{fig-wqw}b.  In this model, one can transform 
to the layer-basis (right and left layers) by 
taking $|L\rangle=(|S\rangle+|AS\rangle)/\sqrt{2}$ and 
$|R\rangle=(|S\rangle-|AS\rangle)/\sqrt{2}$, cf. Fig.~\ref{fig-wqw}.  

The Hamiltonian for our two-component model systems can be 
written (in the $SAS$-basis) as
\begin{widetext}
\begin{eqnarray}
H&=&\frac{1}{2}\sum_{\{m_i,\sigma_i=S,AS\}}\langle
m_1\sigma_1,m_2\sigma_2|V|m_3\sigma_3,m_4\sigma_4\rangle
c^\dagger_{m_1\sigma_1}c^\dagger_{m_2\sigma_2}c_{m_3\sigma_3}c_{m_4\sigma_4}
\delta_{m_1+m_2,m_3+m_4}\nonumber
\\
&-&\frac{\Delta_{SAS}}{2}\sum_{m}(c^\dagger_{mS}c_{mS} - 
c^\dagger_{mAS}c_{mAS})
+ \frac{\Delta \rho}{2}\sum_m (c^\dagger_{mS}c_{mAS} +
c^\dagger_{mAS}c_{mS})\;.
\label{bilayer-ham}
\end{eqnarray}
The second to last term (the term with the prefactor $\Delta_{SAS}/2$) controls inter-layer
tunneling and, in the $SAS$-basis is represented by the pseudo-spin operator 
$\hat{S}_z$, while the last term (prefactor $\Delta\rho/2$) controls 
charge imbalance and is represented 
by the pseudo-spin operator $\hat{S}_x$.  In our 
convention, positive $\Delta\rho$ drives the electrons into the right ($R$) layer.    
Notice that $\langle m_1\sigma_1,m_2\sigma_2|V|m_3\sigma_3,
m_4\sigma_4\rangle$ is different in the bilayer case than it is in the WQW case, 
in general.  Namely, in the 
bilayer case, the Coulomb matrix element  is found by using the transformations 
given in Eqs.~\ref{s-state} and~\ref{as-state} and the fact that the potential 
energy between 
two electrons a distance $r$ apart in a given layer (the right layer, say) of 
width $w$ is given by the 
Zhang-Das Sarma~\cite{zds} 
potential $V_\mathrm{intra}(r)=1/\sqrt{r^2+w^2}$ and the potential between 
two electrons in two different layers separated 
by a distance $d$ is given by $V_\mathrm{inter}(r)=1/\sqrt{r^2 + d^2}$.  In the 
WQW system the Coulomb 
matrix element is calculated by considering the wavefunctions for 
the $S$ and $AS$ states ($\langle z|S\rangle$ and $\langle z|AS\rangle$) 
and doing the required integrals.

In the (left/right) layer-basis we can write the bilayer Hamiltonian very simply as
\begin{eqnarray}
\hat{H}=\sum_{i<j} [V_\mathrm{intra}(|\mathbf{r}_i-\mathbf{r}_j|) +
V_\mathrm{intra}(|\tilde{\mathbf{r}}_i-\tilde{\mathbf{r}}_j|)
+ V_\mathrm{inter}(|\mathbf{r}_i-\tilde{\mathbf{r}}_j|)] - \Delta_{SAS} \hat{\tilde{S}}_x 
+ \Delta \rho \hat{\tilde{S}}_z\;,
\end{eqnarray}
where operators $\hat{\tilde{O}}$ with a tilde are 
written in the layer-basis and coordinates $\mathbf{r}$ and 
$\tilde\mathbf{r}$ belong to electrons in different layers.  Switching from 
the layer- to the $SAS$-basis is a pseudo-spin 
rotation.  Note that for $d=0$ the Hamiltonian is symmetric between $\Delta_{SAS}
$ and $\Delta\rho$.  
Increasing $d$ destroys this symmetry and it becomes harder to drive the system to be 
one-component in the 
layer-sense.

\end{widetext}

In all of our results we use the \emph{natural} FQHE units:  lengths are given 
in units of the magnetic length $l$ and energies are given in units 
of the Coulomb energy $e^2/\epsilon l$.

(Note that the system we are considering is a rather general 
 two-component Hamiltonian and when the layer 
separation is zero any 
results we find are applicable to SU(2) 
symmetric two-component systems where the constituents interact by a Coulomb-like 
interaction--exactly Coulombic for the lowest Landau level and a slightly modified Coulombic 
interaction for the second Landau level.   
For example, when $d=0$ the Hamiltonian (Eq.~\ref{bilayer-ham}) describes 
spin-full electrons interacting via a Coulomb interaction with both $\hat{S}_z$ and 
$\hat{S}_x$ terms.)

\section{Variational wavefunctions: Moore-Read Pfaffian and Halperin 331}
\label{sec-ansatz}

We consider four variational wavefunctions 
to describe our FQH system for half-filled lowest and 
second Landau levels ($\nu=1/2$ and $\nu=5/2$, respectively) 
even though it will appear at first that we are only 
considering two.  Namely, we consider the one-component Moore-Read Pfaffian~\cite{mr-pf} 
(Pf) and two-component Halperin 331 (331) wavefunctions~\cite{halperin-331}.  (Note that
 work~\cite{macdonald-1,schliemann,nomura-1,park,simon,moller-1,moller-2} has 
 been done on the FQHE at total 
filling factor of unity, i.e., $\nu=1/2+1/2=1$, however, that is not the case that 
we study in this work.)

The Moore-Read Pfaffian state is written as
\begin{eqnarray}
\Psi_{\mathrm{Pf}}=\mathrm{Pf}\left\{\frac{1}{z_{i,a}-z_{j,a}}\right\}\prod_{i<j}^N(z_{i,a}-z_{j,a})^2
\end{eqnarray}
where $z_{i,a}=x_{i,a}-iy_{i,a}$ is the position of the $i$th electron in complex coordinates
with $a$ labeling its state, i.e., $a=S$, $AS$, $R$ or $L$.  The origin of 
this wavefunction can be understood at an intuitive level.  
If one writes down an ansatz wavefunction to describe a BCS 
paired state of Composite Fermions in a 
half-filled Landau level then one arrives at a wavefunction of the Pfaffian form~\cite{footnote1}, i.e., 
\begin{eqnarray}
&&\mathrm{Pf}\{g(r_{ij})\}\prod_{i<j}(z_i-z_j)^2=\nonumber\\
&&\mathcal{A}\{g(r_{12})g(r_{34})\ldots g(r_{N-1,N})\}\prod_{i<j}(z_i-z_j)^2
\end{eqnarray}
where $g(|\mathbf{r}_i-\mathbf{r}_j|)$ is the real space pairing amplitude for a pair of electrons 
located at position $\mathbf{r}_i$ and $\mathbf{r}_j$, $\prod_{i<j}(z_i-z_j)^2$ binds two quantum 
mechanical vortices of the many-body wavefunction to each electron (the Composite Fermion 
transformation and fixes the filling factor in a Landau level 
to 1/2). $\mathcal{A}$ is the anti-symmetrization operator.
  
Despite the intuitive picture that goes along with the Moore-Read Pfaffian wavefunction 
it was originally derived using conformal field theory and, as such, the  
Pf has a low-energy effective conformal field theory description, cf. 
Refs.~\onlinecite{mr-pf},~\onlinecite{mvm-nr} and~\onlinecite{read-green}.  
It is within this conformal 
field theory that the understanding 
of the non-Abelian nature of the quasiparticle excitations was first illuminated.  However, 
whether or not the Pf has anything to do with the physics of the FQHE 
at 5/2 has depended crucially on numerical calculations and comparisons with the results of 
exact diagonalization~\cite{morf,rezayi-haldane,storni,feiguin,mrp-ft-prl,mrp-ft-prb,papic-2}.  As 
mentioned above, the details matter greatly in 
determining the physics of the FQHE at 5/2, such as, finite 
thickness effects~\cite{mrp-ft-prl,mrp-ft-prb}, 
the competition between FQHE states and non-FQHE 
striped phases~\cite{rezayi-haldane} and the compressible (Composite 
Fermion) Fermi sea~\cite{hlr,kalmeyer} (as mentioned, the latter occurs 
in the lowest Landau level at $\nu=1/2$ while the former occurs in the second Landau 
level at $\nu=5/2$), the effects of 
Landau level mixing~\cite{bishara-nayak,simon-rezayi,wojs-toke-jain}, etc.  
Note that most
of the numerical studies mentioned above do not treat the
$\nu = 5/2$ problem in full complexity, i.e. with the lowest
Landau level filled with two spin species and one Landau
level half filled. Instead, because of computational
simplicity, one considers a half filled lowest Landau level
(usually without spin) with an effective interaction
that contains the form factors of the first excited
Landau level--the appropriate Haldane pseudopotentials~\cite{haldane}. 
This is what we mean by ``$\nu = 5/2$Ó in an exact diagonalization context.

Thus, it should be clear that any effective conformal field theory (or 
effective BCS mean field Hamiltonian~\cite{read-green}) written down to describe the 
physics at 5/2 is only as good as its physical predictions and agreement with 
experiments \emph{and} its agreement with numerical calculations.  
This is because the entire reason 
for the existence of the FQHE at all filling factors is due to non-perturbative physics 
arising from the strongly interacting electrons interacting with a Coulomb interaction 
modified by the details of which Landau level is fractionally filled by electrons, the 
thickness the quantum well, the amount of Landau level mixing that is taking 
place, etc.  Hence, any mean field theory that throws away all these details at the first 
step cannot  explain, for example, why the FQHE occurs at 
$\nu=5/2$ and not $\nu=1/2$ (or for that matter, at 9/2 or 13/2...) 
without the aid of both numerical calculations and, most importantly, experiments.  In the end, 
whether a particular candidate wavefunction, e.g. Moore-Read Pfaffian
 or Laughlin or Halperin 331 or Composite Fermion states, etc., describes a real 
FQHE at some filling factor is an energetic question 
delicately depending on competing 
states where all the details of the microscopic Hamiltonian may eventually matter, and therefore, 
extreme caution is necessary in identifying experimental FQHE states with candidate 
incompressible wavefunctions, particularly in higher LLs and/or multi-component systems where 
various competing states are, in general, more viable.

\begin{figure}[]
\begin{center}
\includegraphics[width=8.cm,angle=0]{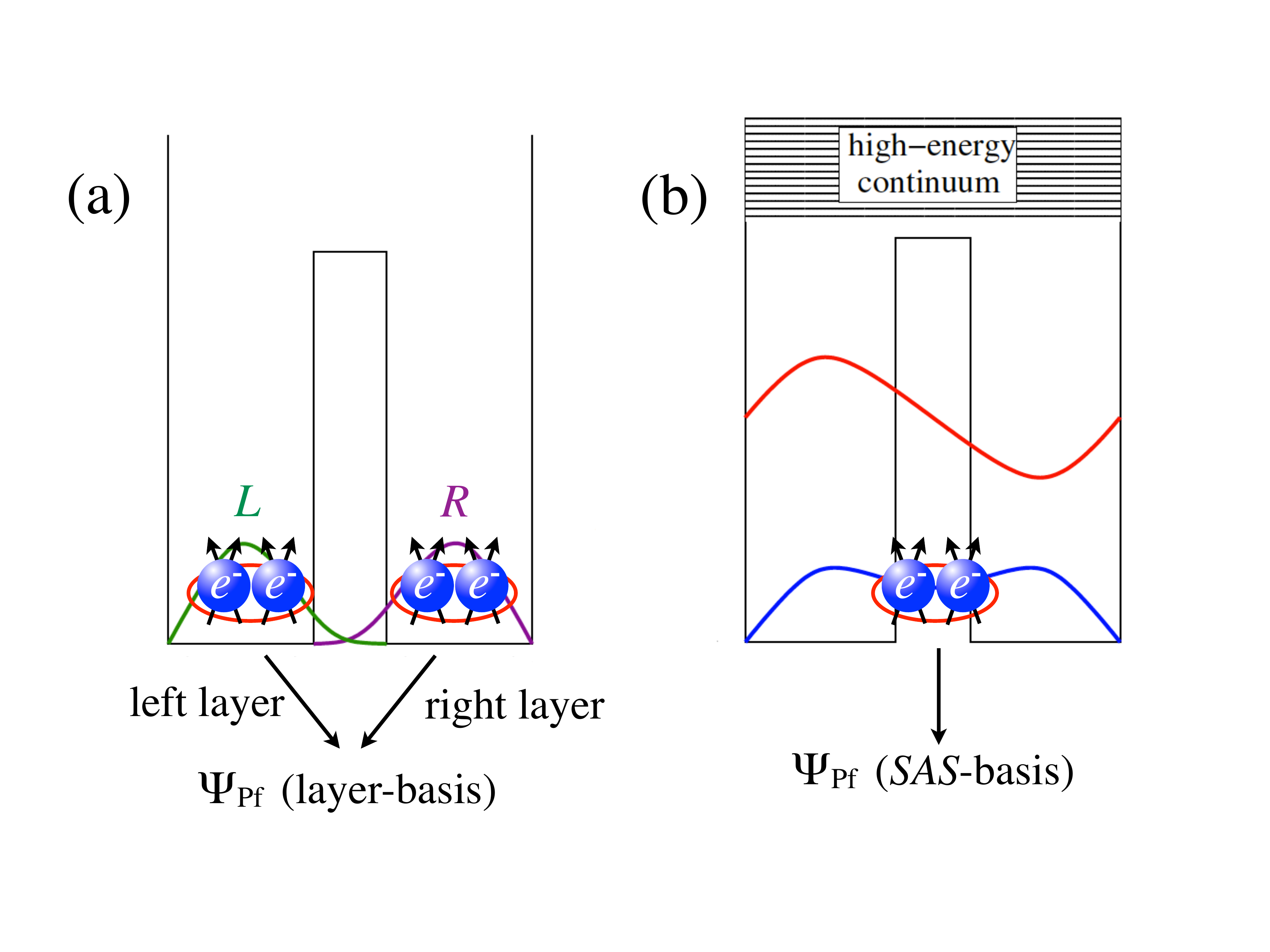}
\includegraphics[width=8.cm,angle=0]{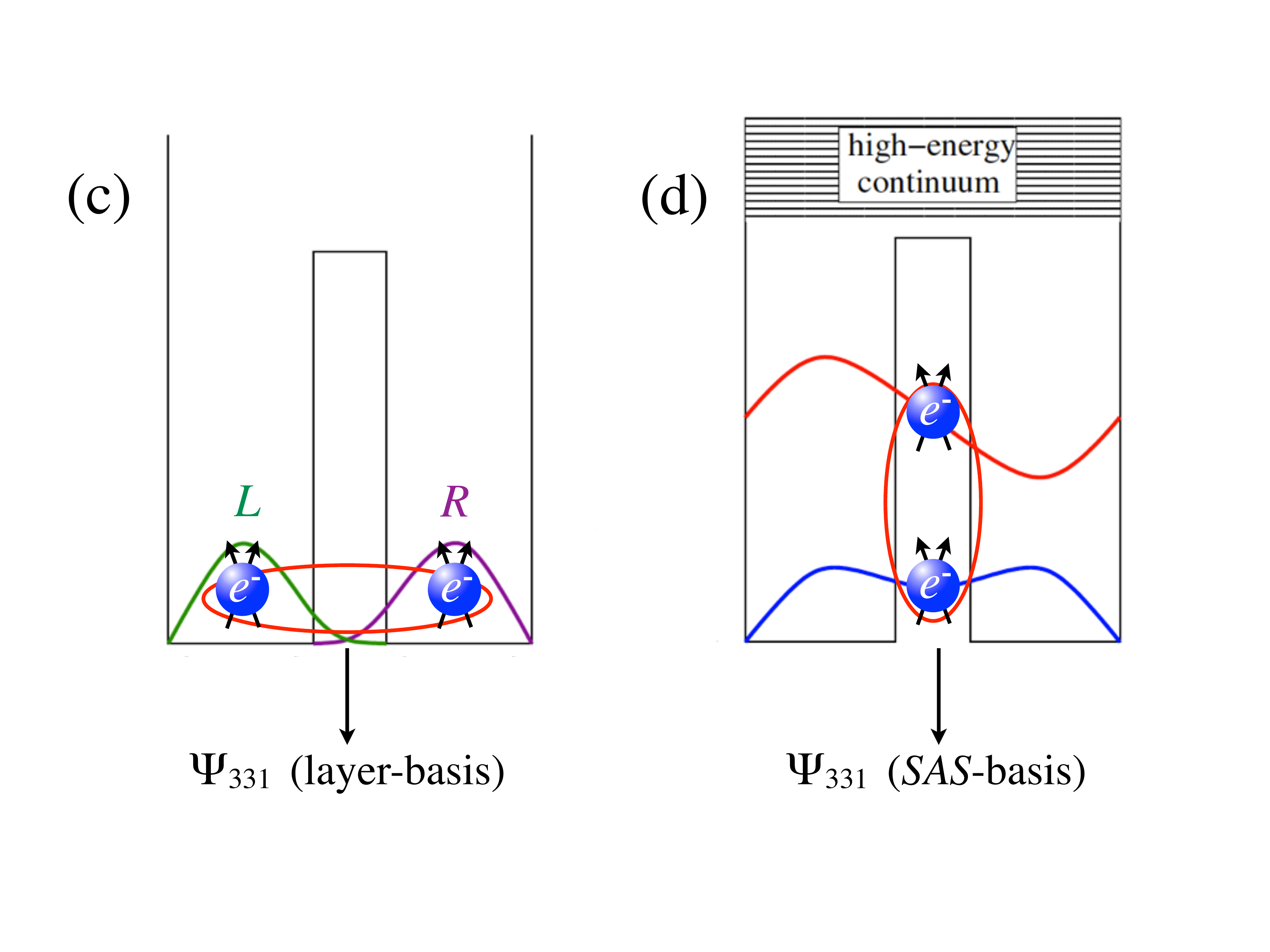}
\caption{(Color online) The Moore-Read Pfaffian 
wavefunction written in the (a) layer- and (b) $SAS$-basis where the $p$-wave pairing 
in the former is between Composite Fermions in the right (or left) layer and the $p$-wave 
pairing in the $SAS$-basis is between electrons in the symmetric ($S$) state.  
The Halperin 331 state is shown in (c) the layer-basis and (d) 
the $SAS$-basis where in the layer-basis the 331 state pairs Composite Fermions between 
layers and in the $SAS$-basis it pairs Composite Fermions in the symmetric ($S$) 
and anti-symmetric ($AS$) states.}
\label{fig-wfs-cartoon}
\end{center}
\end{figure}

The Halperin (two-component) 331 wavefunction is written as
\begin{eqnarray}
&&\Psi_{331}=\nonumber\\
 &&\prod_{i<j}^{N/2}(z_{i,1}-z_{j,1})^3\prod_{i<j}^{N/2}(z_{i,2}-z_{j,2})^3\prod_{i,j}^{N}(z_
{i,1}-z_{j,2}) \;,
\end{eqnarray}
where the subscript 1 (or 2) on $z_{i,1}$ labels the quantum number of the 
electron.  This quantum number could describe spin (up or down), layer (R or L), subband 
($S$ or $AS$). Note that in our work, we consider the bilayer situation so the 
subscript will either denote the layer or the symmetric or anti-symmetric state 
depending on in which basis we choose to work.   Intuitively, the origin of this 
wavefunction is that, in a 
bilayer system in a half-filled Landau level, 
the electrons form $\nu=1/3$ Laughlin states~\cite{laughlin} 
($\prod_{i<j}^{N}(z_i-z_j)^3$) among each electron component and then pair 
among different components (simple 
Jastrow factor $\prod_{i,j}^{N}(z_{i,1}-z_{j,2})$).

(Note that we always drop the Gaussian factors ($\sim \exp (-\sum_i |z_{i,a}|^2)$) that are always 
present when writing wavefunctions describing electrons entirely in (or projected into) the lowest 
Landau level--in fact, the Gaussian is often considered to be part of the measure 
so one is technically not ``dropping" anything at all.)

It is important to realize that the Moore-Read Pfaffian is a one-component state 
that can be one-component in both  the $SAS$-basis sense or in the layer-basis 
sense, see Fig.~\ref{fig-wfs-cartoon}a and~\ref{fig-wfs-cartoon}b.  
That is, the Pf wavefunction either describes pairing among one-component electrons in the 
right (R) layer or left (L) layer (the layer-basis) or pairing among one-component electrons 
in the symmetric ($S$) state (the $SAS$-basis).    The former is the Pfaffian
state that has its origin in the single-layer FQHE, while the latter is the Pfaffian 
understood as the ÒevenÓ (symmetric) channel of the BCS description by
Read and Green~\cite{read-green}.  (We 
choose our charge imbalancing term to drive the electrons 
into the right layer, however, we could have equivalently chosen the charge imbalance term to 
drive the 
electrons into the left layer.)  Thus, there are actually \emph{two} Moore-Read Pfaffian states 
to consider when both tunneling terms ($\Delta_{SAS}$ and $\Delta\rho$) are non-zero (or 
three if we allow the sign of $\Delta\rho$ to change).  When 
the system is SU(2) symmetric, i.e., when the layer separation is zero, the two different 
states are simply related via a rotation in pseudo-spin space.

The Halperin 331 state can also be two-component in two ways: 
it can describe pairing of electrons in the R and L layers (the layer-basis)
or it can describe pairing of electrons in the $S$ and $AS$ states 
(the $SAS$-basis), see 
Fig.~\ref{fig-wfs-cartoon}c and Fig.~\ref{fig-wfs-cartoon}d.  
Hence, there are \emph{two} Halperin 331 states to 
consider when both tunneling terms are finite.  Note, however, that the Halperin 331 state, 
as usually understood, is defined in the layer-basis.  Again, in the SU(2) situation, 
these two 331 states are related via  a pseudo-spin rotation.

Thus, instead of dealing with two variational wavefunctions (Pf and 331) we 
are dealing with four--two Pfs and two 331s.  Even though the two pairs of wavefunctions 
are related via a pseudo-spin rotation, when the SU(2) symmetry is broken this 
is no longer the case and the 
determination of which state describes the physics is non-trivial.

In our calculations we utilize the spherical geometry~\cite{haldane} where the electrons 
are confined to the surface of a sphere of radius $\sqrt{N_\phi/2}$ and a radial magnetic 
field (perpendicular to the surface) is produced by a magnetic monopole of 
strength $N_\phi/2$ placed at the center.  (We 
briefly consider the torus geometry below.)  The total magnetic flux 
piercing the surface is $N_\phi$ and is either an integer or half-integer according 
to Dirac~\cite{wu-yang}.  The filling factor (in the partially filled Landau level) is given 
by $\lim_{N\rightarrow\infty}N/N_\phi$ for $N$ electrons.  In our case, we 
are considering the Moore-Read Pfaffian and Halperin 331 states which 
describe a half-filled Landau level, thus, the relationship between $N$ and $N_\phi$ for 
a finite spherical system is $N_\phi=2N-3$.  Due to computational constraints 
we consider $N=8$ and $N_\phi=13$ (the $N=6$ particle system is 
aliased with a $2/3$ filled Landau level and, thus, produces ambiguous results and the $N=10$ 
electron system is too big to consider while adequately exploring the large 
parameter space inherent when tackling a two-component Hamiltonian with 
both inter-layer and charge imbalancing tunneling).  In the spherical geometry, possible 
FQHE states are uniform states with total angular momentum $L=0$.

In this work we consider the overlap between the exact ground state of the model Hamiltonian 
 and the four variational states: (1) Pf in the 
$SAS$-basis, (2) Pf in the layer-basis, (3) 331 in 
the $SAS$-basis and (4) 331 in the layer-basis.

\begin{figure*}[t]
\begin{center}
\includegraphics[width=5.9cm,angle=0]{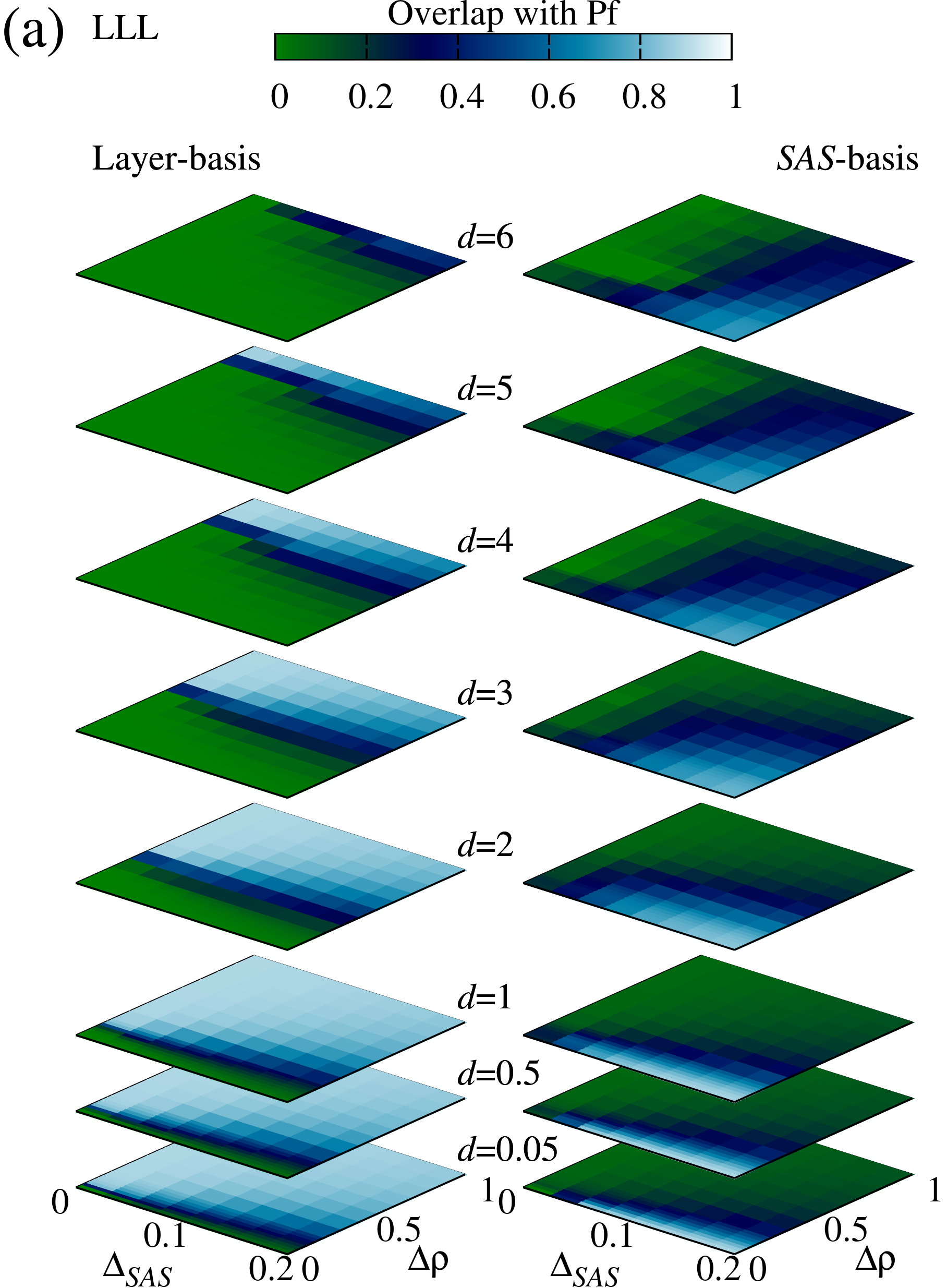}
\includegraphics[width=5.9cm,angle=0]{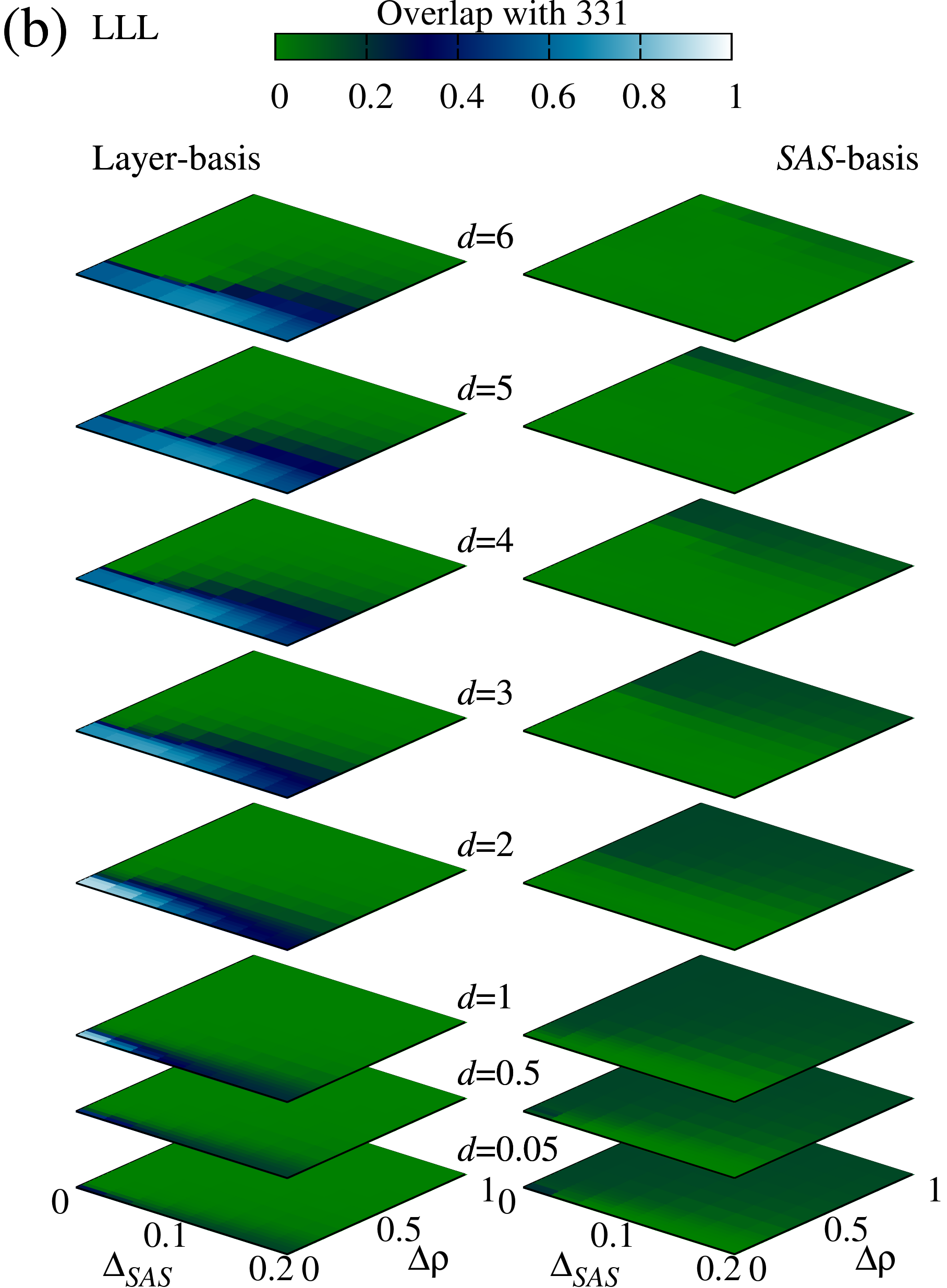}\\
\mbox{}\\
\includegraphics[width=5.9cm,angle=0]{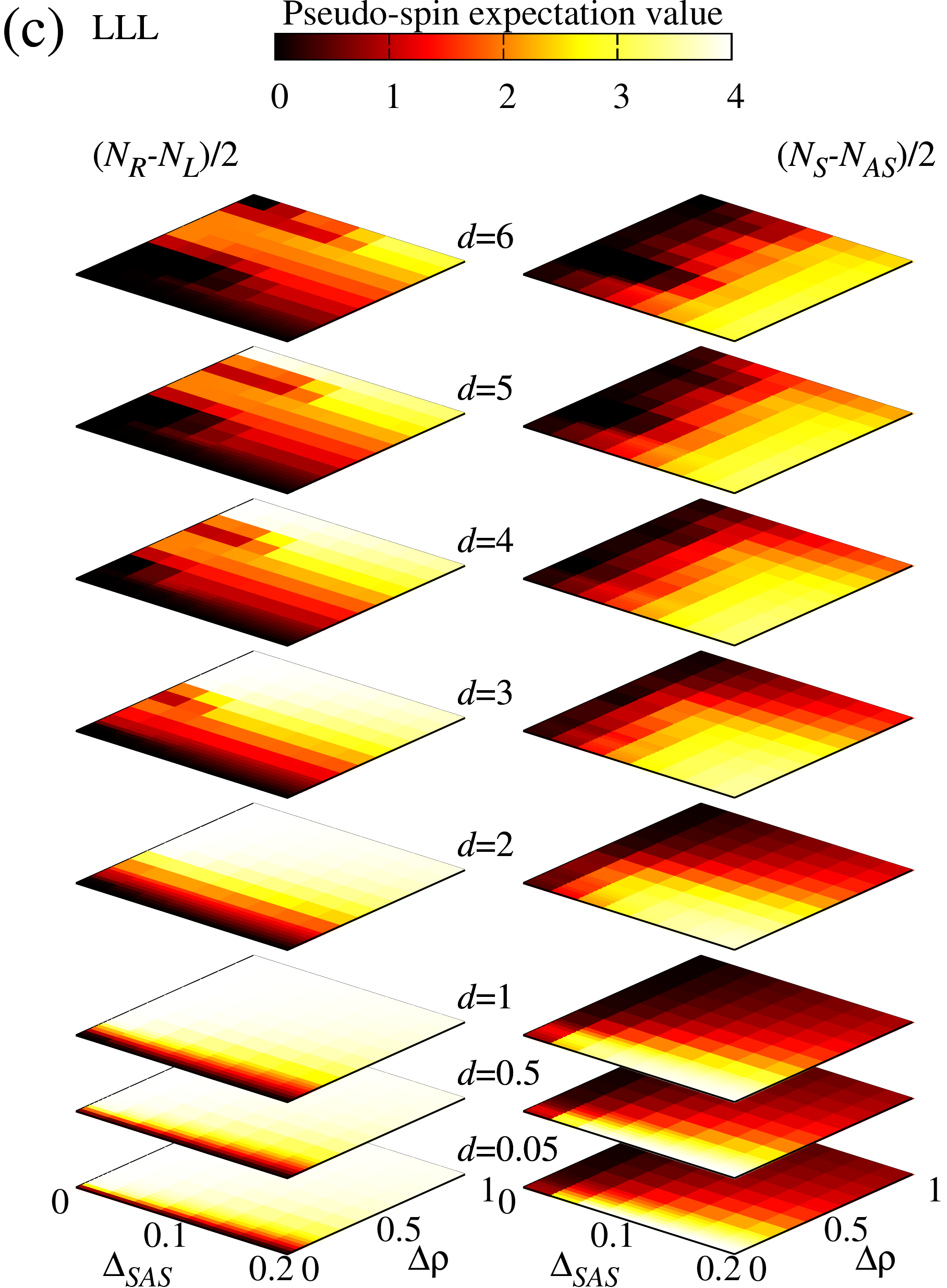}
\includegraphics[width=5.25cm,angle=0]{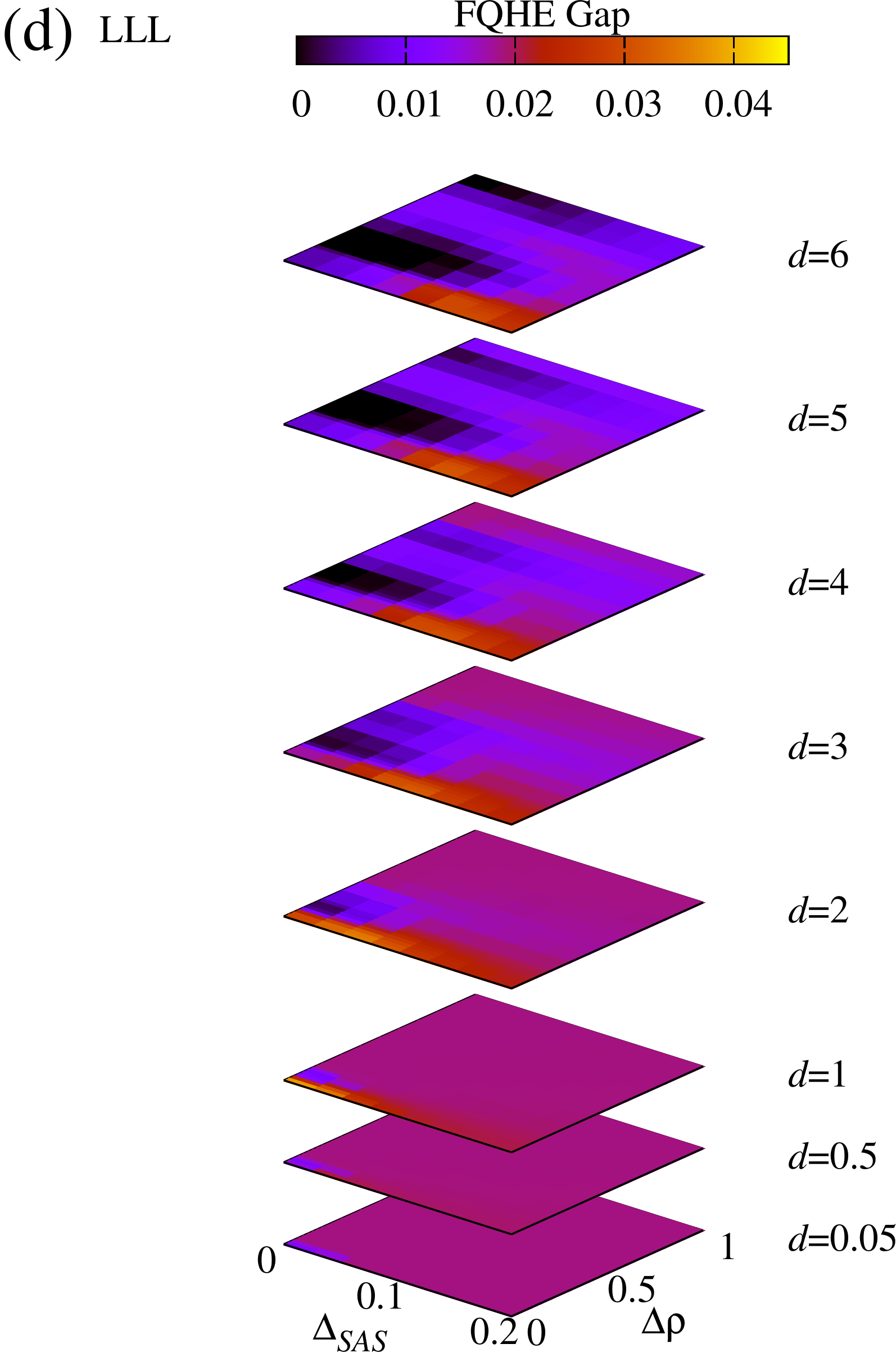}
\caption{(Color online). Lowest Landau level:  (a) Wavefunction overlap between the exact ground 
state of the bilayer Hamiltonian and the Moore-Read Pfaffian 
written in the layer-basis (left column) 
and the $SAS$-basis (right column) shown as a function of inter-layer tunneling $\Delta_{SAS}$ 
and charge imbalance $\Delta\rho$ for different values of layer separation $d$ and zero individual 
layer thickness $w=0$.  (b) Same as (a) but for the Halperin 331 wavefunction.  (c) Pseudo-spin 
expectation value or, more physically, the expectation value of $(N_R-N_L)/2$ (left column) and $
(N_S-N_{AS})/2$ (right column) as a function of $\Delta_{SAS}$ and $\Delta\rho$.  (d) shows the 
FQHE (neutral) energy gap for the exact bilayer Hamiltonian. }
\label{fig-LLL-stacked-Dsas-v-Drho}
\end{center}
\end{figure*}

\begin{figure}
\begin{tabular}{c}
\includegraphics[width=3.5cm,angle=0]{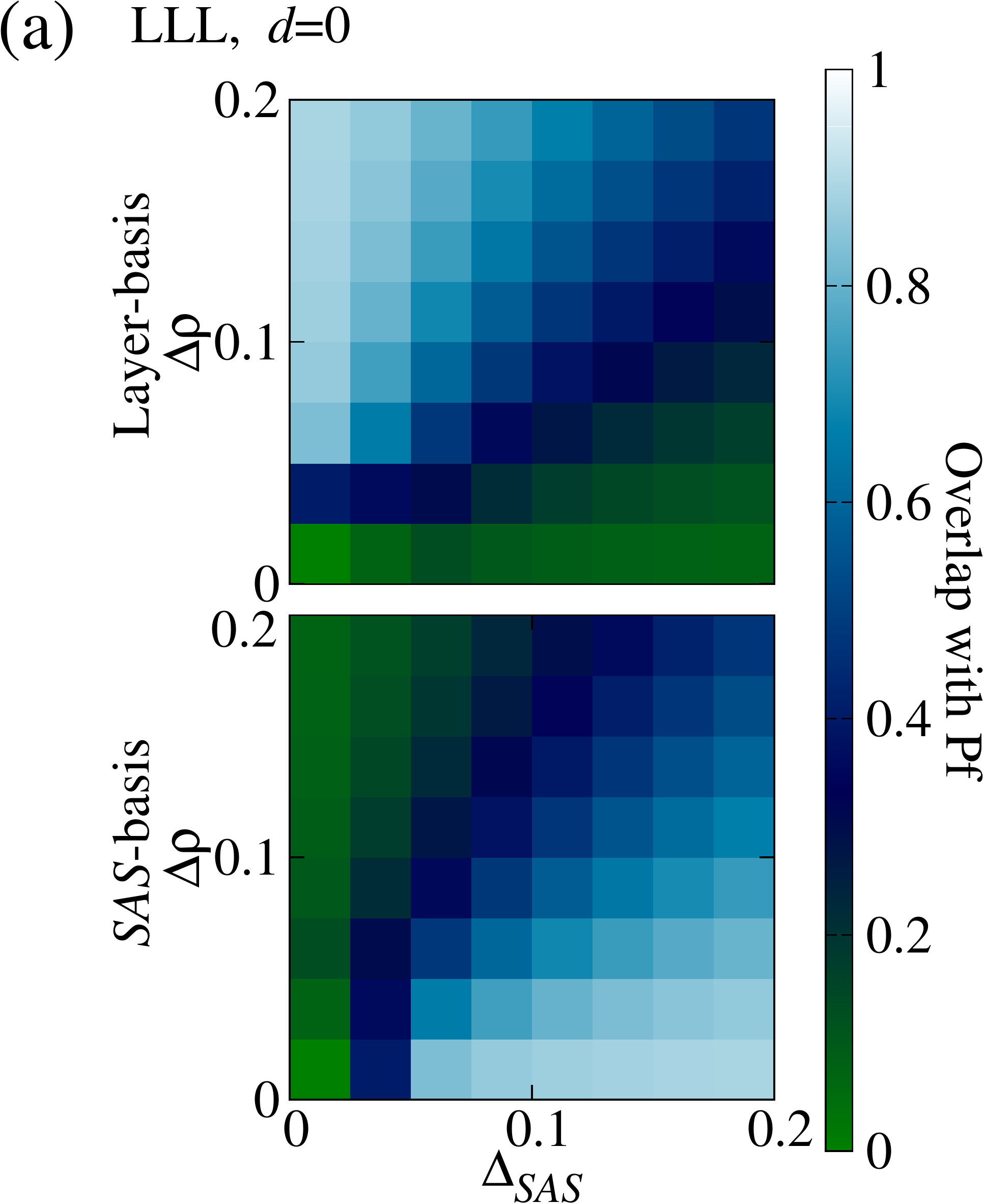} 
\includegraphics[width=3.5cm,angle=0]{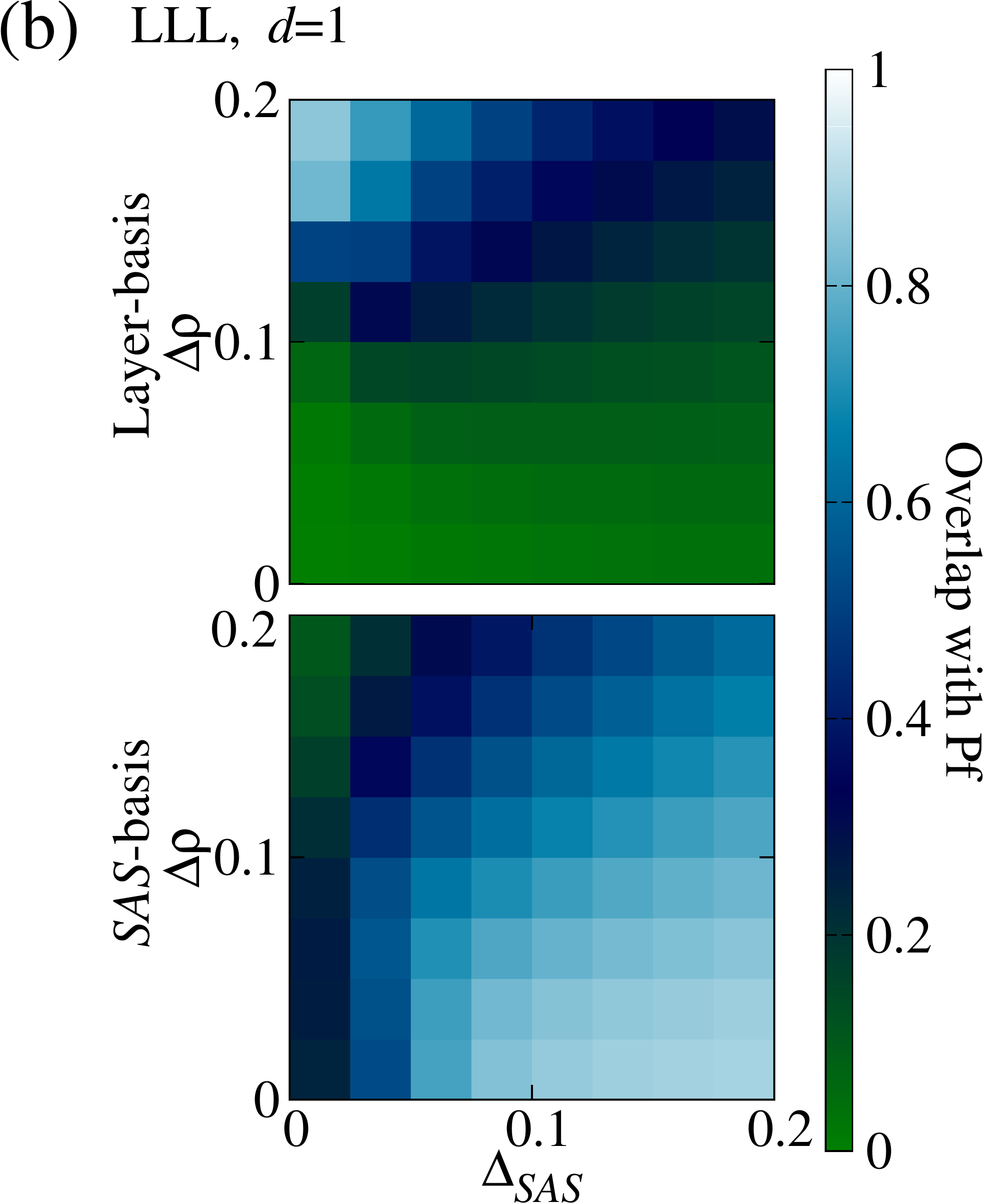} \\
\includegraphics[width=3.5cm,angle=0]{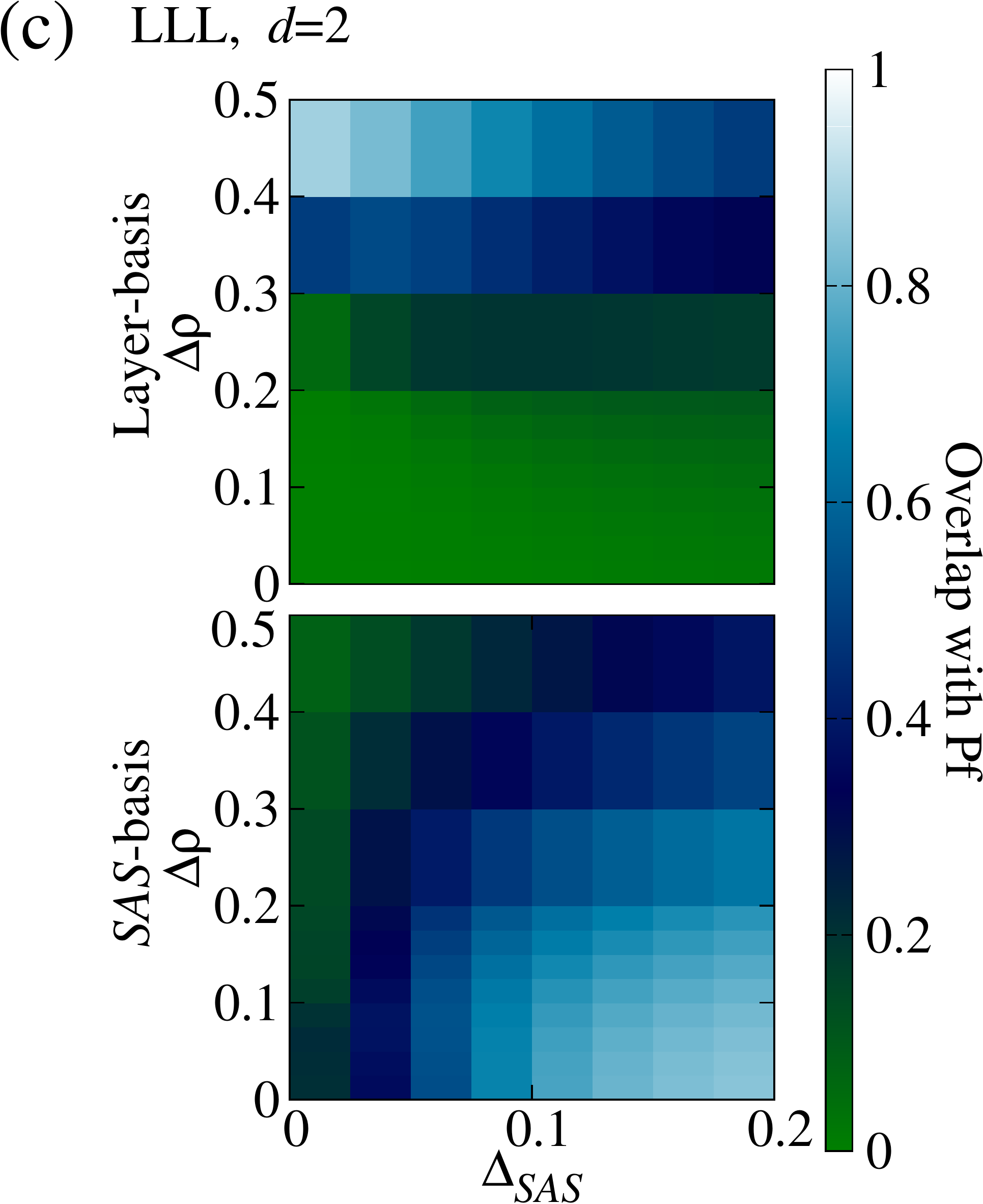}
\includegraphics[width=3.5cm,angle=0]{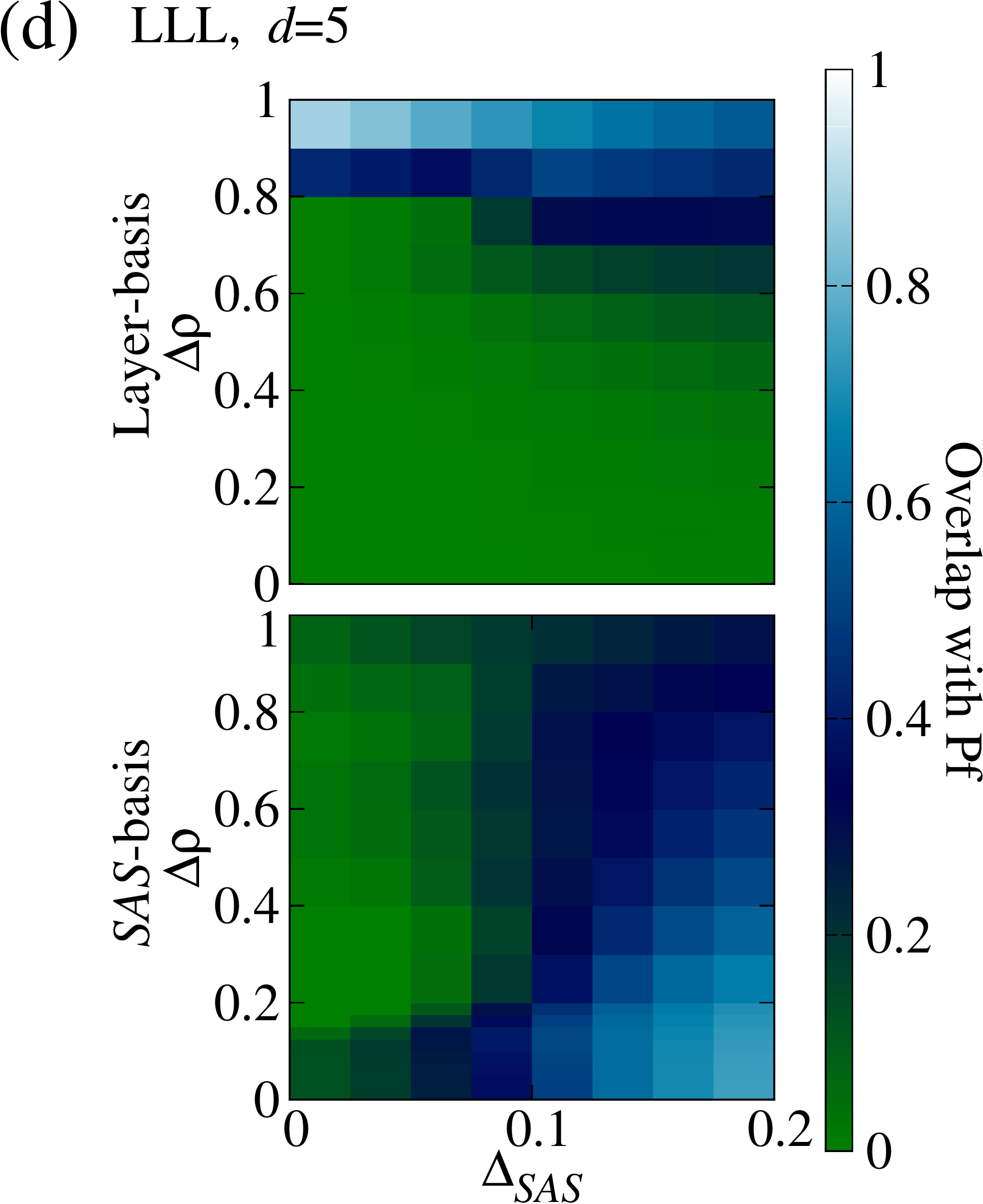}
\end{tabular}
\caption{(Color online) Lowest Landau level:  Wavefunction overlap between the Moore-Read 
Pfaffian wavefunction in the layer-basis (top panel) and the $SAS$-basis (lower panel) and the 
exact ground state for (a) $d=0$, (b) $d=1$, (c) $d=2$ and (d) $d=5$.}
\label{fig-LLL-Dsas-v-Drho-Pf}
\mbox{}\\
\begin{tabular}{c}
\includegraphics[width=3.5cm,angle=0]{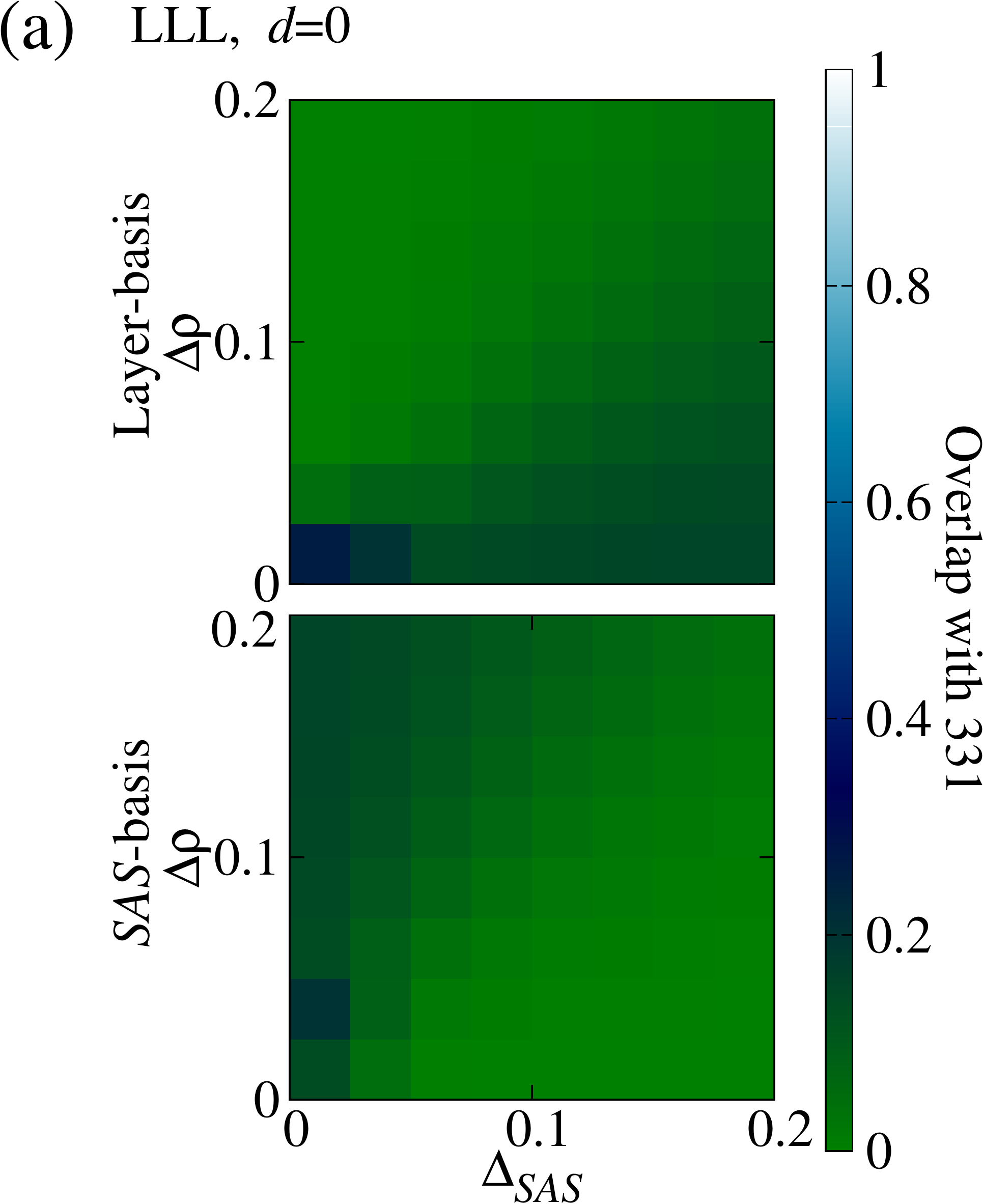} 
\includegraphics[width=3.5cm,angle=0]{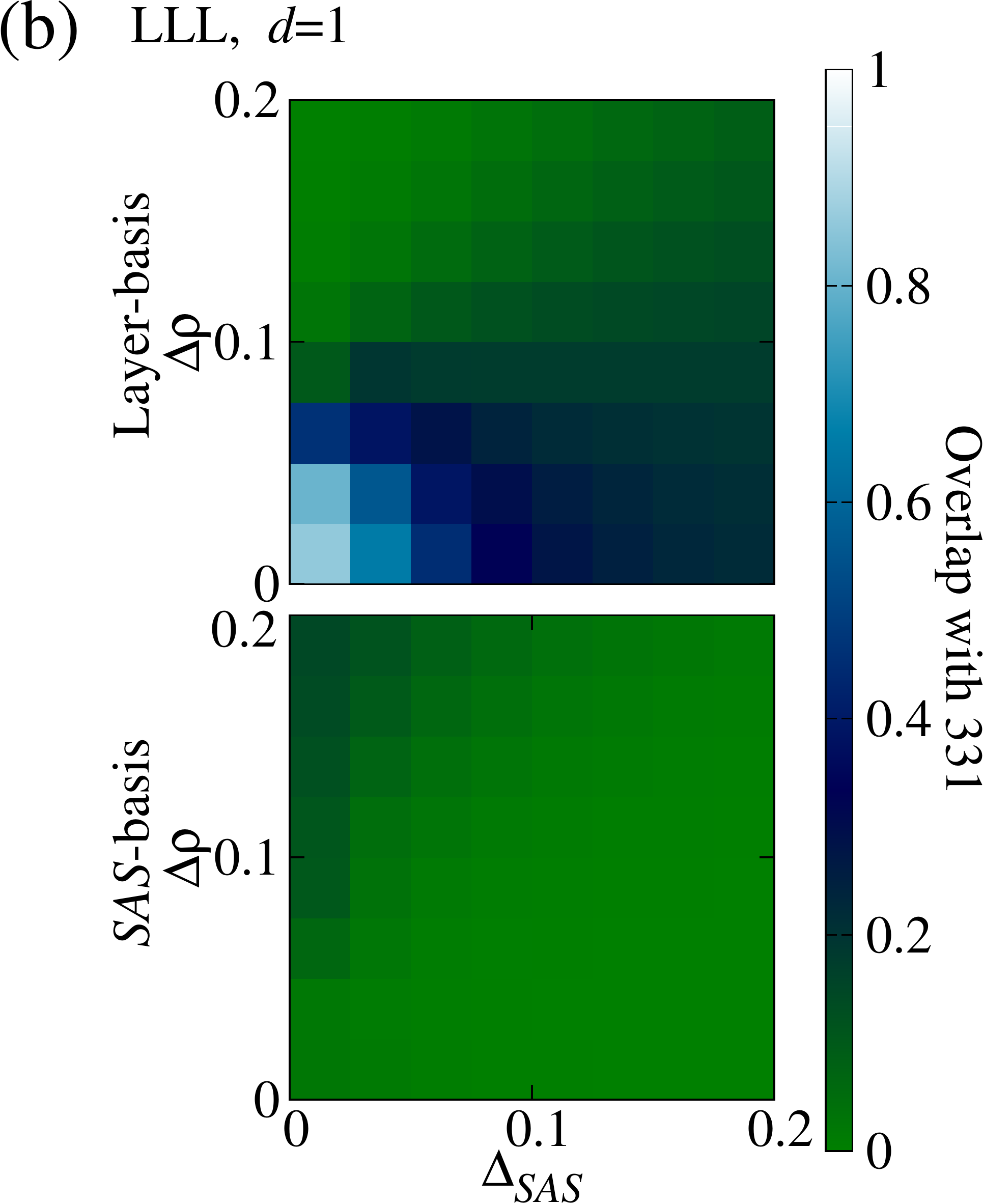} \\
\includegraphics[width=3.5cm,angle=0]{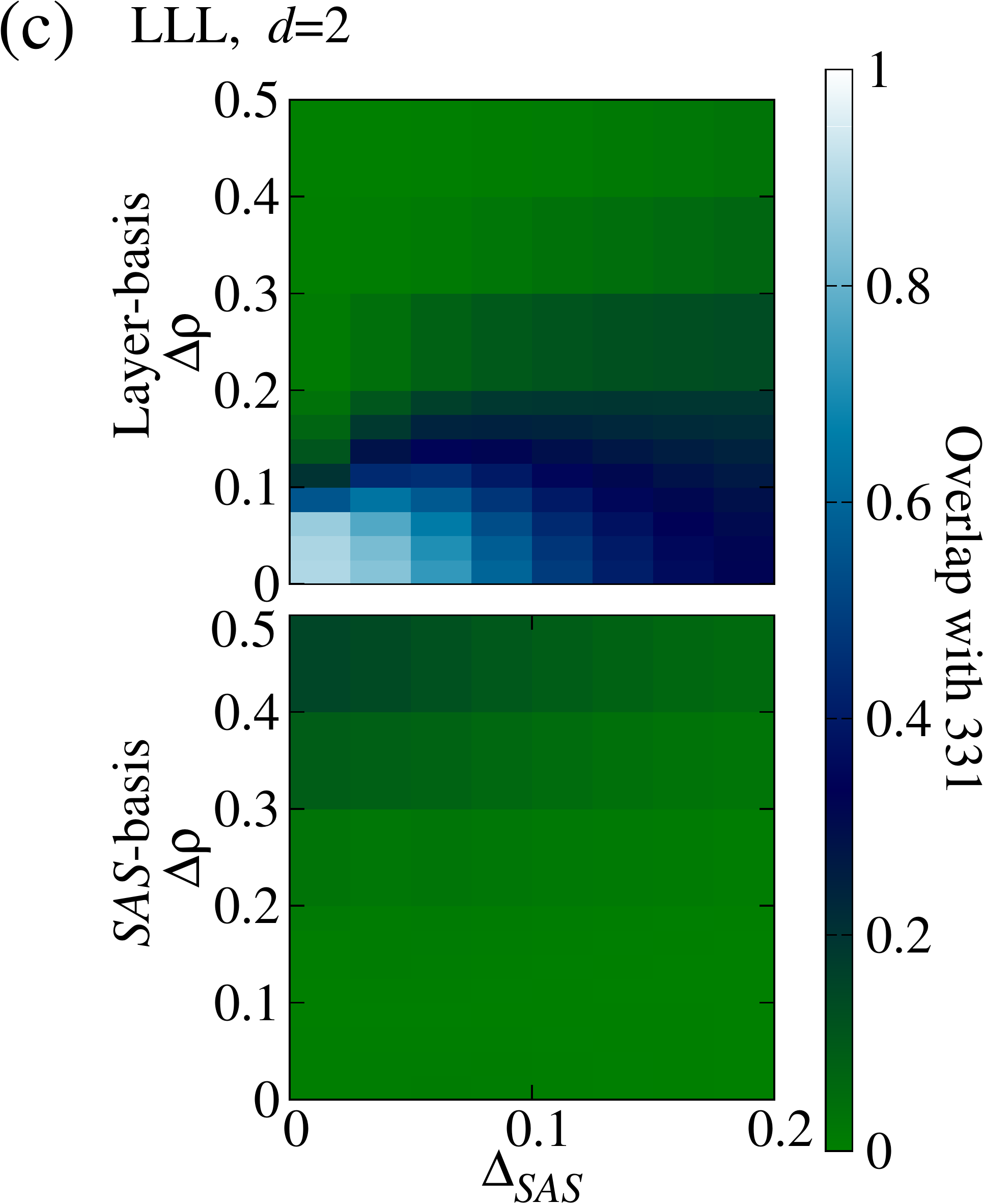}
\includegraphics[width=3.5cm,angle=0]{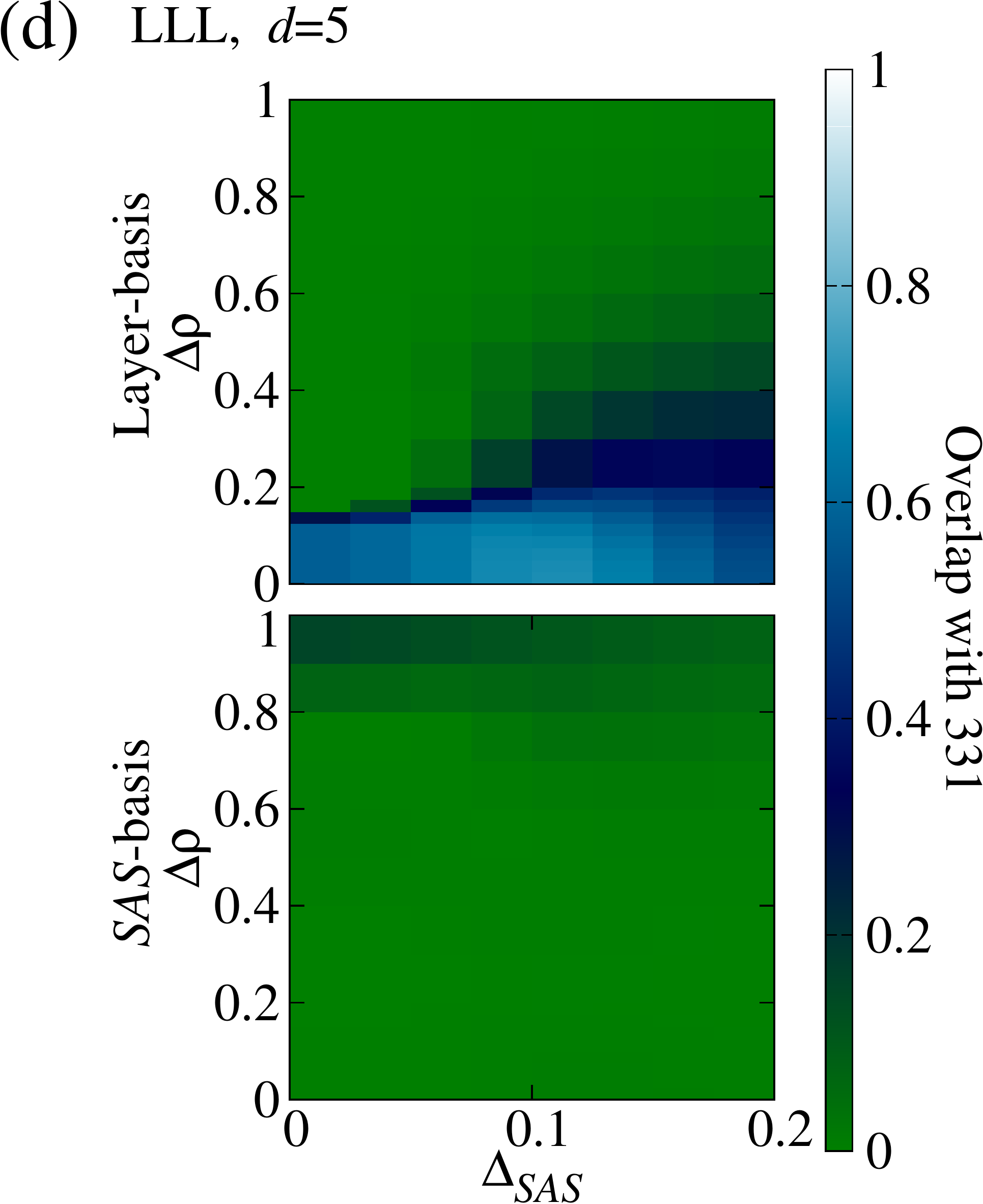}
\end{tabular}
\caption{(Color online) Lowest Landau level:  Wavefunction overlap between the Halperin 331 
wavefunction in the layer-basis (top panel) and the $SAS$-basis (lower panel) and the exact 
ground state for (a) $d=0$, (b) $d=1$, (c) $d=2$ and (d) $d=5$.}
\label{fig-LLL-Dsas-v-Drho-331}
\end{figure}

\begin{figure}
\begin{tabular}{c}
\includegraphics[width=3.5cm,angle=0]{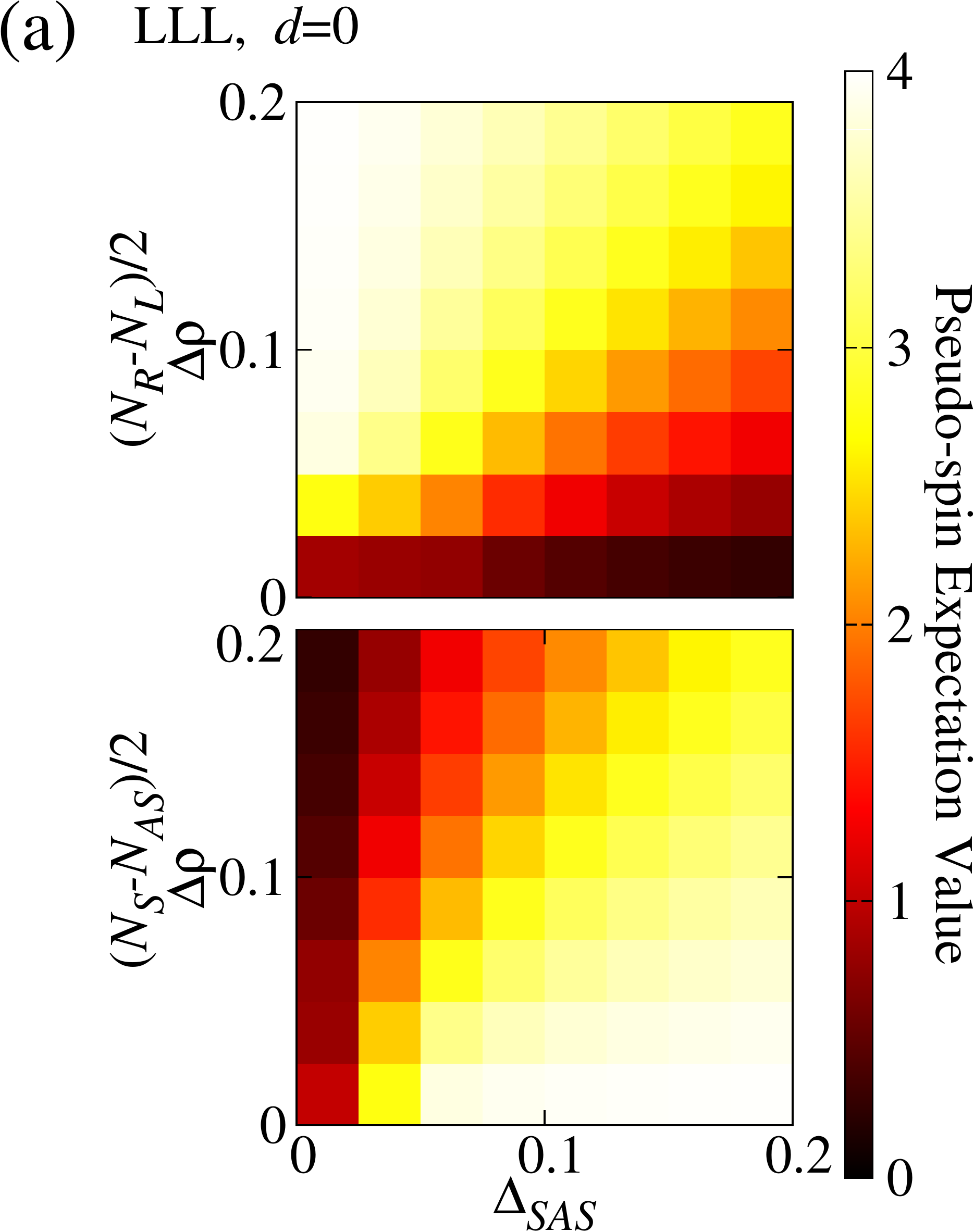}
\includegraphics[width=3.5cm,angle=0]{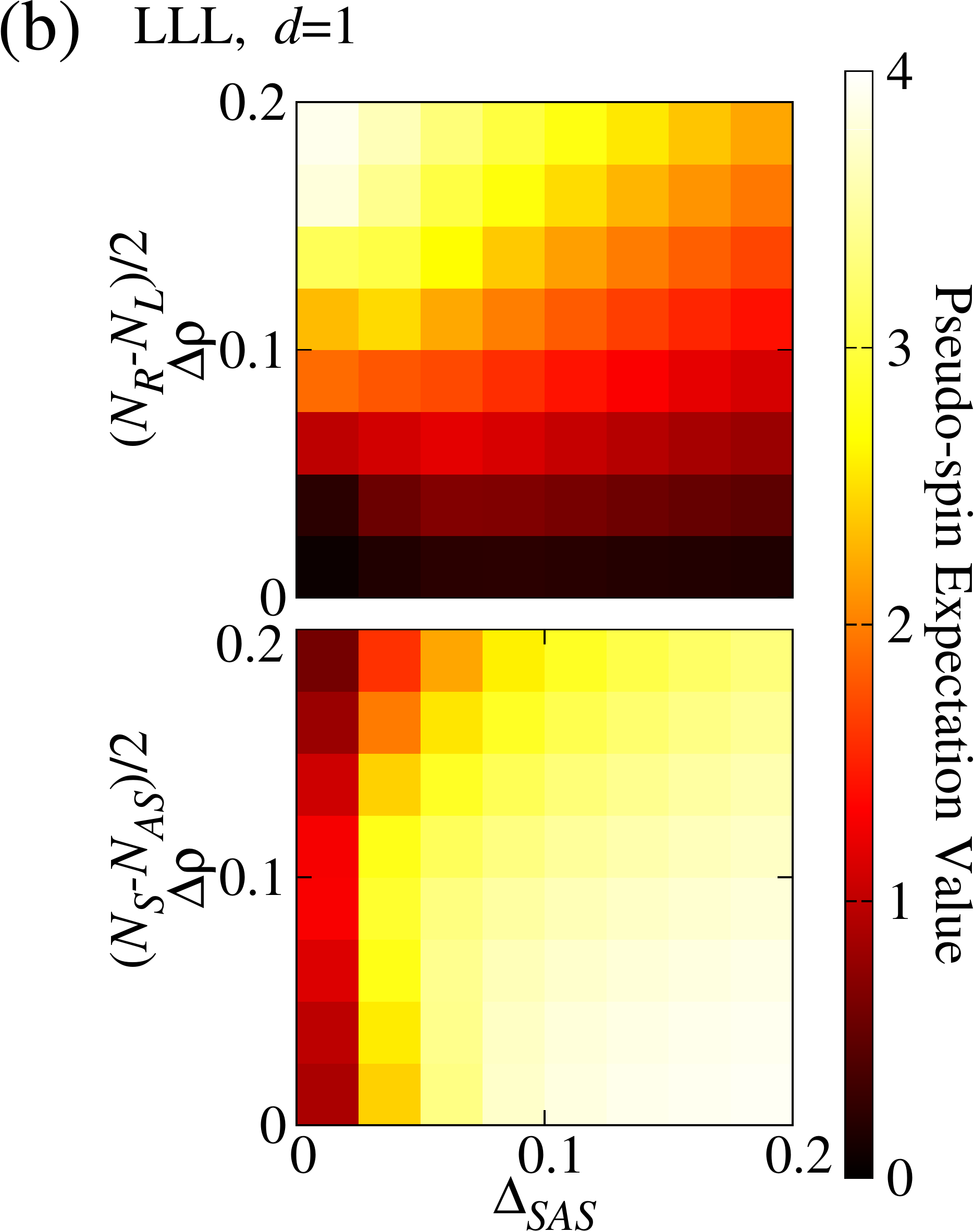}\\
\includegraphics[width=3.5cm,angle=0]{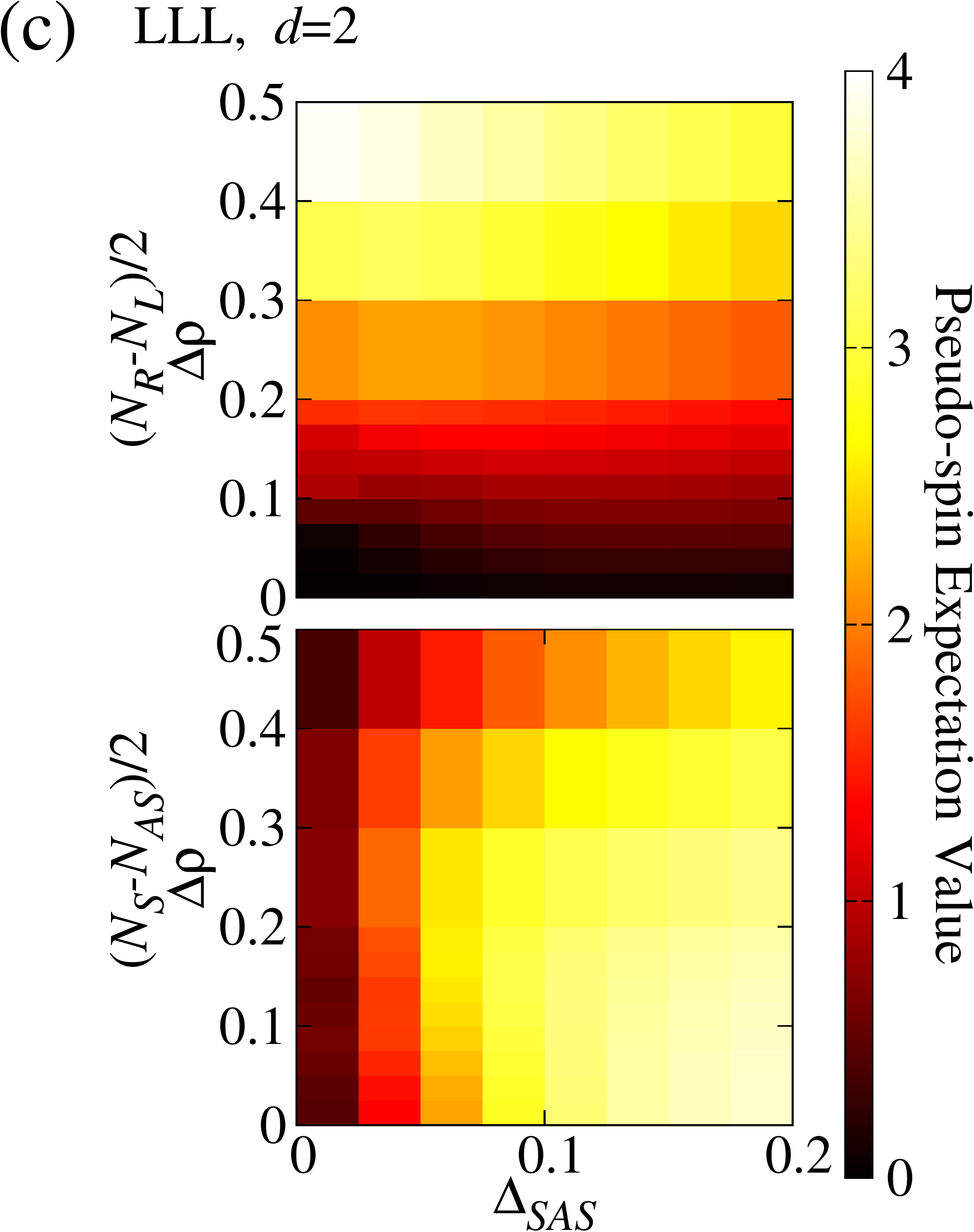}
\includegraphics[width=3.5cm,angle=0]{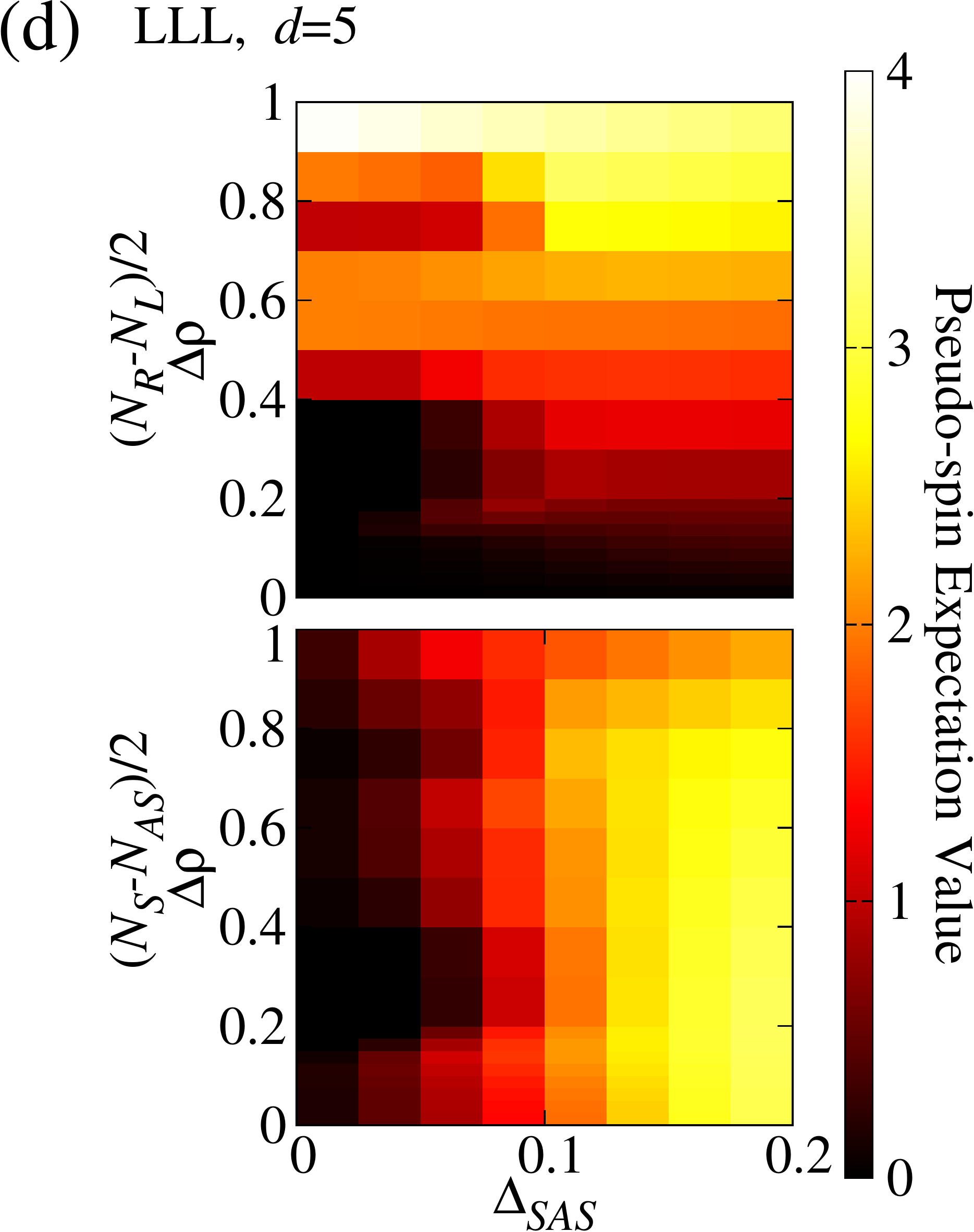}
\end{tabular}
\caption{(Color online) Lowest Landau level: (Pseudo-spin) expectation value of the exact ground 
state of $(N_R-N_L)/2$ (top panel) 
and $(N_S-N_{AS})/2$ (lower panel)  for (a) $d=0$, (b) $d=1$, (c) 
$d=2$ and (d) $d=5$.  Note that these figures (especially for $d=0$) are qualitatively similar 
to our cartoon schematic in Fig.~\ref{fig-phasediagram-cartoon} since the psuedo-spin expectation 
value essentially describes the one- or two-component nature of the ground state.  Also, 
finite $d$ breaks the SU(2) symmetry of reflection across 
the $\Delta_{SAS}=\Delta\rho$ line assumed in Fig.~\ref{fig-phasediagram-cartoon}.}
\label{fig-LLL-pseudospin-Dsas-v-Drho}
\mbox{}\\
\begin{tabular}{c}
\includegraphics[width=3.5cm,angle=0]{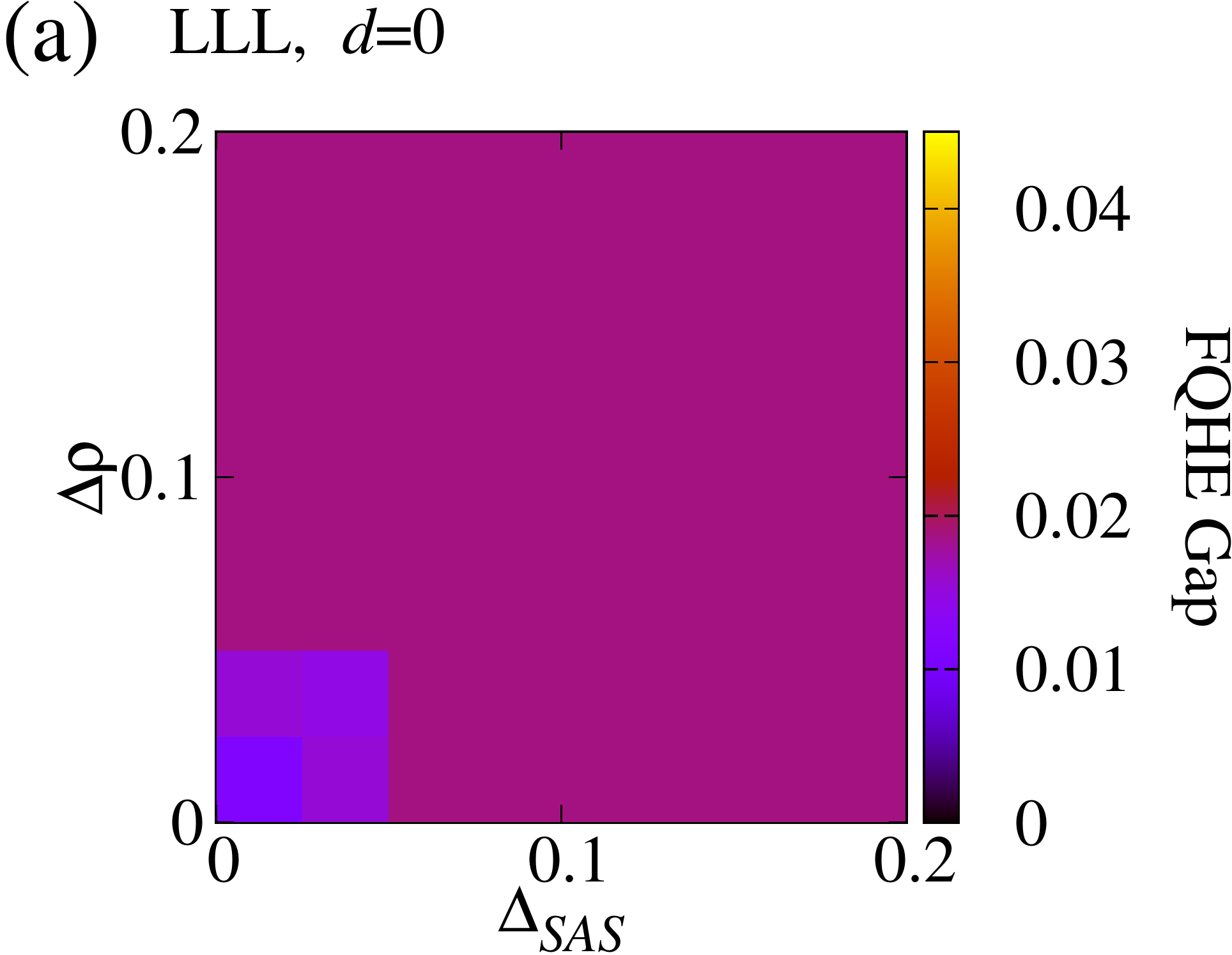}
\includegraphics[width=3.5cm,angle=0]{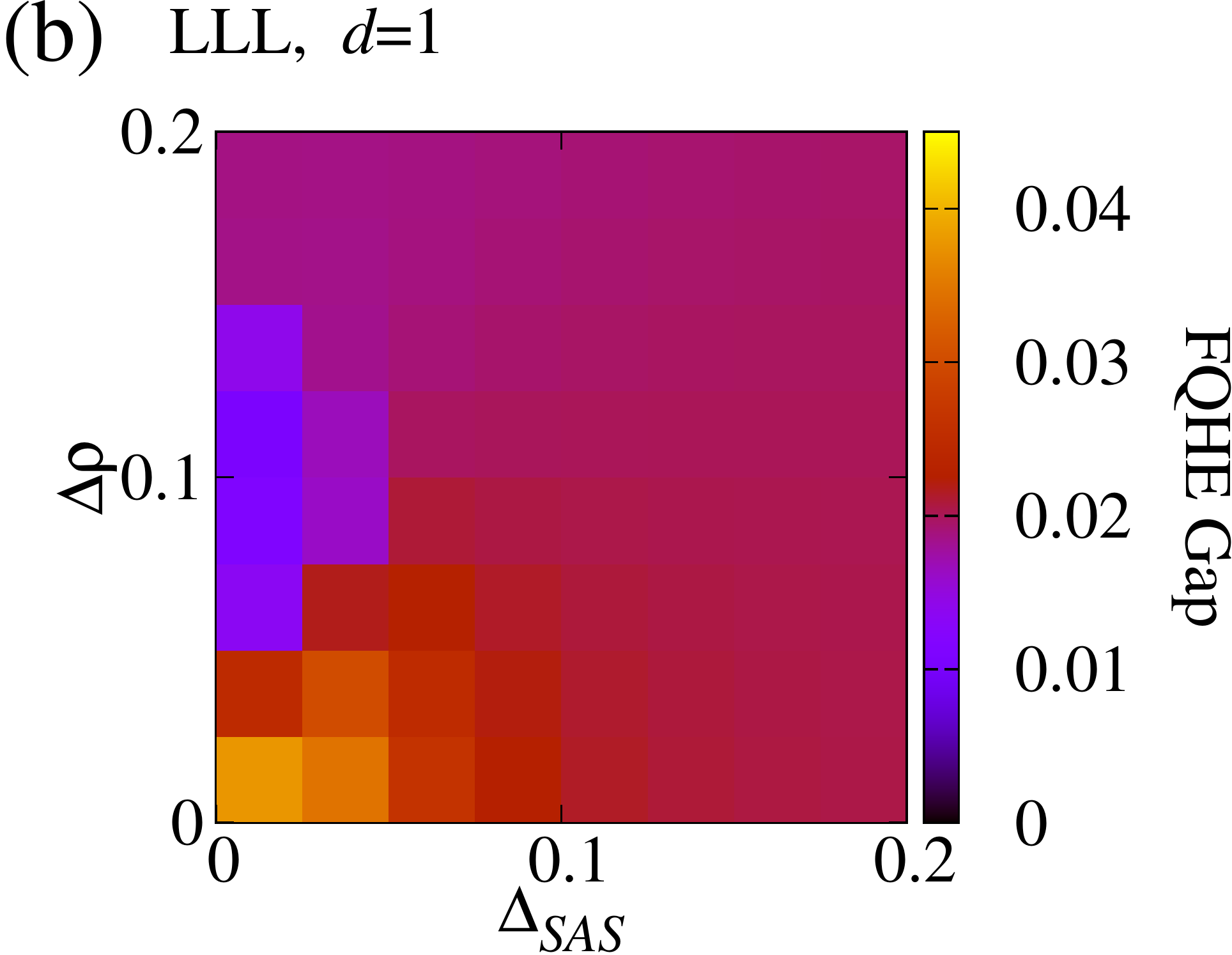}\\
\includegraphics[width=3.5cm,angle=0]{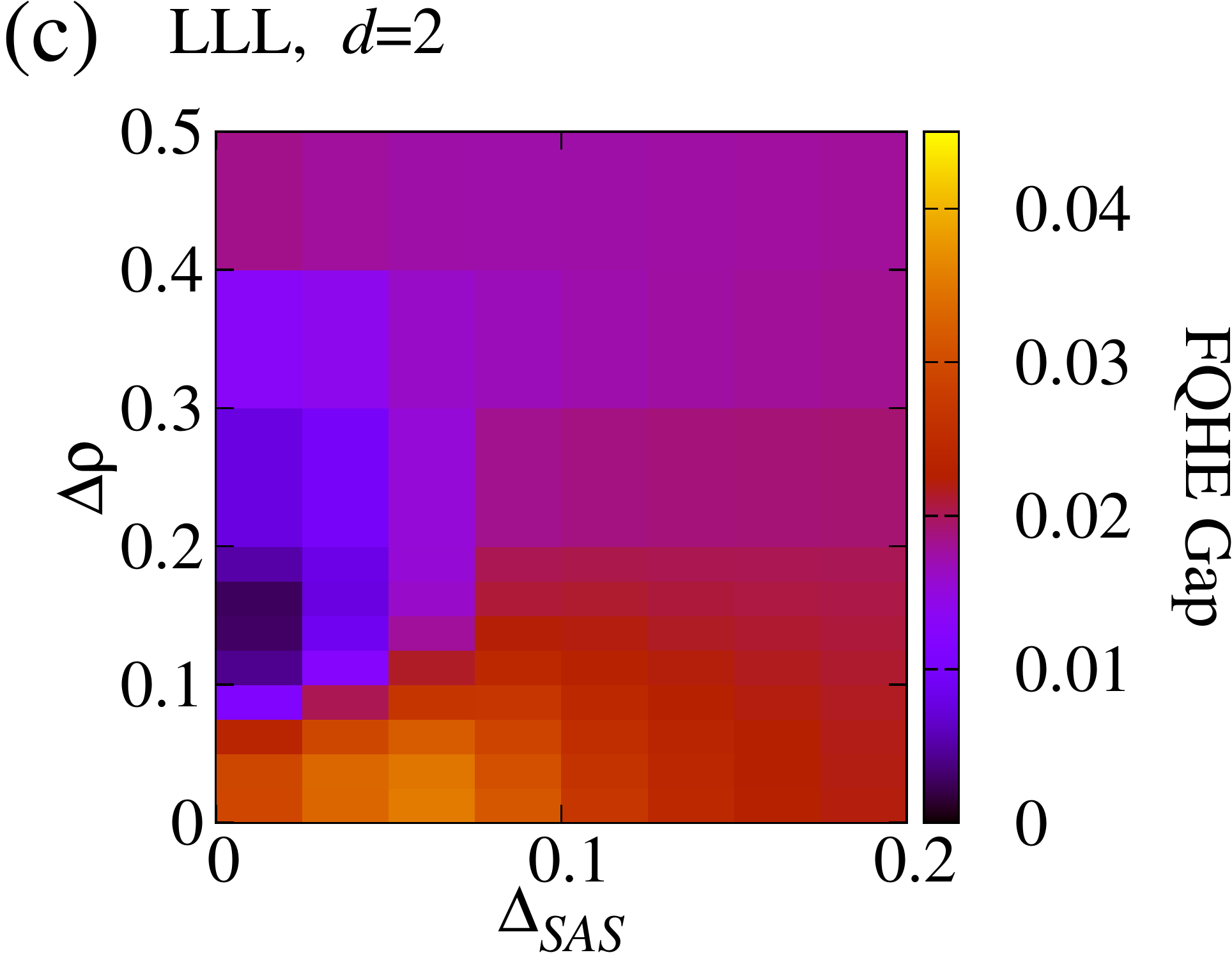}
\includegraphics[width=3.5cm,angle=0]{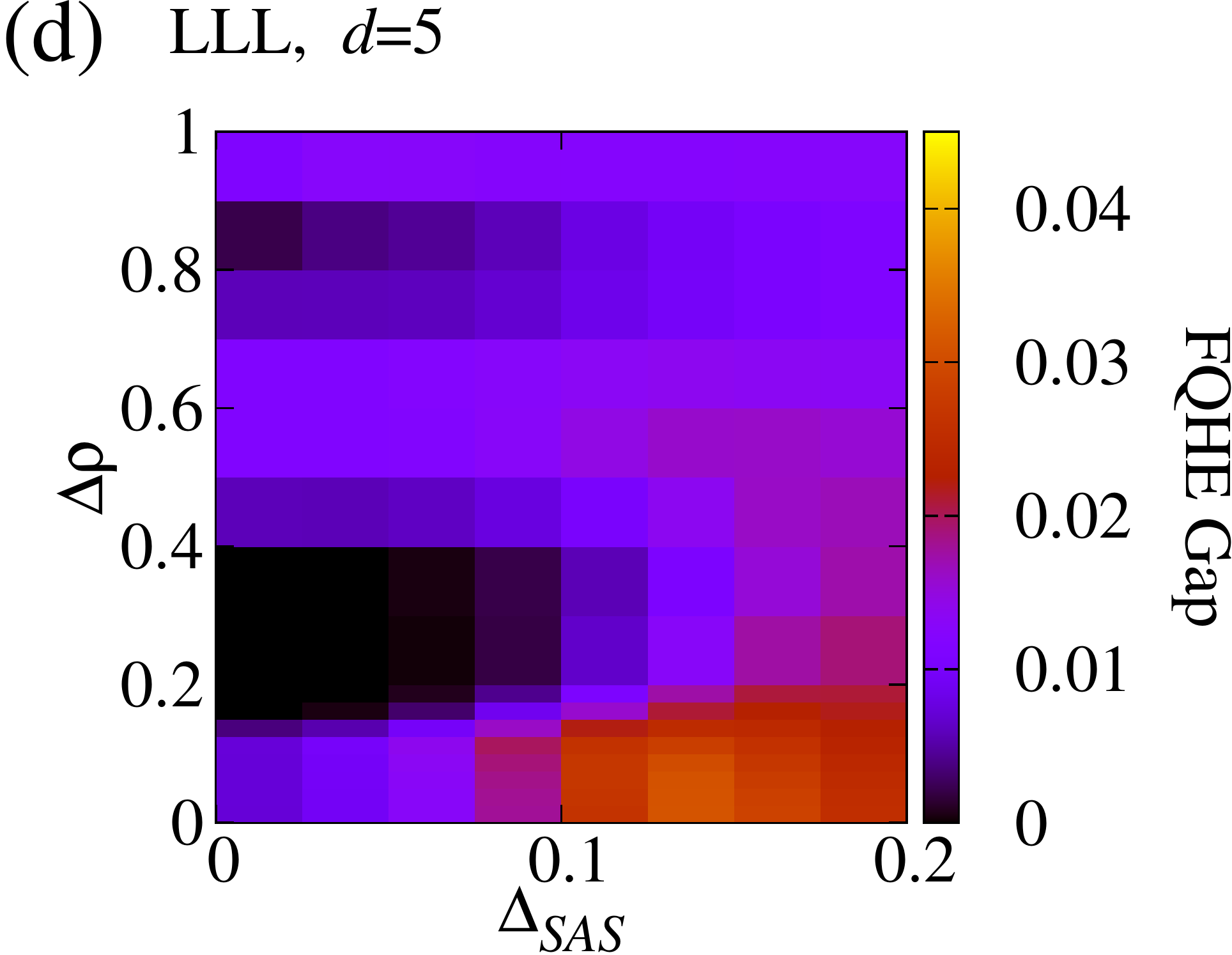}
\end{tabular}
\caption{(Color online) Lowest Landau level:  FQHE energy gap for (a) $d=0$, (b) $d=1$, (c) $d=2$ 
and (d) $d=5$.}
\label{fig-LLL-gap-Dsas-v-Drho}
\end{figure}

\begin{figure}[h!]
\begin{center}
\includegraphics[width=7.cm,angle=0]{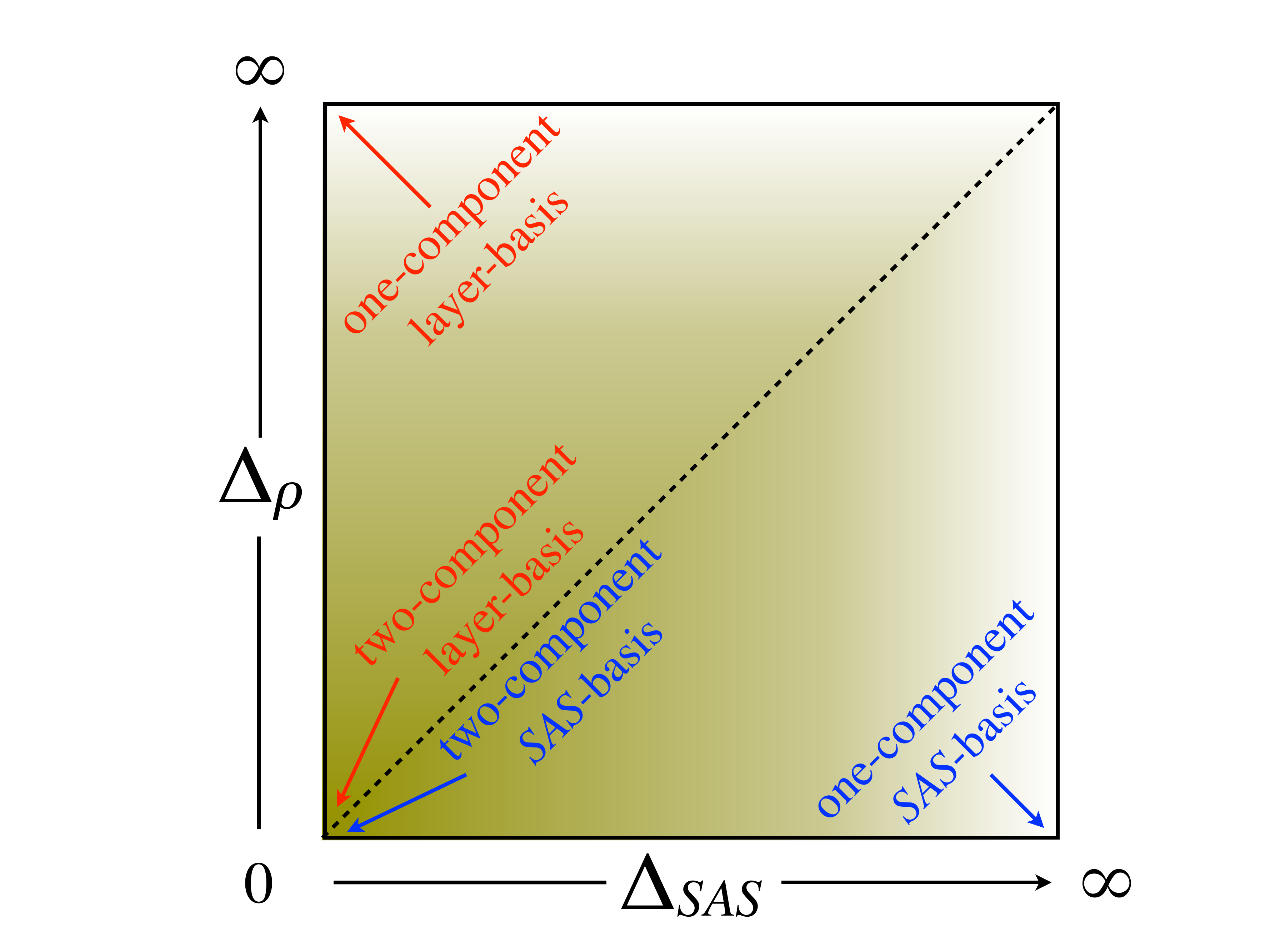}
\caption{(Color online) By varying either $\Delta_{SAS}$ or $\Delta\rho$ the 
system can be two-component (in the $SAS$-basis or layer-basis sense) when 
$\Delta_{SAS}$ and $\Delta\rho$ are small and one-component (again in 
either the $SAS$-basis or layer-basis sense) for large values of either 
$\Delta_{SAS}$ or $\Delta\rho$.  The shading qualitatively indicates 
the one-component to two-component nature of the system with the 
darker shade indicating a two-component system and the lighter 
color a one-component system.}
\label{fig-phasediagram-cartoon}
\end{center}
\end{figure}

\section{Results: lowest Landau level}
\label{sec-LLL}

We first present our results for the lowest Landau level where we consider (i) the wavefunction 
overlap between the four variational states (Pf in the layer- and $SAS$-basis 
and the 331 in the layer- and $SAS$-basis) and the exact ground state of the Hamiltonian--note 
that since all the variational states describe the FQHE 
they have $L=0$ and so if the exact 
ground state of the Hamiltonian does not also have 
total angular momentum $L=0$ the overlaps will be trivially 
zero since they are then of different symmetry.  (ii) the pseudo-spin 
expectation values of the exact ground state, which is more descriptively referred to as 
the expectation value of $(N_S-N_{AS})/2$ or $(N_R-N_L)/2$ where $N_S$, $N_{AS}$, 
$N_R$ and $N_L$ are the expectation values of 
the number of electrons in the symmetric state, 
the anti-symmetric state, the right layer and the left layer, respectively.  Alternatively, 
this is simply 
$\langle \hat{S}_z\rangle$ and $\langle \hat{S}_x\rangle$ in the $SAS$-basis,   
but we prefer to use the more physical  $(N_S-N_{AS})/2$ and $(N_R-N_L)/2$ 
since this does not depend on the basis in which we choose to write the Hamiltonian.  (iii) 
the FQHE energy gap which is \emph{defined} here as the energy difference 
between the $L=0$ ground state and the first excited state (this is often times 
referred to as the ``neutral gap" but we will call it the FQHE gap throughout this 
work) and, lastly, 
(iv) we investigate the energy spectra in the torus geometry.  We show 
all of these quantities from a variety of vantage points to give the clearest picture 
of the physics since things can get complicated rather quickly with so many 
parameters in the Hamiltonian, namely, layer separation $d$ or WQW width $W$, 
$SAS$ energy gap $\Delta_{SAS}$ and the charge imbalancing gap $\Delta\rho$.

\subsection{Bilayer}

We first present our results for the  bilayer Hamiltonian.  
Fig.~\ref{fig-LLL-stacked-Dsas-v-Drho}a shows the overlap between the 
Moore-Read Pfaffian and the exact ground state in the layer-basis (left column) 
and $SAS$-basis (right column) as a function of $\Delta_{SAS}$ 
and $\Delta\rho$ for values of layer separation $d$ ranging from 
$d=0.05$ to 6.  For the layer-basis (left column), it is clear that 
increasing $\Delta_{SAS}$ does not drive the system into the Pf phase, i.e., 
it does not increase the overlap between the layer-basis Pf with the exact 
ground state.  However, increasing the charge imbalance $\Delta\rho$ does 
drive the system into the Pf phase.  For increasing values of $d$ it takes a 
larger value of $\Delta\rho$ to produce a ground state with a high overlap 
with the layer-basis Pf.  This behavior is easy to understand.  Non-zero 
$\Delta\rho$ makes the system one-component in the layer-sense and 
thus a layer-basis Pf wavefunction (which is 
a one-component wavefunction) has a large overlap.  It is 
also understandable that finite layer separation $d$ would make it 
harder to polarize the electrons 
in the layer-sense, i.e., the electrons prefer to remain two-component 
so they can take advantage of the reduction in potential energy between 
electrons in neighboring layers (the interaction is $1/\sqrt{r^2+d^2}$ as opposed 
to $1/r$ for electrons in the same layer).

\begin{figure*}[t]
\begin{center}
\includegraphics[width=5.9cm,angle=0]{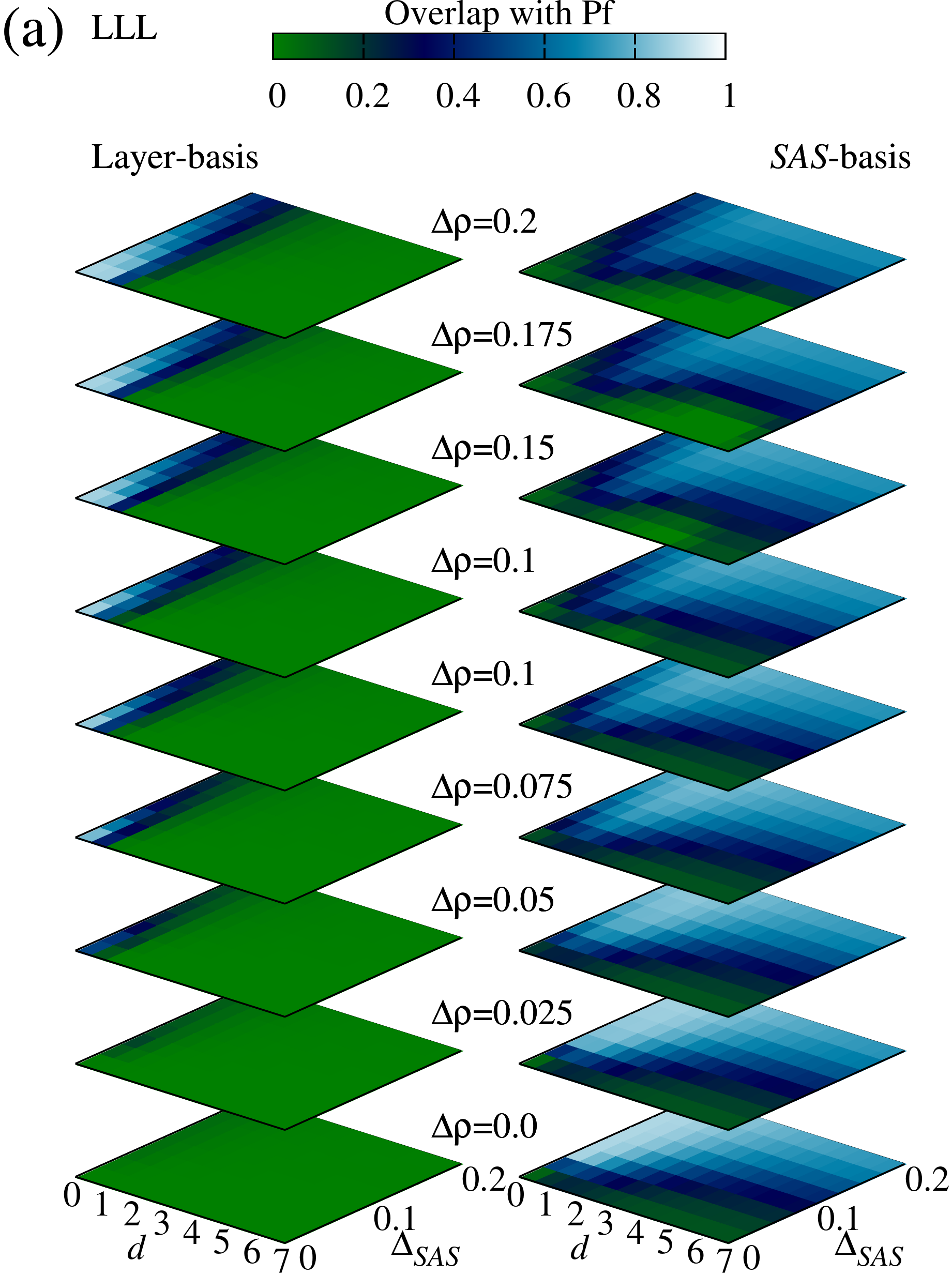}
\includegraphics[width=5.9cm,angle=0]{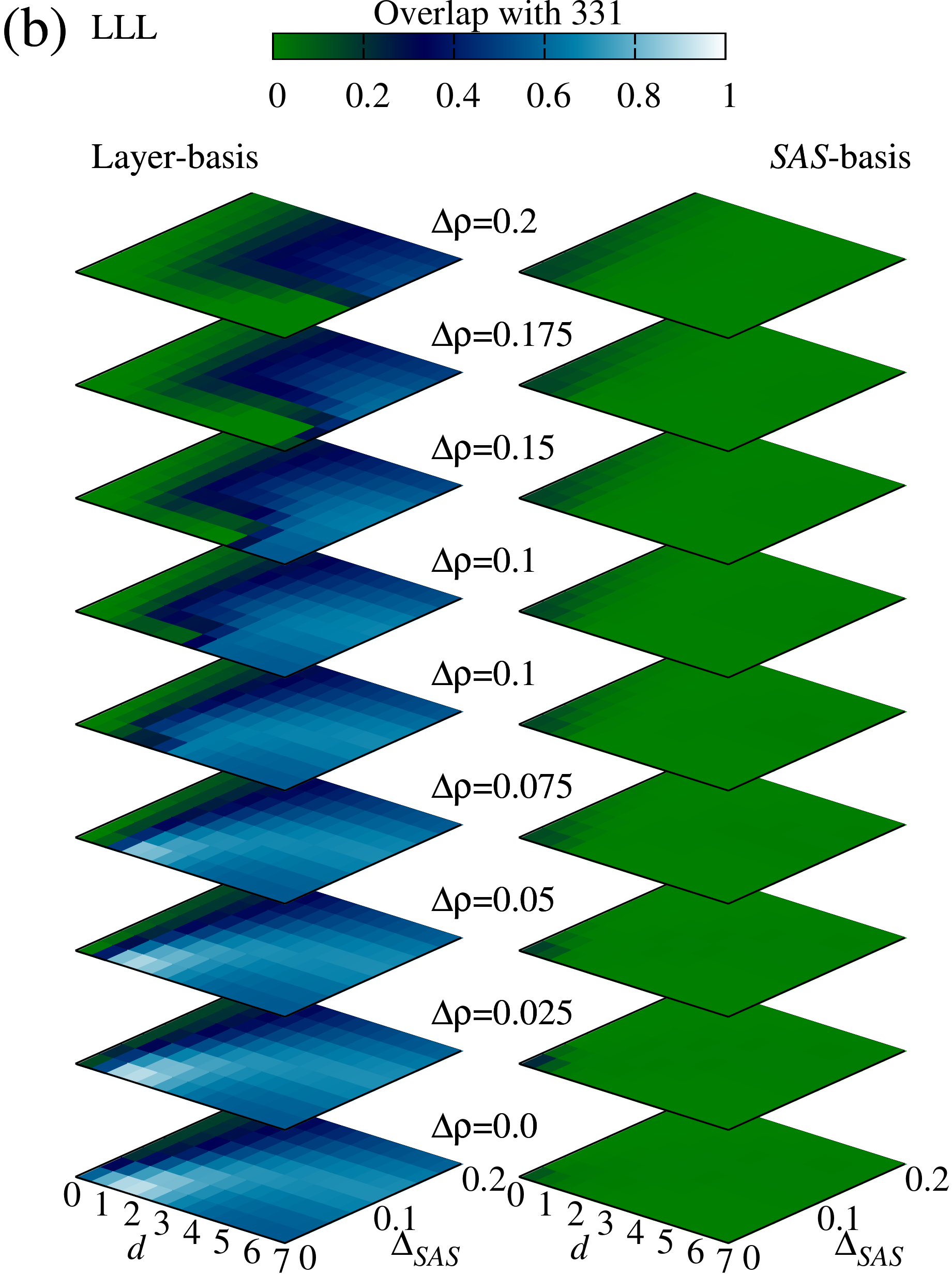}\\
\mbox{}\\
\includegraphics[width=5.9cm,angle=0]{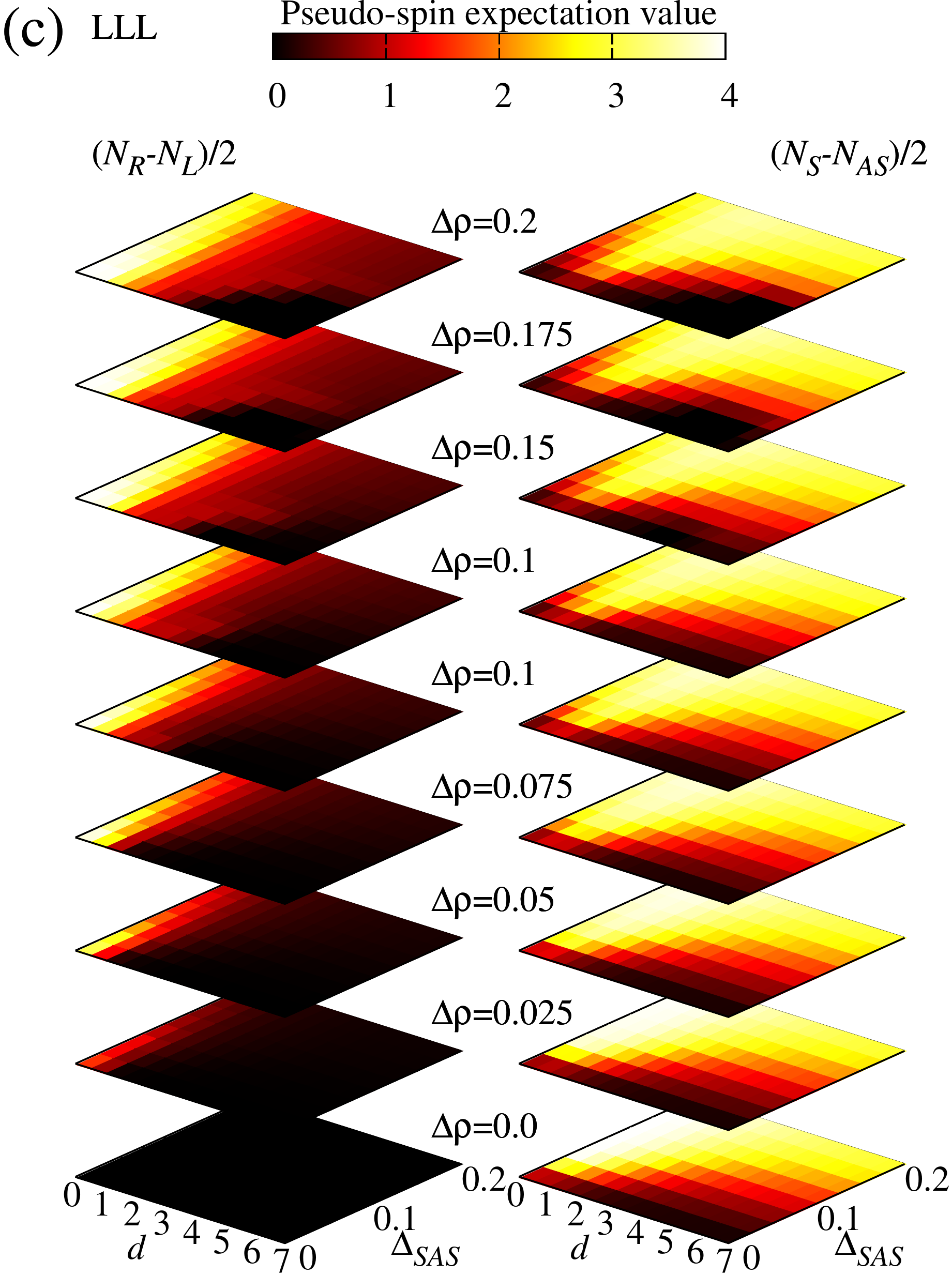}
\includegraphics[width=5.5cm,angle=0]{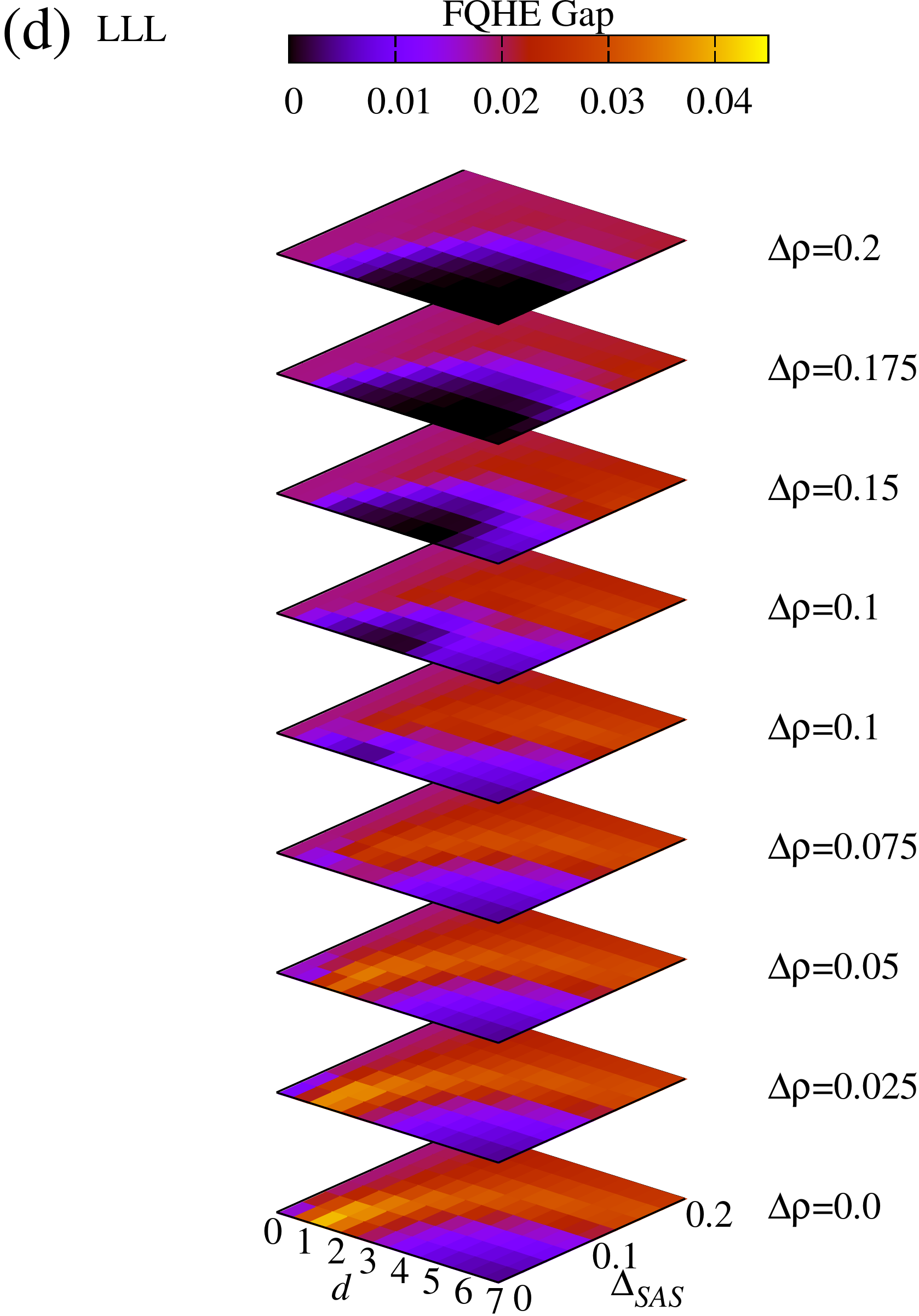}
\caption{(Color online) Lowest Landau level:  Same as Fig.~\ref{fig-LLL-stacked-Dsas-v-Drho} 
except that all plots are shown as a function of inter-layer tunneling $\Delta_{SAS}$ and layer 
separation $d$ for a number of different  charge imbalances $\Delta\rho$ and zero individual layer 
thickness $w=0$.}
\label{fig-LLL-stacked-Dsas-v-d}
\end{center}
\end{figure*}

The overlap between the exact ground state and the Moore-Read Pfaffian in the 
$SAS$-basis shows essentially the opposite behavior.  This time, non-zero 
$\Delta_{SAS}$ drives the system to be one-component.  
Notice that for the smallest value of layer separation $d=0.05$ the system 
is nearly SU(2) symmetric and the two bases (layer and $SAS$) are (nearly) 
pseudo-spin rotations of each other.  The figure is  deceptive because the 
scale on the two axes is different, but the two figures are essentially identical 
upon reflection across the $\Delta_{SAS}=\Delta\rho$ line.  This symmetry is 
quite obviously destroyed upon increasing $d$.

A naive look at Fig.~\ref{fig-LLL-stacked-Dsas-v-Drho}a would lead to a conclusion that
we have found the Pfaffian state in some regions of $\Delta_{SAS}-\Delta\rho$ phase diagram. 
The reader may wonder how robust these conclusions are with increasing system size. 
Although we are unable
to calculate the full phase diagram for $N=10$ electrons, we can gain insight into the limiting 
cases
when either $\Delta\rho$ or $\Delta_{SAS}$ is very large. If we work in the layer-basis 
and $\Delta\rho$ is
very large, we recover the single-layer physics; for our system of $N=8$ electrons, it turns 
out~\cite{papic-2}
that the ground state for the lowest Landau level 
Coulomb interaction is already in the universality class of 
Moore-Read Pfaffian.   However, this no longer holds for larger systems 
such as $N=10$ and $N=12$, leading to the conclusion that
high overlaps with the Pfaffian in the layer-basis in Fig.~\ref{fig-LLL-stacked-Dsas-v-Drho}a are a 
finite size effect for $\nu=1/2$. On the other hand, if we work in the $SAS$-basis and 
increase $\Delta_{SAS}$, all the electrons
will occupy the symmetric subband. This is again a one-component system, but 
with the interaction
slightly softened with respect to pure Coulomb~\cite{papic-3}. It is known~\cite{papic-2} that 
softening the Coulomb interaction via various finite-width corrections can lead to 
a phase transition between compressible
and incompressible (non-Abelian) states, but this also leads to a significant reduction of the 
gap (as we show below).
Therefore, the Pfaffian which is seen in our data for $SAS$ basis is also not a typical 
incompressible state, but rather a ``critical" state with some Pfaffian correlations and 
a small gap. In the remainder of this 
section, we will nonetheless, for the sake of brevity, refer to the one-component phase (fully 
polarized by either $\Delta_{SAS}$ or $\Delta\rho$) as described by 
the appropriate basis ``Pfaffian" state, cautioning the 
reader that such a designation most likely does not survive in the thermodynamic limit.

Next we consider the overlap between the exact ground state and the Halperin 331 
state in the layer-basis (left column) and $SAS$-basis (right column) in  
Fig.~\ref{fig-LLL-stacked-Dsas-v-Drho}b.   For the 
$SAS$-basis 331 state the overlap is very small for all values of 
$\Delta_{SAS}$, $\Delta\rho$ or layer separation $d$.  
On the other hand, the overlap with the layer-basis 331 state is non-zero 
for moderate values of $\Delta_{SAS}$ and for $1\leq d\leq 4$ the overlap is large 
and, despite the fact that non-zero $\Delta\rho$ eventually destroys 
the 331, it is remarkably robust to charge imbalancing.

In Figs.~\ref{fig-LLL-Dsas-v-Drho-Pf} and~\ref{fig-LLL-Dsas-v-Drho-331} we show a clearer 
view of the overlaps.  In particular, for $d=0$ when the system is SU(2) symmetric it is 
obvious in Figs.~\ref{fig-LLL-Dsas-v-Drho-Pf} and~\ref{fig-LLL-Dsas-v-Drho-331} that the 
layer-basis and $SAS$-basis are merely pseudo-spin rotations of one another.  Furthermore, 
it is clear that for $d=0$ there is no region in phase-space were the Halperin 331 wavefunction 
(in either basis) is a good description
and only upon increasing $d$ away from zero do 
we see any sort of reasonably sized overlaps when increasing $\Delta_{SAS}$ (though 
it should be noted that for $d=1$ and 2 the overlap with 331 is sizable for small values 
of both $\Delta_{SAS}$ and $\Delta\rho$).  For the half-filled Landau level at $d=0$, 
theoretically, the system is known to be spin-polarized~\cite{morf,feiguin}, or 
one-component, so it is not surprising that a two-component wavefunction like the 
Halperin 331 state would not be a good description and, therefore, have a 
small overlap with the exact ground state.  Note also that the one-component 
Moore-Read Pfaffian is also not a good description for the $d=0$ case 
since, in the lowest Landau level, the ground state is most likely a non-FQHE 
(Composite Fermion) Fermi sea~\cite{rezayi-haldane,rezayi-read}.

Generally, on the example of our $N=8$ system 
we can conclude that (i) when the system is largely two-component 
in nature then the overlap between the exact ground state and the Halperin 331 state 
is large, and (ii) when the system is largely one-component in nature the overlap between the 
exact ground state and the Pf is large.   
This is more clearly shown in 
Fig.~\ref{fig-LLL-stacked-Dsas-v-Drho}c where in the left and right columns 
we show the expectation value of $(N_R-N_L)/2$ and $(N_S-N_{AS})/2$, 
respectively.  When the system is largely one-component, 
either in the layer sense or $SAS$ sense, the system concomitantly 
has a sizable overlap with the one-component Pf state and when the system 
is largely two-component the overlap with the Halperin 331 state is large.  Again, 
Fig.~\ref{fig-LLL-pseudospin-Dsas-v-Drho} shows the pseudo-spin 
expectation values clearer for a few specific examples of layer separation $d$--note 
again how non-zero $d$ destroys the symmetry between the layer- and $SAS$-basis 
obtained when $d=0$ and the system is SU(2) symmetric.
We emphasize that the first conclusion (i)
is not affected by finite-size effects, whereas the conclusion (ii) likely
holds only for a few special systems such as $N=8$ and it may be a finite-size artifact.

In Fig.~\ref{fig-phasediagram-cartoon} we show a schematic diagram that 
encapsulates broad features of the bilayer model (and WQW model) for 
small layer separation $d$ (small WQW width $W$).  
For large $\Delta_{SAS}$ and small $\Delta\rho$ the system is one-component 
in the $SAS$-basis and for large $\Delta\rho$ and small $\Delta_{SAS}$ the 
system is one-component in the layer-basis.  When, both tunneling 
strengths approach zero, the system is two-component in the layer-basis, 
the $SAS$-basis, or both.  
The diagram would be topologically similar for non-zero (and even large) 
values of layer separation $d$ and WQW width $W$.  The only difference is
 that the diagram would not be symmetric between $\Delta_{SAS}$ and 
 $\Delta\rho$ since the system is not SU(2) symmetric.

Lastly, in Fig.~\ref{fig-LLL-stacked-Dsas-v-Drho}d we show the FQHE energy gap.  For
$d\lesssim1$ the FQHE energy gap is relatively constant as $\Delta_{SAS}$ and 
$\Delta\rho$ vary.  When $d$ is increased past approximately $d\sim 2$ the FQHE 
gap shows interesting features as a function of the strength of the two 
tunneling terms.  Somewhat surprisingly, the FQHE gap obtains 
a maximum for non-zero $\Delta_{SAS}$ and finite $d$ (this is related to 
similar results found recently in Ref.~\onlinecite{mrp-sds-bilayer} where the 
FQHE gap is maximum on the ``ridge" as a function of $\Delta_{SAS}$ and 
separation $d$ for zero charge imbalance $\Delta\rho$).  This ``ridge" 
basically separates the regions in the quantum phase diagram where 
the system is either in the Pf or 331 phase.  
Fig.~\ref{fig-LLL-gap-Dsas-v-Drho} is a more detailed depiction of 
the FQHE energy gap.

Before moving on to results for the second Landau level we show essentially 
the same results as Fig.~\ref{fig-LLL-stacked-Dsas-v-Drho}a-d but as a function of 
$\Delta_{SAS}$ and layer separation $d$ for several values 
of $\Delta\rho$ in Fig.~\ref{fig-LLL-stacked-Dsas-v-d}a-d.  This 
presentation is to more easily compare with previous bilayer works~\cite{mrp-sds-bilayer,papic} 
where the data was presented this way.  When considering the overlap between 
the exact ground state and the Pf we see that increasing imbalance (increasing 
$\Delta\rho$) eventually produces a high overlap for small values of $d$ in the 
layer-basis.  In the $SAS$-basis, charge imbalance pushes the maximum overlap to 
larger values of $\Delta_{SAS}$ and slightly larger values of $d$ while generally 
decreasing the overlap along the way.   
For the overlap with the Halperin 331 wavefunction it is clear 
that there is no region in parameter space that produces a sizable overlap in 
the $SAS$-basis.  For the layer-basis 331, the overlap is large for moderate 
layer separation ($d\sim1-2$) and $\Delta_{SAS}$.  The position of the maximum 
overlap is relatively constant as a function of $\Delta\rho$ until $\Delta\rho\sim0.1$ 
when the maximum overlap shifts to higher values of $d$ and decreases markedly 
in magnitude.  The expectation values of the pseudo-spin 
operators (Fig.~\ref{fig-LLL-stacked-Dsas-v-d}c) mirrors the results of the overlap 
calculations.

For the FQHE gap shown in Fig.~\ref{fig-LLL-stacked-Dsas-v-d}d we see the 
familiar picture found in Ref.~\onlinecite{mrp-sds-bilayer}.  We first 
reiterate that result.  For $\Delta\rho=0$, the FQHE gap is largest along a 
``ridge" in $\Delta_{SAS}$-$d$ space and the overlap between the $SAS$-basis 
Halperin 331 state is largest in the region of phase space where the system is 
two-component and the overlap between the $SAS$-basis Pf state 
is largest in the region of phase space where the system is largely 
one-component.  The FQHE gap ``ridge" seems to function as the phase 
boundary between the two states in the quantum phase diagram, however, 
the maximum FQHE gap is slightly on the Halperin 331 side of the phase diagram 
and the experimentally observed~\cite{suen-1,suen-2,suen-3,eisenstein-bilayer} 
FQHE at $\nu=1/2$ is  
of that nature.  This is strongly thought to be the case because of recent 
work by Storni \emph{et al}.~\cite{storni} that showed that in the lowest Landau level the 
FQHE gap for one-component systems vanishes in the thermodynamic limit--even 
though it is non-zero in our finite size calculation.  Thus, our non-zero gap in the 
Pf region of the phase diagram for the lowest Landau level 
is most likely a finite size effect.

For non-zero $\Delta\rho$ 
the above picture changes.  The ``ridge" is weakened and a maximum in the FQHE 
gap starts to appear for large $\Delta_{SAS}$ and large $d$ until eventually 
weakening further.  In the large $\Delta\rho$ limit we see a $SAS$-basis Pf
state for large $\Delta_{SAS}$ and  non-zero $d$ and a layer-basis Pf 
for moderate $\Delta_{SAS}$ and very small $d$.  However, we caution that the 
FQHE gap is globally weakened upon inclusion of $\Delta\rho$ and taking into 
account recent results~\cite{storni}, an experimental system would most likely 
not exhibit the FQHE in that region of parameter space.

To summarize our results from the calculations on the sphere in the 
lowest Landau level, we find a robust
layer-basis 331 state that has high overlap with the exact ground state and dominant gap in the 
phase diagram. Furthermore, we find a one-component state that has some properties of the 
Moore-Read Pfaffian, but it is likely to show up as a compressible state in experiments. These 
conclusions are in 
agreement with the results of Ref. \onlinecite{papic-3} (and consistent with 
Ref.~\onlinecite{storni}) where transitions between 331 state, Pfaffian and 
Composite Fermion Fermi sea were
studied in a bilayer model with tunneling $\Delta_{SAS}$ (in the layer-basis) 
using exact diagonalization and effective mean field BCS theory of 
Read-Green~\cite{read-green}. 
There it was found that the increase of inter-layer tunneling converts the
331 state into a Composite Fermion Fermi sea because the effective chemical
potential of ``even"-channel electrons becomes very large. Similar analysis is
expected to apply when $\Delta\rho$ is large and the system is viewed in $SAS$-basis.

\subsection{Wide-quantum-well}

We now turn to the wide-quantum-well (WQW) model  which 
 largely mirrors the results
presented for the bilayer model.  
However, there are some differences between the two models 
which we point out below.

\begin{figure}
\begin{tabular}{c}
\includegraphics[width=3.5cm,angle=0]{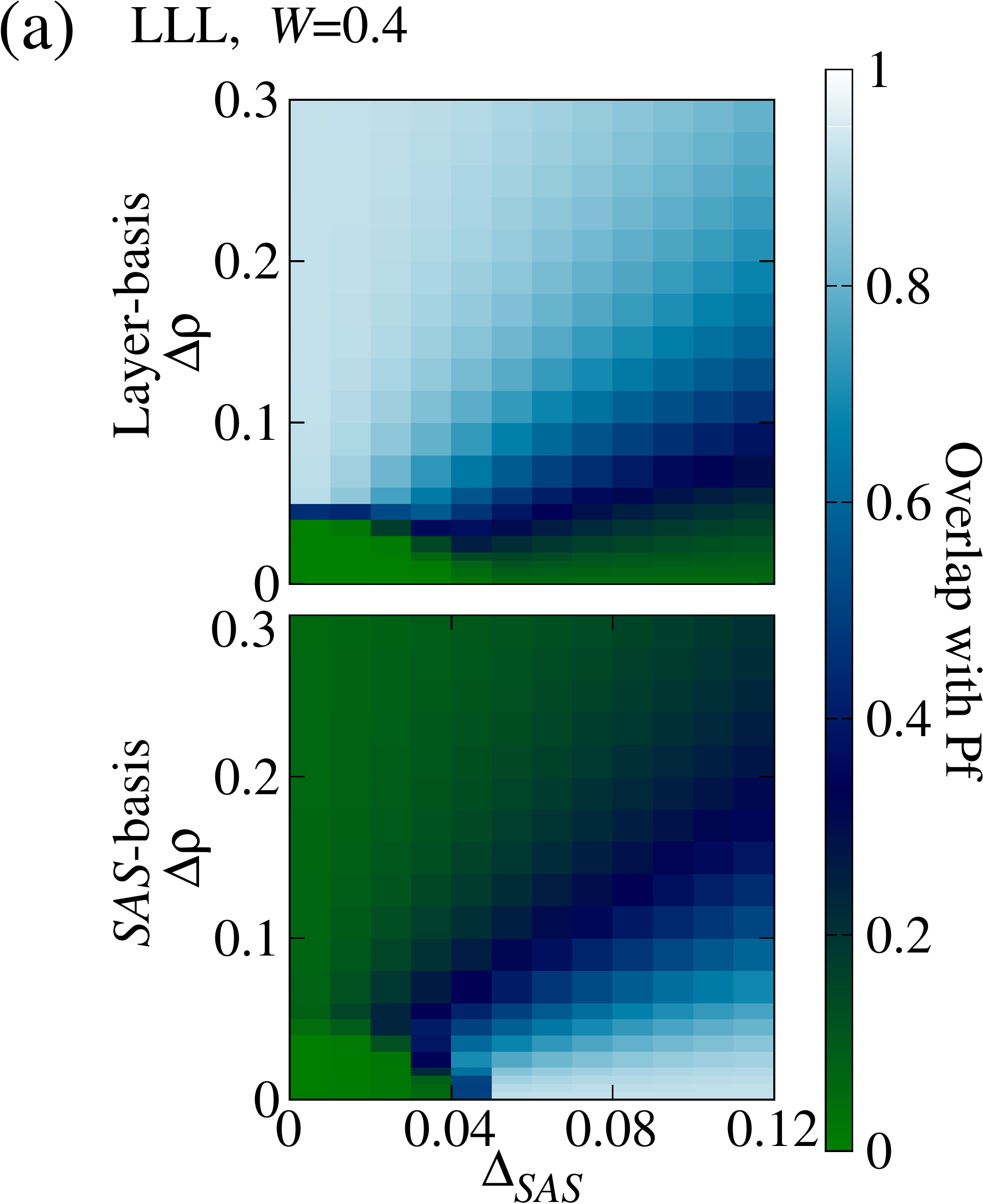}
\includegraphics[width=3.5cm,angle=0]{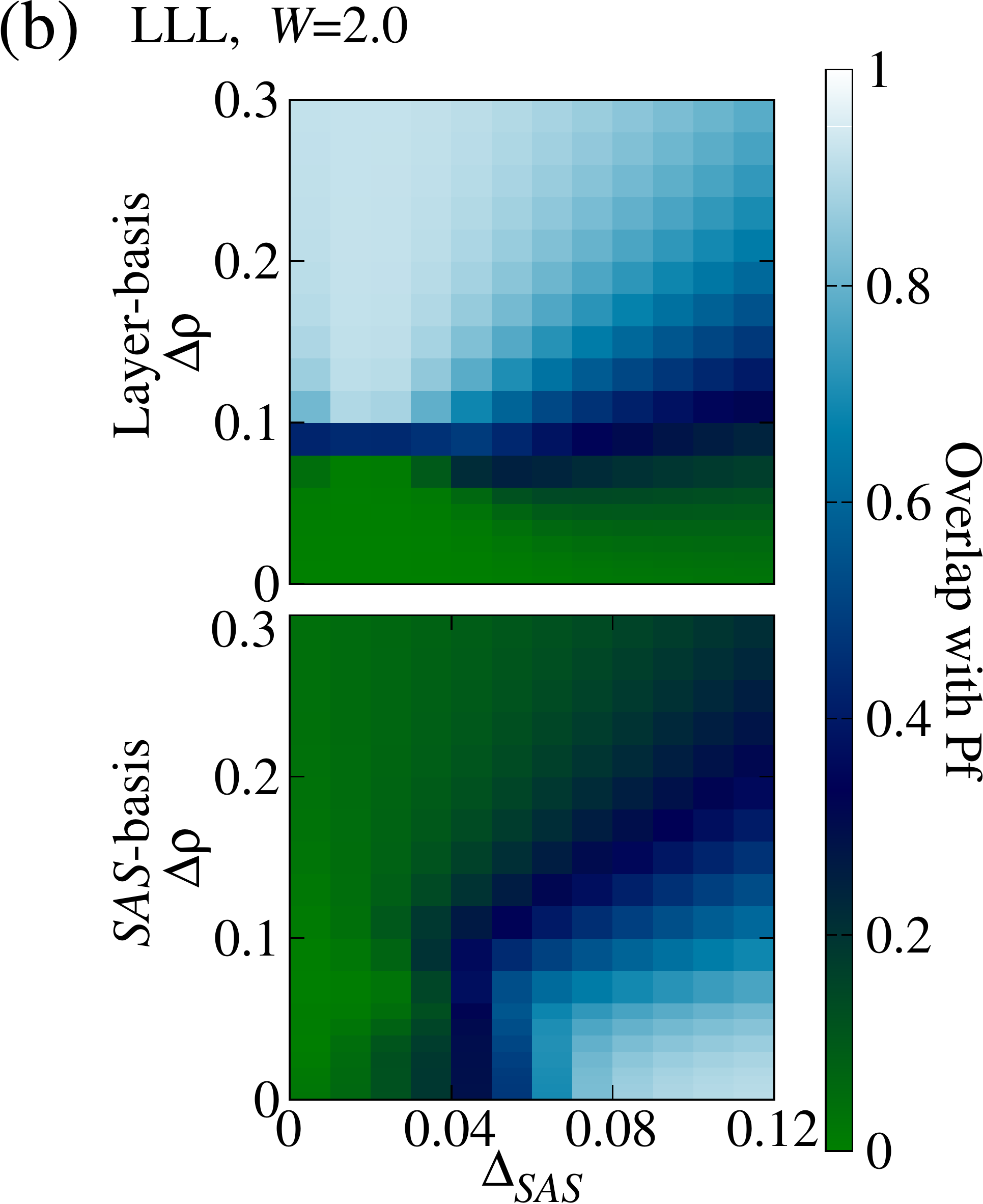}\\
\includegraphics[width=3.5cm,angle=0]{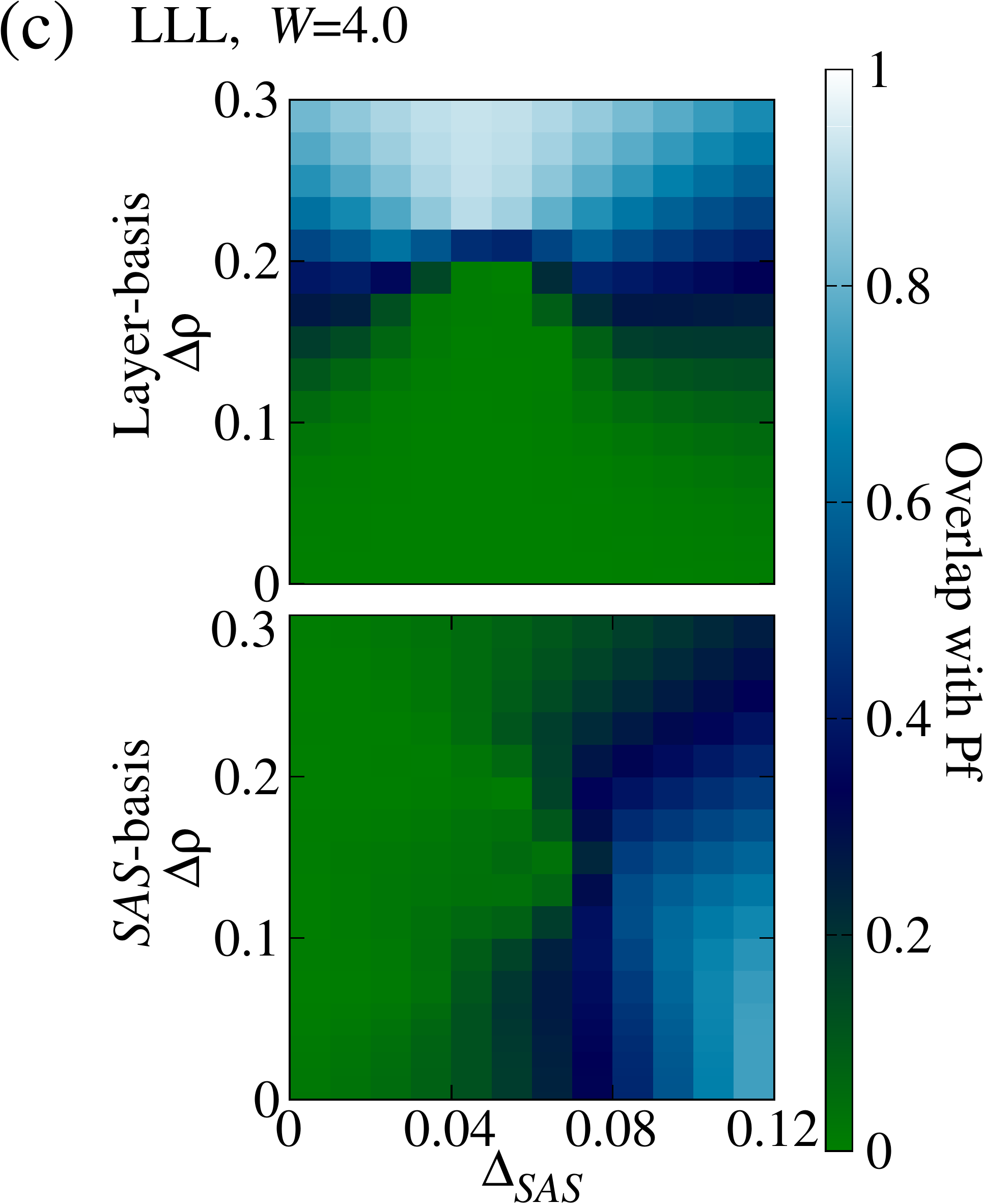}
\includegraphics[width=3.5cm,angle=0]{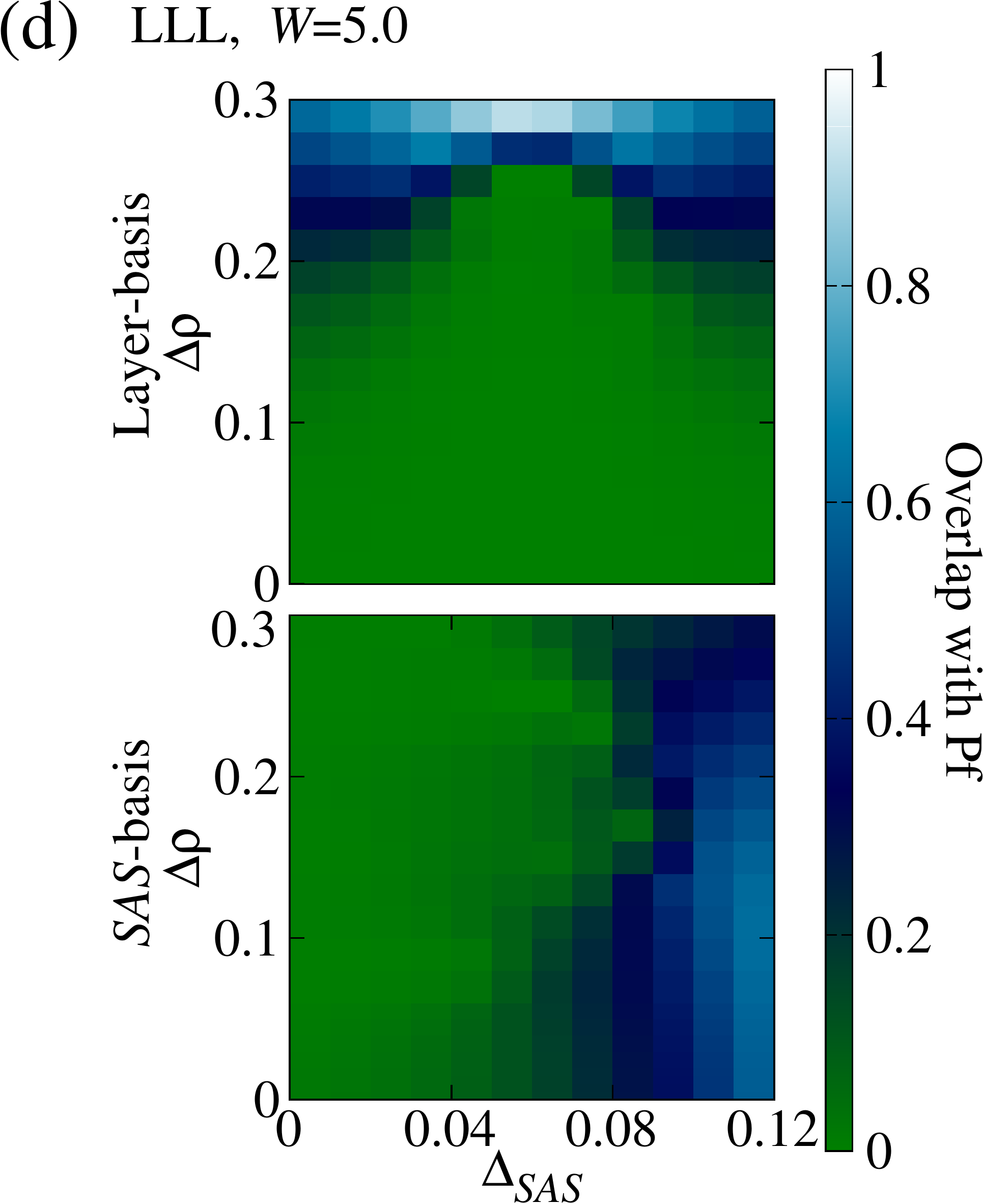}
\end{tabular}
\caption{(Color online) Lowest Landau level: Wavefunction overlap
  between the Moore-Read Pfaffian wavefunction in the layer-basis (top
  panel) and the $SAS$-basis (lower panel) and the exact ground state
  for the WQW for (a) $W=0.4$, (b) $W=2$, (c) $W=4$ and (d) $W=5$.}
\label{fig-LLL-Dsas-v-Drho-Pf-wqw}
\mbox{}\\
\begin{tabular}{c}
\includegraphics[width=3.5cm,angle=0]{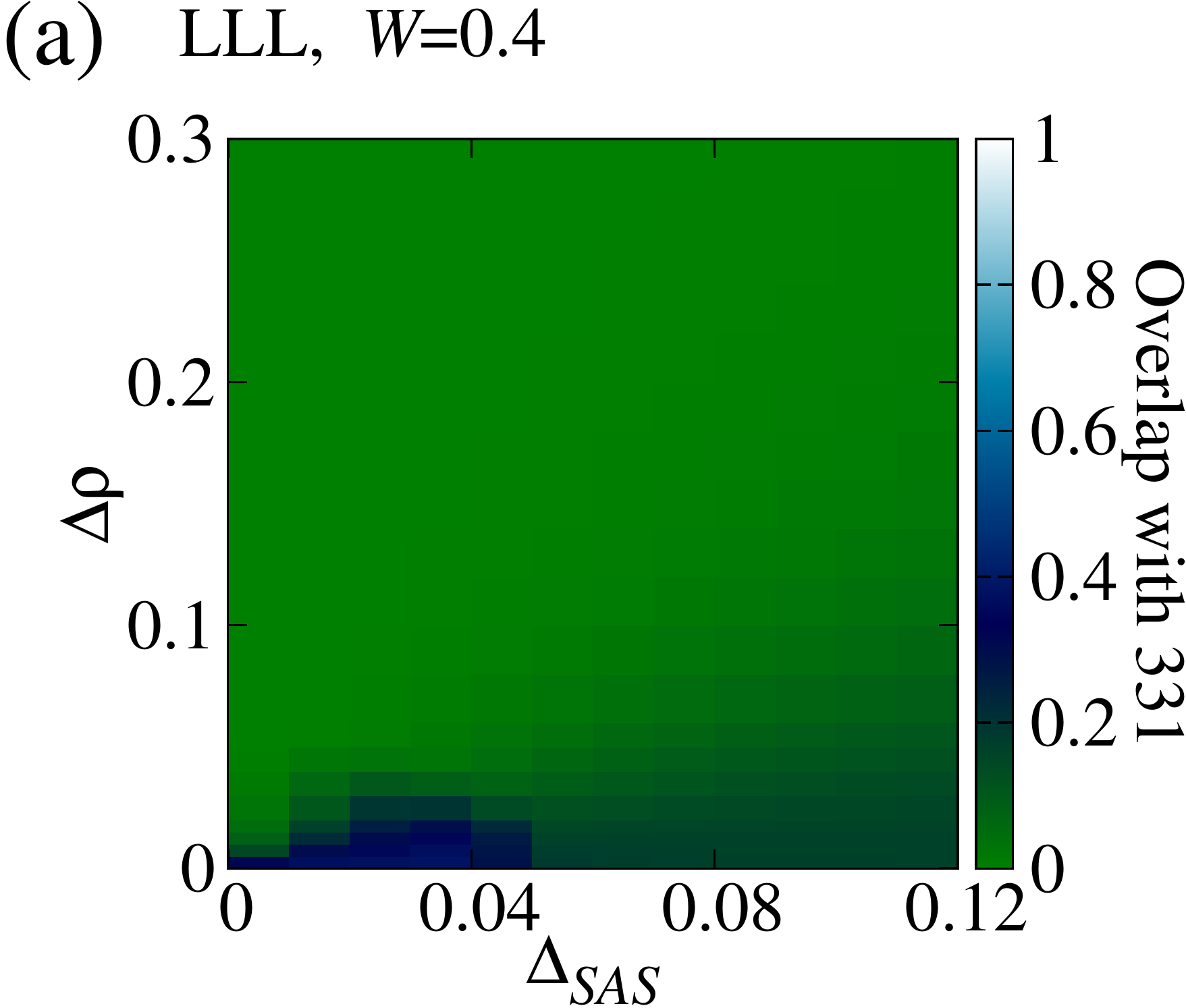}
\includegraphics[width=3.5cm,angle=0]{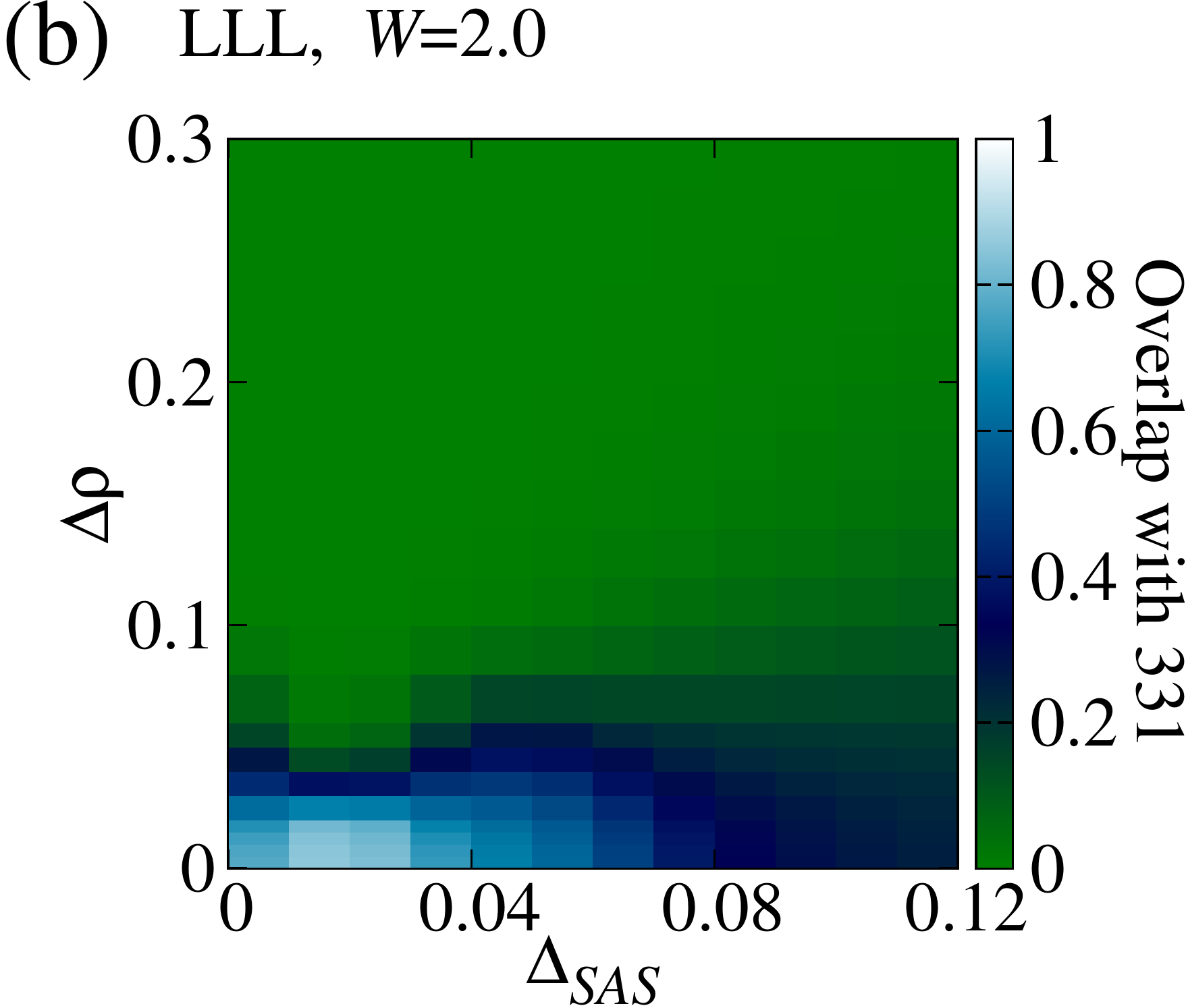}\\
\includegraphics[width=3.5cm,angle=0]{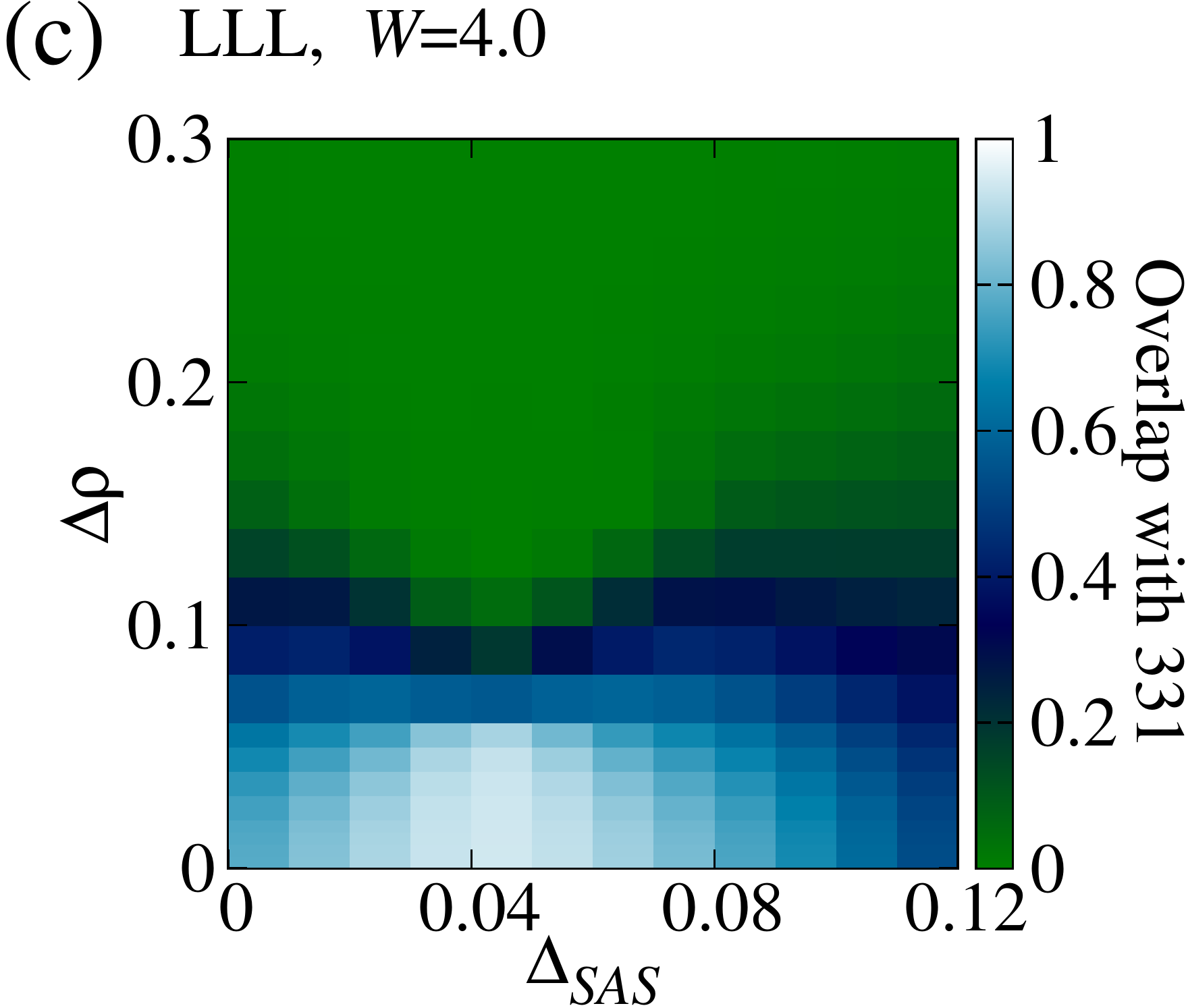}
\includegraphics[width=3.5cm,angle=0]{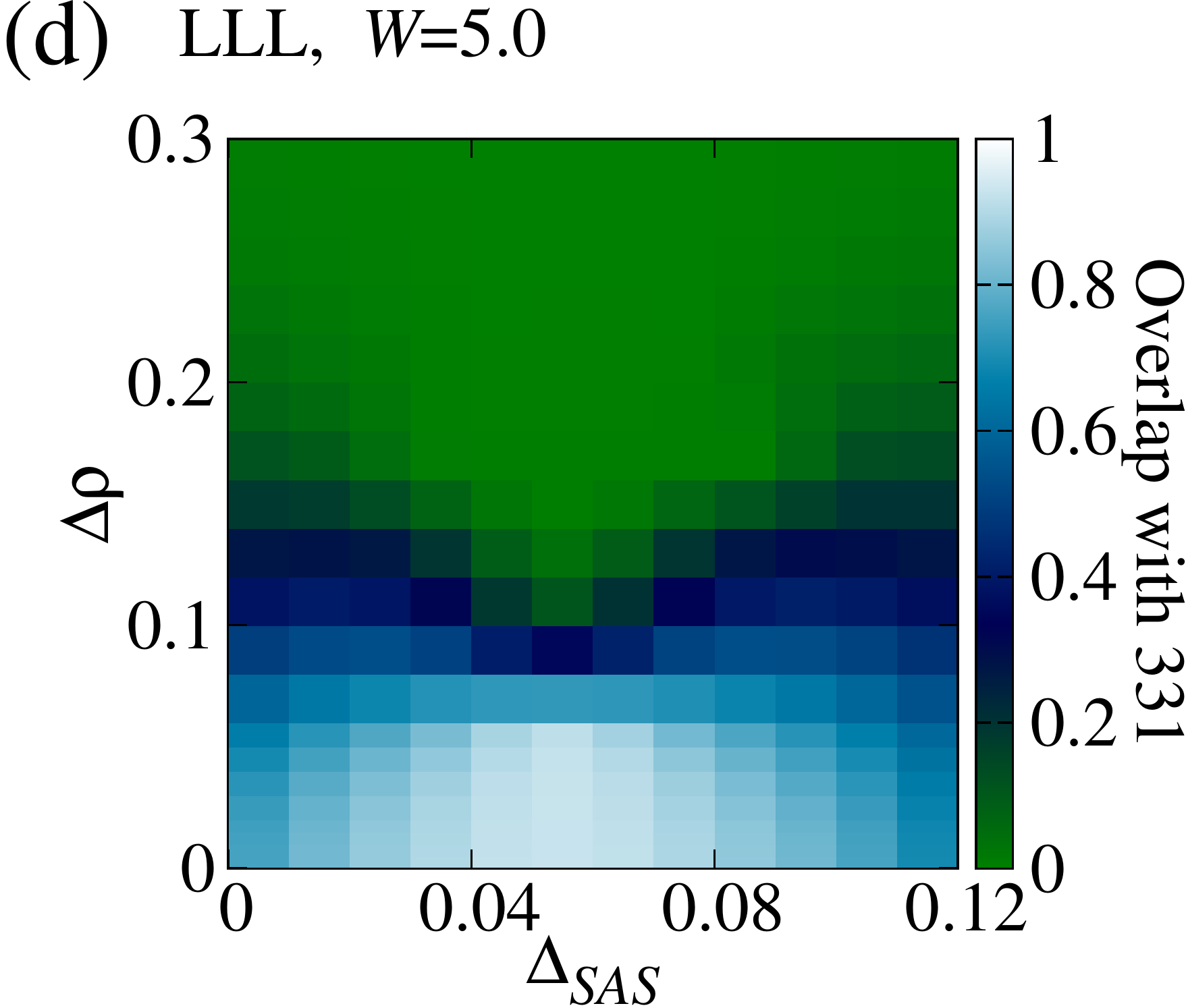}
\end{tabular}
\caption{(Color online) Lowest Landau level:  Wavefunction overlap
  between the Halperin 331 wavefunction in the layer-basis 
  and the exact ground state
  for the WQW for (a) $W=0.4$, (b) $W=2$, (c) $W=4$ and (d) $W=5$.}
\label{fig-LLL-331-Dsas-v-Drho-wqw}
\end{figure}

Fig.~\ref{fig-LLL-Dsas-v-Drho-Pf-wqw} shows the overlap between the 
Moore-Read Pfaffian wavefunction (in both the $SAS$-basis and the 
layer-basis) and the exact ground state of the WQW model as a function of 
$\Delta_{SAS}$ and $\Delta\rho$ for a few values of the WQW width $W$.  
This figure should be compared with Fig.~\ref{fig-LLL-Dsas-v-Drho-Pf} 
for the bilayer model.  \emph{Qualitatively}, the two models produce 
very similar results.  Of course, a layer separation of $d$ in the 
bilayer model is not equivalent to the WQW width $W$ and any similarities 
between the two at $d\sim W$ is coincidental.  However, 
the behavior for the bilayer model as $d$ increases is the same as the 
behavior for the WQW model as $W$ increases.  That is, for small 
$W=0.4$, the overlap between the exact ground state and the 
one-component Pf in the layer-basis becomes 
large when $\Delta\rho$ is increased  since the 
system is being driven to be more one-component in the layer sense.  
When $\Delta_{SAS}$ is increased, the overlap between the exact 
ground state and the  Pf in the $SAS$-basis 
becomes large since the system is being driven to be one-component 
in the $SAS$-basis.  

\begin{figure}
\begin{tabular}{c}
\includegraphics[width=3.5cm,angle=0]{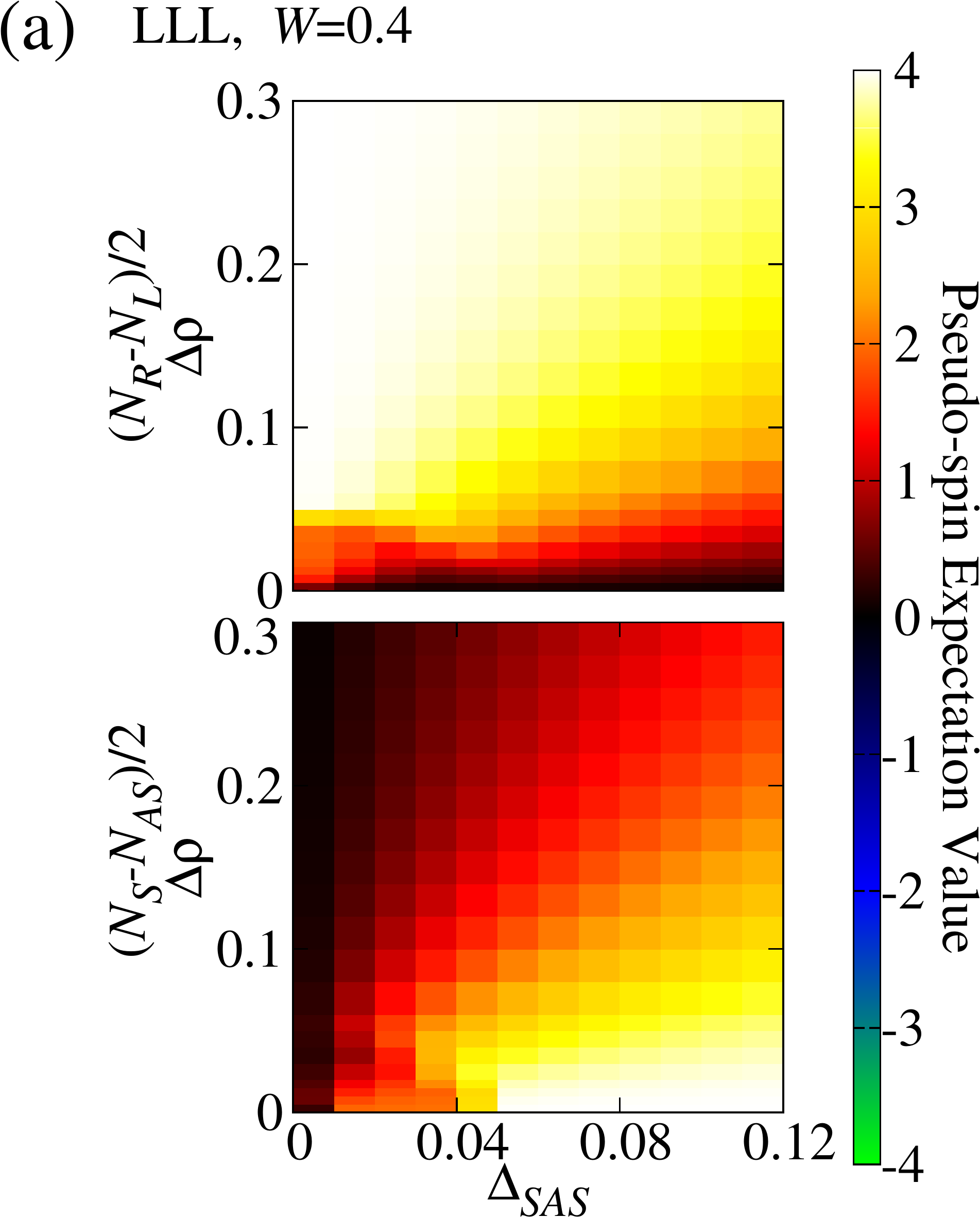}
\includegraphics[width=3.5cm,angle=0]{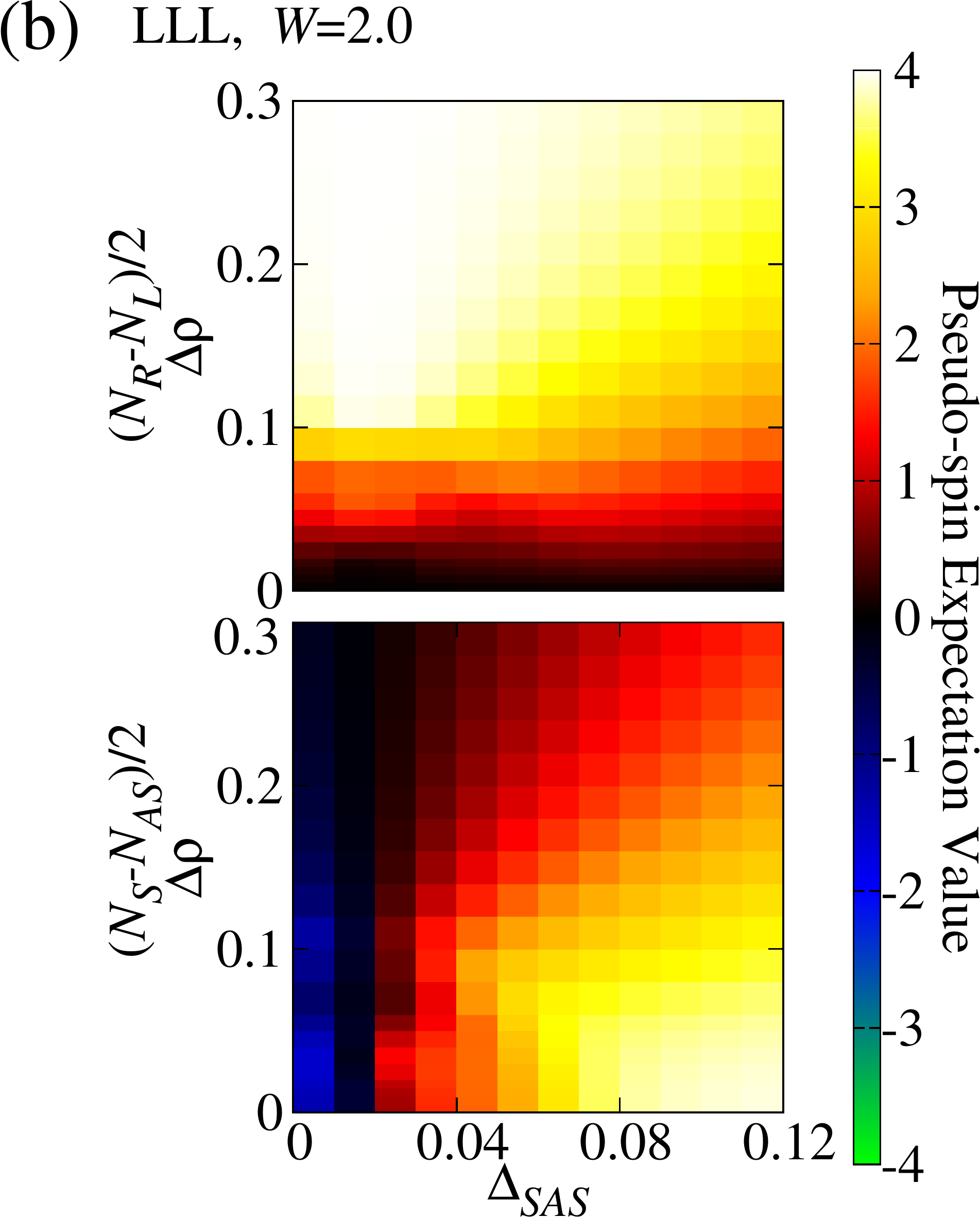}\\
\includegraphics[width=3.5cm,angle=0]{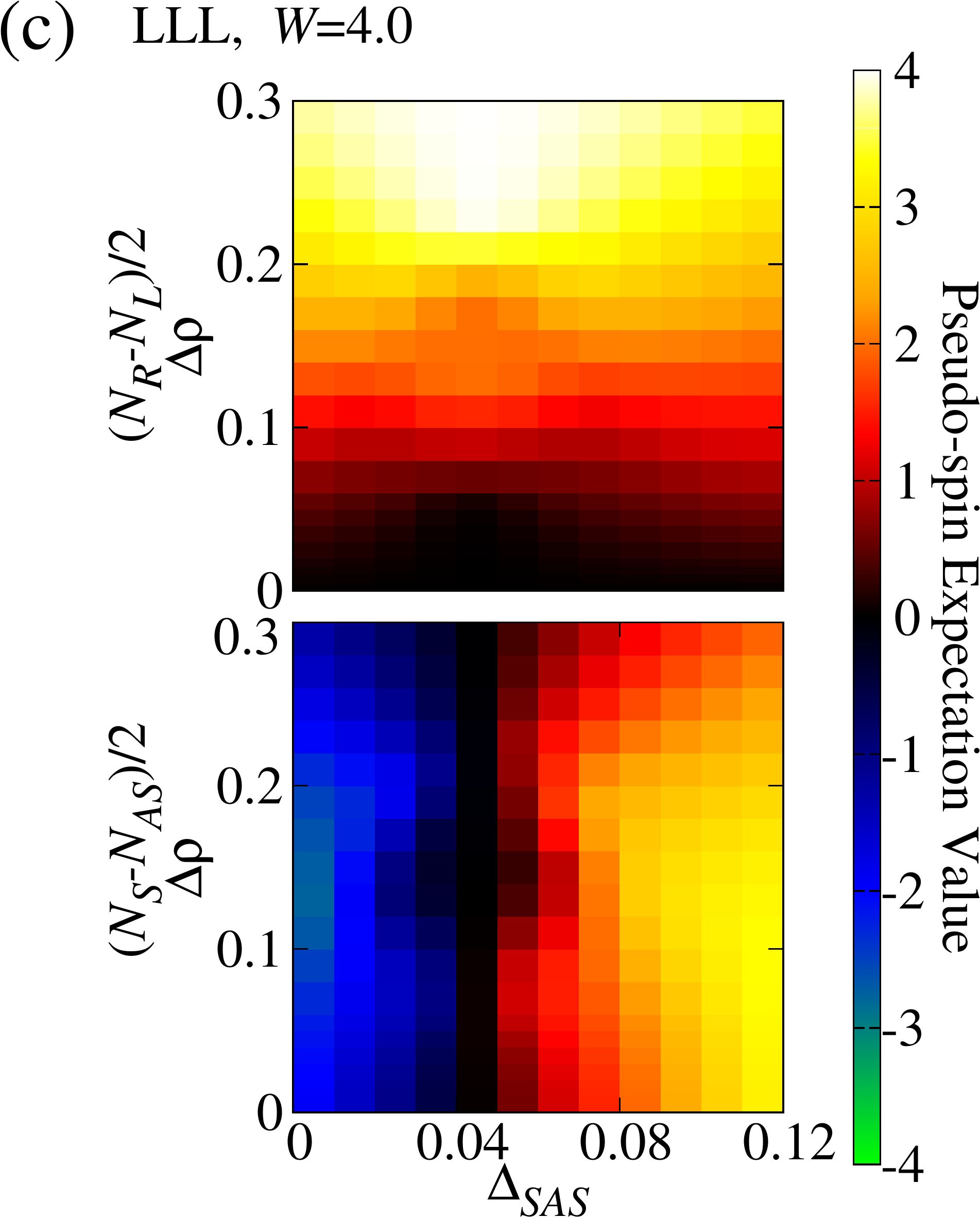}
\includegraphics[width=3.5cm,angle=0]{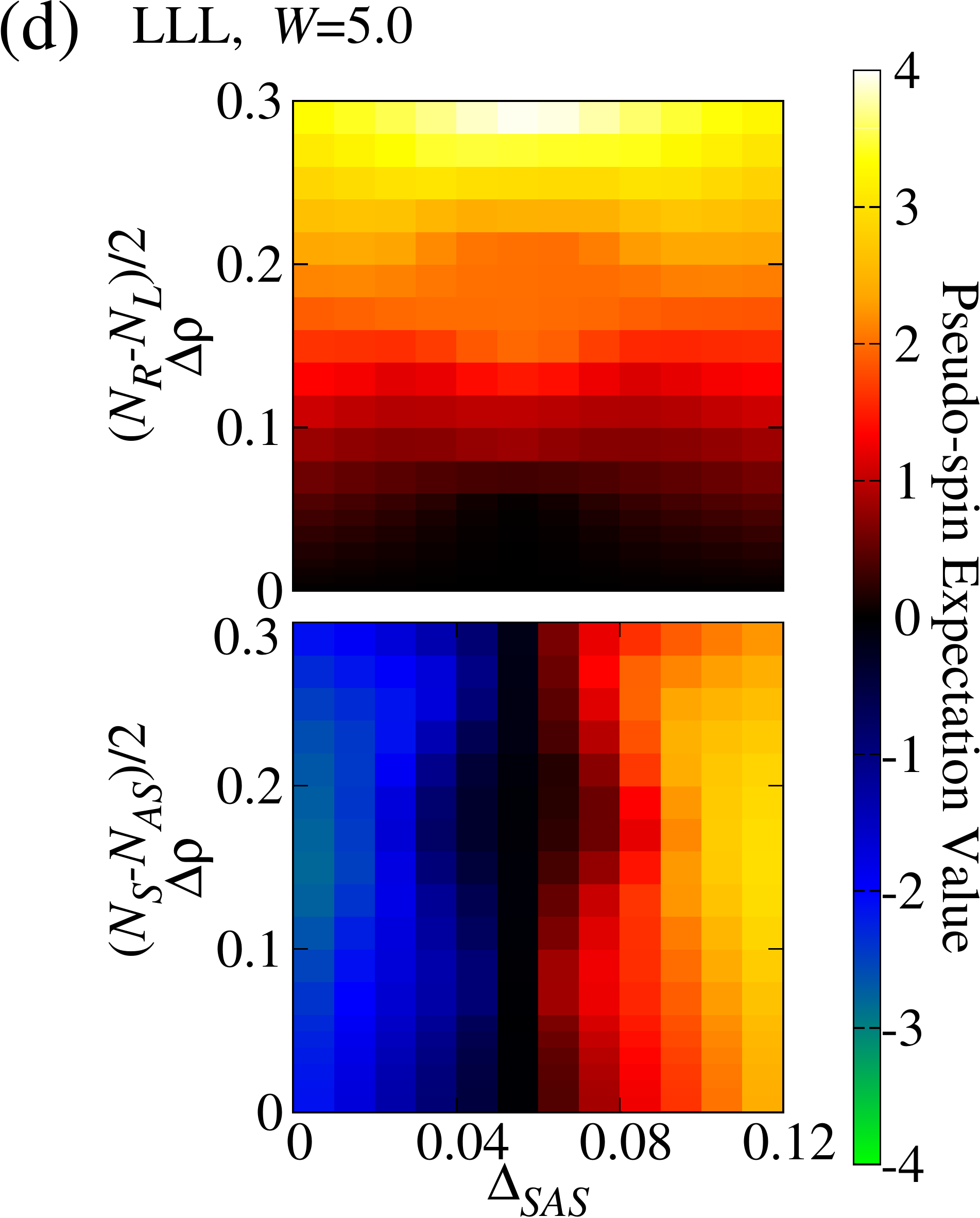}
\end{tabular}
\caption{(Color online)  Lowest Landau level:  (Pseudo-spin) 
expectation value of the exact ground 
state (for the WQW model) of $(N_R-N_L)/2$ (top panel) 
and $(N_S-N_{AS})/2$ (lower panel)  for (a) $W=0.4$, (b) $W=2$, (c) 
$W=4$ and (d) $W=5$.  Note that the WQW model always breaks SU(2) 
symmetry even for small $W$.}
\label{fig-LLL-pseudospin-Dsas-v-Drho-wqw}
\mbox{}\\
\begin{tabular}{c}
\includegraphics[width=3.5cm,angle=0]{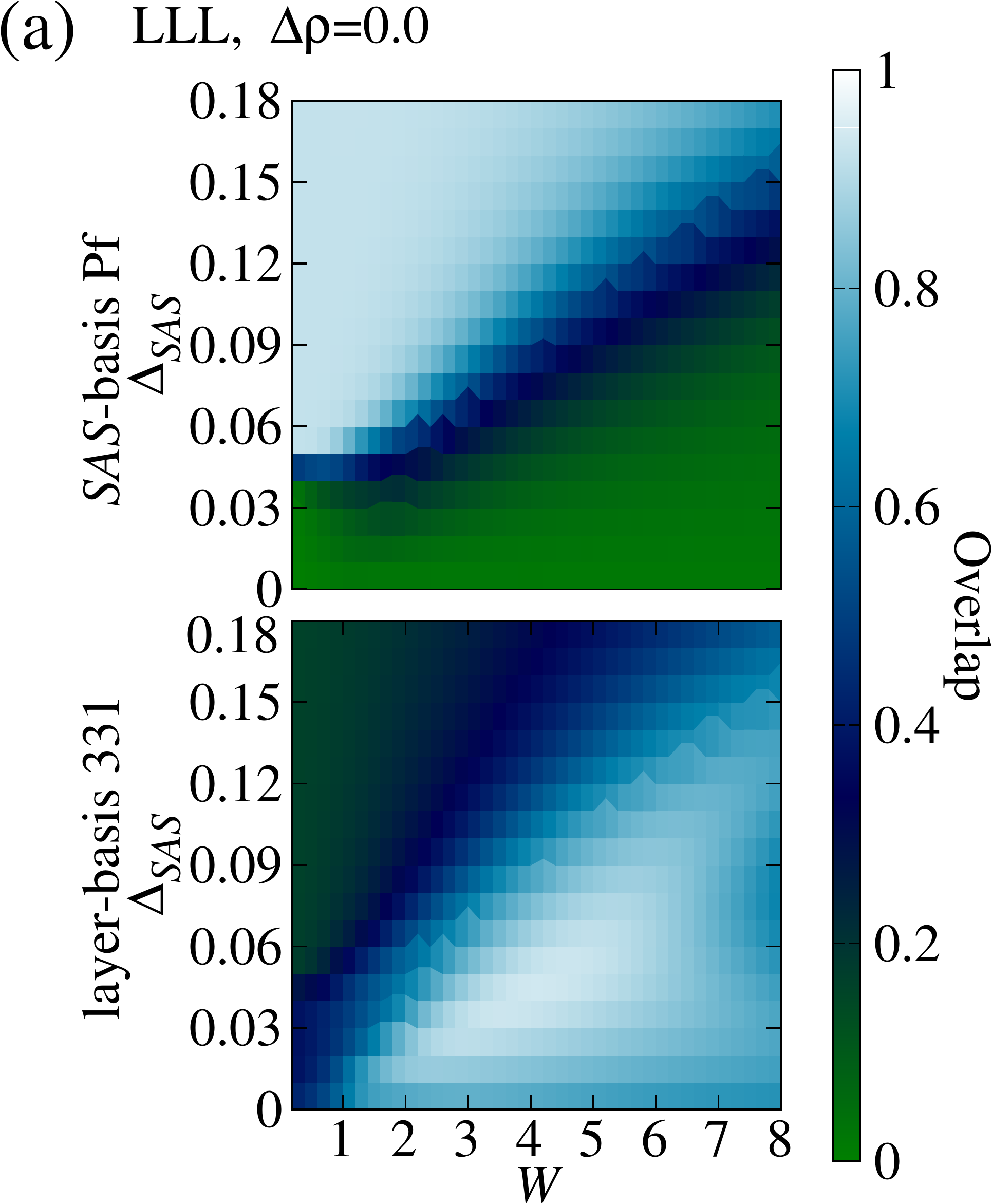}
\includegraphics[width=3.5cm,angle=0]{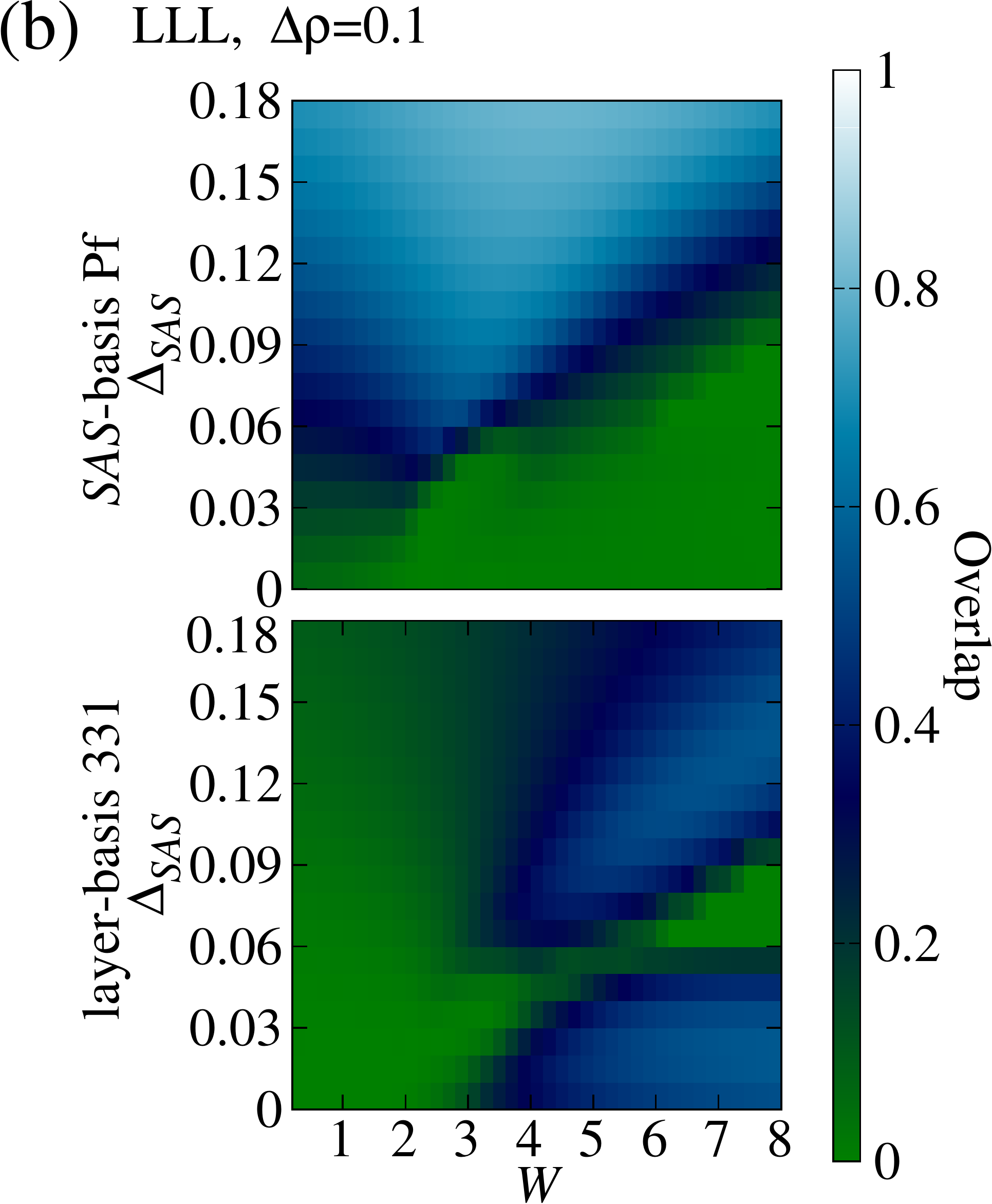}\\
\includegraphics[width=3.5cm,angle=0]{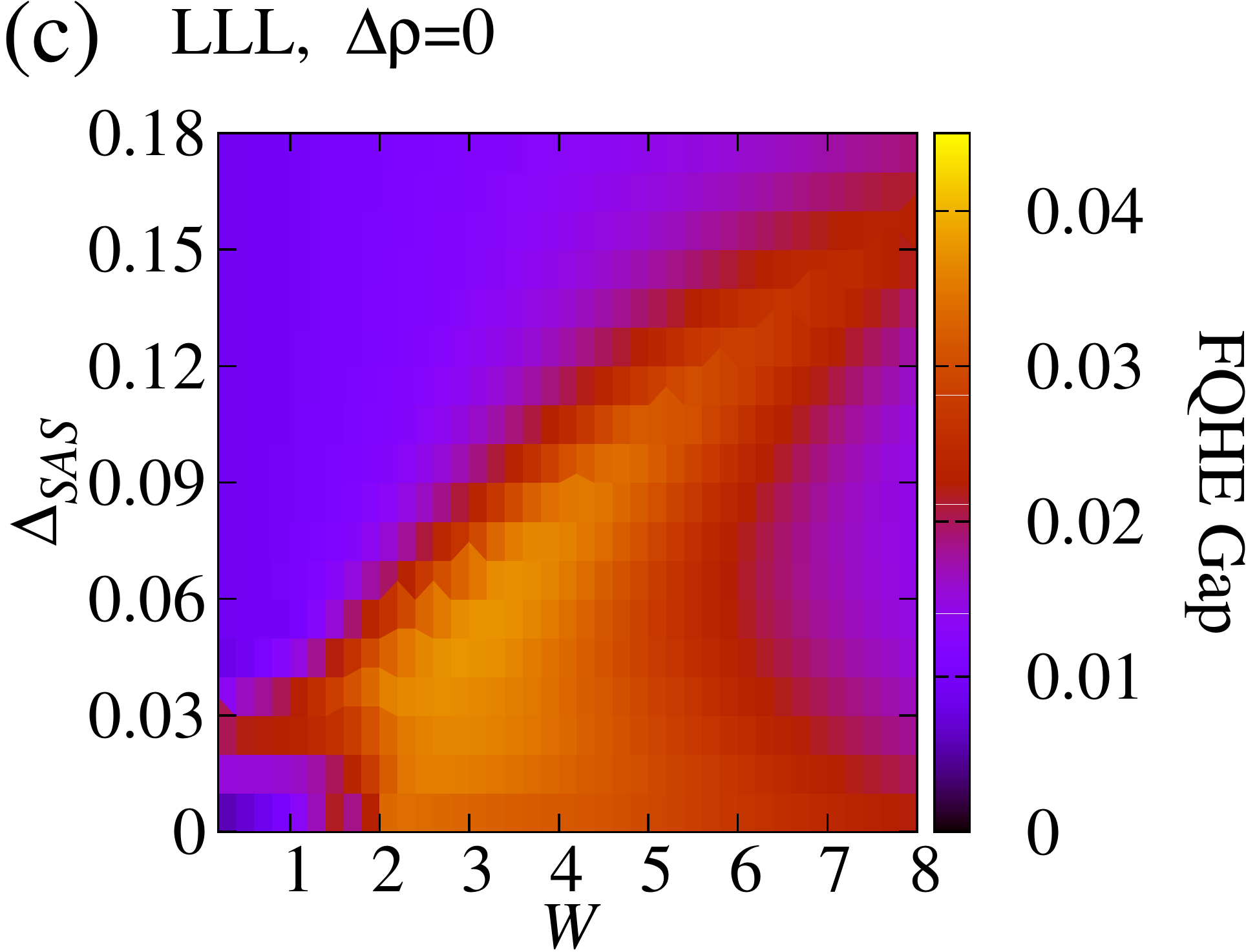}
\includegraphics[width=3.5cm,angle=0]{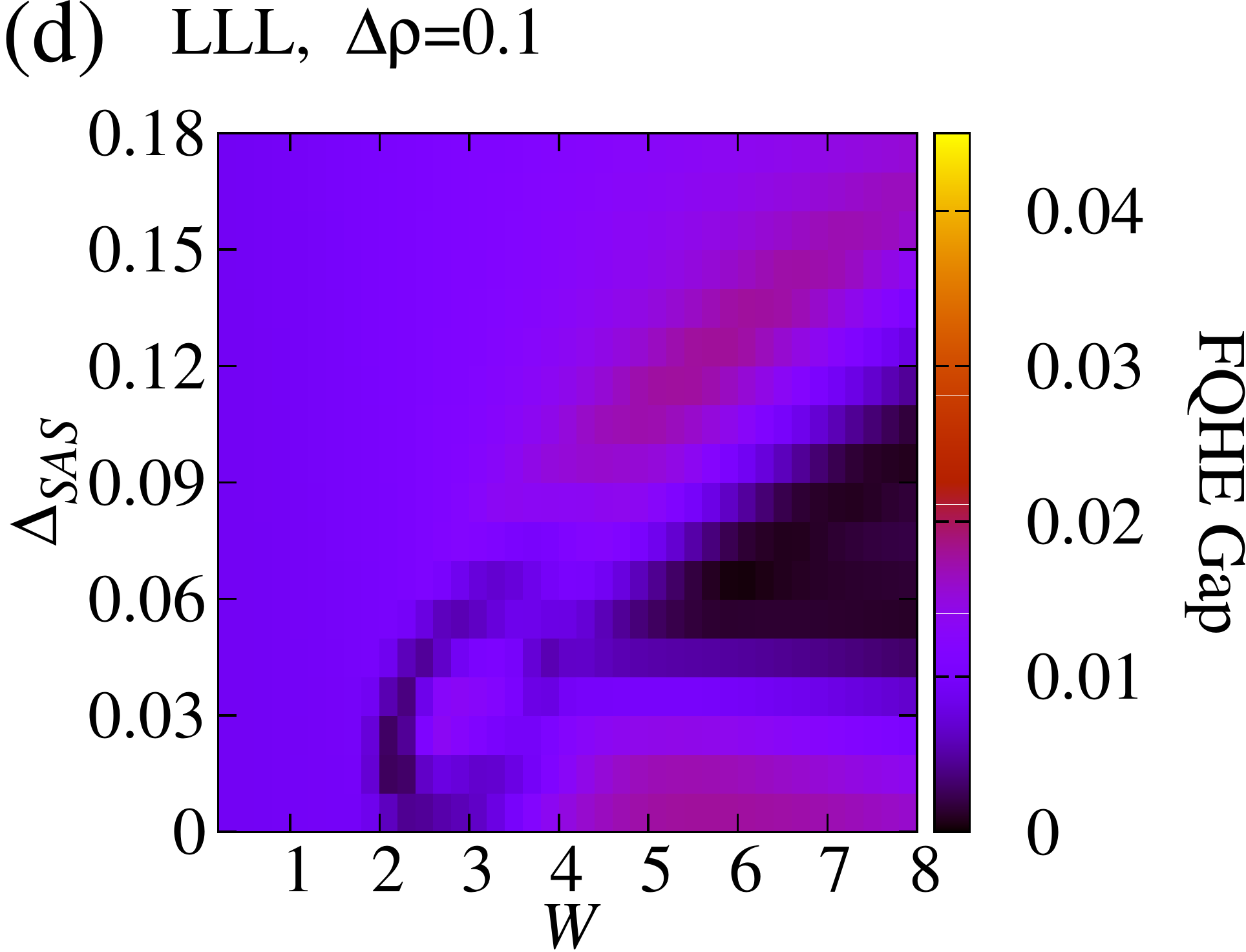}
\end{tabular}
\caption{(Color online) Lowest Landau level:  FQHE energy gap for the 
WQW model as a function of $W$ and $\Delta_{SAS}$ 
with (c) $\Delta\rho=0$ (left column) and (d) $\Delta\rho=0.1$ (right column).  
Also shown is the overlap between the $SAS$-basis Moore-Read Pfaffian (top panel)  
and the layer-basis Halperin 331 (bottom panel) 
wavefunctions and the exact ground state for (a) $\Delta\rho=0$ and (b) $\Delta\rho=0.1$.}
\label{fig-LLL-gap-Dsas-v-Drho-wqw}
\end{figure}

Note, however, that for $W=0.4$ (Fig.~\ref{fig-LLL-Dsas-v-Drho-Pf-wqw}a) 
the system is not as symmetric upon  $\Delta_{SAS}\leftrightarrow
\Delta\rho$ as it was for the $d=0$ bilayer model case where 
the model was actually SU(2) symmetric.  This is because the WQW model 
is not SU(2) symmetric for small $W$ because even as $W$ becomes 
very small the Coulomb energy between electrons in the $S$ and 
$AS$ states is never equal, i.e., the $AS$ state always has a node and 
therefore higher kinetic energy.  That being said, the models produce 
very similar results.

In Fig.~\ref{fig-LLL-331-Dsas-v-Drho-wqw} we show the overlap 
between the exact ground state of the WQW model and the Halperin 
331 state.  This time, as opposed to Fig.~\ref{fig-LLL-Dsas-v-Drho-331} 
we only show the overlap with the 331 state written in the layer-basis 
since, as we learned previously by studying the bilayer model results, 
and by confirming this with the WQW model, the overlap between 
the ground state of the WQW model and the $SAS$-basis  331 
state is always nearly zero, hence, 
we do not bother to show these results explicitly.  For the 
layer-basis 331, however, we again see qualitatively similar 
behavior compared to the bilayer model.  For small $W$ the overlap 
with the 331 state is very small and it can be increased by 
increasing $W$.  One interesting difference between the results 
of the two models is that for $W=5$ the maximum overlap with 331 has a 
maximum 
for non-zero $\Delta_{SAS}$ for the WQW model.  Furthermore, 
non-zero charge imbalance $\Delta\rho$ eventually destroys 
the Halperin 331 state by driving the system to 
be one-component, however, the 331 state is remarkably 
robust to charge imbalancing.

Similar to the bilayer model, the pseudo-spin expectation 
value for the WQW model in 
Fig.~\ref{fig-LLL-pseudospin-Dsas-v-Drho-wqw} mirrors the 
behavior of the overlaps.  When the system 
is largely one-component, either in the layer- or $SAS$-basis 
(large value of $(N_R-N_L)/2$ or $(N_S-N_{AS})/2$, respectively),
the overlap with the appropriate basis Pf state 
is large.  When the system is two-component the overlap 
with the 331 is large.  In investigating the pseudo-spin expectation value for the WQW model 
we see the most serious discrepancy between the two models--this effect 
was also seen previously in Ref.~\onlinecite{papic}.   For $W\geq 2$, and 
relatively small values of $\Delta_{SAS}$, the ground has 
a negative value of $(N_S - N_{AS})/2$.  This means that the electrons occupy 
the $AS$ state compared to the $S$ state.  
If we recall the overlap between the exact ground state and the 
Moore-Read Pfaffian in the $SAS$-basis (lower panel 
of Fig.~\ref{fig-LLL-Dsas-v-Drho-Pf-wqw})  the overlap is large for 
large $\Delta_{SAS}$ and this behavior mirrors the large value of $(N_S-N_{AS})/2$ 
shown in the lower panel of Fig.~\ref{fig-LLL-pseudospin-Dsas-v-Drho-wqw}.  However, 
the one-component Moore-Read Pfaffian we consider pairs electrons in the $S$ state 
(shown schematically in Fig.~\ref{fig-wfs-cartoon}b).  We have not 
checked the overlap with the one-component Pf written such that 
the pairing occurs between electrons in the $AS$ state, thus, adding another 
version of the Pf.  By examining the value of $(N_S-N_{AS})/2$ for $W\geq 2$ in 
the small $\Delta_{SAS}$ region of phase space we would expect a reasonable 
value of the overlap with a one-component Pf pairing electrons in the $AS$ state.

\begin{figure}[]
\includegraphics[scale=0.5,angle=0]{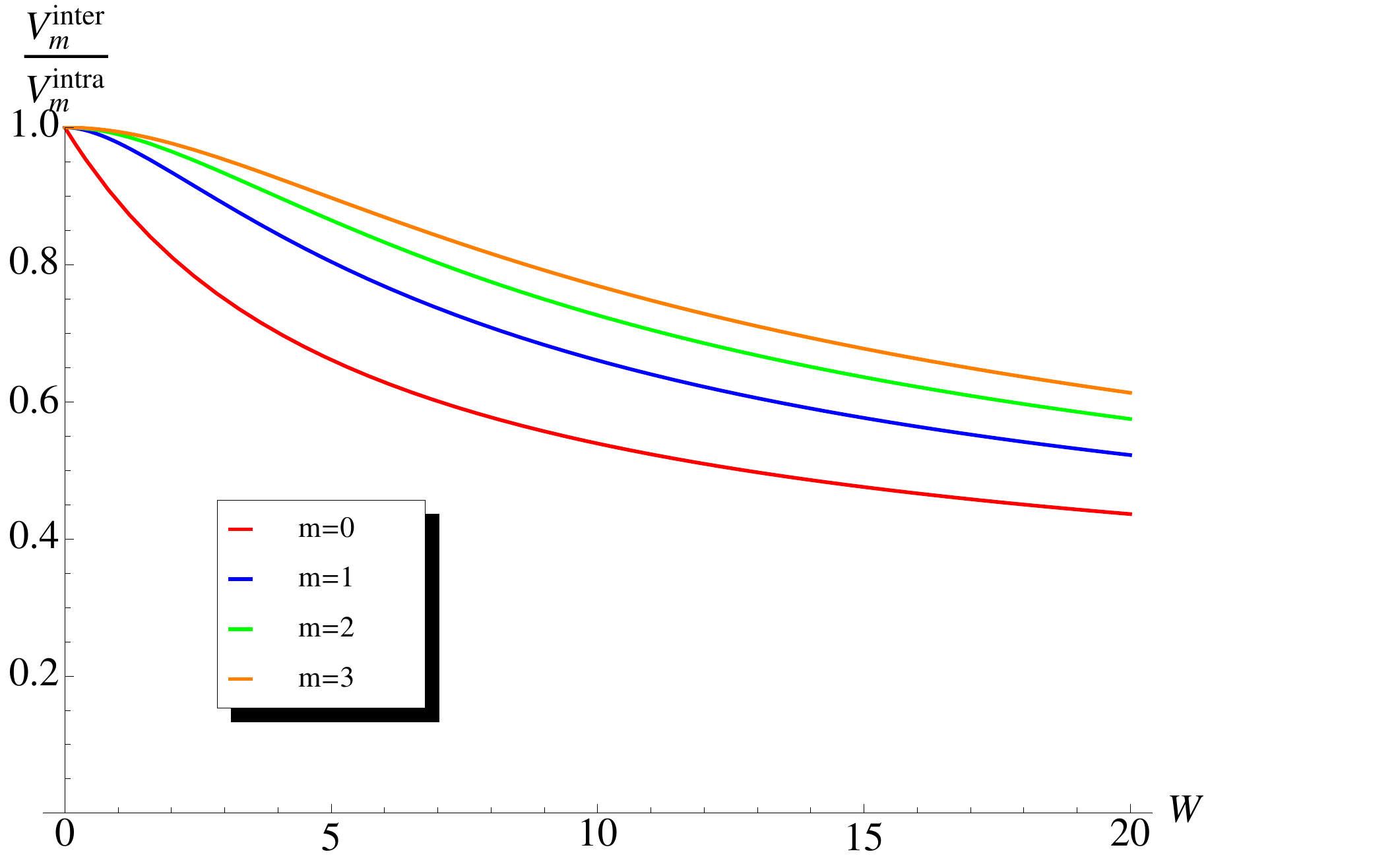}
\caption{(Color online) Ratio of the few strongest, inter-layer 
and intra-layer (bilayer) pseudopotentials 
$V_m^{\rm inter}/V_m^{\rm intra}$ derived using the WQW model~\cite{papic}
as a function of width $W$. Note the saturation of the ratios for large widths $W$,
illustrating the limit of validity of the WQW to describe bilayers.}
\label{fig-wqwpp}
\end{figure}

The WQW model is more general than the usual bilayer model and the latter can be
derived from it (see Section IV-B of Ref.~\onlinecite{papic} for a detailed discussion).  
However, if the well width $W$ becomes very large, it is unjustified to restrict
the model to only the lowest two subbands (and the inclusion of higher subbands cannot be
treated exactly because of the computational complexity). On the other hand, as we map the
WQW Hamiltonian to an effective bilayer, we find that the effective
bilayer distance $\tilde{d}$ saturates for large $W$. To see this, we can calculate the Haldane
pseudopotentials $V_m^{\sigma_1\sigma_2\sigma_3\sigma_4}$, 
where $m$ is the relative angular moment
of the two electrons and $\sigma=S,AS$. If we denote 
by $\mathcal{F}^{\sigma_1\sigma_2\sigma_3\sigma_4}(\mathbf{r})$
the effective interaction in the plane, we have
\begin{eqnarray}
\nonumber \mathcal{F}{\sigma_1\sigma_2\sigma_3\sigma_4} (\mathbf{r}) &=& \int_0^W dz_1 
\int_0^W dz_2  
\langle \mathbf{r}|\sigma_1\rangle (z_1)  \langle \mathbf{r}|\sigma_2\rangle (z_2) \\
&&\!\!\!\!\!\!\!\!\!\!
\frac{1}{\sqrt{r^2+|z_1-z_2|^2}}  \langle \mathbf{r}|\sigma_3\rangle(z_1)  \langle \mathbf{r}|\sigma_4\rangle(z_2)\;.
\end{eqnarray}
The Haldane pseudopotentials for
the effective interaction, written on the disk for simplicity (for the lowest Landau level), are
$$
V_m^{\sigma_1\sigma_2\sigma_3\sigma_4} = \int \frac{d^2\mathbf{k}}{(2\pi)^2} e^{-k^2} \mathcal
{L}_m(k^2) \mathcal{F}_\mathbf{k}^{\sigma_1\sigma_2\sigma_3\sigma_4},
$$
where $\mathcal{L}_m$ is the Laguerre polynomial and the Fourier 
transform $\mathcal{F}_\mathbf{k}^{\sigma_1\sigma_2\sigma_3\sigma_4}$ can be evaluated 
analytically. Furthermore, we can construct linear combinations~\cite{papic} 
of $V_m^{\sigma_1\sigma_2\sigma_3\sigma_4}$ to get the effective bilayer 
pseudopotentials, $V_m^{\rm intra}$ and $V_m^{\rm inter}$. We plot the ratios of the few 
strongest $V_m^{\rm intra}$ and $V_m^{\rm inter}$ in Fig.~\ref{fig-wqwpp} as a function 
of $W$.   Notice that the limits saturate for 
large $W$, indicating that the inter-layer repulsion decreases very 
slowly with respect to intra-repulsion for larger $W$, thus suggesting that the model becomes 
unrealistic in this regime. Also that the pseudopotentials involving a node in the $z$-wave 
function (the $AS$ single particle wavefunciton has one node) 
can be smaller in absolute value than those without a node, indicating a possibility for 
some $W$ to have a depopulation of the lowest subband and negative 
polarization $(N_S-N_{AS})$, as seen in the data.

Fig.~\ref{fig-LLL-gap-Dsas-v-Drho-wqw} shows the FQHE energy gap 
for the WQW model for $\Delta\rho=0$ and $\Delta\rho=0.1$, respectively.  For 
$\Delta\rho=0$, we see similar behavior to the results of the bilayer model and 
shown previously~\cite{mrp-sds-bilayer}, i.e., Fig.~\ref{fig-LLL-stacked-Dsas-v-d}d for 
$\Delta\rho=0$.  Of course, there are quantitative differences between the 
two models but the FQHE energy gap 
still shows a prominent ``ridge" as a function of $\Delta_{SAS}$ 
and $W$.  The ridge marks the transition line between the 331 and the Pfaffian, or
perhaps compressible states.
Also, note that the maximum FQHE gap (the ``ridge") is slightly 
on the Halperin 331 side of the quantum phase diagram if the phase boundary is taken 
to be the place at which the 331 overlap is larger than the Pfaffian overlap, i.e., the 331 phase 
is the region where the Halperin 331 overlap is larger than the Pfaffian and vice versa. 
cf. Figs.~\ref{fig-LLL-gap-Dsas-v-Drho-wqw}a and b.
When the charge imbalancing term is increased to $\Delta\rho=0.1$ the FQHE 
gap is markedly reduced, as it was in the bilayer model.  Again, as for $\Delta\rho=0$ 
the maximum gap is in the Halperin 331 part of the phase diagram.  Interestingly, 
there appear to be two ``ridges" forming for the $\Delta\rho=0.1$ situation in 
the FQHE gap and both ``ridge" maxima are mirrored  peaks in the 
331 overlap--although, note that the overlaps (both 331 and Pf) 
for the $\Delta\rho=0.1$ situation are never very large and in a real experimental 
system the phase would most likely have been taken over by some other 
competing phase by the time the charge imbalancing strength reaches $\Delta\rho=0.1$, 
namely a striped phase or, perhaps, a (Composite Fermion) Fermi sea.

\subsubsection{Torus geometry}

\begin{figure*}[t]
\begin{center}
\includegraphics[width=6.5cm,angle=0]{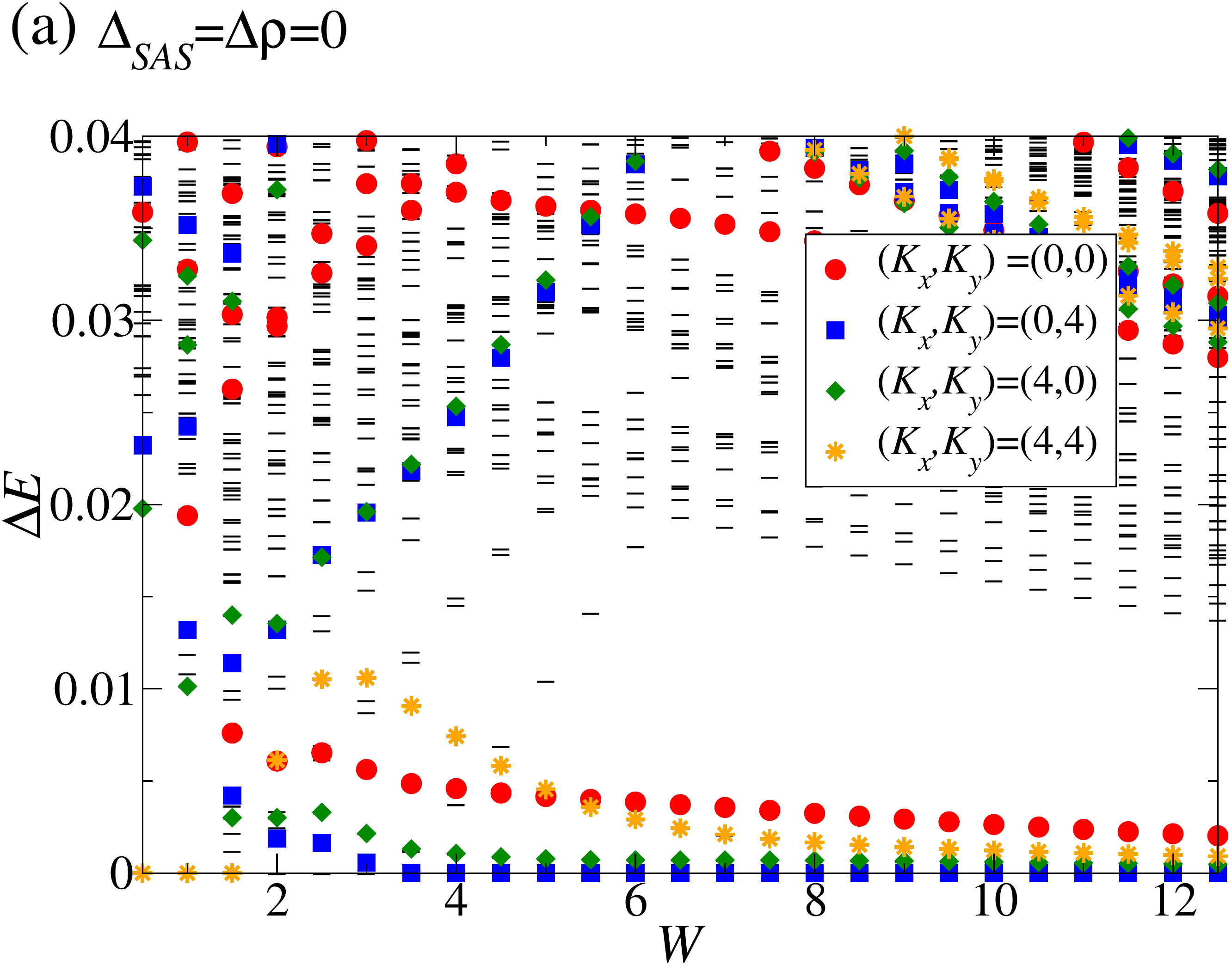}
\includegraphics[width=6.75cm,angle=0]{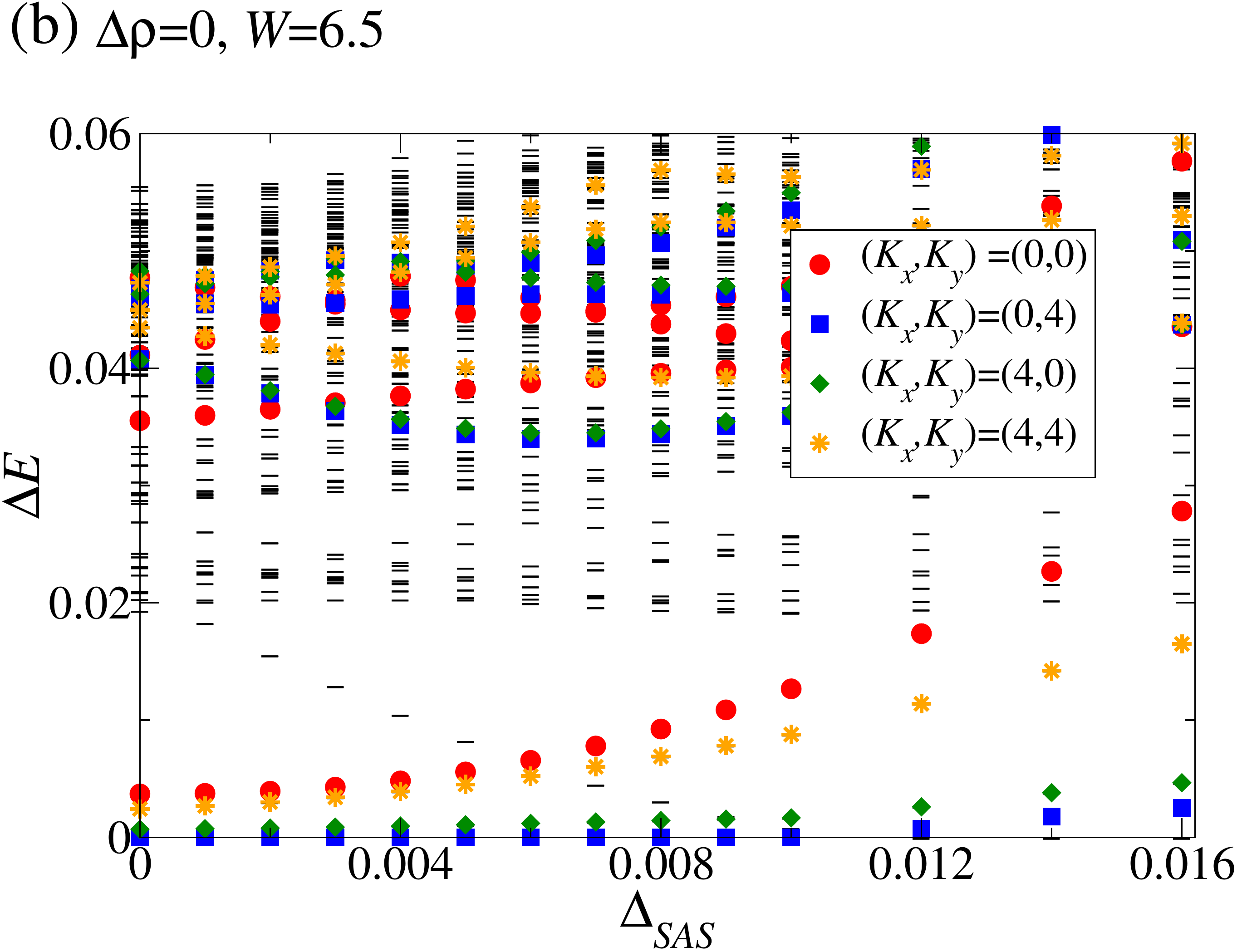}
\includegraphics[width=6.75cm,angle=0]{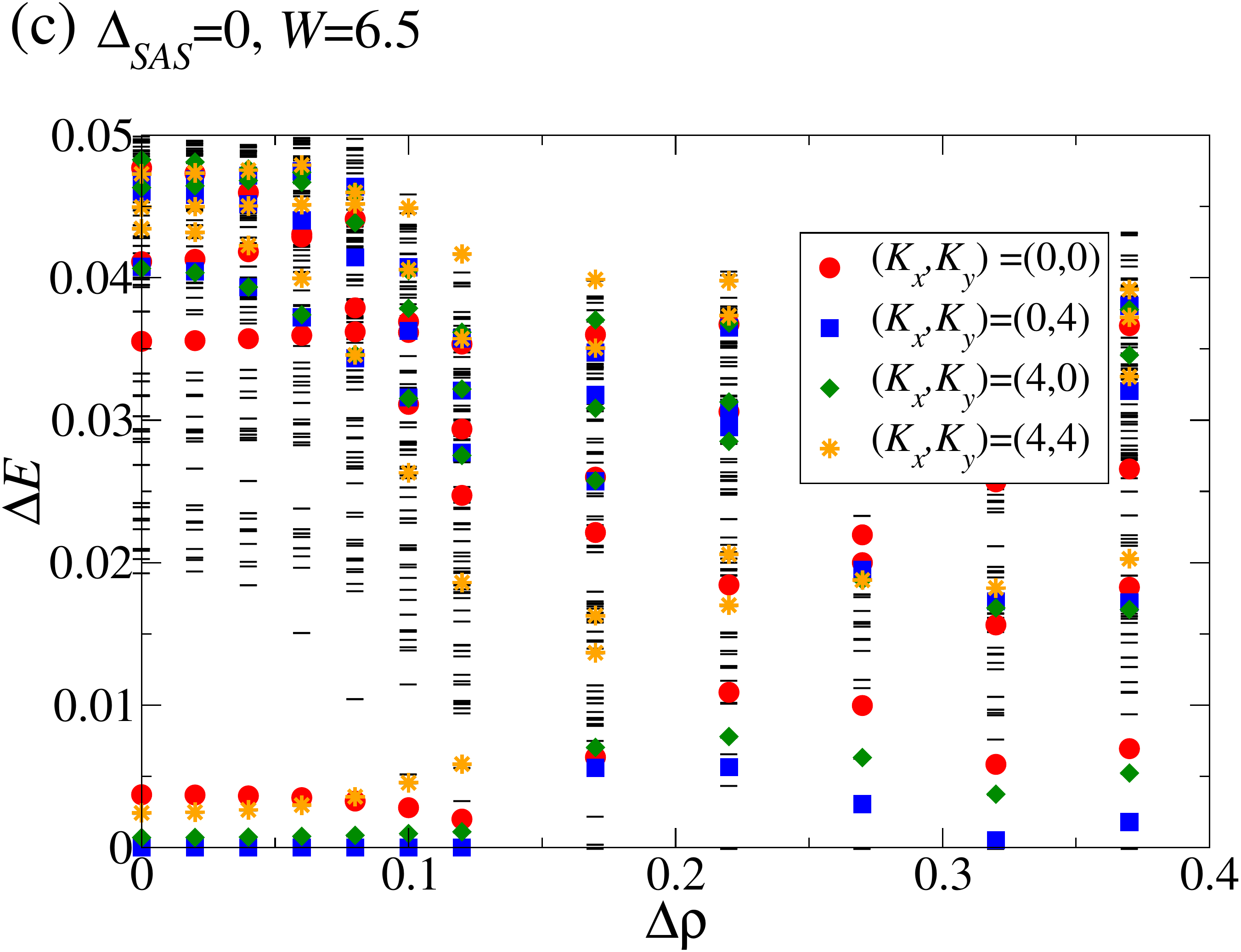}
\caption{(Color online) Lowest Landau level:  
Energy spectra (given  as the energy $\Delta E$ with respect to the 
ground state) for different pseudo-momenta (the different levels) with 
(a) $\Delta_{SAS}=\Delta\rho=0$ as a function of WQW width $W$,  
(b) $W = 6.5$ and $\Delta\rho=0$ as a function of $\Delta_{SAS}$ and 
 (c) $W = 6.5$ and $\Delta_{SAS}=0$ when
the imbalance $\Delta\rho$ is increased from zero.  The pseudo-momenta 
 corresponding to the Halperin 331 state and Moore-Read Pfaffian 
 state ($(K_x,K_y)=(0,0)$ (solid red circle), (0,4) (solid blue square), (4,0) 
 (solid green diamond) and (4,4) (orange star) 
 and $(K_x,K_y)=(0,4)$, (4,0) and (4,4), respectively.  The black dashes
 correspond to states with pseudo-momenta not belonging to 
 either the 331 or Pf states.}
\label{fig-LLL-torus-wqw}
\end{center}
\end{figure*}

In this section we consider the wide-quantum-well model in the torus 
geometry~\cite{yhl,yoshioka-1,haldane-torus} which has one distinct 
advantage to the spherical geometry (of course, 
at the cost of other disadvantages).  Recall in the spherical geometry, the filling factor is 
defined as $\lim_{N\rightarrow\infty}N/N_\phi$ and the relationship between $N$ and 
$N_\phi$ for the Moore-Read Pfaffian and Halperin 331 is $N_\phi=2N-3$ where 
the ``-3" is a consequence of the curvature of a spherical surface.  Other competing 
phases for a half-filled Landau level, such as a (Composite Fermion) Fermi sea, have 
a different shift and comparing different states on an equal footing requires great 
care and extrapolation to the thermodynamic limit.    
However, on the torus, the filling factor is uniquely defined as 
simply $N/N_\phi$ and, therefore, different states can be directly compared 
for finite size systems--of course, this does not mean a conclusion made for a finite 
sized system will maintain as the thermodynamic limit is approached.  (We note that it 
makes no sense to consider overlaps extrapolated to the thermodynamic limit since 
they trivially vanish.)

The torus geometry is generally defined using a domain with sides of 
 length $a$ and $b$ with $a\neq b$.  
The aspect ratio of the toroidal system is $\tau=a/b$.  The magnetic field does not allow 
the use of the usual translation operators but many-body states can be 
found~\cite{haldane-torus} using 
the so-called magnetic translation operators with conserved pseudo-momenta $(K_x,K_y)$.  
The pseudo-momenta belong to a Brillouin zone with $(K_x=2\pi s/a, K_y=2\pi t/b)$ 
with $s,t=0,\ldots,N_0-1$ with $N_0$ being the greatest common divider of $N$ 
and $N_\phi$.  

Different  FQHE states on the torus can be specified by their 
ground state (topological) degeneracy.  Generically, for a Landau 
level filling factor of $p/q$ there is always a center-of-mass degeneracy equal to $q$ 
which is invariant to the form of the Hamiltonian and, thus, of no physical significance--we 
shall ignore it.  Also there can be additional degeneracies that occur at 
special points in the Brillouin zone, such as at certain points in a hexagonal Brillouin 
zone, and we will consider these trivial.  Finally, there can be degeneracies that 
are related to the specific topological nature of certain ground states and, 
therefore, non-trivial.    For 
the two-component (Abelian) 
Halperin 331 state~\cite{wz} we expect a quadruplet of states (up to the center of mass 
degeneracy, which 
in our case is 2) one of which belongs in the $(K_x,K_y)=(0,0)$ sector and the remaining 
three are at the Brillouin zone corners $(K_x,K_y)=(0,N_0/2)$, $(N_0/2,0)$, and 
$(N_0/2,N_0/2)$.  The non-Abelian Moore-Read Pfaffian~\cite{mr-pf,read-rezayi} 
has only a \emph{three-fold} 
degeneracy~\cite{read-green} of $(K_x,K_y)=(0,N_0/2)$, $(N_0/2,0)$, and 
$(N_0/2,N_0/2)$, i.e., the $(K_x,K_y)=(0,0)$ state is missing.  
Compressible states, such as the (Composite Fermion) Fermi sea, generally do not have clearly 
defined degeneracies--they may have accidental degeneracies that are strong functions 
of the aspect ratio $\tau$, particle number $N$, or other Hamiltonian parameters.  The 
\emph{exact} degeneracies are expectations based on the (analytic) form of the 
variational ansatz wavefunctions and their respective conformal field 
theories~\cite{mr-pf, mvm-nr}.  The ground 
state(s) of an actual Hamiltonian will not display exact 
degeneracies~\cite{wang-sheng-haldane, rezayi-haldane,mrp-ft-prl,mrp-ft-prb} (perhaps they 
do in the thermodynamic limit?) but the ground state(s) should qualitatively show the 
ground state topological degeneracy corresponding to the variational ansatz 
if they are to be thought of as being in that ``phase".  Many 
states are sensitive to changes in $\tau$ while others are not.  One 
should really investigate the properties of the system with regard to changes in $\tau$, 
however, in the present case, a somewhat involved analytical calculation regarding the 
``background" charge is needed, and so we focus here on a fixed aspect ration of $\tau=0.97$.

All results on the torus correspond to a system of $N=8$ electrons in the 
half-filled lowest Landau level--$\nu=1/2$. 
Fig.~\ref{fig-LLL-torus-wqw}a shows the energy spectrum of the low-lying states for the WQW 
model in the absence of any tunneling terms, i.e., $\Delta_{SAS}=\Delta\rho=0$, as 
a function of the WQW width $W$.  For each $W$, the lowest-lying energies 
are plotted relative to the ground state and shape- and color-coded to emphasize which 
states belong to which pseudo-momenta sectors expected when considering the 
two-component Halperin 331 and one-component Moore-Read Pfaffian wavefunctions.  Namely, 
we track the pseudo-momenta sectors $(K_x,K_y)=(0,4)$, $(4,0)$ and $(4,4)$ for the 
Pf with the addition of $(K_x,K_y)=(0,0)$ for the 331 state.  
For large widths $W>4.5$, the spectrum is characterized by a large energy gap separating 
a high-energy (quasi-)continuum of states from a low-energy, nearly degenerate 
manifold of states at the pseudo-momenta corresponding to the 331 state.  As 
the width is decreased towards zero the $(K_x,K_y)=(0,0)$ state that specifies the 
Halperin 331 state from the Moore-Read Pfaffian state goes up in energy joining the 
continuum.  However, the $(K_x,K_y)=(0,4)$ and $(4,0)$ states of the Moore-Read Pfaffian 
also rise in energy into the continuum while at the same time a few states from the 
continuum drop down in energy mixing with states belonging to the 331 or Pf states.  
Finally, at very small $W<1$ there is a single ground state of $(K_x,K_y)=(4,4)$ and a 
large energy gap.  However, there is no three-fold degeneracy characteristic of the 
Moore-Read Pfaffian state present.

In Fig.~\ref{fig-LLL-torus-wqw}b we fix $W=6.5$ and $\Delta\rho=0$ and vary 
the inter-layer tunneling strength $\Delta_{SAS}$.  For small $\Delta_{SAS}=0$ we 
are clearly in the Halperin 331 part of the phase diagram since the spectra has 
a quasi-four-fold degeneracy characteristic of the 331 phase separated from the 
continuum by a large energy gap.  Upon increasing $\Delta_{SAS}$ in an 
attempt to drive the system into the Moore-Read Pfaffian phase we see that 
while the $(K_x,K_y)=(0,0)$ state of the 331 phase rise in energy and joins the 
continuum, the $(K_x,K_y)=(4,4)$ state of the Pfaffian phase goes with it and a 
state from the high-energy continuum drops down into a quasi-three-fold degeneracy.  However, 
this quasi-three-fold degeneracy does not contain the right pseudo-momenta 
to describe the non-Abelian Moore-Read Pfaffian phase.

Lastly, in Fig.~\ref{fig-LLL-torus-wqw}c we again set $W=6.5$ but now fix $\Delta_{SAS}=0$ 
and vary the charge imbalance $\Delta\rho$.  As before, for 
small $\Delta\rho$ we see a clear signature 
of the Halperin 331 phase.  However, as $\Delta\rho$ is increased the energy gap 
essentially collapses giving way to a phase that would most likely not exhibit the 
FQHE.

These conclusions corroborate our previous work using the spherical geometry
and agree
with the previous study~\cite{papic-3} of bilayer model on the torus for $\nu=1/2$.  Namely, 
the Halperin 331 state is a good ansatz for the FQHE in the right parameter regions 
for bilayer and WQW systems.  However, when the system is driven to be one-component 
in the hopes of producing a FQHE described by the Moore-Read Pfaffian state the 
Hamiltonian details and Haldane pseudopotential~\cite{haldane} values are such that the 
Moore-Read Pfaffian phase loses out to a different, most likely, non-FQHE state such 
as a striped phase or (Composite Fermion) Fermi sea.

\section{Results: second Landau level}
\label{sec-SLL}

\begin{figure*}[]
\begin{center}
\includegraphics[width=5.9cm,angle=0]{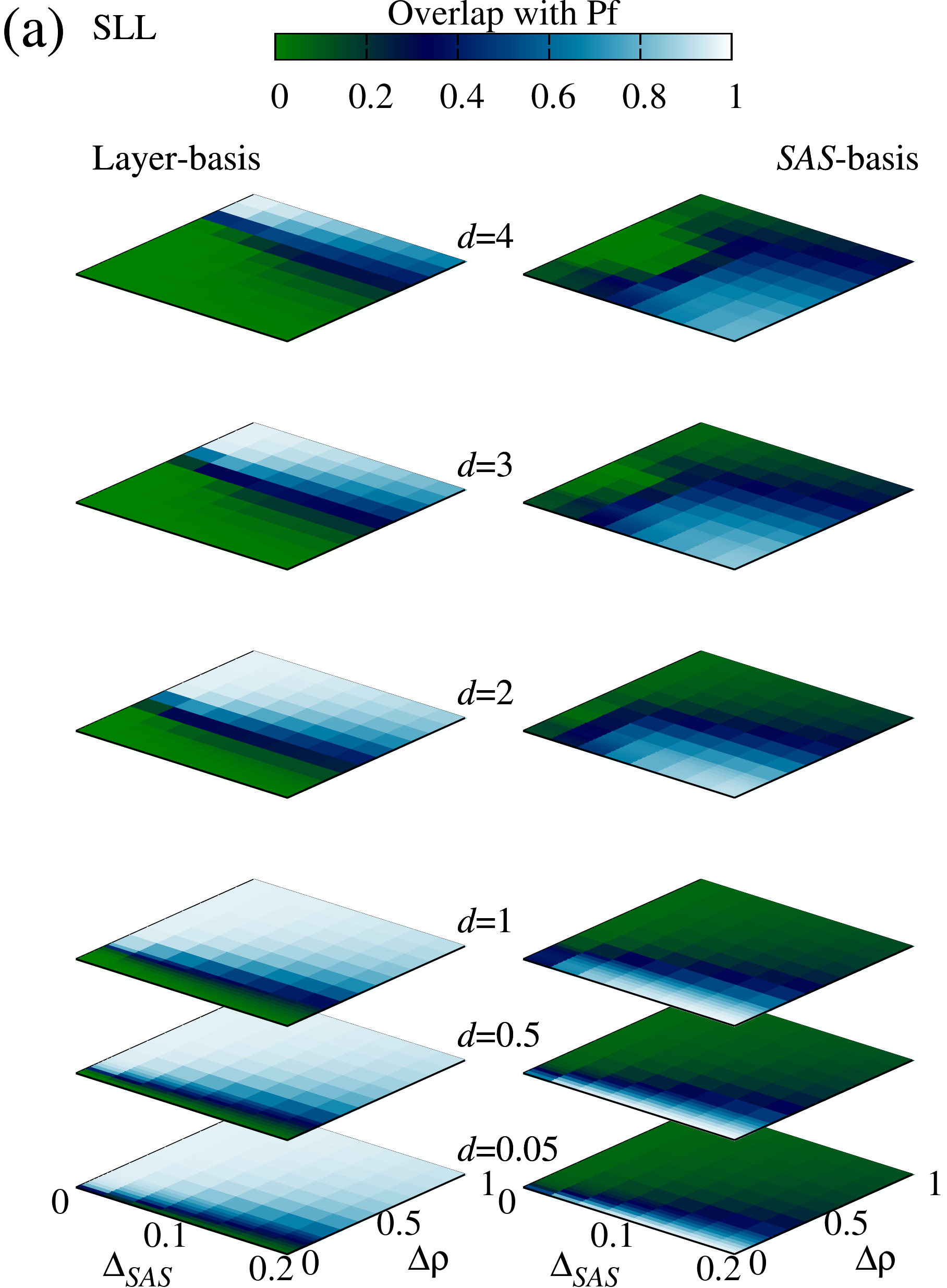}
\includegraphics[width=5.9cm,angle=0]{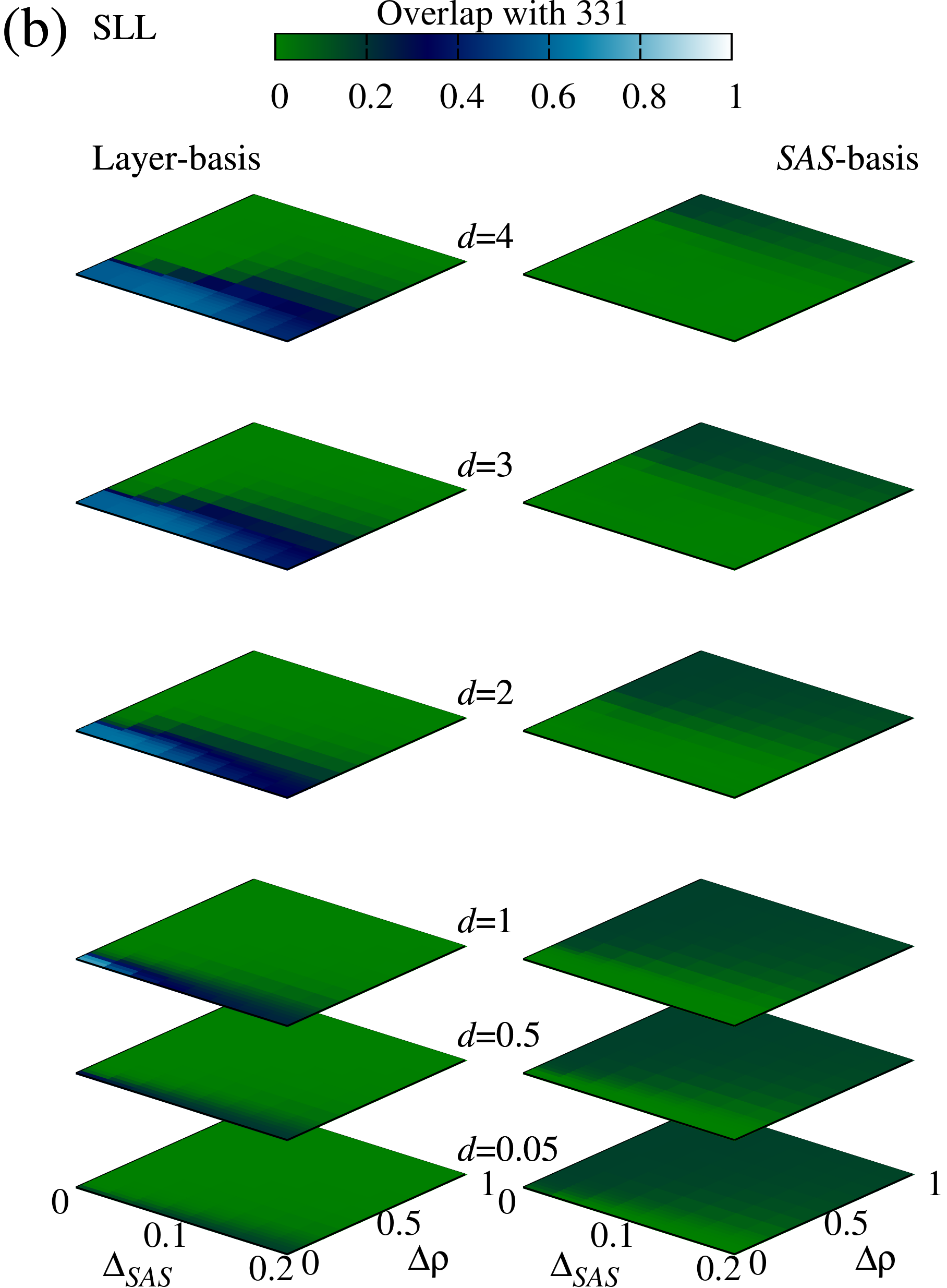}\\
\mbox{}\\
\includegraphics[width=5.9cm,angle=0]{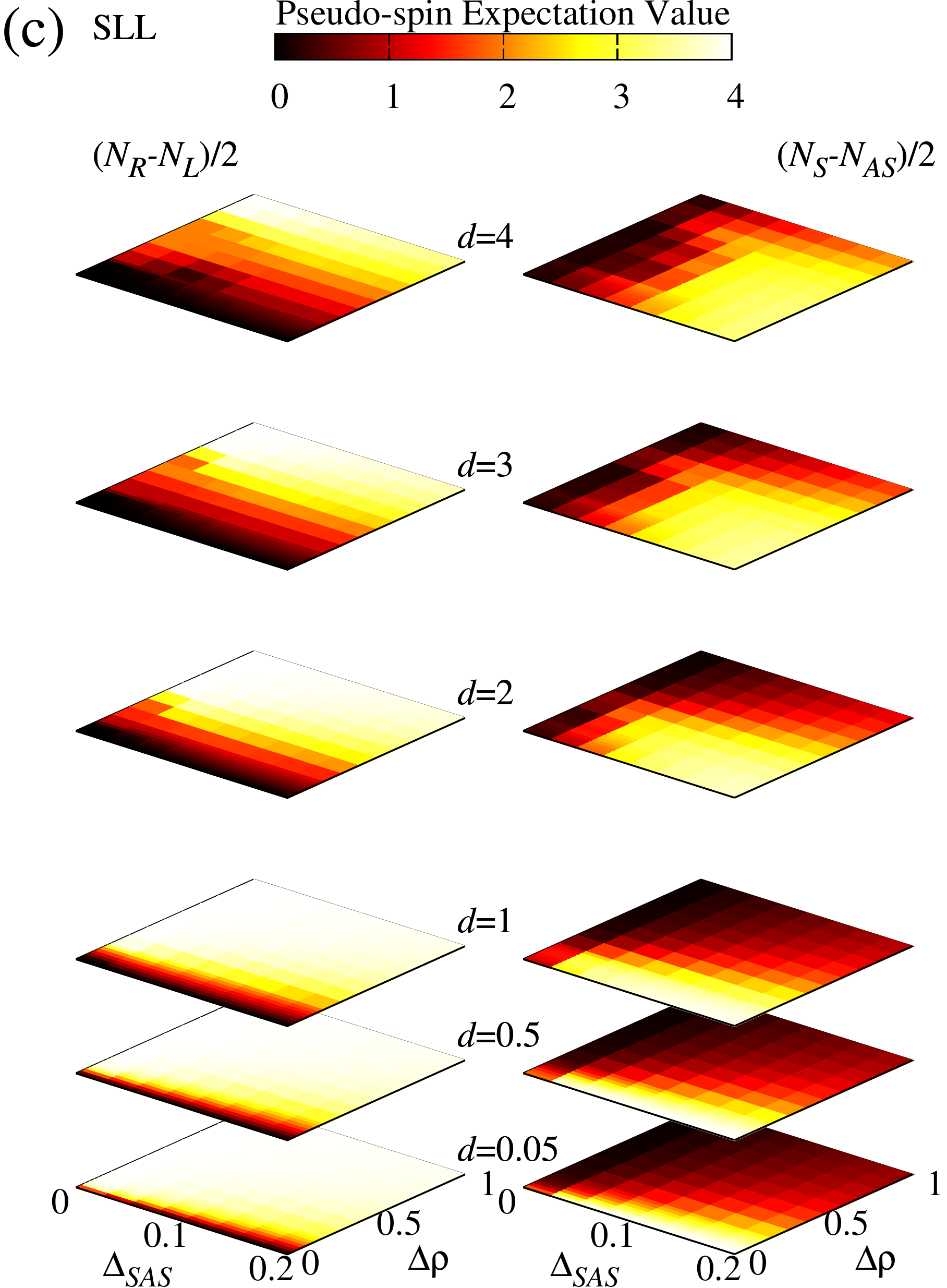}
\includegraphics[width=5.25cm,angle=0]{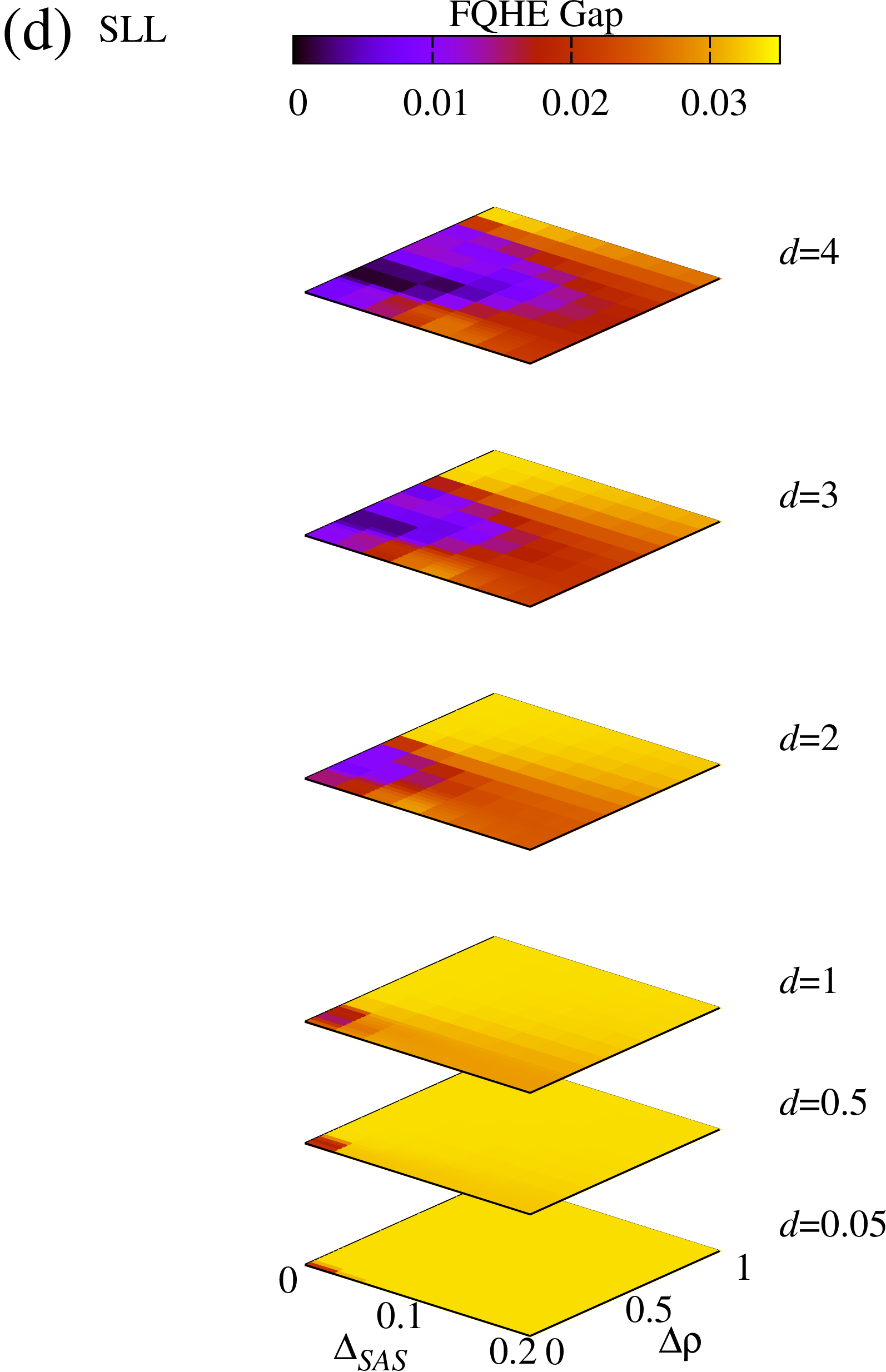}
\caption{(Color online) Second Landau level:  (a) Wavefunction overlap between the exact ground 
state of the bilayer Hamiltonian and the Moore-Read Pfaffian written in the layer-basis (left column) 
and the $SAS$-basis (right column) shown as a function of inter-layer tunneling $\Delta_{SAS}$ 
and charge imbalance $\Delta\rho$ for different values of layer separation $d$ and zero individual 
layer thickness $w=0$.  (b) Same as (a) but for the Halperin 331 wavefunction.  (c) Pseudo-spin 
expectation value or, more physically, the expectation value of $(N_R-N_L)/2$ (left column) and $
(N_S-N_{AS})/2$ (right column) as a function of $\Delta_{SAS}$ and $\Delta\rho$.  (d) The 
FQHE energy gap for the exact bilayer Hamiltonian.}
\label{fig-SLL-stacked-Dsas-v-Drho}
\end{center}
\end{figure*}

In this section we present results of our calculations for the second Landau level, that 
is, bilayer FQHE at $\nu=5/2$.  (Work~\cite{shi} has been done 
considering bilayer FQHE where the total filling factor is $\nu=5/2+5/2=5$ but 
it is unrelated to our work.)  We 
are neglecting Landau level mixing and considering the electrons occupying 
the second Landau level to be spin-polarized.  Operationally, we have projected 
the half-filled electrons in the second Landau level into the lowest Landau level using the 
Haldane pseudopotentials~\cite{haldane}.  These effective pseudopotentials 
take into account the screening of the Coulomb interaction that occurs 
between the electrons in the SLL due to the (taken to be) inert electrons 
in the LLL.  We will not detail the procedure of using Haldane 
pseudopotentials in the FQHE as this procedure has been given in 
many places~\cite{cf-book,haldane}.  Also, we (Peterson and Das 
Sarma~\cite{mrp-sds-bilayer}) 
have recently studied the FQHE in the SLL \emph{without} 
the presence of a charge imbalancing term and will compare our results 
here extensively with the ones given previously.

\begin{figure}
\begin{tabular}{c}
\includegraphics[width=3.5cm,angle=0]{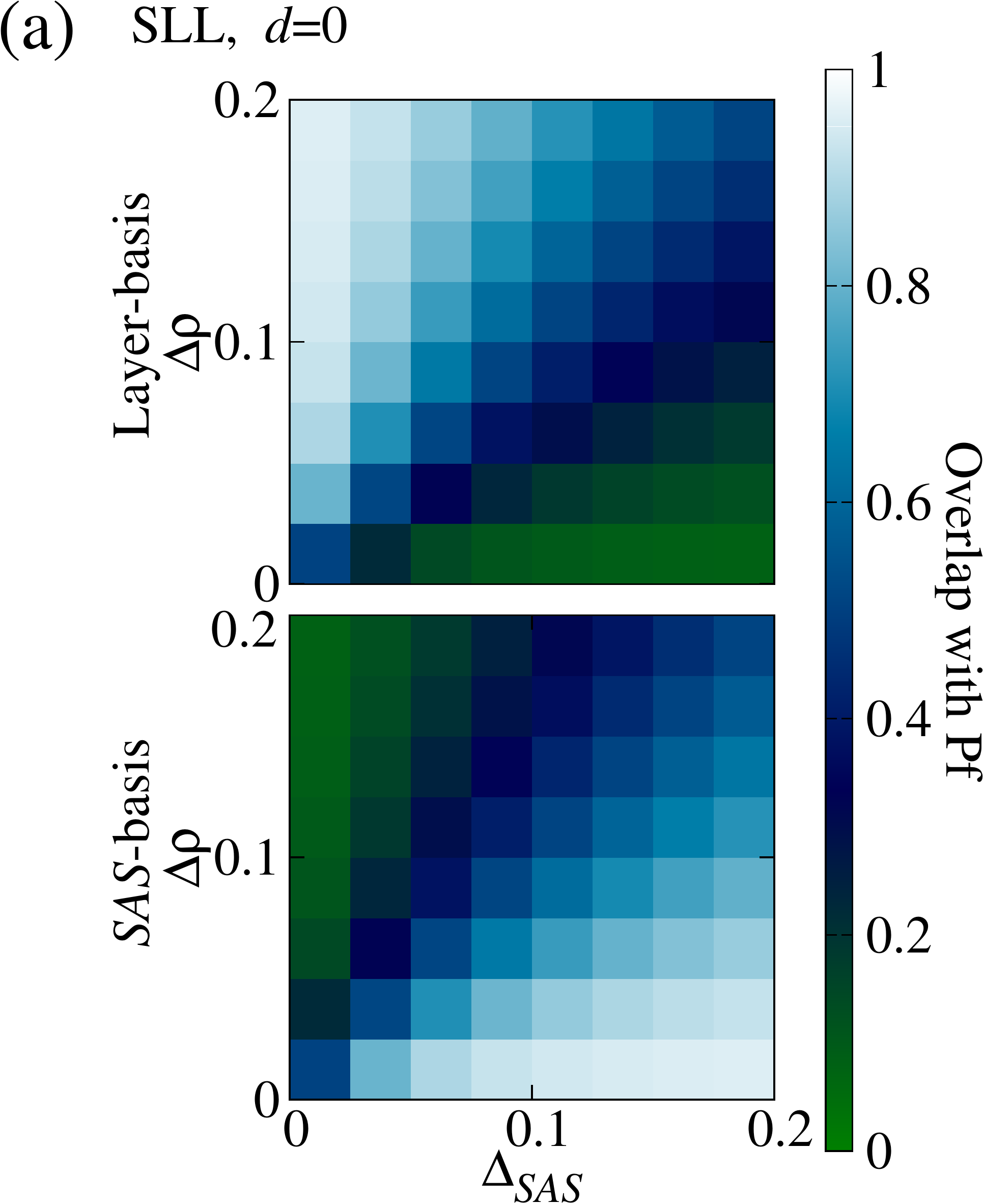} 
\includegraphics[width=3.5cm,angle=0]{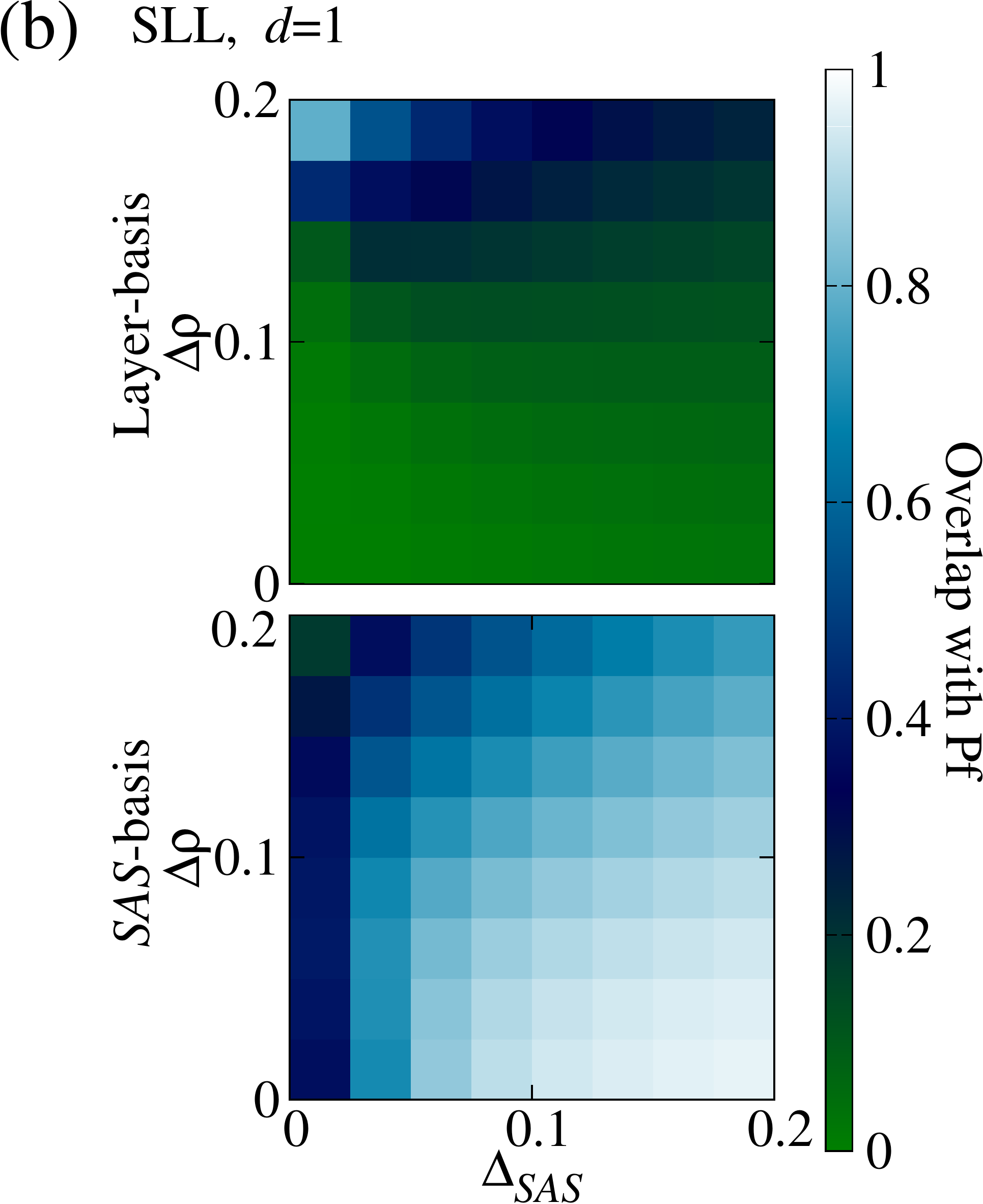} \\
\includegraphics[width=3.5cm,angle=0]{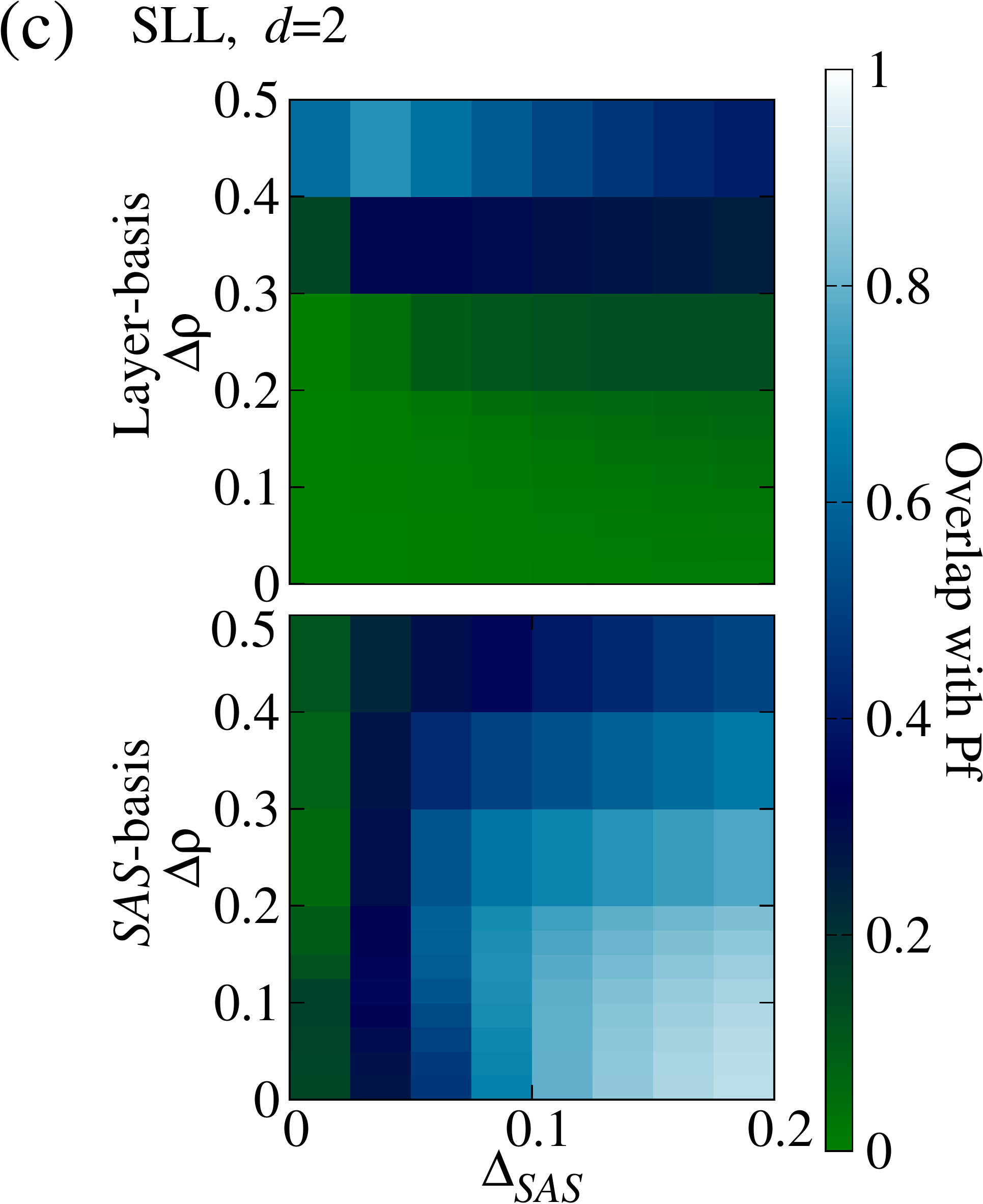}
\includegraphics[width=3.5cm,angle=0]{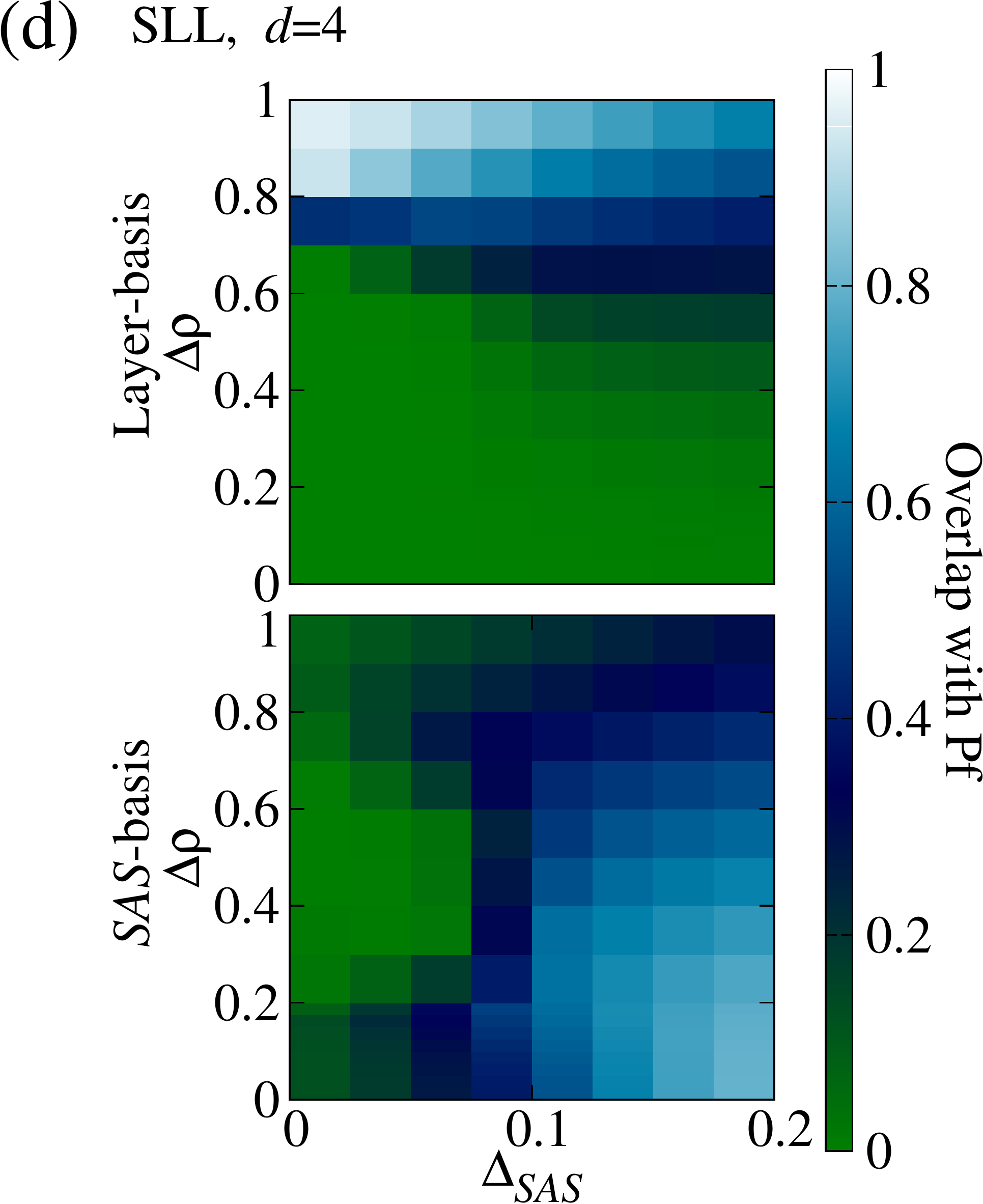}
\end{tabular}
\caption{(Color online) Second Landau level:  Wavefunction overlap between the Moore-Read 
Pfaffian wavefunction in the layer-basis (top panel) and the $SAS$-basis (lower panel) and the 
exact ground state for (a) $d=0$, (b) $d=1$, (c) $d=2$ and (d) $d=4$.}
\label{fig-SLL-Dsas-v-Drho-Pf}
\mbox{}\\
\begin{tabular}{c}
\includegraphics[width=3.5cm,angle=0]{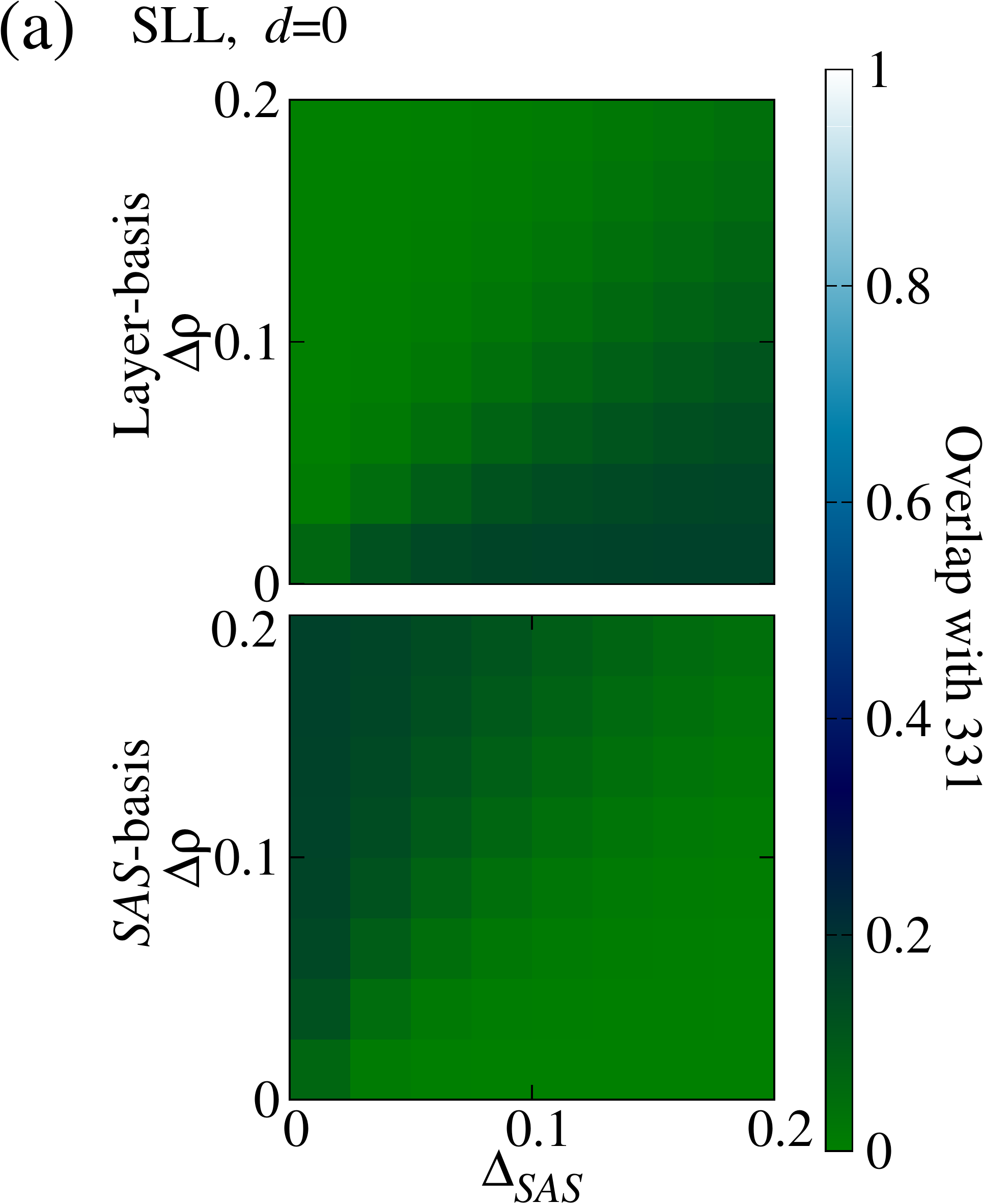} 
\includegraphics[width=3.5cm,angle=0]{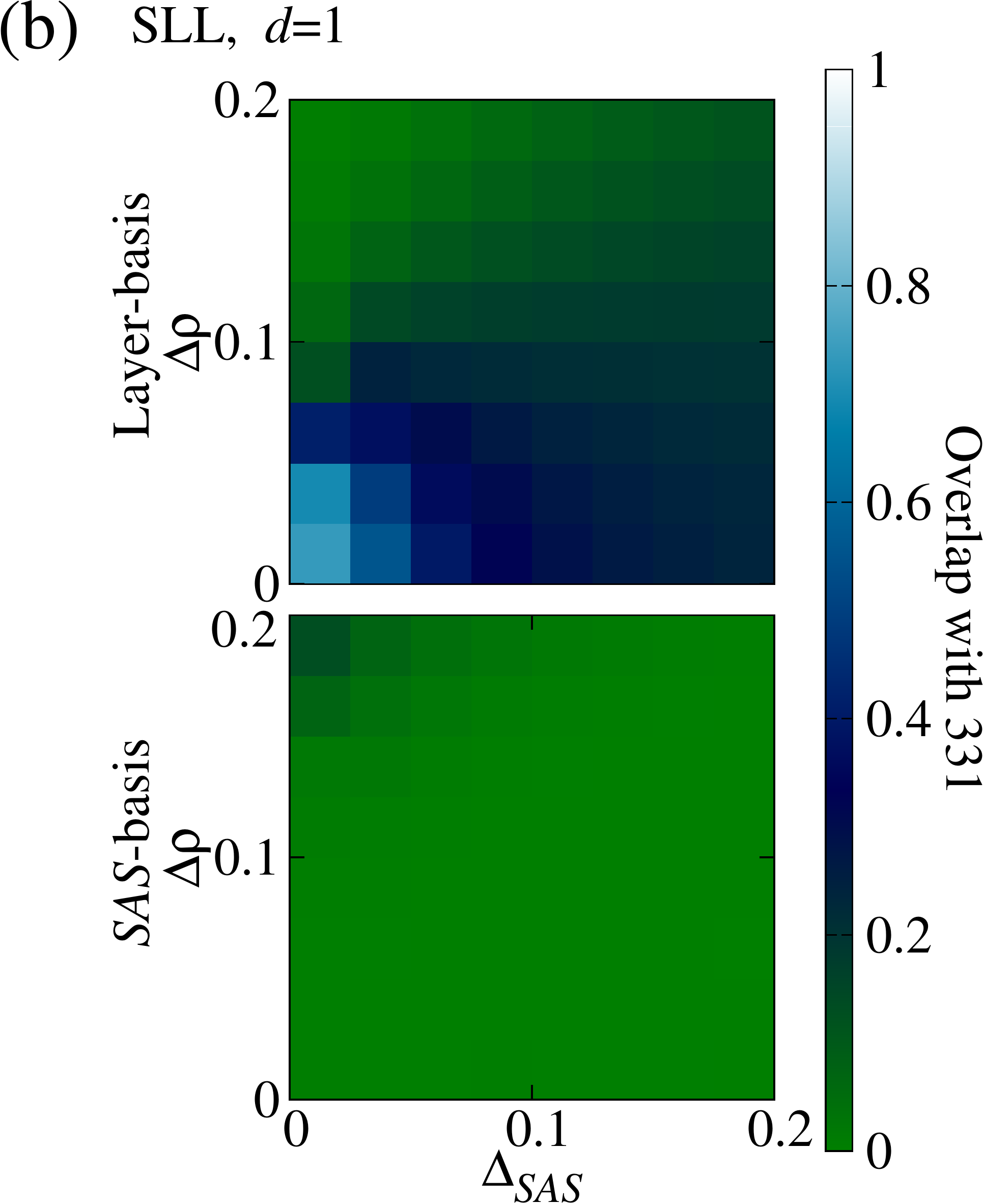} \\
\includegraphics[width=3.5cm,angle=0]{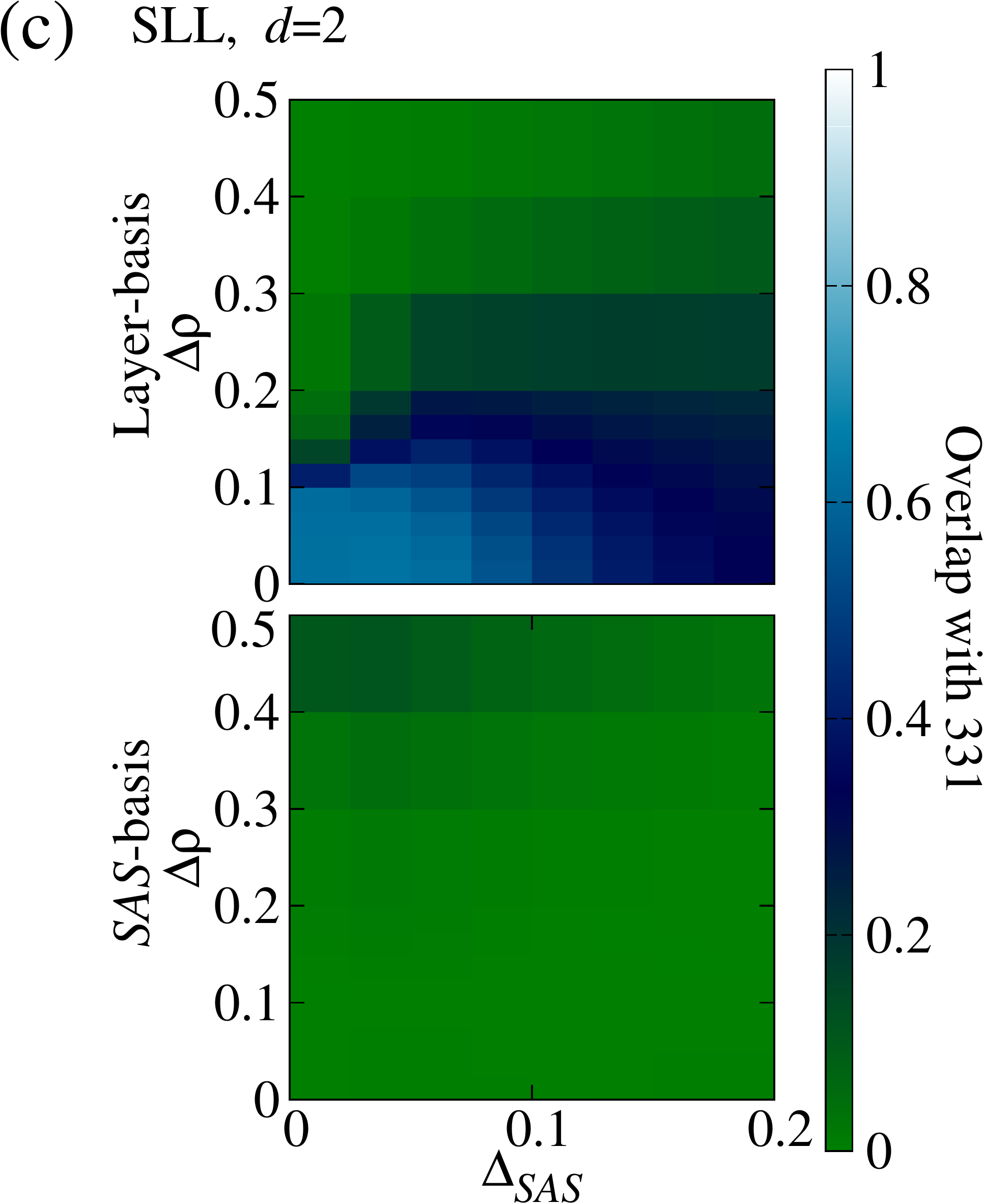}
\includegraphics[width=3.5cm,angle=0]{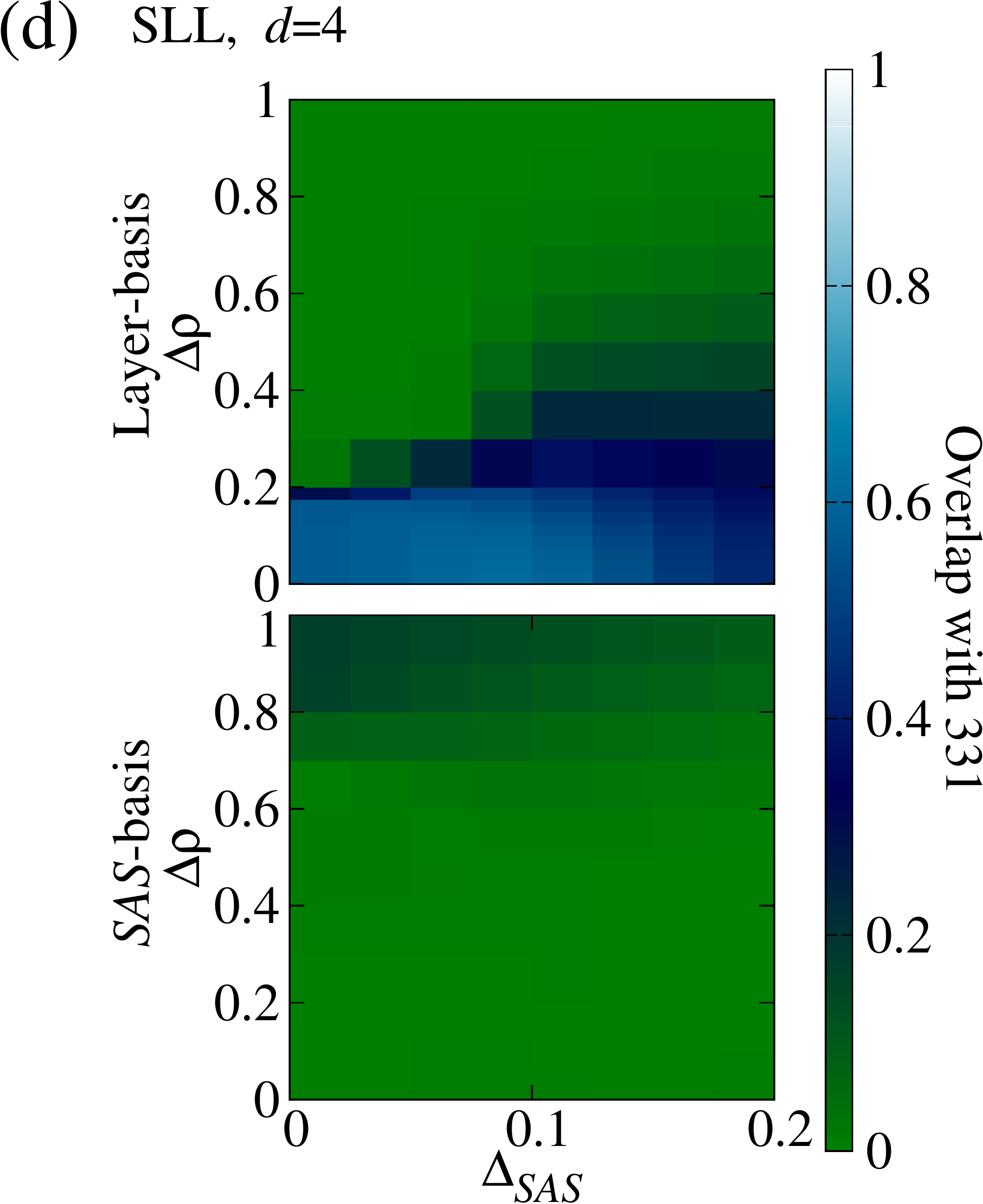}
\end{tabular}
\caption{(Color online) Second Landau level:  Wavefunction overlap between the Halperin 331 
wavefunction in the layer-basis (top panel) and the $SAS$-basis (lower panel) and the exact 
ground state for (a) $d=0$, (b) $d=1$, (c) $d=2$ and (d) $d=4$.}
\label{fig-SLL-Dsas-v-Drho-331}
\end{figure}

\begin{figure}
\begin{tabular}{c}
\includegraphics[width=3.5cm,angle=0]{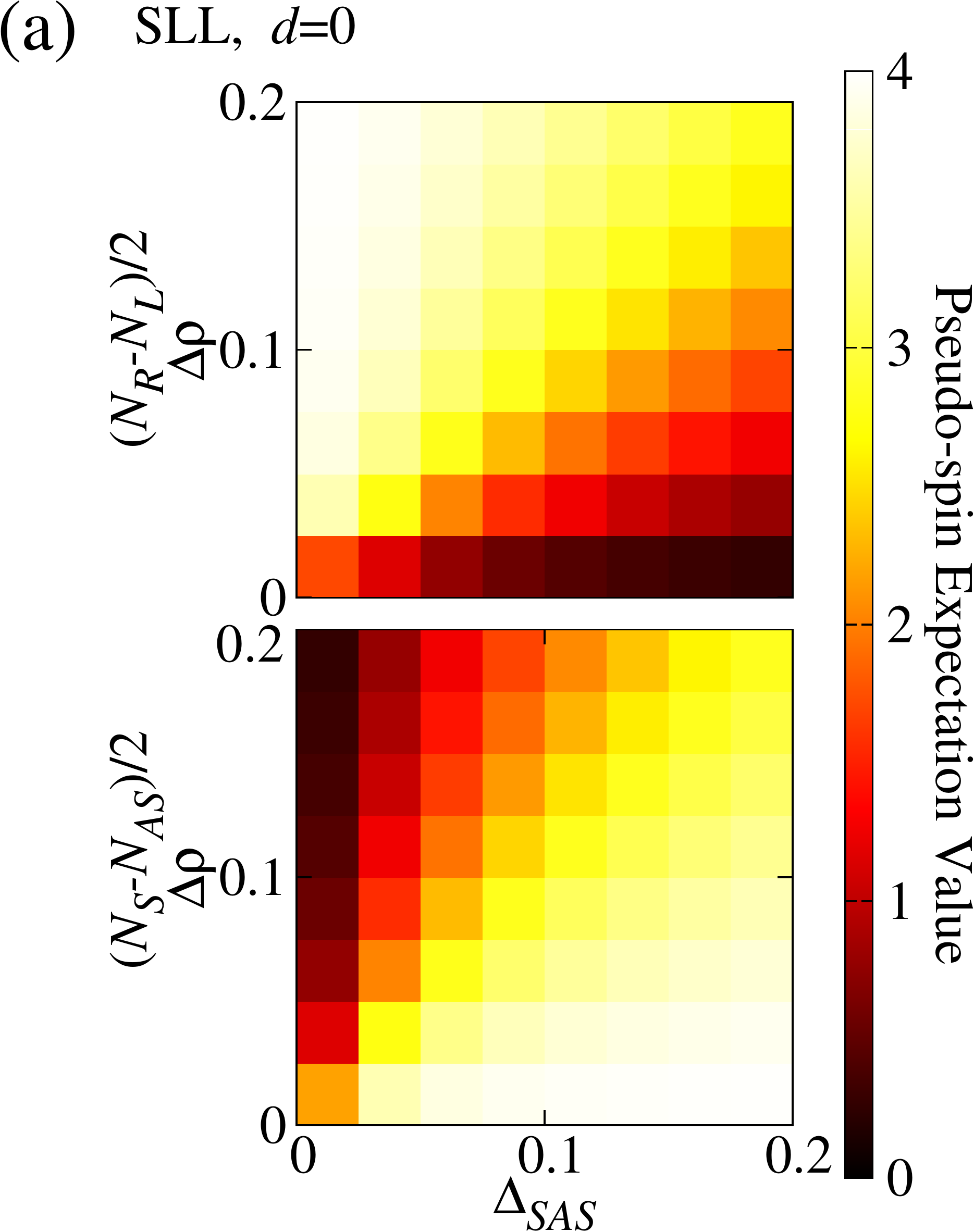}
\includegraphics[width=3.5cm,angle=0]{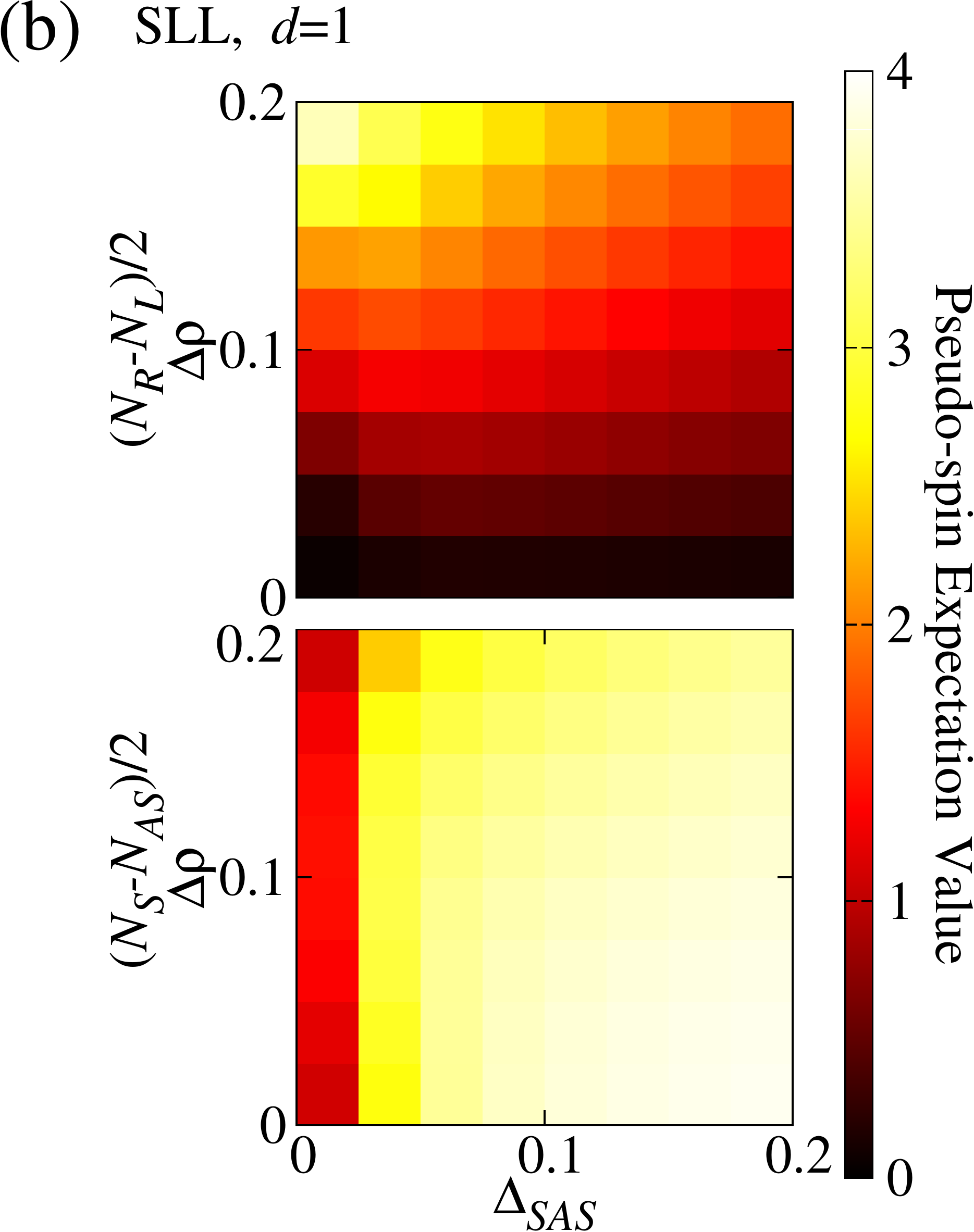}\\
\includegraphics[width=3.5cm,angle=0]{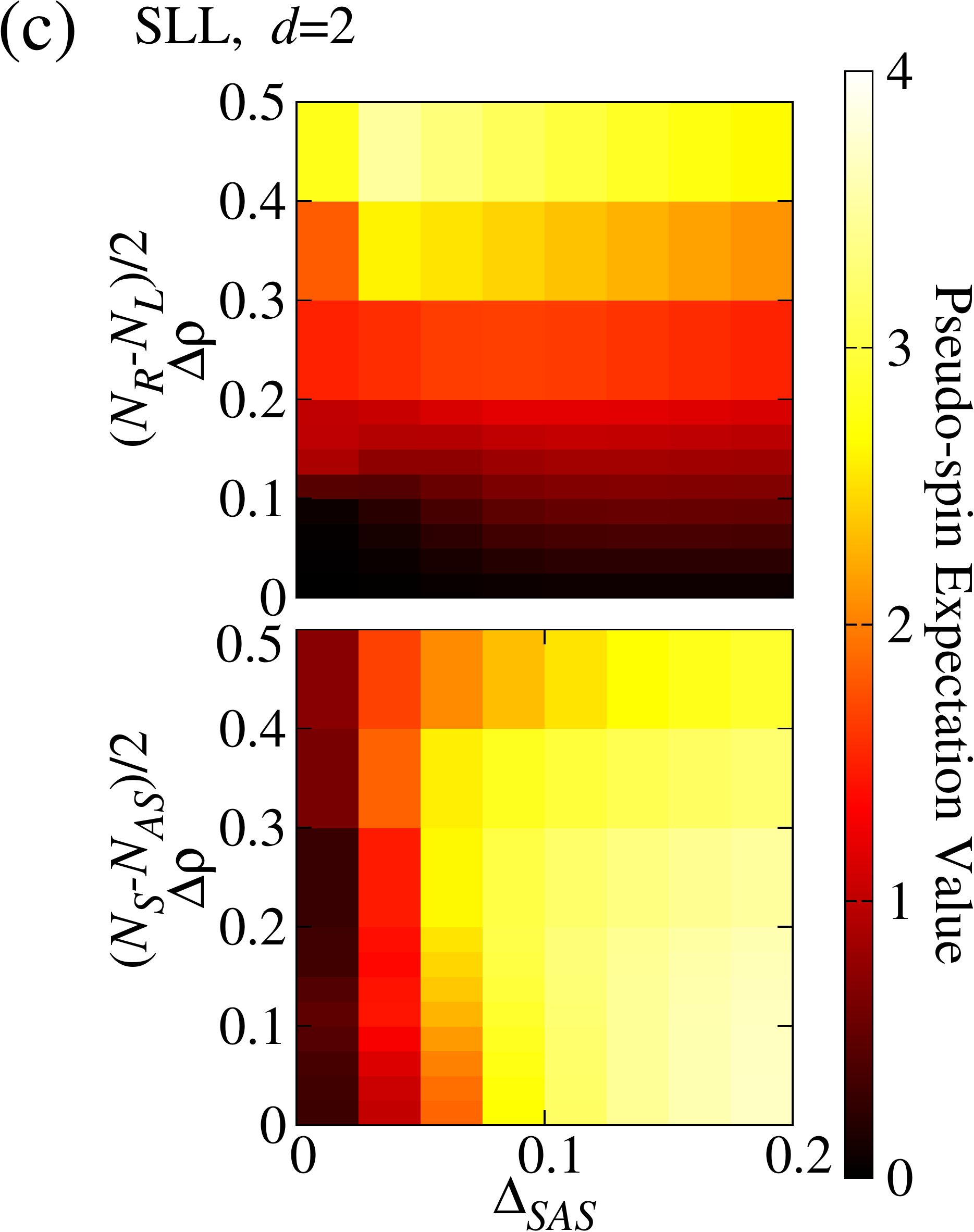}
\includegraphics[width=3.5cm,angle=0]{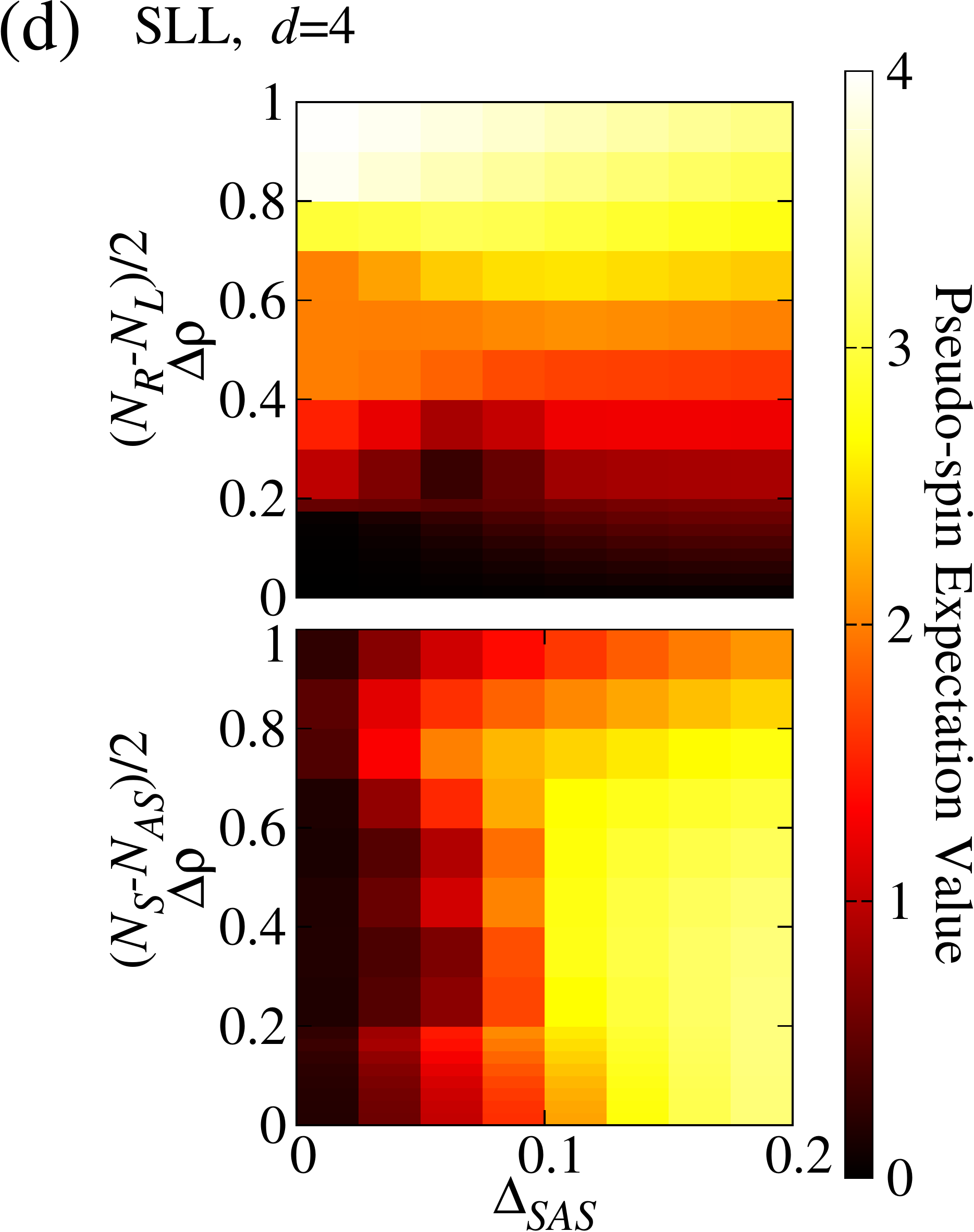}
\end{tabular}
\caption{(Color online) Second Landau level:  (Pseudo-spin) 
expectation value of the exact ground 
state of $(N_R-N_L)/2$ (top panel) and $(N_S-N_{AS})/2$ (lower panel)  
for (a) $d=0$, (b) $d=1$, (c) 
$d=2$ and (d) $d=4$.}
\label{fig-SLL-Dsas-v-Drho-pseudospin}
\mbox{}\\
\begin{tabular}{c}
\includegraphics[width=3.5cm,angle=0]{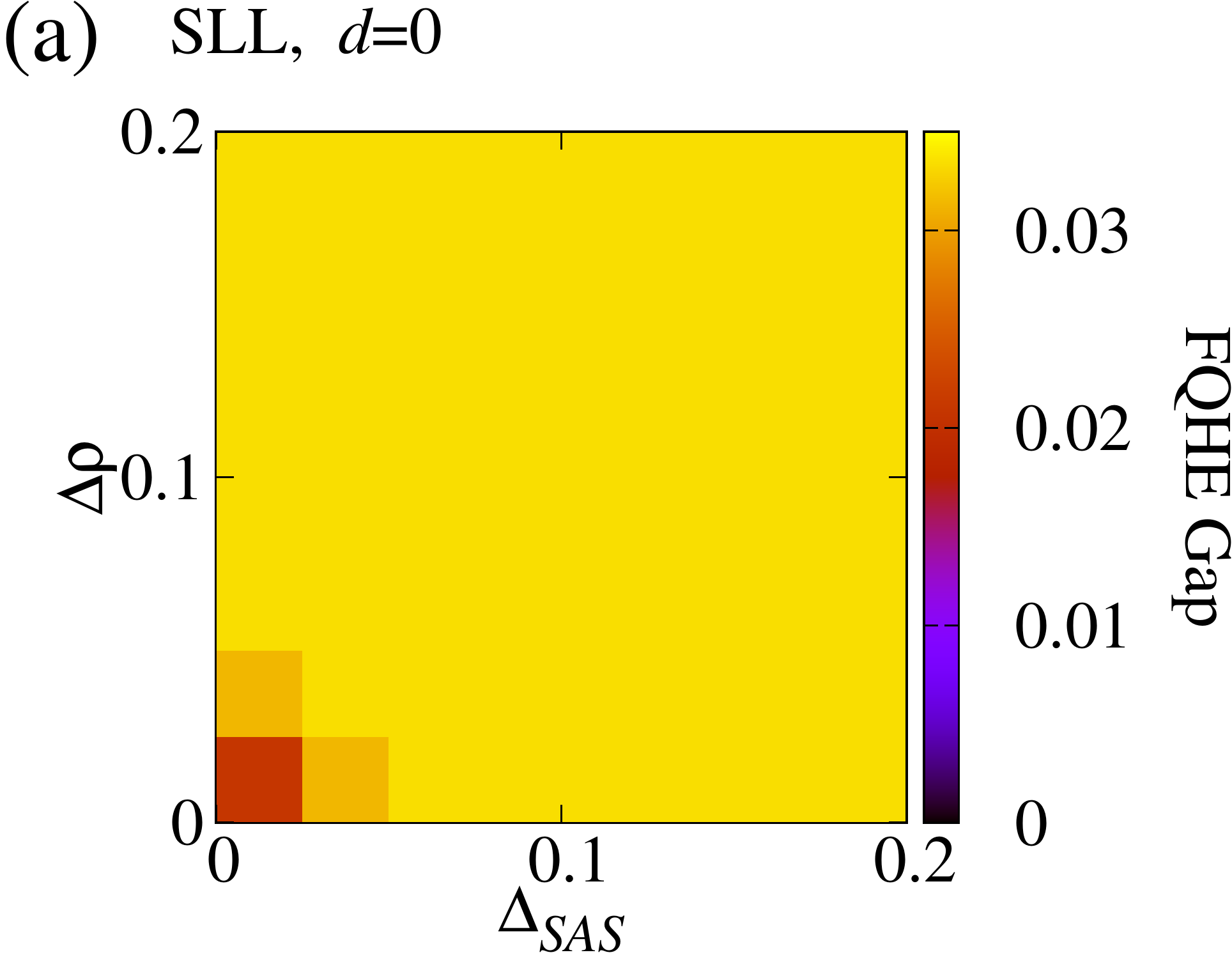}
\includegraphics[width=3.5cm,angle=0]{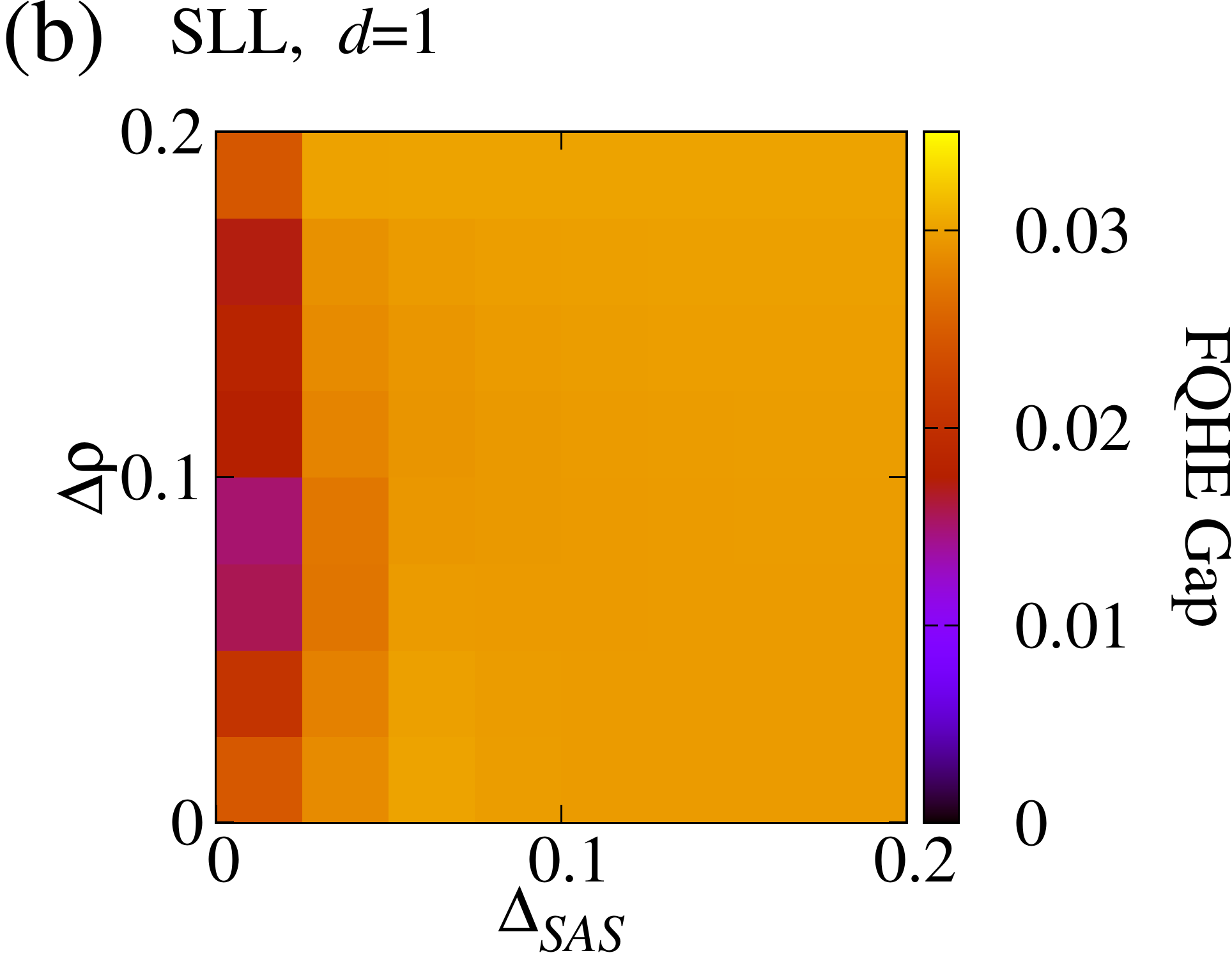}\\
\includegraphics[width=3.5cm,angle=0]{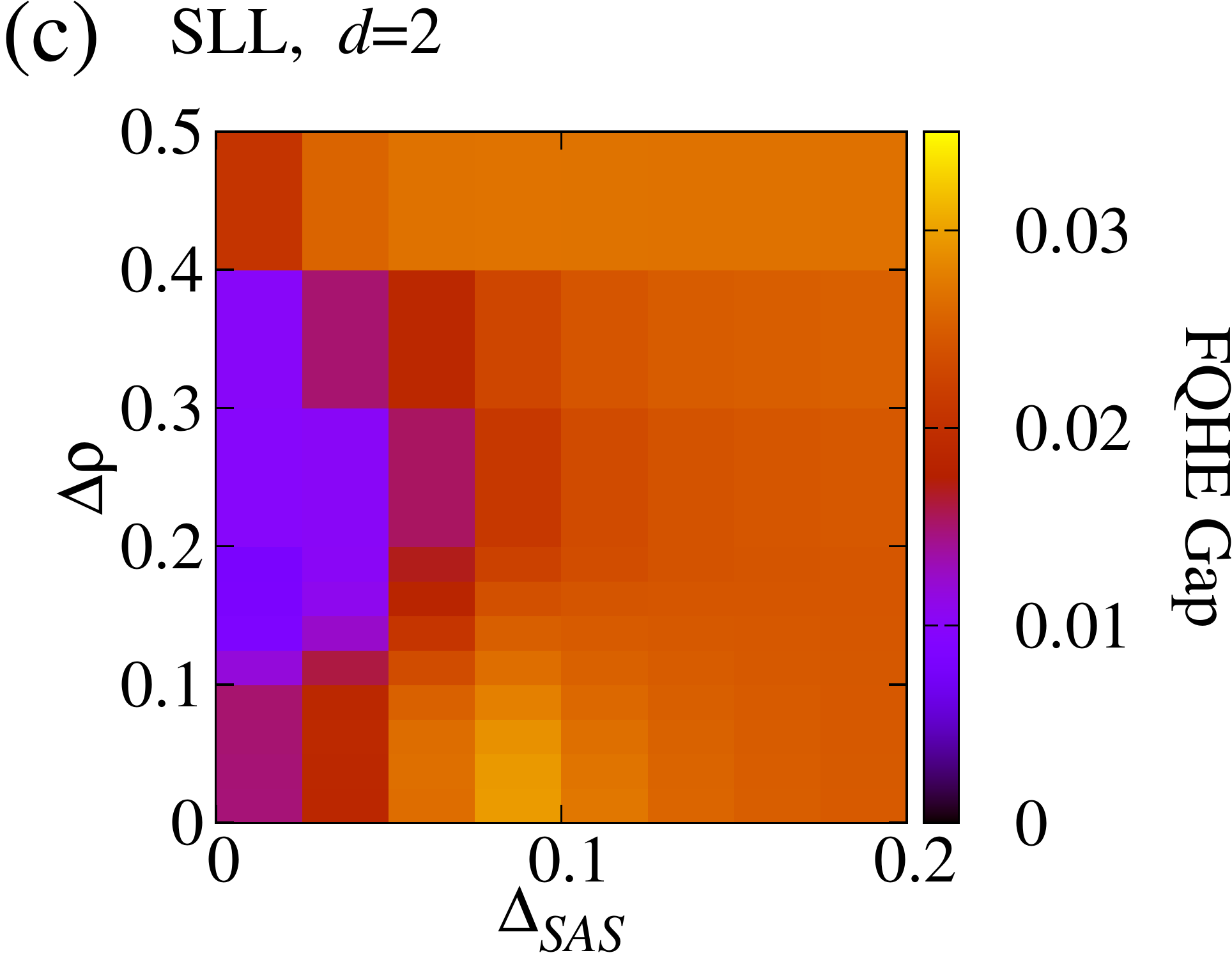}
\includegraphics[width=3.5cm,angle=0]{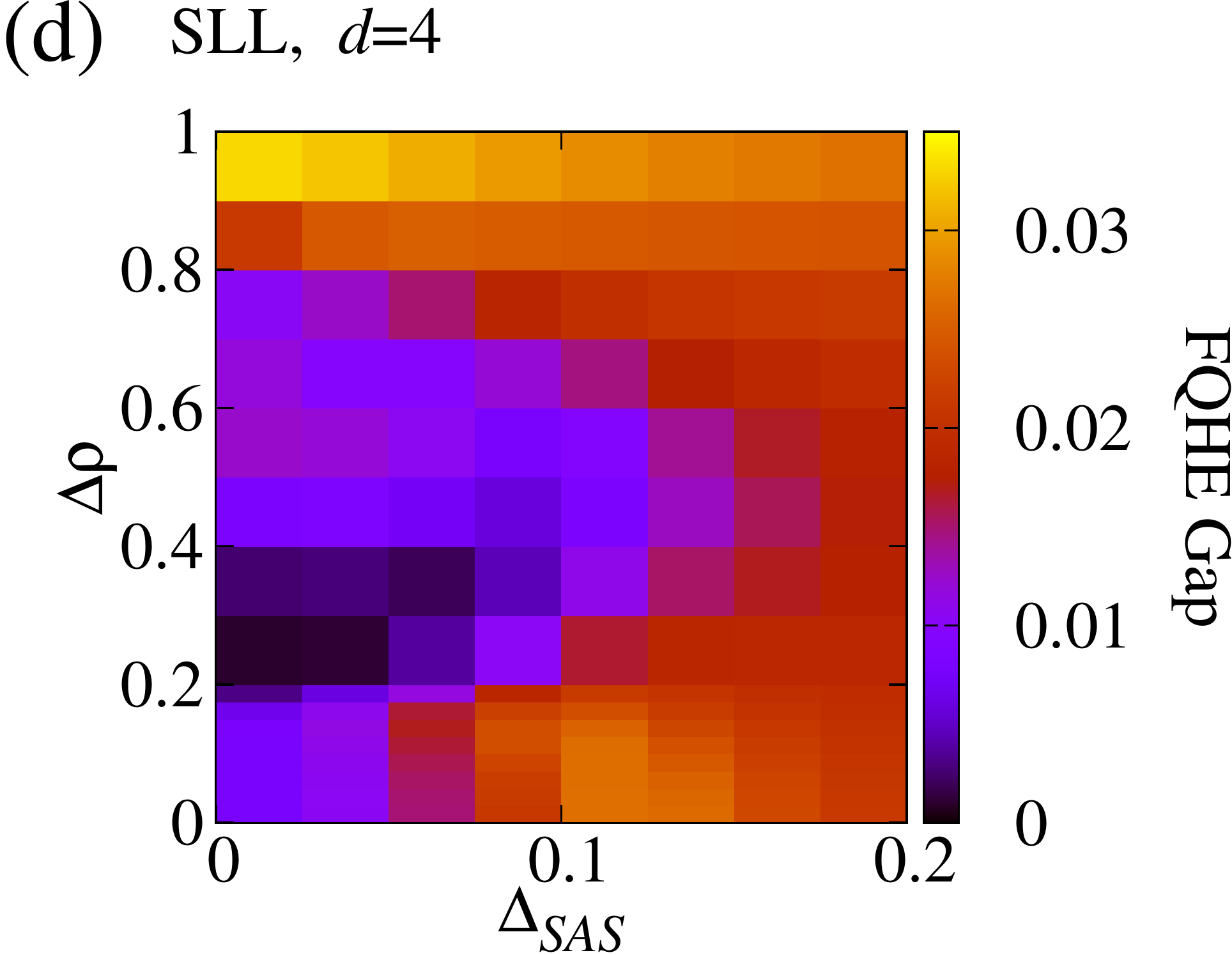}
\end{tabular}
\caption{(Color online) Second Landau level:  FQHE energy gap for (a) $d=0$, (b) $d=1$, (c) 
$d=2$ and (d) $d=4$.}
\label{fig-SLL-Dsas-v-Drho-gap}
\end{figure}

Before we tackle the results for the second Landau level we briefly note that 
bilayer systems in higher Landau levels are quite subtle and non-trivial and 
actually provide some conceptual difficulty.  As 
recently discussed in Ref.~\onlinecite{mrp-sds-bilayer}, it is not obvious what happens to 
the electrons in the lowest spin-up and spin-down Landau levels 
when driving a one-component system at $\nu=5/2$ to a two-component (bilayer)
system at total filling $\nu=5/2$.  For this 
work, however, we take the conceptually well-defined and straightforward 
solution~\cite{mrp-sds-bilayer}:
we assume full spin-polarization, hence the system is essentially spin-less with 
each Landau level having only one spin-index.  Then the $\nu=5/2$ (balanced) 
two-component system is equivalent to one where each layer has $1+1/4$ filling with 
the lowest spin Landau level being completely full and the second Landau level 
being 1/4 full.  In this way, the incompressible FQHE states (Halperin 331 or Moore-Read 
Pfaffian) form completely in the second Landau level.  This ``solution" provides a 
completely well-defined mathematical problem.  Of course, the real physical 
system, i.e., bilayer FQHE systems in higher Landau levels, 
could be more complicated and produce rich physics.  In fact, theoretically, the 
\emph{full} solution is out of reach for any conceivable computer--the Hilbert space is 
too vast when including many (or at least three) Landau levels, spin-up and spin-down, and 
layer index.  Thus, theoretical and experimental efforts in studying bilayer FQHE systems 
in higher Landau levels is likely a fertile ground for new discoveries.

\subsection{Bilayer}

In Fig.~\ref{fig-SLL-stacked-Dsas-v-Drho} we show overlaps between the exact 
ground state and the layer-basis and $SAS$-basis Pf and 331 wavefunctions, 
pseudo-spin expectation values, and the FQHE gap for the second Landau level.   The 
difference between the LLL and the SLL are subtle but important.

First we focus on the overlaps between the Pf and 331 states and the 
exact ground state in both the layer- and $SAS$-bases, cf. 
Fig.~\ref{fig-SLL-stacked-Dsas-v-Drho}a-b.  
The main difference between the results in the LLL and the SLL is that, as 
has been shown previously~\cite{mrp-ft-prl,mrp-ft-prb,mrp-sds-bilayer} for the 
case of zero charge imbalance $\Delta\rho=0$, 
the overlap between the 
exact state and the Pf state(s) is higher in the SLL than it is in the LLL.  In the SLL, 
the Pfaffian overlap is approximately $\sim 0.97$ at its largest while in the LLL it is 
approximately $\sim 0.9$ at its largest.   This can be seen more clearly 
in Fig.~\ref{fig-SLL-Dsas-v-Drho-Pf}  where, compared to the results of the 
LLL, the overlap with the Pf state is higher in the SLL.  (Again, for $d=0$ 
the system is SU(2) symmetric and the layer-basis and $SAS$-basis 
results are related via a pseudo-spin rotation.)  
We note that by increasing the width of the individual quantum well $w$, 
the Moore-Read Pfaffian overlap can be increased to nearly 
unity~\cite{mrp-ft-prl,mrp-ft-prb,mrp-sds-bilayer}.  This is a  
consequence of the differences between the Haldane pseudopotentials corresponding to the 
LLL versus those of the SLL with the SLL pseudopotentials being 
more amenable to pairing into a $p$-wave paired 
BCS state (of Composite Fermions), i.e., the Moore-Read 
Pfaffian~\cite{rezayi-haldane,morf,scarola-park-jain,moller,mrp-ft-prl,mrp-ft-prb}.

There is another striking difference between the results in the LLL versus the 
SLL when considering the overlap between the Halperin 331 state.  
In the LLL, the 331 state provides a 
very good description of the FQHE for two-component systems 
when the separation $d$ (or wide-well width $W$ for the wide-quantum-well) 
and the tunneling strengths are appropriate--this is evident by the value high 
value of the overlap that is obtained ($\sim0.99$).  In the SLL, the qualitative 
behavior of the overlap as the system parameters are varied is similar 
to the LLL but the overlap does not obtain as high a value (only $\sim 0.8$ 
for $d\sim 1$ and $\Delta_{SAS}=\Delta\rho=0$).  This, again, can be 
seen more clearly in Fig.~\ref{fig-SLL-Dsas-v-Drho-331}.

Looking at the pseudo-spin expectation 
value $\langle \hat{S}_z\rangle=(N_S-N_{AS})/2$ and $\langle\hat{S}_z\rangle=
(N_R-N_L)/2$ in Fig.~\ref{fig-SLL-stacked-Dsas-v-Drho}c (as well 
as Fig.~\ref{fig-SLL-Dsas-v-Drho-pseudospin}) we see very little 
difference between the lowest and second Landau levels.  In fact, visually 
it is difficult distinguishing the two.

Lastly, we consider the FQHE gap  (Fig.~\ref{fig-SLL-stacked-Dsas-v-Drho}d 
and Fig.~\ref{fig-SLL-Dsas-v-Drho-gap}), 
which does display some behavior that is 
qualitatively different from that of the LLL.  This qualitatively different 
behavior manifests itself for values of layer separation $d>2$.  In this region 
of parameter space we see that for increasing values of $\Delta\rho$ the 
FQHE gap has a maximum for a wide range of 
$\Delta_{SAS}$.  This is in stark contrast to the 
FQHE gap in the LLL (cf. Fig.~\ref{fig-LLL-stacked-Dsas-v-Drho}d) where 
the gap shows some indication of increasing for increasing $\Delta\rho$ 
but not to a \emph{maximum}.  

\begin{figure*}[t]
\begin{center}
\includegraphics[width=5.9cm,angle=0]{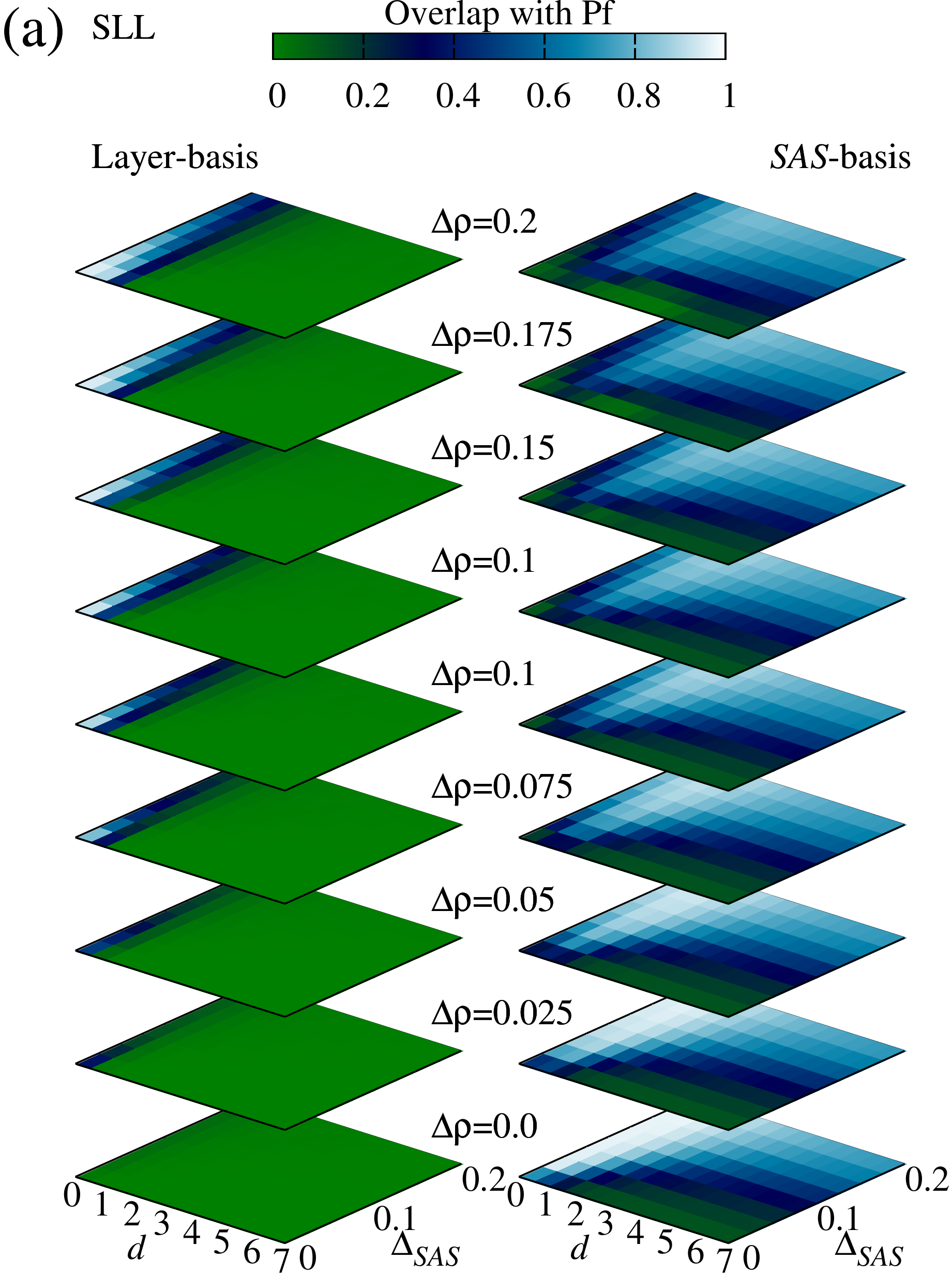}
\includegraphics[width=5.9cm,angle=0]{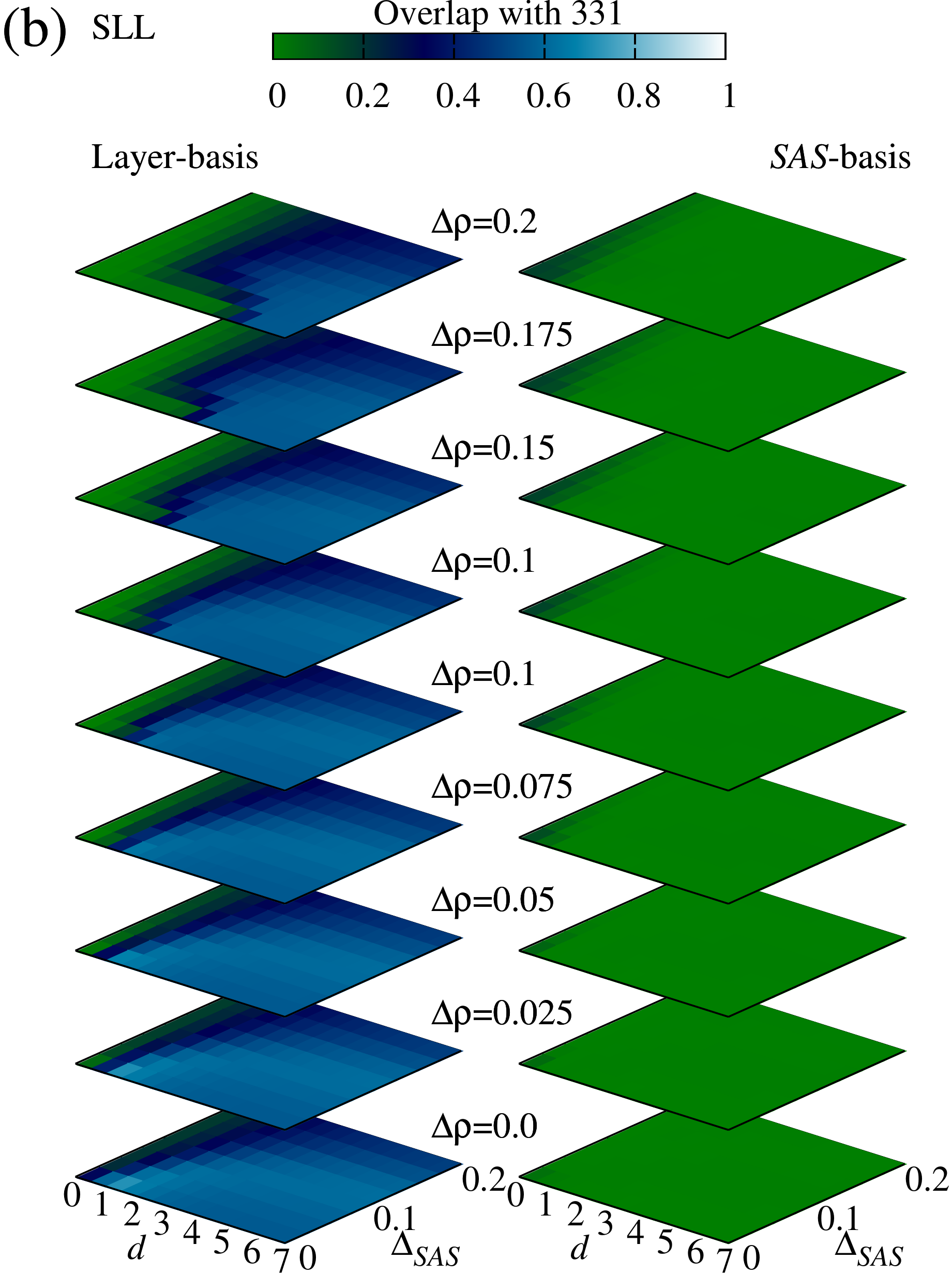}\\
\mbox{}\\
\includegraphics[width=5.9cm,angle=0]{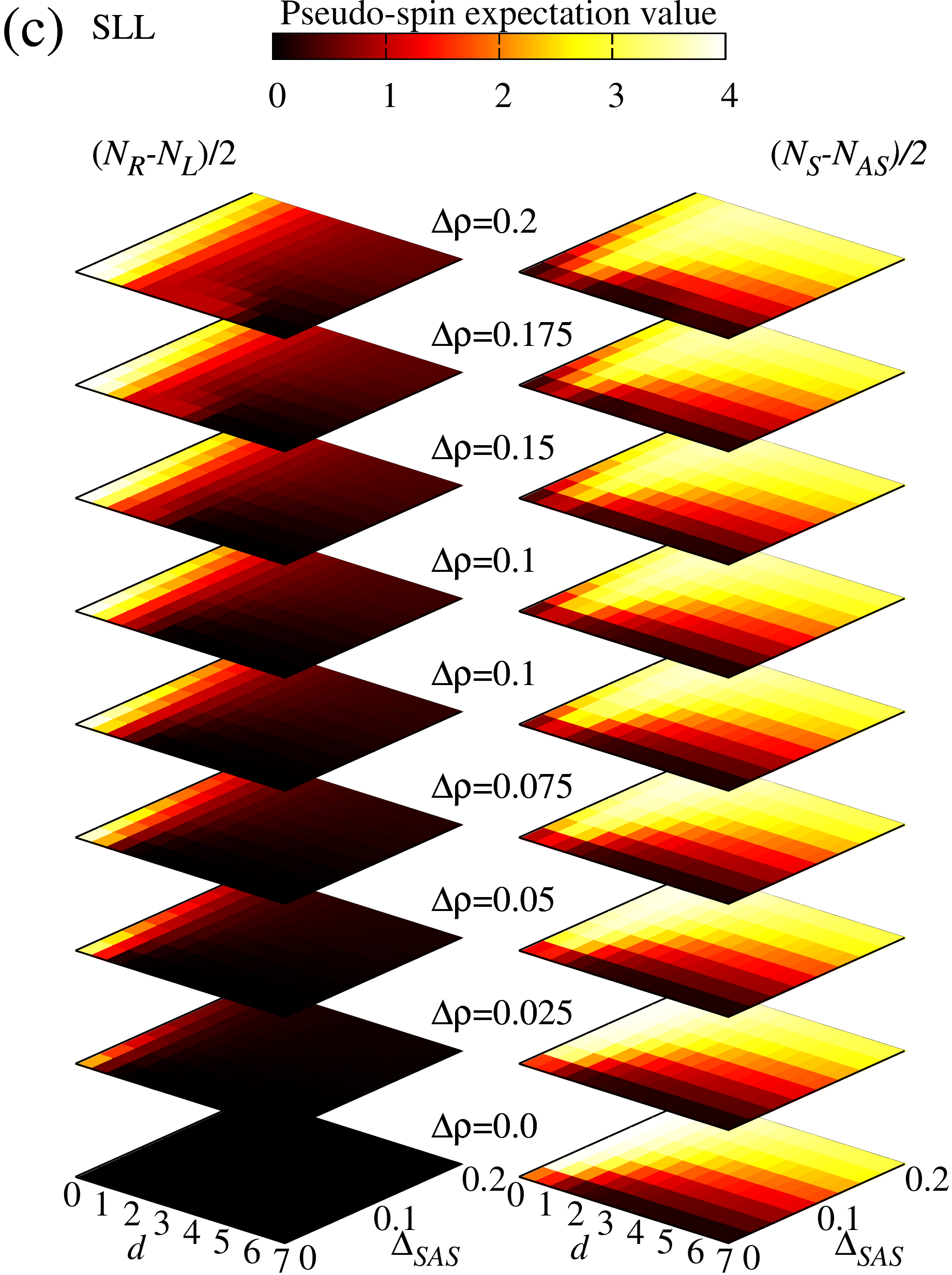}
\includegraphics[width=5.2cm,angle=0]{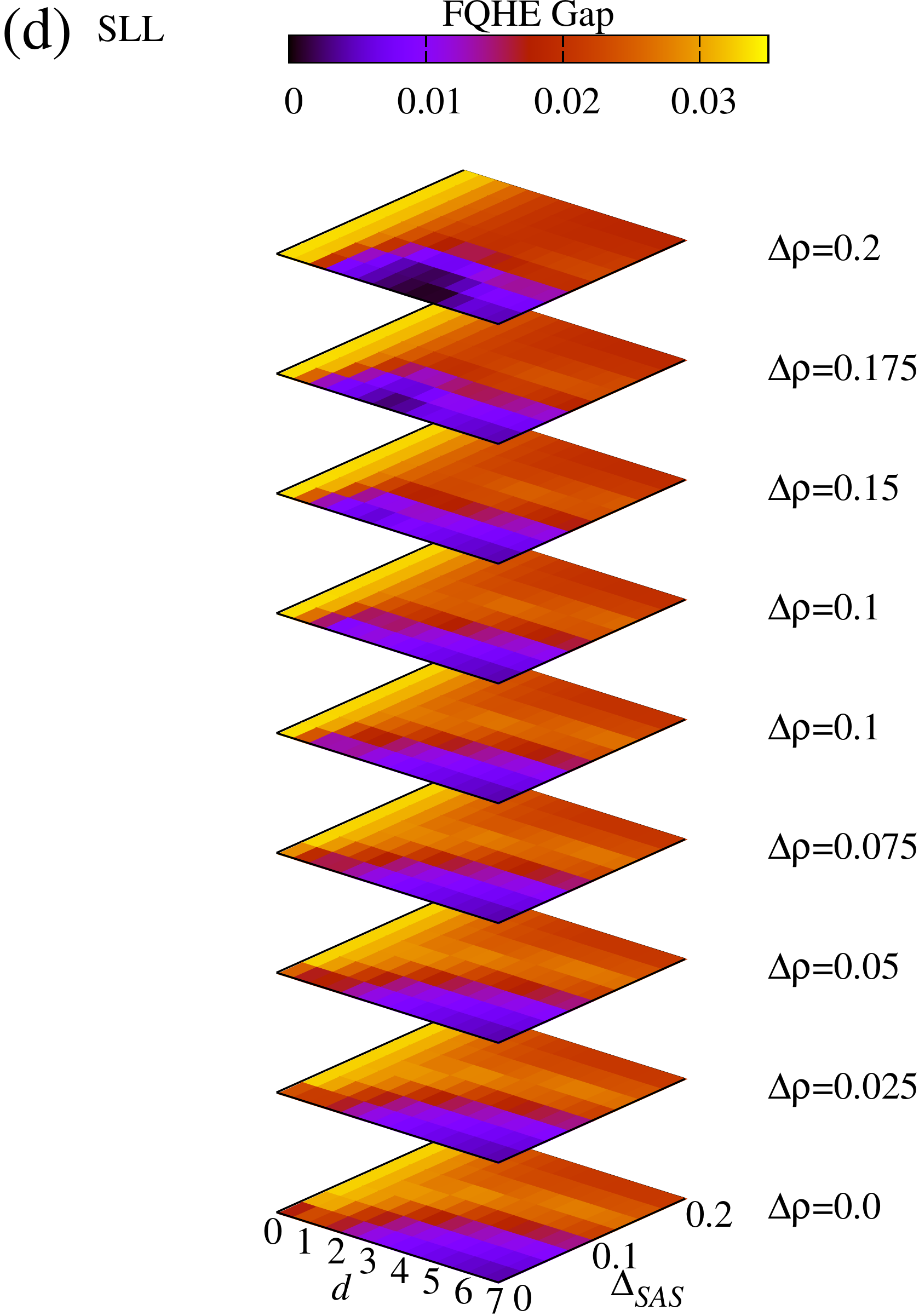}
\caption{(Color online) Second Landau level:  Same as Fig.~\ref{fig-SLL-stacked-Dsas-v-Drho} 
except that all plots are shown as a function of inter-layer tunneling $\Delta_{SAS}$ and layer 
separation $d$ for a number of different  charge imbalances $\Delta\rho$ and zero individual layer 
thickness $w=0$.}
\label{fig-SLL-stacked-Dsas-v-d}
\end{center}
\end{figure*}

Fig.~\ref{fig-SLL-stacked-Dsas-v-d} shows the overlaps, pseudo-spin 
expectation values, and FQHE gap as a function of layer separation 
$d$ and tunneling strength $\Delta_{SAS}$ for several values of 
charge imbalance tunneling strength $\Delta\rho$.  Qualitatively, the 
overlaps (Fig.~\ref{fig-SLL-stacked-Dsas-v-d}a and b) 
are similar in the SLL and LLL.  However, comparing them 
side-by-side, the overlap with the Halperin 331 state achieves a 
higher value in the LLL than it does in the SLL.  The opposite 
is true for the overlap with the Moore-Read Pfaffian state where 
the overlap achieves a higher value in the SLL than it does in the  
LLL.  These differences are manifest even while the 
pseudo-spin expectation values (Fig.~\ref{fig-SLL-stacked-Dsas-v-d}c) 
in the SLL, compared to the LLL, cannot visually be distinguished.  Thus, 
while the values of the Haldane pseudopotentials in the SLL are not 
different enough from the LLL to change pseudo-spin expectation 
values, i.e., the system is one-component or two-component 
at almost exactly the same place in parameter space in both the 
LLL and SLL, the differences are enough to change the character 
of the overlaps and the FQHE energy gap.

Fig.~\ref{fig-SLL-stacked-Dsas-v-d}d shows the FQHE energy gap, and again, 
there is a marked difference between the result in the LLL versus SLL.  Similar 
to the higher overlap value between the exact ground state and the Moore-Read 
Pfaffian state in the SLL, the FQHE energy gap has a maximum in the Pf 
region of the approximate quantum phase diagram especially 
along the $\Delta_{SAS}$ axis for small $d$.  (The opposite is 
true in the LLL--the FQHE energy gap is largest in the Halperin 331 region 
of the quantum phase diagram.)    We see this behavior also in the 
wide-quantum-well results.

\begin{figure}
\begin{tabular}{c}
\includegraphics[width=3.5cm,angle=0]{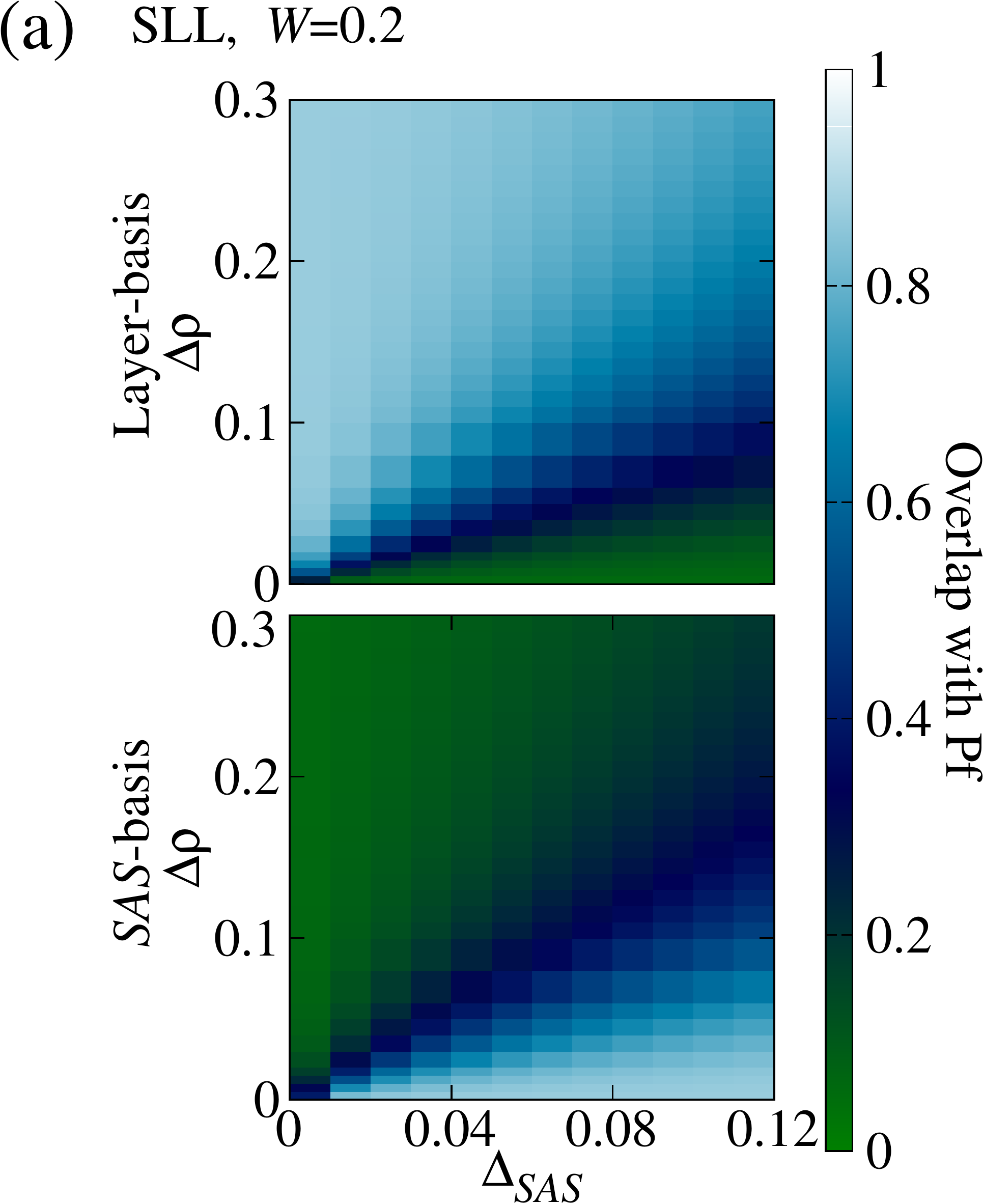}
\includegraphics[width=3.5cm,angle=0]{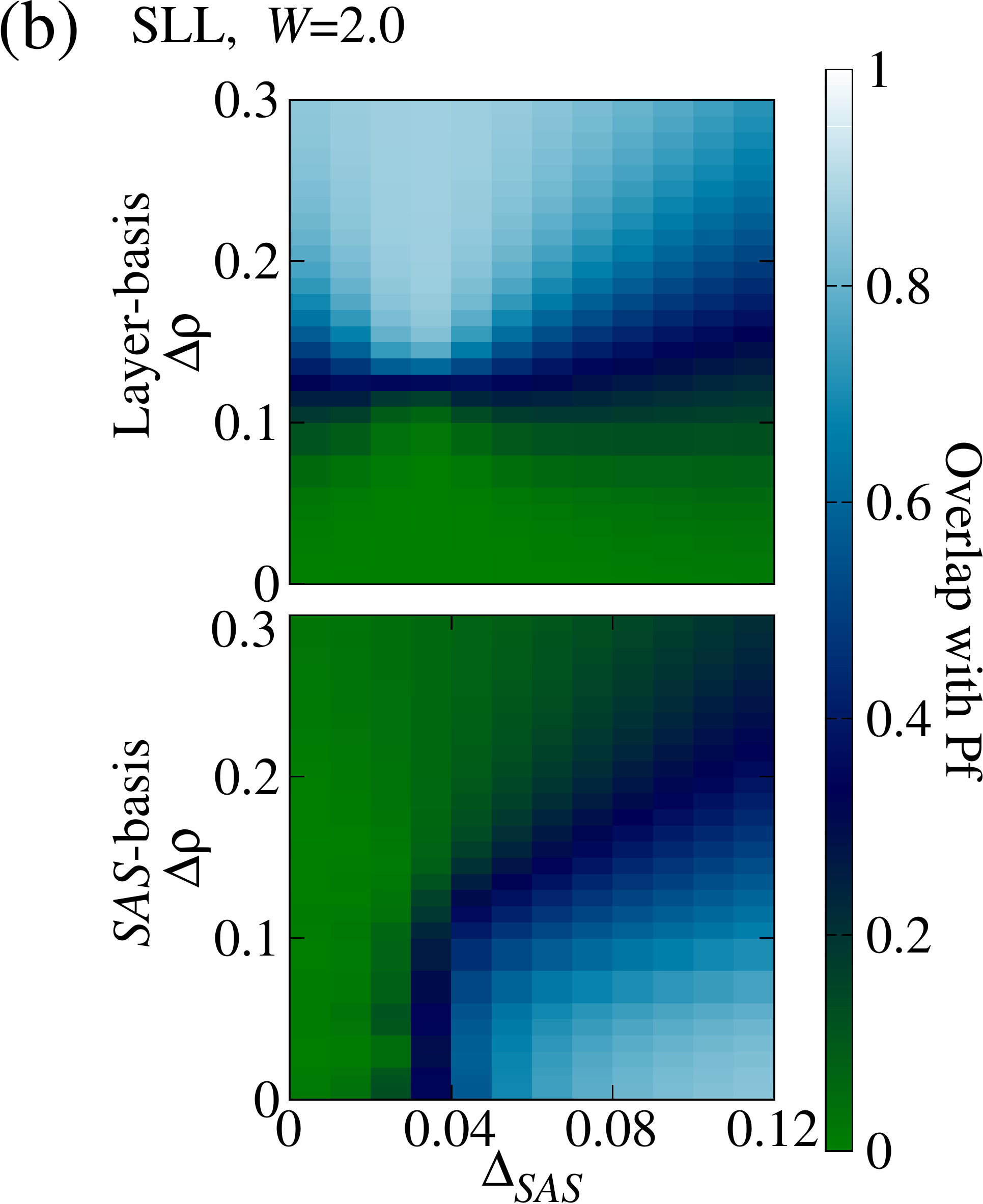}\\
\includegraphics[width=3.5cm,angle=0]{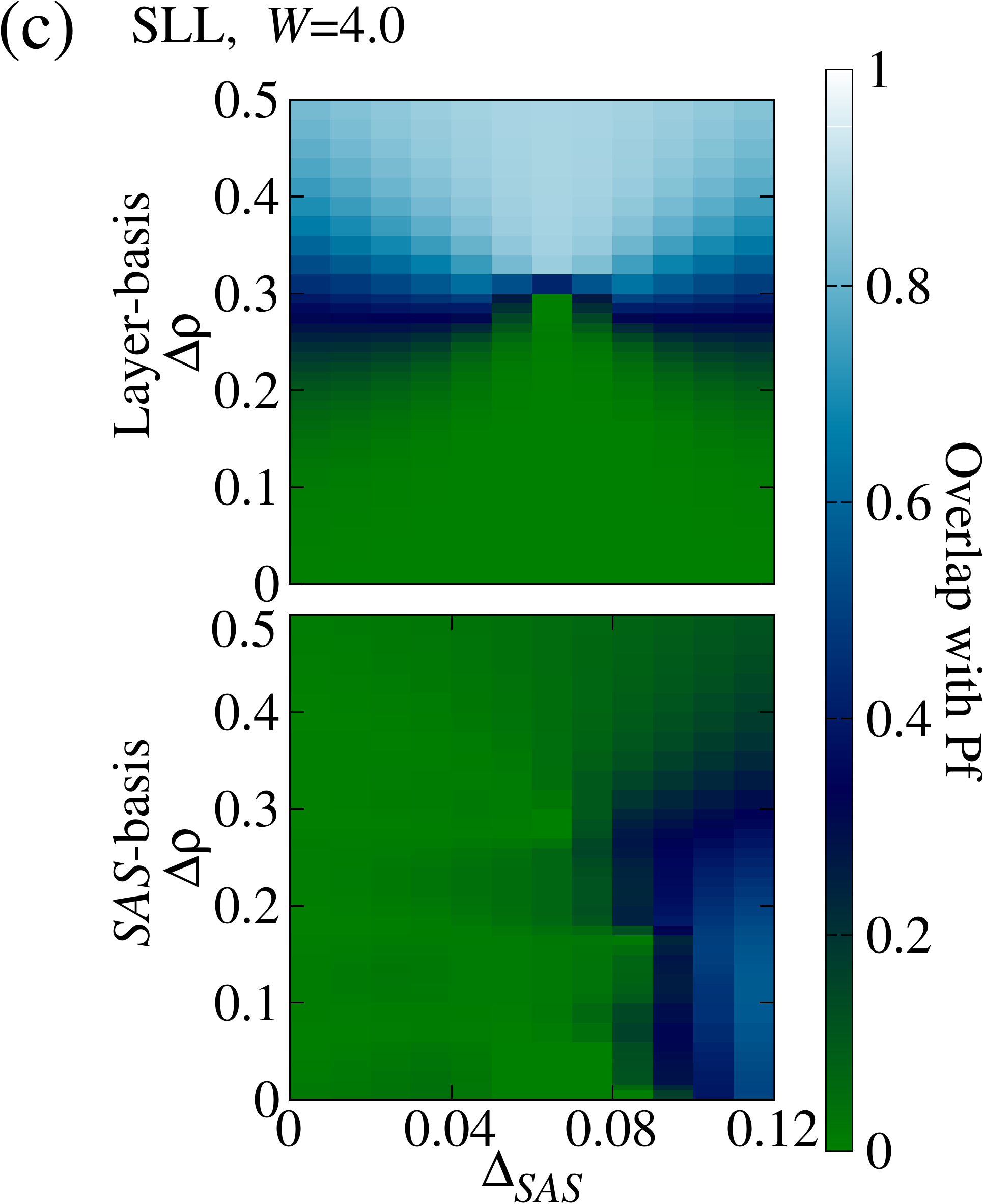}
\includegraphics[width=3.5cm,angle=0]{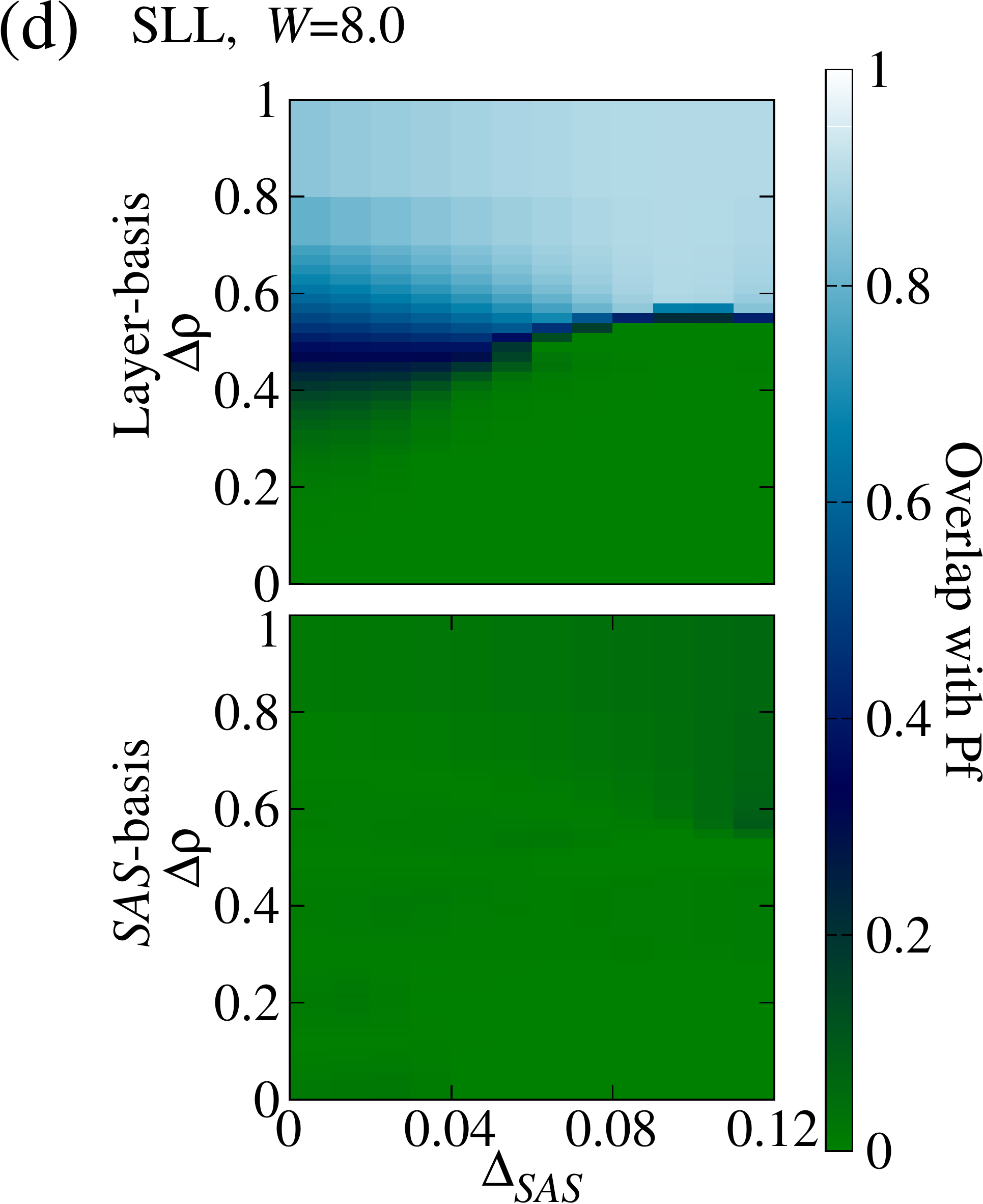}
\end{tabular}
\caption{(Color online) Second Landau level: Wavefunction overlap
  between the Moore-Read  Pfaffian wavefunction in the layer-basis (top
  panel) and the $SAS$-basis (lower panel) and the exact ground state
  for the WQW for (a) $W=0.2$, (b) $W=2$, (c) $W=4$ and (d) $W=8$ 
  as a function of $\Delta_{SAS}$ and $\Delta\rho$.}
\label{fig-SLL-Dsas-v-Drho-Pf-wqw}
\mbox{}\\

\begin{tabular}{c}
\includegraphics[width=3.5cm,angle=0]{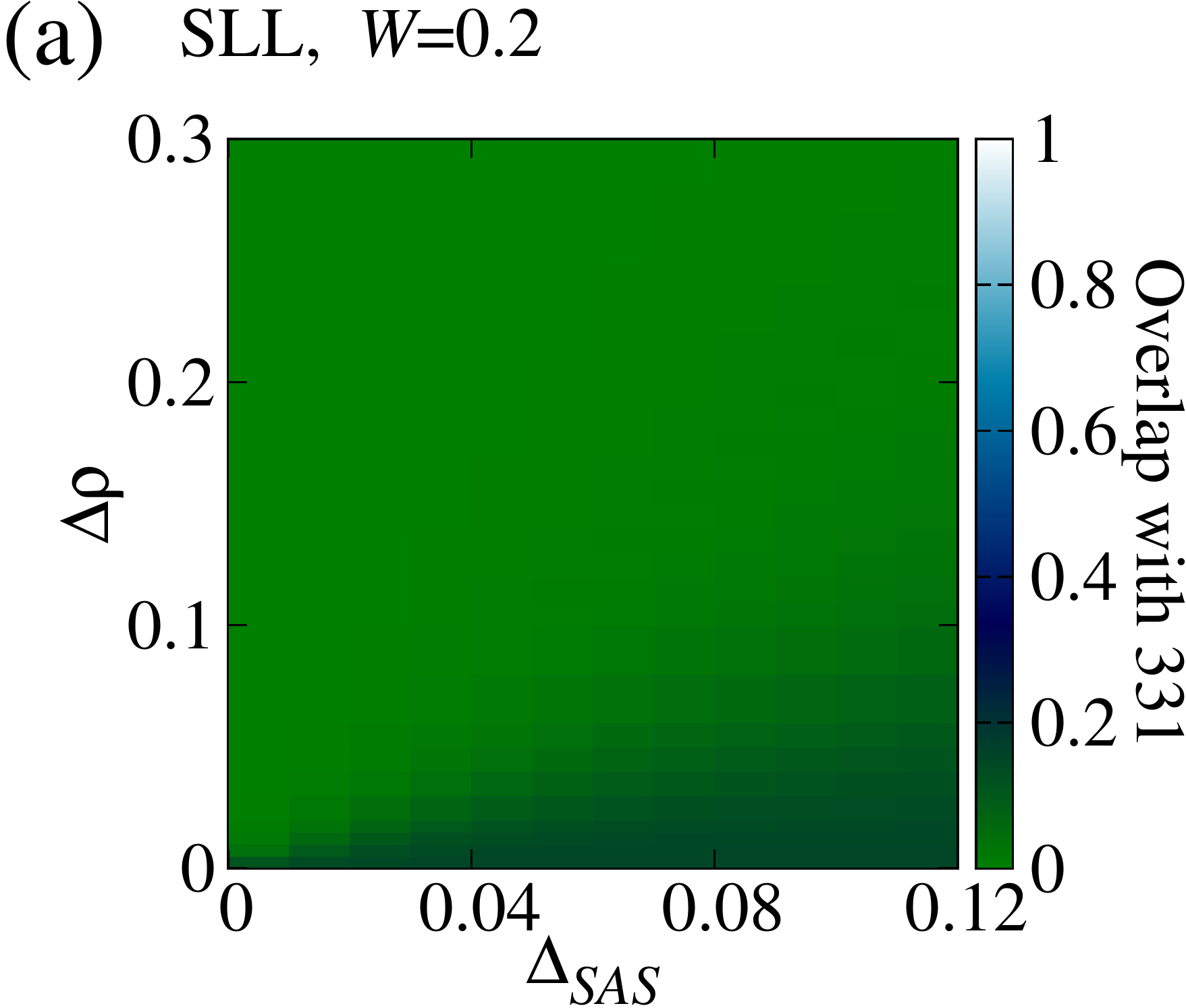}
\includegraphics[width=3.5cm,angle=0]{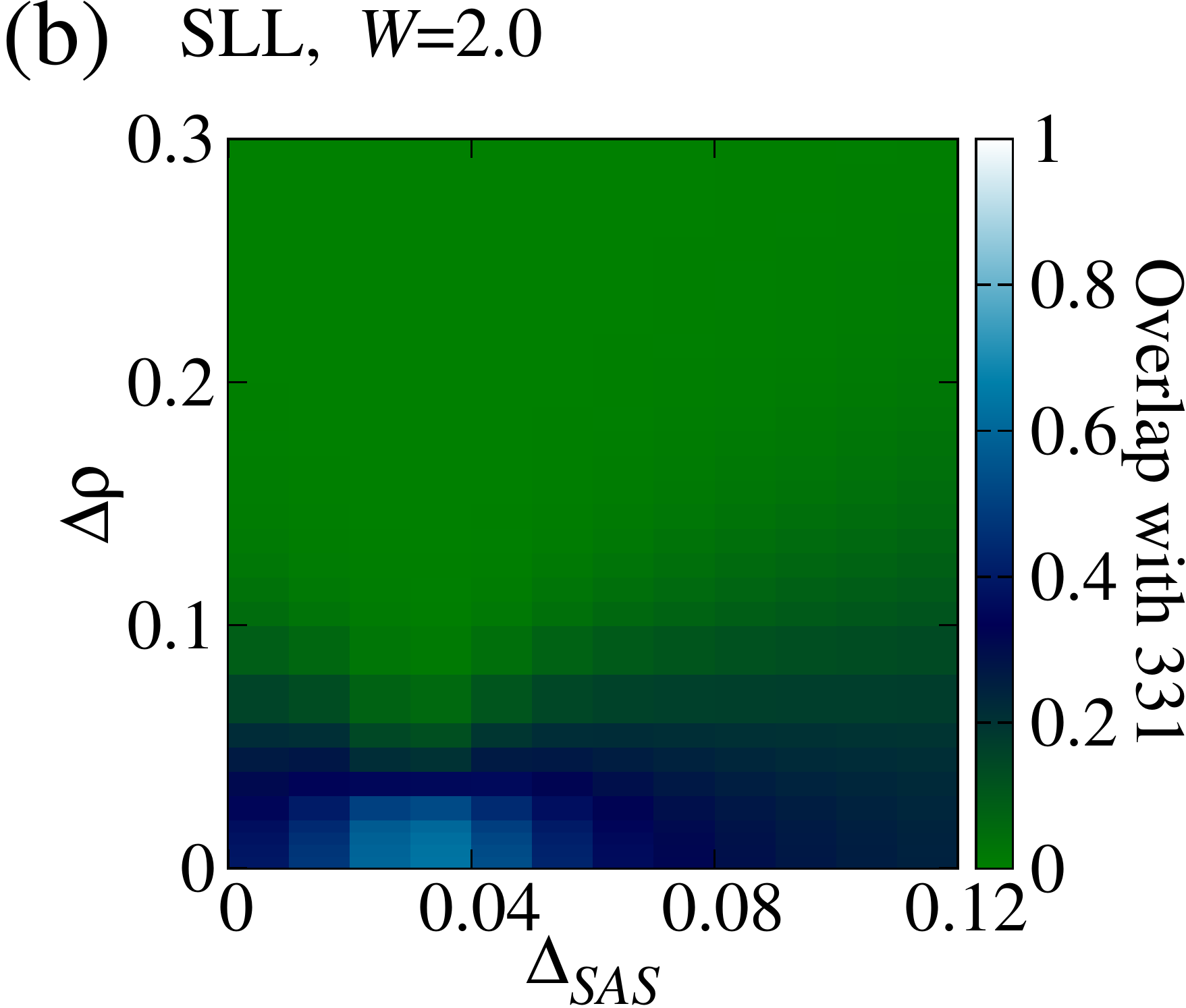}\\
\includegraphics[width=3.5cm,angle=0]{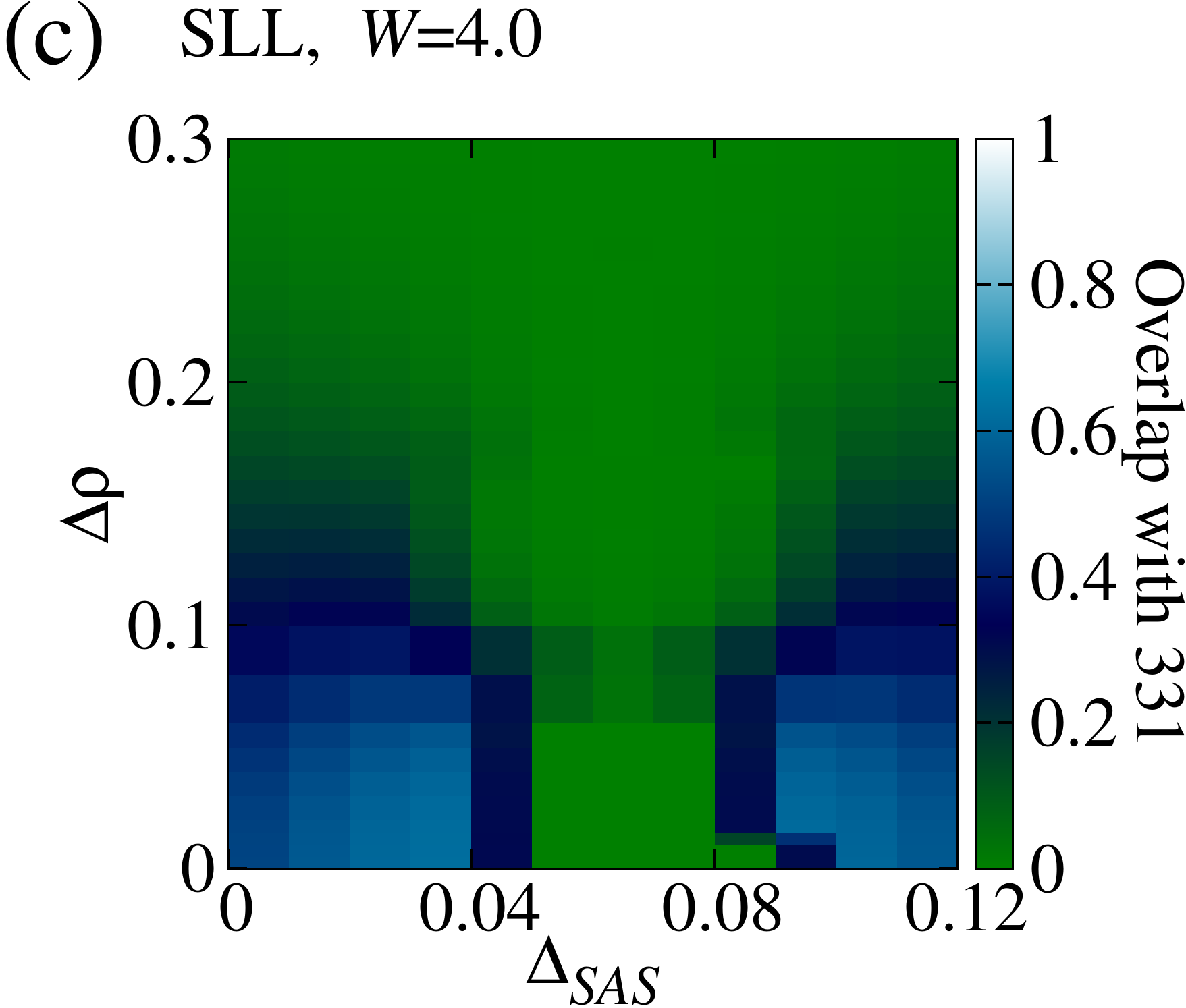}
\includegraphics[width=3.5cm,angle=0]{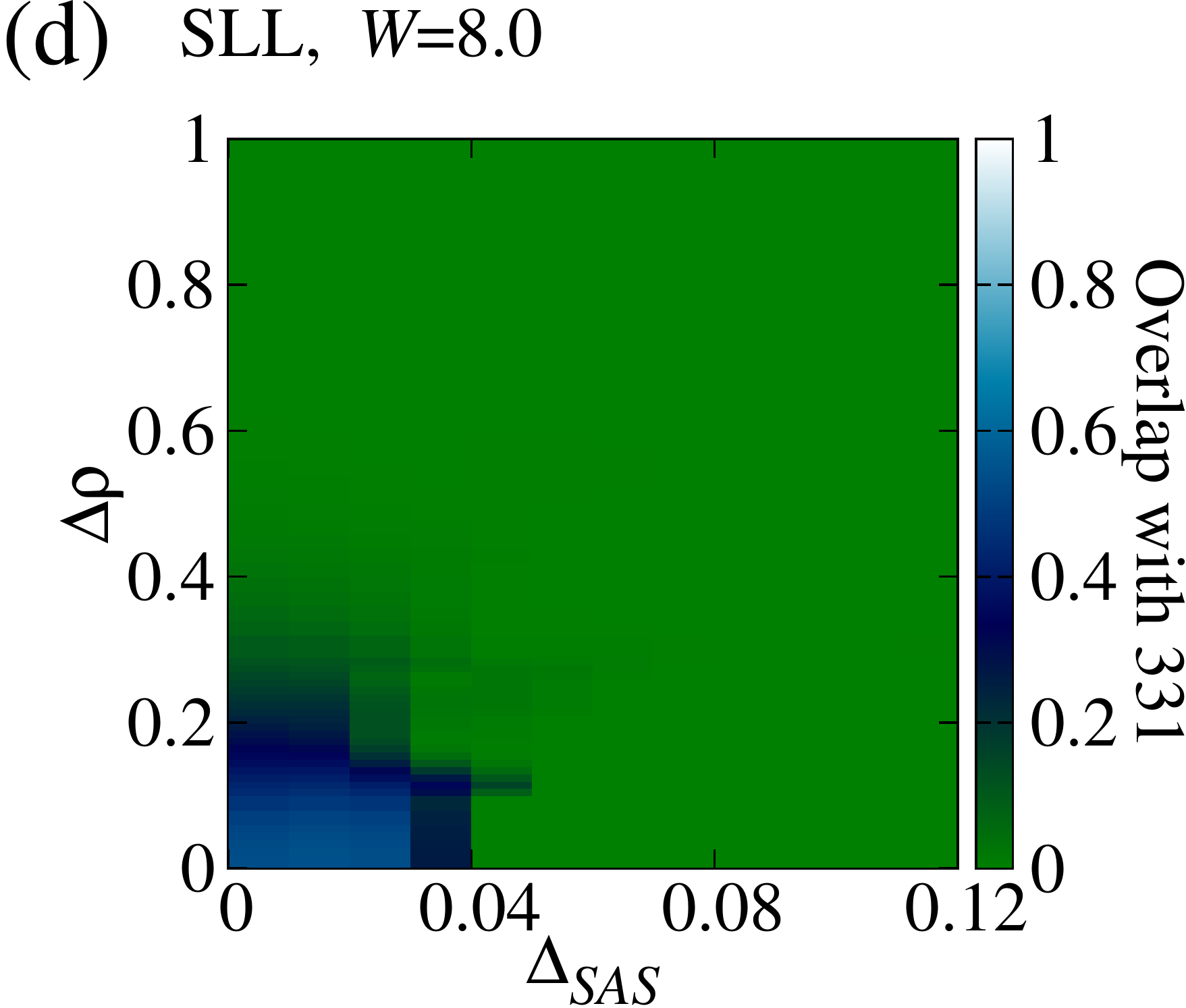}
\end{tabular}
\caption{(Color online) Second Landau level: Wavefunction overlap
  between the Halperin 331 wavefunction in the layer-basis and the exact ground state
  for the WQW for (a) $W=0.2$, (b) $W=2$, (c) $W=4$ and (d) $W=8$
    as a function of $\Delta_{SAS}$ and $\Delta\rho$.  Note that the 
    overlap with the layer-basis Halperin 331 state is not shown since 
    it is always nearly zero.}
\label{fig-SLL-Dsas-v-Drho-331-wqw}
\end{figure}

\begin{figure}
\begin{tabular}{c}
\includegraphics[width=3.5cm,angle=0]{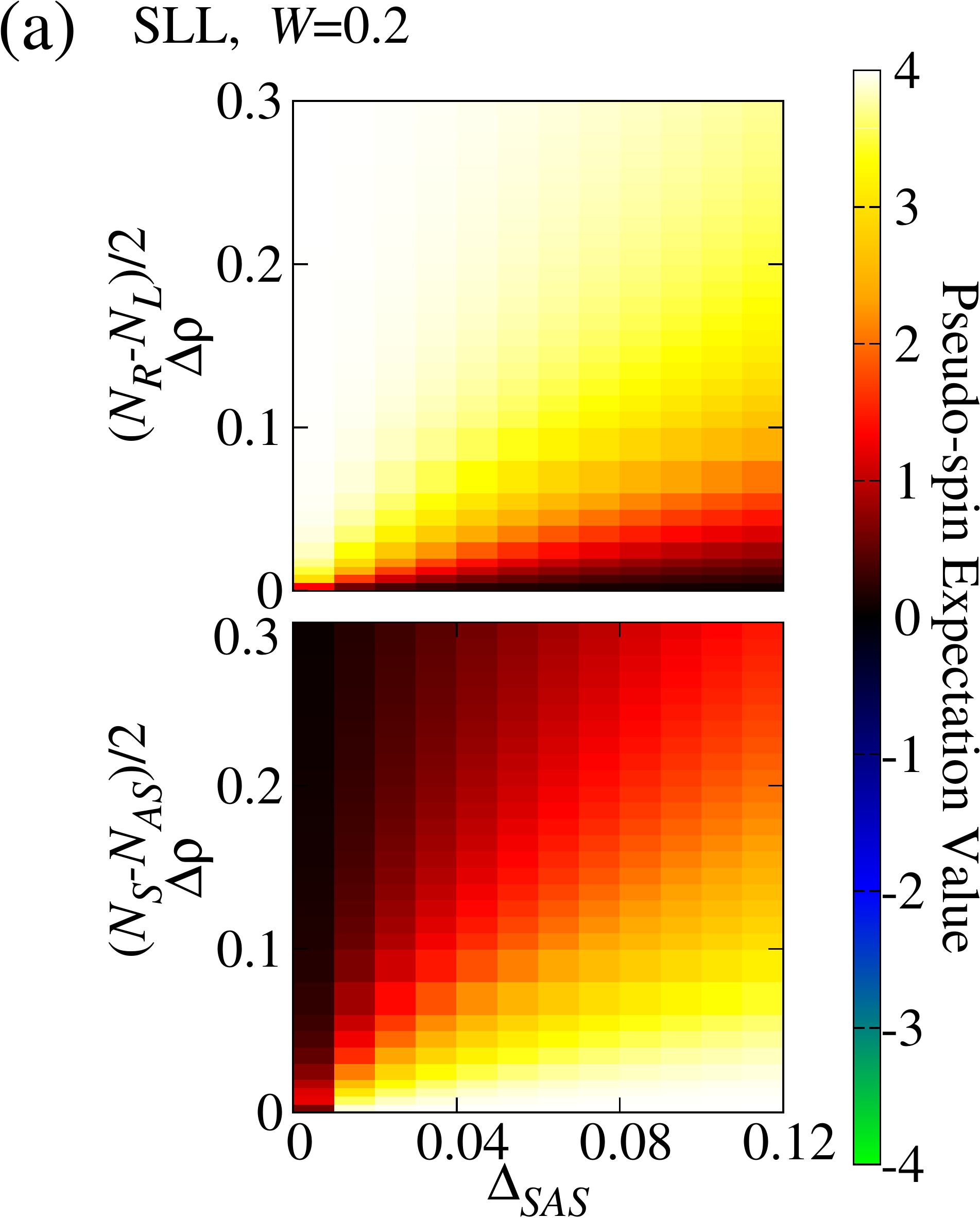}
\includegraphics[width=3.5cm,angle=0]{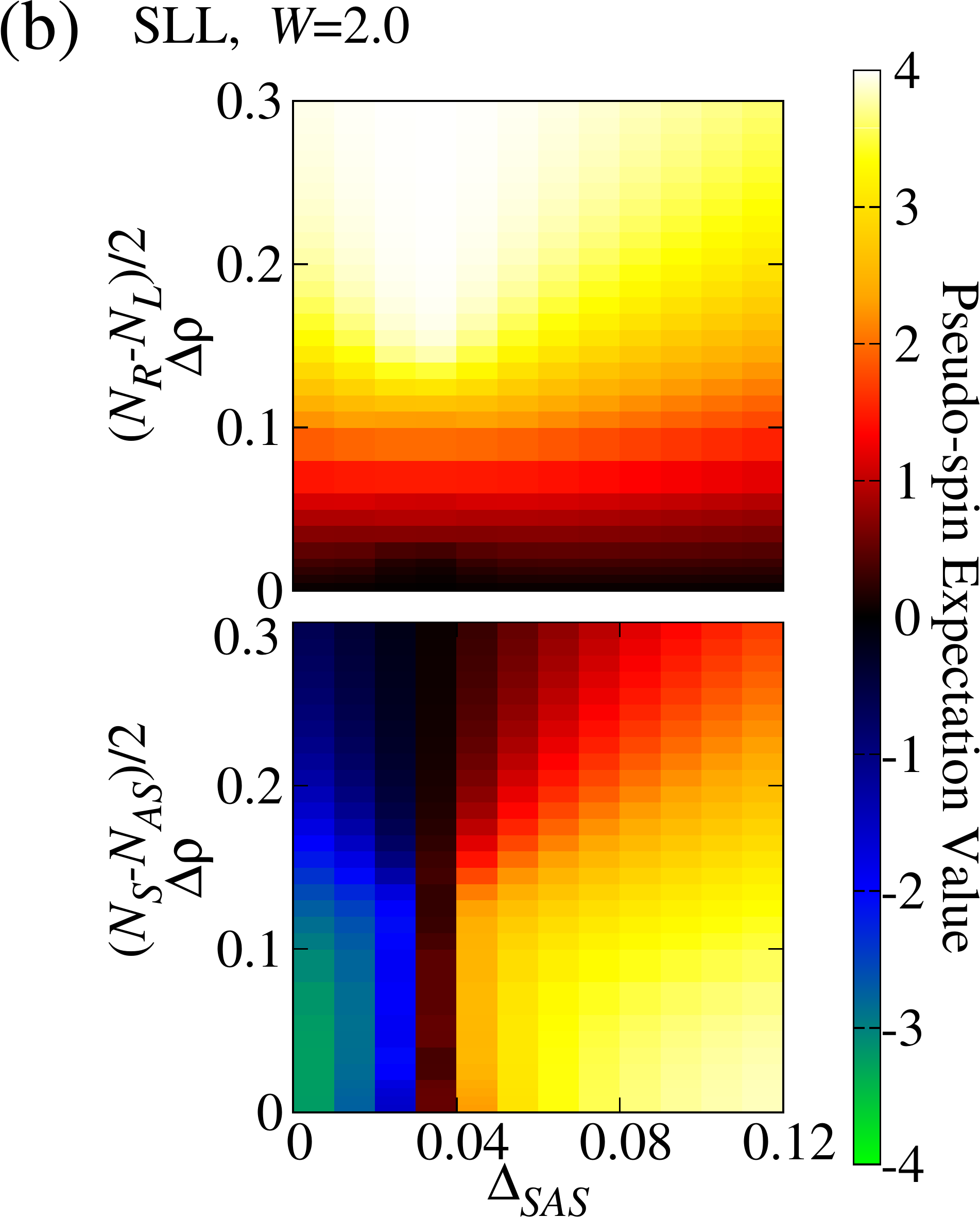}\\
\includegraphics[width=3.5cm,angle=0]{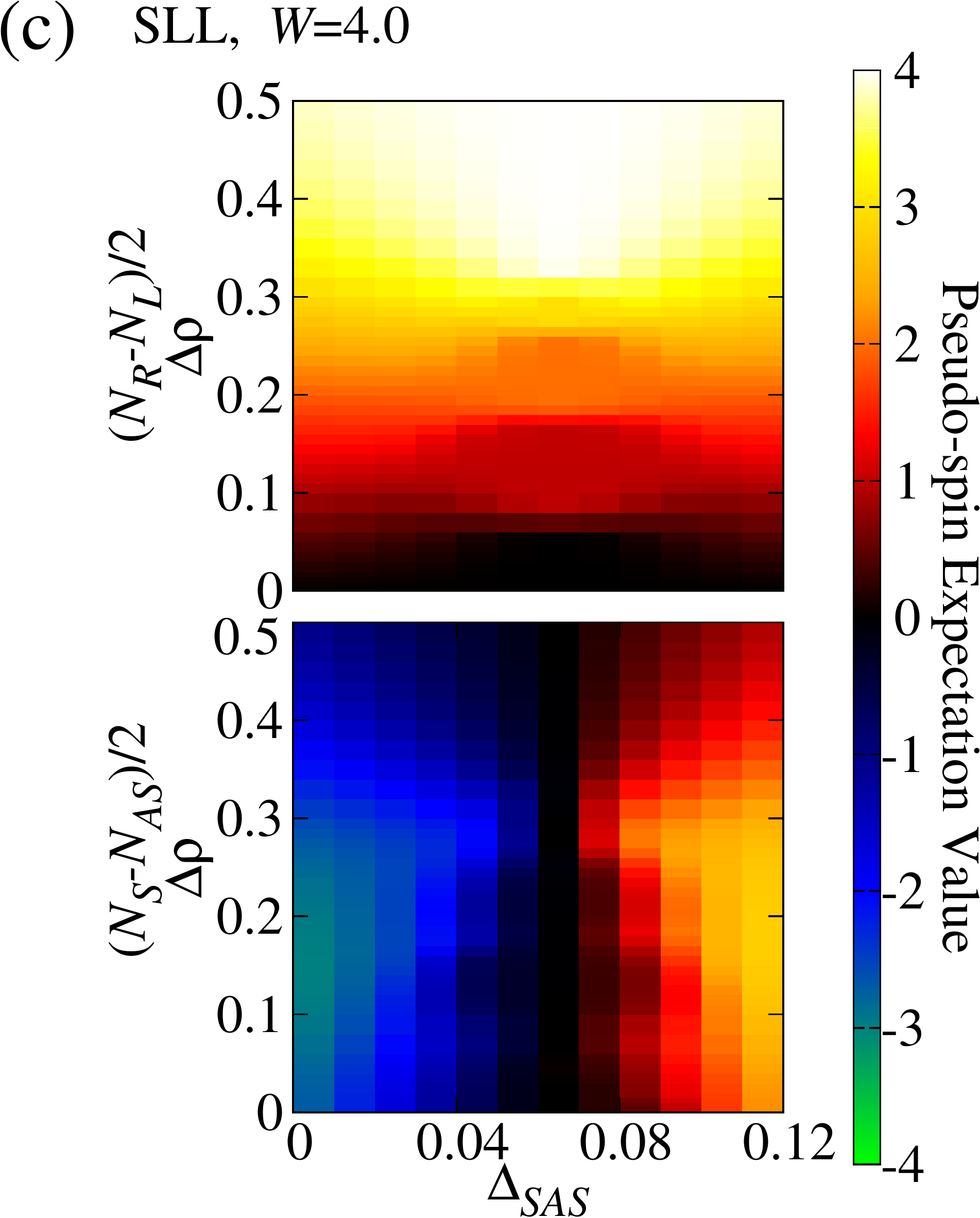}
\includegraphics[width=3.5cm,angle=0]{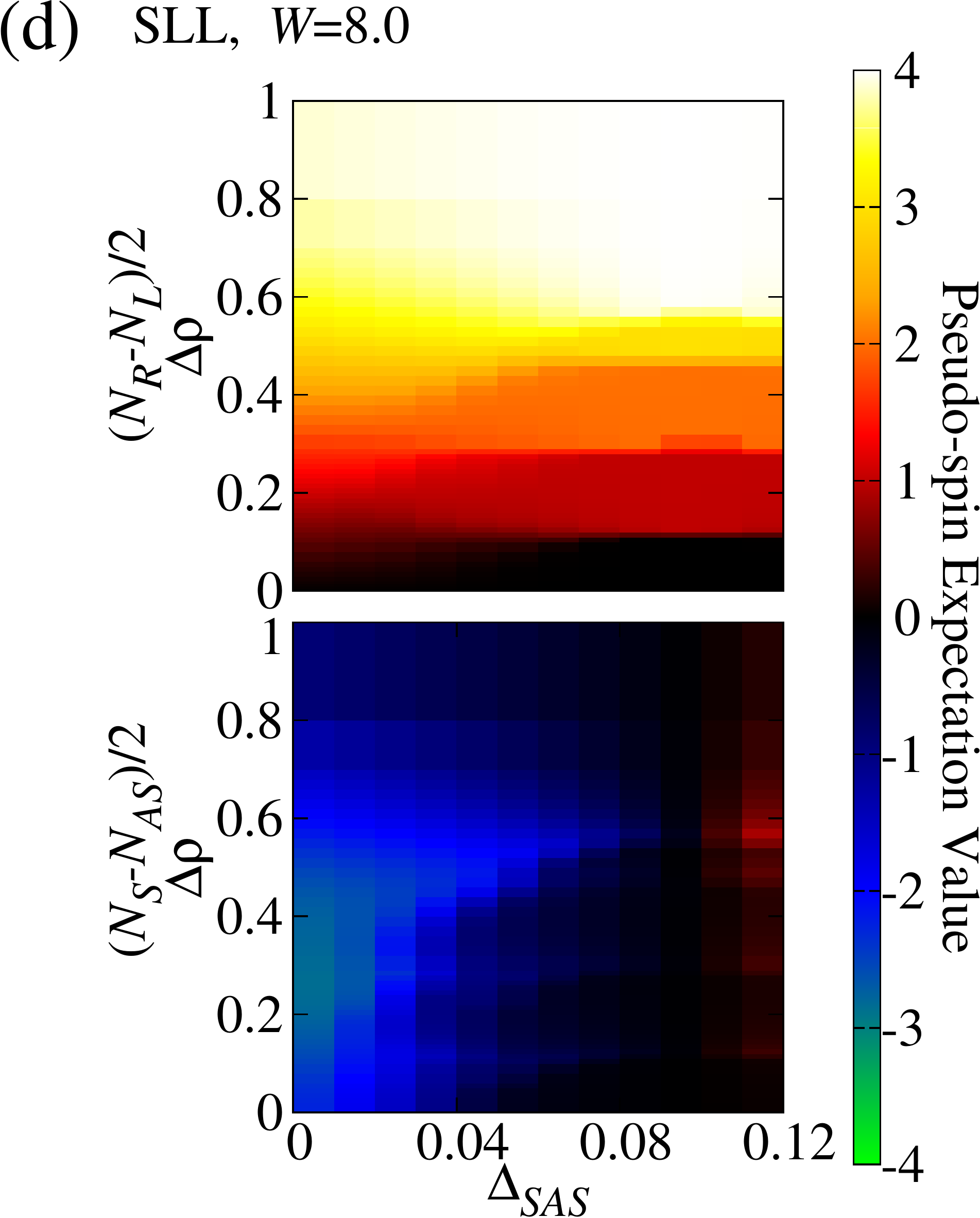}
\end{tabular}
\caption{(Color online)  Second Landau level:  Pseudo-spin expectation value of the exact ground 
state (for the WQW model) of $(N_R-N_L)/2$ (top panel) 
and $(N_S-N_{AS})/2$ (lower panel)  for (a) $W=0.2$, (b) $W=2$, (c) 
$W=4$ and (d) $W=8$.  Note that the WQW model always breaks SU(2) 
symmetry even for small $W$ and that for $W\geq 2$ there are negative 
values of $(N_S-N_{AS})/2$ meaning the system prefers, in some 
regions of parameter space, to have more electrons in the $AS$ 
state compared to the $S$ state even for positive and non-zero $\Delta_{SAS}$.}
\label{fig-SLL-Dsas-v-Drho-pseudospin-wqw}
\mbox{}\\
\begin{tabular}{c}
\includegraphics[width=3.5cm,angle=0]{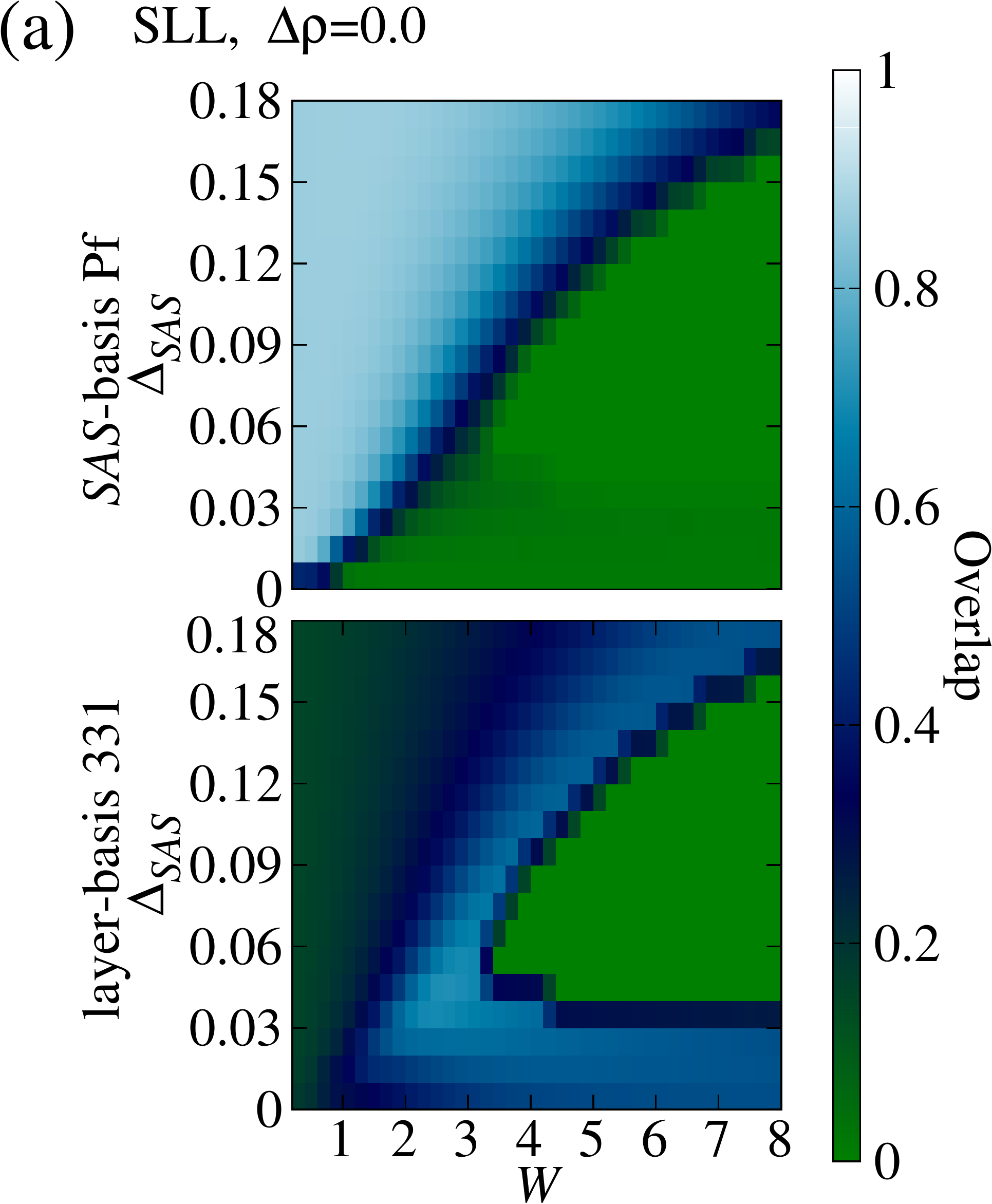}\\
\includegraphics[width=3.5cm,angle=0]{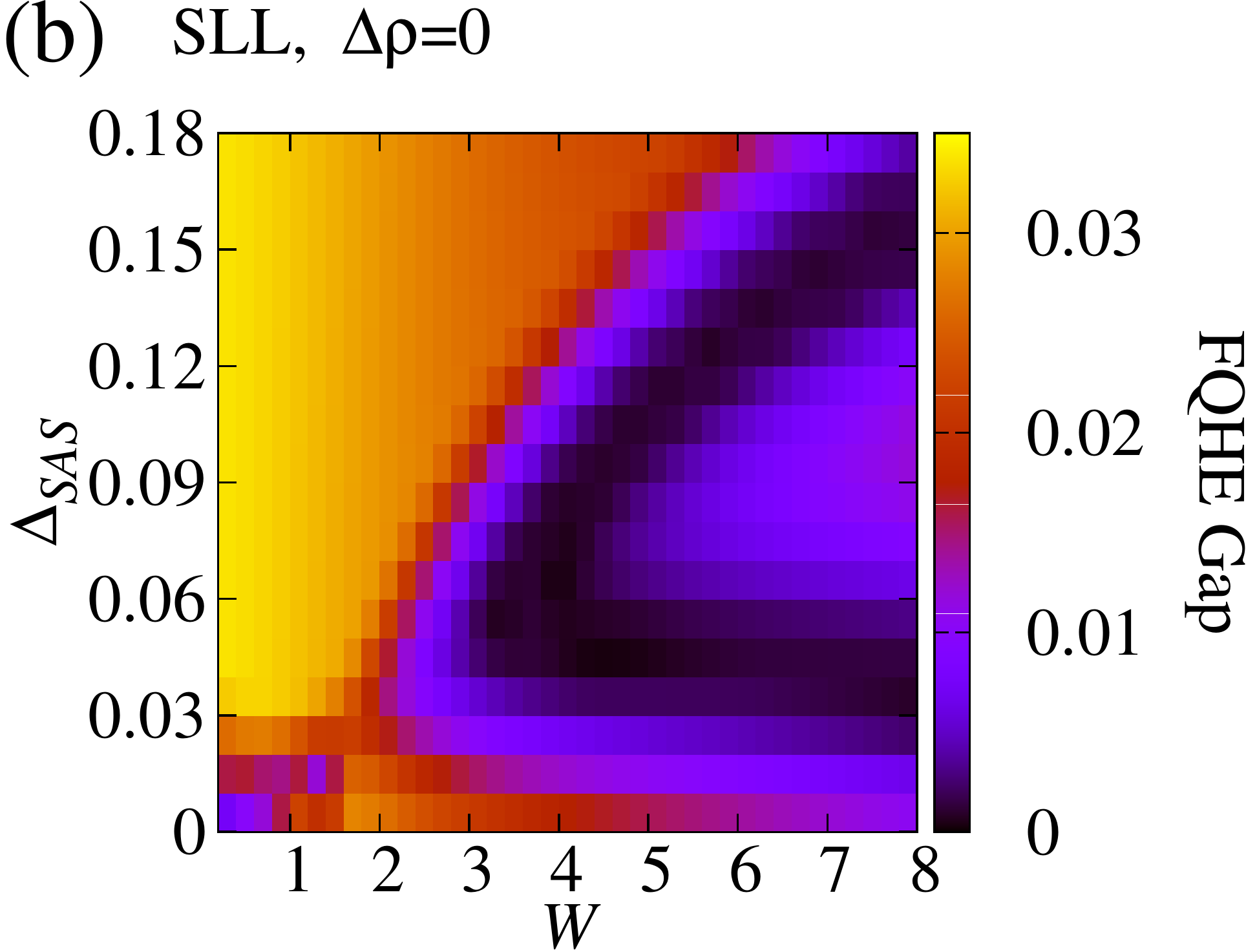}
\end{tabular}
\caption{(Color online) Second Landau level:  (b) FQHE energy gap for the 
WQW model as a function of $W$ and $\Delta_{SAS}$ 
with $\Delta\rho=0$.  Also shown is the (a) overlap between the $SAS$-basis Moore-Read 
Pfaffian (top panel)  
and the layer-basis Halperin 331 (bottom panel) 
wavefunctions and the exact ground state for $\Delta\rho=0$.}
\label{fig-SLL-gap-Dsas-v-Drho-wqw}
\end{figure}

\subsection{Wide-quantum-well}

In Fig.~\ref{fig-SLL-Dsas-v-Drho-Pf-wqw} we plot the overlap between the exact 
ground state of the WQW model and the Moore-Read Pfaffian wavefunction 
written in the layer-basis (top panel) and $SAS$-basis (lower panel).  This figure is, 
of course, similar to, and should be compared to, Fig.~\ref{fig-SLL-Dsas-v-Drho-Pf} 
showing the same thing for the bilayer model.  Qualitatively, the same behavior 
is manifest in the two models.   As the WQW width $W$ is increased it takes a 
larger and larger tunneling strength to increase the overlap with the Pf.  

The overlap between the ground state and the Halperin 331 state is shown 
in Fig.~\ref{fig-SLL-Dsas-v-Drho-331-wqw}.  Again, we only show the overlap 
with the Halperin 331 state written in the layer-basis since the overlap with the 
$SAS$-basis 331 state is almost zero throughout parameter space.  The results 
are qualitatively similar to those of the bilayer model.  However, 
for $W=2$ (Fig.~\ref{fig-SLL-Dsas-v-Drho-331-wqw}b) 
there is a slight maximum in the overlap as $\Delta_{SAS}$ is increased 
from zero at $\Delta_{SAS}\sim0.03$.  This is an interesting result that is 
harder to understand--increasing $\Delta_{SAS}$ drives the system to be 
one-component and one would surmise that increasing this parameter 
would lower the overlap with the 331 state monotonically.  For $W=4$ we 
see that two peaks in the overlap appear; one for small 
values of $\Delta_{SAS}$ and another for a larger value with a strong minimum 
in-between; behavior is particular to the WQW model.  
Similar to the lowest Landau level, the expectation value of $(N_S-N_{AS})/2$ 
for $W\geq2$ is negative for small $\Delta_{SAS}$ and $\Delta\rho$.  In fact, this 
negative value means that the system prefers to have more electrons in the $AS$ state 
than the $S$ state.  However, the system is still more one-component in those regions, such 
as the region in Fig.~\ref{fig-SLL-Dsas-v-Drho-pseudospin-wqw}b along the $\Delta\rho=0$ 
axis near $\Delta_{SAS}\sim0.3$, which correspond to a maximum in the overlap 
with the 331 state.  At any rate, this behavior marks a qualitative difference between the 
bilayer and WQW models.

Lastly, we show the FQHE energy gap for $\Delta\rho=0$ in 
Fig.~\ref{fig-SLL-gap-Dsas-v-Drho-wqw}b along with the overlap between the 
$SAS$-basis Moore-Read Pfaffian state (top panel of Fig.~\ref{fig-SLL-gap-Dsas-v-Drho-wqw}a) 
and the layer-basis Halperin 331 state 
(bottom panel of Fig.~\ref{fig-SLL-gap-Dsas-v-Drho-wqw}a), respectively.  When the 
FQHE energy gap is non-zero, and the system therefore would be expected to 
exhibit the FQHE, the overlap with the Pfaffian state is large.  On the other hand, in the 
region of parameter space where the FQHE energy gap is small, the Halperin 331 
state has a larger overlap.  Thus, independent of our model of choice, the 
details of the electron-electron interaction in the second Landau level, compared to the 
lowest Landau level, make a profound difference.  In the second Landau 
level the Pfaffian wavefunction 
is a good description of the physics when the system is largely one-component.  
In the lowest Landau level, the Halperin 331 is a good description of the physics when the 
system is largely two-component.

Note that there is a peak in the FQHE energy gap for $W\sim2$ and $\Delta_{SAS}\approx0$ 
which is well within the Halperin 331 part of the approximate quantum phase diagram.  Even 
though the overlap with the 331 state is small in that region of parameter space it 
could correspond to a two-component Abelian FQHE for bilayer systems 
in the second Landau level at total $\nu=5/2$ that is in the same universality class 
as the Halperin 331 state.

\section{Conclusions}
\label{sec-conc}

In conclusion, we have investigated the FQHE in two-component systems 
for the half-filled lowest and second Landau levels ($\nu=1/2$ and 
$\nu=5/2$, respectively) as a function of both 
tunneling strengths; inter-layer tunneling and 
charge imbalancing.  This work was motivated by the recent 
experimental systems investigated in Refs.~\onlinecite{shabani-1} 
and~\onlinecite{shabani-2}.  

Our main results are as follows:

(i) The FQHE at $\nu=1/2$ is described by the (Abelian) Halperin 331
two-component state 
which is remarkably robust 
to charge imbalance tunneling.  When the system is driven into the 
one-component region of  parameter space, we find that 
the (non-Abelian) Moore-Read Pfaffian 
state is most likely beaten out 
by other competing non-FQHE phases (cf. striped phase or (Composite Fermion) 
Fermi sea).   The reason we suspect this is because even 
though the overlap between the exact ground state and the Pf (in the $SAS$- 
or layer-basis, depending on the region of parameter space, i.e., whether 
$\Delta_{SAS}$ is large and $\Delta\rho$ is small or vice-versa) is large, the FQHE energy 
gap has a peak that is (slightly) on the Halperin 331 side of the approximate 
quantum phase diagram, where the quantum phase diagram determination 
is described in Sec.~\ref{sec-LLL}.  This result, along with recent numerical 
calculations by Storni \emph{et al.}~\cite{storni} (as well as 
Refs.~\onlinecite{papic-2} and~\onlinecite{papic-3}), 
leads us to conclude that the one-component $\nu=1/2$ is likely to be a non-FQHE striped 
phase or (Composite Fermion) Fermi 
sea.   In the effective BCS description, the increase of tunneling (in the layer-basis) or
charge imbalance (in the $SAS$-basis) leads also to the increase of the 
effective chemical potential
of the ``even" channel which drives the system into a compressible 
phase~\cite{papic-3}.

We also find that our calculations are unable to explain the recent results 
of Shabani \emph{et al.}~\cite{shabani-1,shabani-2}, which consider \emph{only} 
the competing Moore-Read Pfaffian and Halperin 331 states.  In that work, they observe 
no FQHE at $\nu=1/2$ when the charge imbalance is zero, an emerging 
FQHE at $\nu=1/2$ for non-zero charge imbalance, and finally, that the 
FQHE is eventually destroyed upon increasing charge imbalance.  While we 
find that the Halperin 331 FQHE is quite robust to a charge imbalance 
tunneling term, our results would suggest that the FQHE would monotonically 
decrease in strength with increasing charge imbalance (that is, the 
measured activation gap would decrease in strength, the minimum 
in $R_{xx}$ would weaken, and/or the ``quality" of the plateau in $R_{xy}$ 
would deteriorate)--we would expect experiments 
to observe a FQHE at $\nu=1/2$ with zero charge imbalance that is 
eventually destroyed upon further imbalancing.  The fact that this does not happen in the 
experiment indicates that the observed state at large imbalance must be something 
different from the 331 or the Pfaffian state, at least within our model calculations.

(ii) The FQHE in the second Landau level at $\nu=5/2$ is most likely the spin-polarized 
and one-component (non-Abelian) Moore-Read Pfaffian 
state.  This state, similar to the Halperin 331 state in the lowest Landau level, is remarkably
robust to charge imbalancing.  We also find that in regions of 
parameter space when the system is 
largely two-component, i.e., for small $\Delta_{SAS}$ and $\Delta\rho$, the 
system might display a FQHE described by the (Abelian) Halperin 331 state.  This 
suggests~\cite{mrp-sds-bilayer} the exciting possibility of experimentally 
tuning parameters to drive an Abelian FQHE state (331) at $\nu=5/2$ into a 
non-Abelian FQHE state (Pf) at $\nu=5/2$ by way of a quantum 
phase transition. 

(iii) One of our surprising theoretical findings is that the Halperin 
331 two-component bilayer Abelian paired FQHE is extremely robust, and survives, 
not only substantial inter-layer tunneling, but also 
substantial charge imbalance.  We suspect that this result is quite 
general and possibly also applies to other Abelian paired 
states~\cite{scarola-jain} in the same universality class of the 331 state.

The physics of the bilayer FQHE at $\nu=5/2$ is extremely rich with many 
unanswered and fascination questions (not to mention the fact that the bilayer 
problem in higher Landau levels is conceptually difficult and 
non-trivial (see Sec.~\ref{sec-SLL})) that, 
in our opinion, are awaiting experimental answers.

\section{Acknowledgements}
MRP and SDS are grateful to Microsoft Q  for support. ZP was supported by the
Serbian Ministry of Science under Grant No. 141035 and by the Agence Nationale
de la Recherche under Grant No. ANR-JCJC-0003-01. 
MRP thanks Vito Scarola for helpful comments. ZP most gratefully acknowledges
helpful discussions and former collaboration with M. Goerbig and N. Regnault
on setting up the wide-quantum-well model.


\begin{thebibliography}{90}
\expandafter\ifx\csname natexlab\endcsname\relax\def\natexlab#1{#1}\fi
\expandafter\ifx\csname bibnamefont\endcsname\relax
  \def\bibnamefont#1{#1}\fi
\expandafter\ifx\csname bibfnamefont\endcsname\relax
  \def\bibfnamefont#1{#1}\fi
\expandafter\ifx\csname citenamefont\endcsname\relax
  \def\citenamefont#1{#1}\fi
\expandafter\ifx\csname url\endcsname\relax
  \def\url#1{\texttt{#1}}\fi
\expandafter\ifx\csname urlprefix\endcsname\relax\def\urlprefix{URL }\fi
\providecommand{\bibinfo}[2]{#2}
\providecommand{\eprint}[2][]{\url{#2}}

\bibitem[{\citenamefont{He et~al.}(1993)\citenamefont{He, Das~Sarma, and
  Xie}}]{he-1}
\bibinfo{author}{\bibfnamefont{S.}~\bibnamefont{He}},
  \bibinfo{author}{\bibfnamefont{S.}~\bibnamefont{Das~Sarma}},
  \bibnamefont{and} \bibinfo{author}{\bibfnamefont{X.~C.} \bibnamefont{Xie}},
  \bibinfo{journal}{Phys. Rev. B} \textbf{\bibinfo{volume}{47}},
  \bibinfo{pages}{4394} (\bibinfo{year}{1993}).

\bibitem[{\citenamefont{He et~al.}(1991)\citenamefont{He, Xie, Das~Sarma, and
  Zhang}}]{he-2}
\bibinfo{author}{\bibfnamefont{S.}~\bibnamefont{He}},
  \bibinfo{author}{\bibfnamefont{X.~C.} \bibnamefont{Xie}},
  \bibinfo{author}{\bibfnamefont{S.}~\bibnamefont{Das~Sarma}},
  \bibnamefont{and} \bibinfo{author}{\bibfnamefont{F.~C.} \bibnamefont{Zhang}},
  \bibinfo{journal}{Phys. Rev. B} \textbf{\bibinfo{volume}{43}},
  \bibinfo{pages}{9339} (\bibinfo{year}{1991}).

\bibitem[{\citenamefont{Nomura and Yoshioka}(2004)}]{nomura}
\bibinfo{author}{\bibfnamefont{K.}~\bibnamefont{Nomura}} \bibnamefont{and}
  \bibinfo{author}{\bibfnamefont{D.}~\bibnamefont{Yoshioka}},
  \bibinfo{journal}{J. Phys. Soc. Jpn.} \textbf{\bibinfo{volume}{73}},
  \bibinfo{pages}{2612} (\bibinfo{year}{2004}).

\bibitem[{\citenamefont{Papi\'{c} et~al.}(2009)\citenamefont{Papi\'{c},
  M\"{o}ller, Milovanovi\'{c}, Regnault, and Goerbig}}]{papic}
\bibinfo{author}{\bibfnamefont{Z.}~\bibnamefont{Papi\'{c}}},
  \bibinfo{author}{\bibfnamefont{G.}~\bibnamefont{M\"{o}ller}},
  \bibinfo{author}{\bibfnamefont{M.~V.} \bibnamefont{Milovanovi\'{c}}},
  \bibinfo{author}{\bibfnamefont{N.}~\bibnamefont{Regnault}}, \bibnamefont{and}
  \bibinfo{author}{\bibfnamefont{M.~O.} \bibnamefont{Goerbig}},
  \bibinfo{journal}{Phys. Rev. B} \textbf{\bibinfo{volume}{79}},
  \bibinfo{eid}{245325} (\bibinfo{year}{2009}).

\bibitem[{\citenamefont{Peterson and Das~Sarma}(2010)}]{mrp-sds-bilayer}
\bibinfo{author}{\bibfnamefont{M.~R.} \bibnamefont{Peterson}} \bibnamefont{and}
  \bibinfo{author}{\bibfnamefont{S.}~\bibnamefont{Das~Sarma}},
  \bibinfo{journal}{Phys. Rev. B} \textbf{\bibinfo{volume}{81}},
  \bibinfo{pages}{165304} (\bibinfo{year}{2010}).

\bibitem[{\citenamefont{Shabani
  et~al.}(2009{\natexlab{a}})\citenamefont{Shabani, Gokmen, and
  Shayegan}}]{shabani-1}
\bibinfo{author}{\bibfnamefont{J.}~\bibnamefont{Shabani}},
  \bibinfo{author}{\bibfnamefont{T.}~\bibnamefont{Gokmen}}, \bibnamefont{and}
  \bibinfo{author}{\bibfnamefont{M.}~\bibnamefont{Shayegan}},
  \bibinfo{journal}{Phys. Rev. Lett.} \textbf{\bibinfo{volume}{103}},
  \bibinfo{pages}{046805} (\bibinfo{year}{2009}{\natexlab{a}}).

\bibitem[{\citenamefont{Shabani
  et~al.}(2009{\natexlab{b}})\citenamefont{Shabani, Gokmen, Chiu, and
  Shayegan}}]{shabani-2}
\bibinfo{author}{\bibfnamefont{J.}~\bibnamefont{Shabani}},
  \bibinfo{author}{\bibfnamefont{T.}~\bibnamefont{Gokmen}},
  \bibinfo{author}{\bibfnamefont{Y.~T.} \bibnamefont{Chiu}}, \bibnamefont{and}
  \bibinfo{author}{\bibfnamefont{M.}~\bibnamefont{Shayegan}},
  \bibinfo{journal}{Phys. Rev. Lett.} \textbf{\bibinfo{volume}{103}},
  \bibinfo{pages}{256802} (\bibinfo{year}{2009}{\natexlab{b}}).

\bibitem[{\citenamefont{Tsui et~al.}(1982)\citenamefont{Tsui, Stormer, and
  Gossard}}]{tsui-stormer-gossard}
\bibinfo{author}{\bibfnamefont{D.~C.} \bibnamefont{Tsui}},
  \bibinfo{author}{\bibfnamefont{H.~L.} \bibnamefont{Stormer}},
  \bibnamefont{and} \bibinfo{author}{\bibfnamefont{A.~C.}
  \bibnamefont{Gossard}}, \bibinfo{journal}{Phys. Rev. Lett.}
  \textbf{\bibinfo{volume}{48}}, \bibinfo{pages}{1559} (\bibinfo{year}{1982}).

\bibitem[{\citenamefont{Laughlin}(1983)}]{laughlin}
\bibinfo{author}{\bibfnamefont{R.~B.} \bibnamefont{Laughlin}},
  \bibinfo{journal}{Phys. Rev. Lett.} \textbf{\bibinfo{volume}{50}},
  \bibinfo{pages}{1395} (\bibinfo{year}{1983}).

\bibitem[{per()}]{perspectives}
\bibinfo{note}{\textit{Perspectives in Quantum Hall Effects}, edited by S. Das
  Sarma and A. Pinczuk (Wiley, New York, 1997).}

\bibitem[{hei()}]{heinonen}
\bibinfo{note}{\textit{Composite Fermions: A Unified View of the Quantum Hall
  Regime}, edited by O. Heinonen (World Scientific, New Jersey, 1998).}

\bibitem[{\citenamefont{Jain}(2007)}]{cf-book}
\bibinfo{author}{\bibfnamefont{J.~K.} \bibnamefont{Jain}},
  \emph{\bibinfo{title}{Composite Fermions}} (\bibinfo{publisher}{Cambridge
  University Press}, \bibinfo{address}{New York}, \bibinfo{year}{2007}).

\bibitem[{\citenamefont{Jain}(1989)}]{jain-prl}
\bibinfo{author}{\bibfnamefont{J.~K.} \bibnamefont{Jain}},
  \bibinfo{journal}{Phys. Rev. Lett.} \textbf{\bibinfo{volume}{63}},
  \bibinfo{pages}{199} (\bibinfo{year}{1989}).

\bibitem[{\citenamefont{Wen and Niu}(1990)}]{wen}
\bibinfo{author}{\bibfnamefont{X.~G.} \bibnamefont{Wen}} \bibnamefont{and}
  \bibinfo{author}{\bibfnamefont{Q.}~\bibnamefont{Niu}},
  \bibinfo{journal}{Phys. Rev. B} \textbf{\bibinfo{volume}{41}},
  \bibinfo{pages}{9377} (\bibinfo{year}{1990}).

\bibitem[{\citenamefont{Willett et~al.}(1987)\citenamefont{Willett, Eisenstein,
  St\"ormer, Tsui, Gossard, and English}}]{willett}
\bibinfo{author}{\bibfnamefont{R.}~\bibnamefont{Willett}},
  \bibinfo{author}{\bibfnamefont{J.~P.} \bibnamefont{Eisenstein}},
  \bibinfo{author}{\bibfnamefont{H.~L.} \bibnamefont{St\"ormer}},
  \bibinfo{author}{\bibfnamefont{D.~C.} \bibnamefont{Tsui}},
  \bibinfo{author}{\bibfnamefont{A.~C.} \bibnamefont{Gossard}},
  \bibnamefont{and} \bibinfo{author}{\bibfnamefont{J.~H.}
  \bibnamefont{English}}, \bibinfo{journal}{Phys. Rev. Lett.}
  \textbf{\bibinfo{volume}{59}}, \bibinfo{pages}{1776} (\bibinfo{year}{1987}).

\bibitem[{\citenamefont{Halperin et~al.}(1993)\citenamefont{Halperin, Lee, and
  Read}}]{hlr}
\bibinfo{author}{\bibfnamefont{B.~I.} \bibnamefont{Halperin}},
  \bibinfo{author}{\bibfnamefont{P.~A.} \bibnamefont{Lee}}, \bibnamefont{and}
  \bibinfo{author}{\bibfnamefont{N.}~\bibnamefont{Read}},
  \bibinfo{journal}{Phys. Rev. B} \textbf{\bibinfo{volume}{47}},
  \bibinfo{pages}{7312} (\bibinfo{year}{1993}).

\bibitem[{\citenamefont{Kalmeyer and Zhang}(1992)}]{kalmeyer}
\bibinfo{author}{\bibfnamefont{V.}~\bibnamefont{Kalmeyer}} \bibnamefont{and}
  \bibinfo{author}{\bibfnamefont{S.-C.} \bibnamefont{Zhang}},
  \bibinfo{journal}{Phys. Rev. B} \textbf{\bibinfo{volume}{46}},
  \bibinfo{pages}{9889} (\bibinfo{year}{1992}).

\bibitem[{\citenamefont{Rezayi and Read}(1994)}]{rezayi-read}
\bibinfo{author}{\bibfnamefont{E.}~\bibnamefont{Rezayi}} \bibnamefont{and}
  \bibinfo{author}{\bibfnamefont{N.}~\bibnamefont{Read}},
  \bibinfo{journal}{Phys. Rev. Lett.} \textbf{\bibinfo{volume}{72}},
  \bibinfo{pages}{900} (\bibinfo{year}{1994}).

\bibitem[{\citenamefont{Willett et~al.}(1993)\citenamefont{Willett, Ruel, West,
  and Pfeiffer}}]{willett-1}
\bibinfo{author}{\bibfnamefont{R.~L.} \bibnamefont{Willett}},
  \bibinfo{author}{\bibfnamefont{R.~R.} \bibnamefont{Ruel}},
  \bibinfo{author}{\bibfnamefont{K.~W.} \bibnamefont{West}}, \bibnamefont{and}
  \bibinfo{author}{\bibfnamefont{L.~N.} \bibnamefont{Pfeiffer}},
  \bibinfo{journal}{Phys. Rev. Lett.} \textbf{\bibinfo{volume}{71}},
  \bibinfo{pages}{3846} (\bibinfo{year}{1993}).

\bibitem[{\citenamefont{Du et~al.}(1993)\citenamefont{Du, Stormer, Tsui,
  Pfeiffer, and West}}]{du}
\bibinfo{author}{\bibfnamefont{R.~R.} \bibnamefont{Du}},
  \bibinfo{author}{\bibfnamefont{H.~L.} \bibnamefont{Stormer}},
  \bibinfo{author}{\bibfnamefont{D.~C.} \bibnamefont{Tsui}},
  \bibinfo{author}{\bibfnamefont{L.~N.} \bibnamefont{Pfeiffer}},
  \bibnamefont{and} \bibinfo{author}{\bibfnamefont{K.~W.} \bibnamefont{West}},
  \bibinfo{journal}{Phys. Rev. Lett.} \textbf{\bibinfo{volume}{70}},
  \bibinfo{pages}{2944} (\bibinfo{year}{1993}).

\bibitem[{\citenamefont{Kang et~al.}(1993)\citenamefont{Kang, Stormer,
  Pfeiffer, Baldwin, and West}}]{kang-1}
\bibinfo{author}{\bibfnamefont{W.}~\bibnamefont{Kang}},
  \bibinfo{author}{\bibfnamefont{H.~L.} \bibnamefont{Stormer}},
  \bibinfo{author}{\bibfnamefont{L.~N.} \bibnamefont{Pfeiffer}},
  \bibinfo{author}{\bibfnamefont{K.~W.} \bibnamefont{Baldwin}},
  \bibnamefont{and} \bibinfo{author}{\bibfnamefont{K.~W.} \bibnamefont{West}},
  \bibinfo{journal}{Phys. Rev. Lett.} \textbf{\bibinfo{volume}{71}},
  \bibinfo{pages}{3850} (\bibinfo{year}{1993}).

\bibitem[{\citenamefont{Eisenstein et~al.}(1988)\citenamefont{Eisenstein,
  Willett, Stormer, Tsui, Gossard, and English}}]{eisenstein-52-1}
\bibinfo{author}{\bibfnamefont{J.~P.} \bibnamefont{Eisenstein}},
  \bibinfo{author}{\bibfnamefont{R.}~\bibnamefont{Willett}},
  \bibinfo{author}{\bibfnamefont{H.~L.} \bibnamefont{Stormer}},
  \bibinfo{author}{\bibfnamefont{D.~C.} \bibnamefont{Tsui}},
  \bibinfo{author}{\bibfnamefont{A.~C.} \bibnamefont{Gossard}},
  \bibnamefont{and} \bibinfo{author}{\bibfnamefont{J.~H.}
  \bibnamefont{English}}, \bibinfo{journal}{Phys. Rev. Lett.}
  \textbf{\bibinfo{volume}{61}}, \bibinfo{pages}{997} (\bibinfo{year}{1988}).

\bibitem[{\citenamefont{Gammel et~al.}(1988)\citenamefont{Gammel, Bishop,
  Eisenstein, English, Gossard, Ruel, and Stormer}}]{gammel}
\bibinfo{author}{\bibfnamefont{P.~L.} \bibnamefont{Gammel}},
  \bibinfo{author}{\bibfnamefont{D.~J.} \bibnamefont{Bishop}},
  \bibinfo{author}{\bibfnamefont{J.~P.} \bibnamefont{Eisenstein}},
  \bibinfo{author}{\bibfnamefont{J.~H.} \bibnamefont{English}},
  \bibinfo{author}{\bibfnamefont{A.~C.} \bibnamefont{Gossard}},
  \bibinfo{author}{\bibfnamefont{R.}~\bibnamefont{Ruel}}, \bibnamefont{and}
  \bibinfo{author}{\bibfnamefont{H.~L.} \bibnamefont{Stormer}},
  \bibinfo{journal}{Phys. Rev. B} \textbf{\bibinfo{volume}{38}},
  \bibinfo{pages}{10128} (\bibinfo{year}{1988}).

\bibitem[{\citenamefont{Pan et~al.}(1999)\citenamefont{Pan, Xia, Shvarts,
  Adams, Stormer, Tsui, Pfeiffer, Baldwin, and West}}]{pan-prl}
\bibinfo{author}{\bibfnamefont{W.}~\bibnamefont{Pan}},
  \bibinfo{author}{\bibfnamefont{J.-S.} \bibnamefont{Xia}},
  \bibinfo{author}{\bibfnamefont{V.}~\bibnamefont{Shvarts}},
  \bibinfo{author}{\bibfnamefont{D.~E.} \bibnamefont{Adams}},
  \bibinfo{author}{\bibfnamefont{H.~L.} \bibnamefont{Stormer}},
  \bibinfo{author}{\bibfnamefont{D.~C.} \bibnamefont{Tsui}},
  \bibinfo{author}{\bibfnamefont{L.~N.} \bibnamefont{Pfeiffer}},
  \bibinfo{author}{\bibfnamefont{K.~W.} \bibnamefont{Baldwin}},
  \bibnamefont{and} \bibinfo{author}{\bibfnamefont{K.~W.} \bibnamefont{West}},
  \bibinfo{journal}{Phys. Rev. Lett.} \textbf{\bibinfo{volume}{83}},
  \bibinfo{pages}{3530} (\bibinfo{year}{1999}).

\bibitem[{\citenamefont{Eisenstein et~al.}(2002)\citenamefont{Eisenstein,
  Cooper, Pfeiffer, and West}}]{eisenstein-52-2}
\bibinfo{author}{\bibfnamefont{J.~P.} \bibnamefont{Eisenstein}},
  \bibinfo{author}{\bibfnamefont{K.~B.} \bibnamefont{Cooper}},
  \bibinfo{author}{\bibfnamefont{L.~N.} \bibnamefont{Pfeiffer}},
  \bibnamefont{and} \bibinfo{author}{\bibfnamefont{K.~W.} \bibnamefont{West}},
  \bibinfo{journal}{Phys. Rev. Lett.} \textbf{\bibinfo{volume}{88}},
  \bibinfo{pages}{076801} (\bibinfo{year}{2002}).

\bibitem[{\citenamefont{Xia et~al.}(2004)\citenamefont{Xia, Pan, Vicente,
  Adams, Sullivan, Stormer, Tsui, Pfeiffer, Baldwin, and West}}]{xia}
\bibinfo{author}{\bibfnamefont{J.~S.} \bibnamefont{Xia}},
  \bibinfo{author}{\bibfnamefont{W.}~\bibnamefont{Pan}},
  \bibinfo{author}{\bibfnamefont{C.~L.} \bibnamefont{Vicente}},
  \bibinfo{author}{\bibfnamefont{E.~D.} \bibnamefont{Adams}},
  \bibinfo{author}{\bibfnamefont{N.~S.} \bibnamefont{Sullivan}},
  \bibinfo{author}{\bibfnamefont{H.~L.} \bibnamefont{Stormer}},
  \bibinfo{author}{\bibfnamefont{D.~C.} \bibnamefont{Tsui}},
  \bibinfo{author}{\bibfnamefont{L.~N.} \bibnamefont{Pfeiffer}},
  \bibinfo{author}{\bibfnamefont{K.~W.} \bibnamefont{Baldwin}},
  \bibnamefont{and} \bibinfo{author}{\bibfnamefont{K.~W.} \bibnamefont{West}},
  \bibinfo{journal}{Phys. Rev. Lett.} \textbf{\bibinfo{volume}{93}},
  \bibinfo{pages}{176809} (\bibinfo{year}{2004}).

\bibitem[{\citenamefont{Cs\'athy et~al.}(2005)\citenamefont{Cs\'athy, Xia,
  Vicente, Adams, Sullivan, Stormer, Tsui, Pfeiffer, and West}}]{csathy}
\bibinfo{author}{\bibfnamefont{G.~A.} \bibnamefont{Cs\'athy}},
  \bibinfo{author}{\bibfnamefont{J.~S.} \bibnamefont{Xia}},
  \bibinfo{author}{\bibfnamefont{C.~L.} \bibnamefont{Vicente}},
  \bibinfo{author}{\bibfnamefont{E.~D.} \bibnamefont{Adams}},
  \bibinfo{author}{\bibfnamefont{N.~S.} \bibnamefont{Sullivan}},
  \bibinfo{author}{\bibfnamefont{H.~L.} \bibnamefont{Stormer}},
  \bibinfo{author}{\bibfnamefont{D.~C.} \bibnamefont{Tsui}},
  \bibinfo{author}{\bibfnamefont{L.~N.} \bibnamefont{Pfeiffer}},
  \bibnamefont{and} \bibinfo{author}{\bibfnamefont{K.~W.} \bibnamefont{West}},
  \bibinfo{journal}{Phys. Rev. Lett.} \textbf{\bibinfo{volume}{94}},
  \bibinfo{pages}{146801} (\bibinfo{year}{2005}).

\bibitem[{\citenamefont{Choi et~al.}(2008)\citenamefont{Choi, Kang, Das~Sarma,
  Pfeiffer, and West}}]{choi}
\bibinfo{author}{\bibfnamefont{H.~C.} \bibnamefont{Choi}},
  \bibinfo{author}{\bibfnamefont{W.}~\bibnamefont{Kang}},
  \bibinfo{author}{\bibfnamefont{S.}~\bibnamefont{Das~Sarma}},
  \bibinfo{author}{\bibfnamefont{L.~N.} \bibnamefont{Pfeiffer}},
  \bibnamefont{and} \bibinfo{author}{\bibfnamefont{K.~W.} \bibnamefont{West}},
  \bibinfo{journal}{Phys. Rev. B} \textbf{\bibinfo{volume}{77}},
  \bibinfo{pages}{081301} (\bibinfo{year}{2008}).

\bibitem[{\citenamefont{Nuebler et~al.}(2010)\citenamefont{Nuebler, Umansky,
  Morf, Heiblum, von Klitzing, and Smet}}]{nuebler}
\bibinfo{author}{\bibfnamefont{J.}~\bibnamefont{Nuebler}},
  \bibinfo{author}{\bibfnamefont{V.}~\bibnamefont{Umansky}},
  \bibinfo{author}{\bibfnamefont{R.}~\bibnamefont{Morf}},
  \bibinfo{author}{\bibfnamefont{M.}~\bibnamefont{Heiblum}},
  \bibinfo{author}{\bibfnamefont{K.}~\bibnamefont{von Klitzing}},
  \bibnamefont{and} \bibinfo{author}{\bibfnamefont{J.}~\bibnamefont{Smet}},
  \bibinfo{journal}{Phys. Rev. B} \textbf{\bibinfo{volume}{81}},
  \bibinfo{pages}{035316} (\bibinfo{year}{2010}).

\bibitem[{\citenamefont{Dean et~al.}(2008{\natexlab{a}})\citenamefont{Dean,
  Piot, Hayden, Das~Sarma, Gervais, Pfeiffer, and West}}]{dean-2}
\bibinfo{author}{\bibfnamefont{C.~R.} \bibnamefont{Dean}},
  \bibinfo{author}{\bibfnamefont{B.~A.} \bibnamefont{Piot}},
  \bibinfo{author}{\bibfnamefont{P.}~\bibnamefont{Hayden}},
  \bibinfo{author}{\bibfnamefont{S.}~\bibnamefont{Das~Sarma}},
  \bibinfo{author}{\bibfnamefont{G.}~\bibnamefont{Gervais}},
  \bibinfo{author}{\bibfnamefont{L.~N.} \bibnamefont{Pfeiffer}},
  \bibnamefont{and} \bibinfo{author}{\bibfnamefont{K.~W.} \bibnamefont{West}},
  \bibinfo{journal}{Phys. Rev. Lett.} \textbf{\bibinfo{volume}{101}},
  \bibinfo{pages}{186806} (\bibinfo{year}{2008}{\natexlab{a}}).

\bibitem[{sds()}]{sds}
\bibinfo{note}{S.~Das Sarma, G. Gervais and X. Zho, arXiv:1007.1688
  (unpublished).}

\bibitem[{\citenamefont{Dean et~al.}(2008{\natexlab{b}})\citenamefont{Dean,
  Piot, Hayden, Sarma, Gervais, Pfeiffer, and West}}]{dean}
\bibinfo{author}{\bibfnamefont{C.~R.} \bibnamefont{Dean}},
  \bibinfo{author}{\bibfnamefont{B.~A.} \bibnamefont{Piot}},
  \bibinfo{author}{\bibfnamefont{P.}~\bibnamefont{Hayden}},
  \bibinfo{author}{\bibfnamefont{S.~D.} \bibnamefont{Sarma}},
  \bibinfo{author}{\bibfnamefont{G.}~\bibnamefont{Gervais}},
  \bibinfo{author}{\bibfnamefont{L.~N.} \bibnamefont{Pfeiffer}},
  \bibnamefont{and} \bibinfo{author}{\bibfnamefont{K.~W.} \bibnamefont{West}},
  \bibinfo{journal}{Phys. Rev. Lett.} \textbf{\bibinfo{volume}{100}},
  \bibinfo{eid}{146803} (\bibinfo{year}{2008}{\natexlab{b}}).

\bibitem[{\citenamefont{Dean et~al.}(2009)\citenamefont{Dean, Piot, Gervais,
  Pfeiffer, and West}}]{dean-3}
\bibinfo{author}{\bibfnamefont{C.~R.} \bibnamefont{Dean}},
  \bibinfo{author}{\bibfnamefont{B.~A.} \bibnamefont{Piot}},
  \bibinfo{author}{\bibfnamefont{G.}~\bibnamefont{Gervais}},
  \bibinfo{author}{\bibfnamefont{L.~N.} \bibnamefont{Pfeiffer}},
  \bibnamefont{and} \bibinfo{author}{\bibfnamefont{K.~W.} \bibnamefont{West}},
  \bibinfo{journal}{Phys. Rev. B} \textbf{\bibinfo{volume}{80}},
  \bibinfo{pages}{153301} (\bibinfo{year}{2009}).

\bibitem[{rho()}]{rhone}
\bibinfo{note}{T.~D. Rhone, American Physical Society March Meeting Invited
  Talk Y2.00003 (unpublished).}

\bibitem[{ste()}]{stern}
\bibinfo{note}{M. Stern, P. Plochocka, V. Umansky, D.~K. Maude, M. Potemski, I.
  Bar-Joseph, arXiv:1005.3112 (unpublished).}

\bibitem[{\citenamefont{Bishara and Nayak}(2009)}]{bishara-nayak}
\bibinfo{author}{\bibfnamefont{W.}~\bibnamefont{Bishara}} \bibnamefont{and}
  \bibinfo{author}{\bibfnamefont{C.}~\bibnamefont{Nayak}},
  \bibinfo{journal}{Phys. Rev. B} \textbf{\bibinfo{volume}{80}},
  \bibinfo{pages}{121302} (\bibinfo{year}{2009}).

\bibitem[{sim()}]{simon-rezayi}
\bibinfo{note}{E.~H. Rezayi and S.~H. Simon, arXiv:0912.0109 (unpublished).}

\bibitem[{woj()}]{wojs-toke-jain}
\bibinfo{note}{A. Wojs, C. T\H{o}ke and J.~K. Jain, arXiv:1005.4365 (unpublished).}

\bibitem[{\citenamefont{Wang et~al.}(2009)\citenamefont{Wang, Sheng, and
  Haldane}}]{wang-sheng-haldane}
\bibinfo{author}{\bibfnamefont{H.}~\bibnamefont{Wang}},
  \bibinfo{author}{\bibfnamefont{D.~N.} \bibnamefont{Sheng}}, \bibnamefont{and}
  \bibinfo{author}{\bibfnamefont{F.~D.~M.} \bibnamefont{Haldane}},
  \bibinfo{journal}{Phys. Rev. B} \textbf{\bibinfo{volume}{80}},
  \bibinfo{pages}{241311} (\bibinfo{year}{2009}).

\bibitem[{\citenamefont{Pan et~al.}(2001)\citenamefont{Pan, Stormer, Tsui,
  Pfeiffer, Baldwin, and West}}]{pan-ssc}
\bibinfo{author}{\bibfnamefont{W.}~\bibnamefont{Pan}},
  \bibinfo{author}{\bibfnamefont{H.~L.} \bibnamefont{Stormer}},
  \bibinfo{author}{\bibfnamefont{D.~C.} \bibnamefont{Tsui}},
  \bibinfo{author}{\bibfnamefont{L.~N.} \bibnamefont{Pfeiffer}},
  \bibinfo{author}{\bibfnamefont{K.~W.} \bibnamefont{Baldwin}},
  \bibnamefont{and} \bibinfo{author}{\bibfnamefont{K.~W.} \bibnamefont{West}},
  \bibinfo{journal}{Solid State Communications} \textbf{\bibinfo{volume}{119}},
  \bibinfo{pages}{641 } (\bibinfo{year}{2001}), ISSN \bibinfo{issn}{0038-1098}.

\bibitem[{\citenamefont{Zhang et~al.}(2010)\citenamefont{Zhang, Knuuttila, Dai,
  Du, Pfeiffer, and West}}]{zhang-prl}
\bibinfo{author}{\bibfnamefont{C.}~\bibnamefont{Zhang}},
  \bibinfo{author}{\bibfnamefont{T.}~\bibnamefont{Knuuttila}},
  \bibinfo{author}{\bibfnamefont{Y.}~\bibnamefont{Dai}},
  \bibinfo{author}{\bibfnamefont{R.~R.} \bibnamefont{Du}},
  \bibinfo{author}{\bibfnamefont{L.~N.} \bibnamefont{Pfeiffer}},
  \bibnamefont{and} \bibinfo{author}{\bibfnamefont{K.~W.} \bibnamefont{West}},
  \bibinfo{journal}{Phys. Rev. Lett.} \textbf{\bibinfo{volume}{104}},
  \bibinfo{pages}{166801} (\bibinfo{year}{2010}).

\bibitem[{\citenamefont{Morf}(1998)}]{morf}
\bibinfo{author}{\bibfnamefont{R.~H.} \bibnamefont{Morf}},
  \bibinfo{journal}{Phys. Rev. Lett.} \textbf{\bibinfo{volume}{80}},
  \bibinfo{pages}{1505} (\bibinfo{year}{1998}).

\bibitem[{\citenamefont{Feiguin et~al.}(2008)\citenamefont{Feiguin, Rezayi,
  Nayak, and Das~Sarma}}]{feiguin}
\bibinfo{author}{\bibfnamefont{A.~E.} \bibnamefont{Feiguin}},
  \bibinfo{author}{\bibfnamefont{E.}~\bibnamefont{Rezayi}},
  \bibinfo{author}{\bibfnamefont{C.}~\bibnamefont{Nayak}}, \bibnamefont{and}
  \bibinfo{author}{\bibfnamefont{S.}~\bibnamefont{Das~Sarma}},
  \bibinfo{journal}{Phys. Rev. Lett.} \textbf{\bibinfo{volume}{100}},
  \bibinfo{pages}{166803} (\bibinfo{year}{2008}).

\bibitem[{\citenamefont{Kitaev}(2003)}]{kitaev}
\bibinfo{author}{\bibfnamefont{A.~Y.} \bibnamefont{Kitaev}},
  \bibinfo{journal}{Annals of Physics} \textbf{\bibinfo{volume}{303}},
  \bibinfo{pages}{2 } (\bibinfo{year}{2003}), ISSN \bibinfo{issn}{0003-4916}.

\bibitem[{\citenamefont{Nayak et~al.}(2008)\citenamefont{Nayak, Simon, Stern,
  Freedman, and Sarma}}]{tqc-rmp}
\bibinfo{author}{\bibfnamefont{C.}~\bibnamefont{Nayak}},
  \bibinfo{author}{\bibfnamefont{S.~H.} \bibnamefont{Simon}},
  \bibinfo{author}{\bibfnamefont{A.}~\bibnamefont{Stern}},
  \bibinfo{author}{\bibfnamefont{M.}~\bibnamefont{Freedman}}, \bibnamefont{and}
  \bibinfo{author}{\bibfnamefont{S.~D.} \bibnamefont{Sarma}},
  \bibinfo{journal}{Rev. Mod. Phys.} \textbf{\bibinfo{volume}{80}},
  \bibinfo{eid}{1083} (\bibinfo{year}{2008}).

\bibitem[{\citenamefont{Moore and Read}(1991)}]{mr-pf}
\bibinfo{author}{\bibfnamefont{G.}~\bibnamefont{Moore}} \bibnamefont{and}
  \bibinfo{author}{\bibfnamefont{N.}~\bibnamefont{Read}},
  \bibinfo{journal}{Nucl. Phys. B} \textbf{\bibinfo{volume}{360}},
  \bibinfo{pages}{362} (\bibinfo{year}{1991}).

\bibitem[{\citenamefont{Das~Sarma et~al.}(2005)\citenamefont{Das~Sarma,
  Freedman, and Nayak}}]{sds-freedman-nayak}
\bibinfo{author}{\bibfnamefont{S.}~\bibnamefont{Das~Sarma}},
  \bibinfo{author}{\bibfnamefont{M.}~\bibnamefont{Freedman}}, \bibnamefont{and}
  \bibinfo{author}{\bibfnamefont{C.}~\bibnamefont{Nayak}},
  \bibinfo{journal}{Phys. Rev. Lett.} \textbf{\bibinfo{volume}{94}},
  \bibinfo{pages}{166802} (\bibinfo{year}{2005}).

\bibitem[{\citenamefont{Bardeen
  et~al.}(1957{\natexlab{a}})\citenamefont{Bardeen, Cooper, and
  Schrieffer}}]{bcs}
\bibinfo{author}{\bibfnamefont{J.}~\bibnamefont{Bardeen}},
  \bibinfo{author}{\bibfnamefont{L.~N.} \bibnamefont{Cooper}},
  \bibnamefont{and} \bibinfo{author}{\bibfnamefont{J.~R.}
  \bibnamefont{Schrieffer}}, \bibinfo{journal}{Phys. Rev.}
  \textbf{\bibinfo{volume}{106}}, \bibinfo{pages}{162}
  (\bibinfo{year}{1957}{\natexlab{a}}).

\bibitem[{\citenamefont{Bardeen
  et~al.}(1957{\natexlab{b}})\citenamefont{Bardeen, Cooper, and
  Schrieffer}}]{bcs-2}
\bibinfo{author}{\bibfnamefont{J.}~\bibnamefont{Bardeen}},
  \bibinfo{author}{\bibfnamefont{L.~N.} \bibnamefont{Cooper}},
  \bibnamefont{and} \bibinfo{author}{\bibfnamefont{J.~R.}
  \bibnamefont{Schrieffer}}, \bibinfo{journal}{Phys. Rev.}
  \textbf{\bibinfo{volume}{108}}, \bibinfo{pages}{1175}
  (\bibinfo{year}{1957}{\natexlab{b}}).

\bibitem[{\citenamefont{Read and Green}(2000)}]{read-green}
\bibinfo{author}{\bibfnamefont{N.}~\bibnamefont{Read}} \bibnamefont{and}
  \bibinfo{author}{\bibfnamefont{D.}~\bibnamefont{Green}},
  \bibinfo{journal}{Phys. Rev. B} \textbf{\bibinfo{volume}{61}},
  \bibinfo{pages}{10267} (\bibinfo{year}{2000}).

\bibitem[{\citenamefont{Lee et~al.}(2007)\citenamefont{Lee, Ryu, Nayak, and
  Fisher}}]{apf-1}
\bibinfo{author}{\bibfnamefont{S.-S.} \bibnamefont{Lee}},
  \bibinfo{author}{\bibfnamefont{S.}~\bibnamefont{Ryu}},
  \bibinfo{author}{\bibfnamefont{C.}~\bibnamefont{Nayak}}, \bibnamefont{and}
  \bibinfo{author}{\bibfnamefont{M.~P.~A.} \bibnamefont{Fisher}},
  \bibinfo{journal}{Phys. Rev. Lett.} \textbf{\bibinfo{volume}{99}},
  \bibinfo{pages}{236807} (\bibinfo{year}{2007}).

\bibitem[{\citenamefont{Levin et~al.}(2007)\citenamefont{Levin, Halperin, and
  Rosenow}}]{apf-2}
\bibinfo{author}{\bibfnamefont{M.}~\bibnamefont{Levin}},
  \bibinfo{author}{\bibfnamefont{B.~I.} \bibnamefont{Halperin}},
  \bibnamefont{and} \bibinfo{author}{\bibfnamefont{B.}~\bibnamefont{Rosenow}},
  \bibinfo{journal}{Phys. Rev. Lett.} \textbf{\bibinfo{volume}{99}},
  \bibinfo{pages}{236806} (\bibinfo{year}{2007}).

\bibitem[{\citenamefont{Peterson
  et~al.}(2008{\natexlab{a}})\citenamefont{Peterson, Park, and
  Das~Sarma}}]{mrp-park-sds}
\bibinfo{author}{\bibfnamefont{M.~R.} \bibnamefont{Peterson}},
  \bibinfo{author}{\bibfnamefont{K.}~\bibnamefont{Park}}, \bibnamefont{and}
  \bibinfo{author}{\bibfnamefont{S.}~\bibnamefont{Das~Sarma}},
  \bibinfo{journal}{Phys. Rev. Lett.} \textbf{\bibinfo{volume}{101}},
  \bibinfo{pages}{156803} (\bibinfo{year}{2008}{\natexlab{a}}).

\bibitem[{\citenamefont{d'Ambrumenil and Reynolds}(1988)}]{abrumenil}
\bibinfo{author}{\bibfnamefont{N.}~\bibnamefont{d'Ambrumenil}}
  \bibnamefont{and} \bibinfo{author}{\bibfnamefont{A.~M.}
  \bibnamefont{Reynolds}}, \bibinfo{journal}{Journal of Physics C: Solid State
  Physics} \textbf{\bibinfo{volume}{21}}, \bibinfo{pages}{119}
  (\bibinfo{year}{1988}).

\bibitem[{\citenamefont{MacDonald}(1984)}]{macdonald}
\bibinfo{author}{\bibfnamefont{A.~H.} \bibnamefont{MacDonald}},
  \bibinfo{journal}{Phys. Rev. B} \textbf{\bibinfo{volume}{30}},
  \bibinfo{pages}{3550} (\bibinfo{year}{1984}).

\bibitem[{\citenamefont{Scarola et~al.}(2000)\citenamefont{Scarola, Park, and
  Jain}}]{scarola-park-jain}
\bibinfo{author}{\bibfnamefont{V.~W.} \bibnamefont{Scarola}},
  \bibinfo{author}{\bibfnamefont{K.}~\bibnamefont{Park}}, \bibnamefont{and}
  \bibinfo{author}{\bibfnamefont{J.~K.} \bibnamefont{Jain}},
  \bibinfo{journal}{Phys. Rev. B} \textbf{\bibinfo{volume}{62}},
  \bibinfo{pages}{R16259} (\bibinfo{year}{2000}).

\bibitem[{\citenamefont{T\"oke et~al.}(2005)\citenamefont{T\"oke, Peterson,
  Jeon, and Jain}}]{toke-1}
\bibinfo{author}{\bibfnamefont{C.}~\bibnamefont{T\"oke}},
  \bibinfo{author}{\bibfnamefont{M.~R.} \bibnamefont{Peterson}},
  \bibinfo{author}{\bibfnamefont{G.~S.} \bibnamefont{Jeon}}, \bibnamefont{and}
  \bibinfo{author}{\bibfnamefont{J.~K.} \bibnamefont{Jain}},
  \bibinfo{journal}{Phys. Rev. B} \textbf{\bibinfo{volume}{72}},
  \bibinfo{pages}{125315} (\bibinfo{year}{2005}).

\bibitem[{\citenamefont{Storni et~al.}(2010)\citenamefont{Storni, Morf, and
  Das~Sarma}}]{storni}
\bibinfo{author}{\bibfnamefont{M.}~\bibnamefont{Storni}},
  \bibinfo{author}{\bibfnamefont{R.~H.} \bibnamefont{Morf}}, \bibnamefont{and}
  \bibinfo{author}{\bibfnamefont{S.}~\bibnamefont{Das~Sarma}},
  \bibinfo{journal}{Phys. Rev. Lett.} \textbf{\bibinfo{volume}{104}},
  \bibinfo{pages}{076803} (\bibinfo{year}{2010}).

\bibitem[{\citenamefont{Peterson
  et~al.}(2008{\natexlab{b}})\citenamefont{Peterson, Jolicoeur, and
  Sarma}}]{mrp-ft-prl}
\bibinfo{author}{\bibfnamefont{M.~R.} \bibnamefont{Peterson}},
  \bibinfo{author}{\bibfnamefont{T.}~\bibnamefont{Jolicoeur}},
  \bibnamefont{and} \bibinfo{author}{\bibfnamefont{S.~D.} \bibnamefont{Sarma}},
  \bibinfo{journal}{Phys. Rev. Lett.} \textbf{\bibinfo{volume}{101}},
  \bibinfo{eid}{016807} (\bibinfo{year}{2008}{\natexlab{b}}).

\bibitem[{\citenamefont{Peterson
  et~al.}(2008{\natexlab{c}})\citenamefont{Peterson, Jolicoeur, and
  Sarma}}]{mrp-ft-prb}
\bibinfo{author}{\bibfnamefont{M.~R.} \bibnamefont{Peterson}},
  \bibinfo{author}{\bibfnamefont{T.}~\bibnamefont{Jolicoeur}},
  \bibnamefont{and} \bibinfo{author}{\bibfnamefont{S.~D.} \bibnamefont{Sarma}},
  \bibinfo{journal}{Phys. Rev. B)} \textbf{\bibinfo{volume}{78}},
  \bibinfo{eid}{155308} (\bibinfo{year}{2008}{\natexlab{c}}).

\bibitem[{\citenamefont{Papi\ifmmode~\acute{c}\else \'{c}\fi{}
  et~al.}(2009)\citenamefont{Papi\ifmmode~\acute{c}\else \'{c}\fi{}, Regnault,
  and Das~Sarma}}]{papic-2}
\bibinfo{author}{\bibfnamefont{Z.}~\bibnamefont{Papi\ifmmode~\acute{c}\else
  \'{c}\fi{}}}, \bibinfo{author}{\bibfnamefont{N.}~\bibnamefont{Regnault}},
  \bibnamefont{and}
  \bibinfo{author}{\bibfnamefont{S.}~\bibnamefont{Das~Sarma}},
  \bibinfo{journal}{Phys. Rev. B} \textbf{\bibinfo{volume}{80}},
  \bibinfo{pages}{201303} (\bibinfo{year}{2009}).

\bibitem[{\citenamefont{Suen et~al.}(1992{\natexlab{a}})\citenamefont{Suen,
  Engel, Santos, Shayegan, and Tsui}}]{suen-1}
\bibinfo{author}{\bibfnamefont{Y.~W.} \bibnamefont{Suen}},
  \bibinfo{author}{\bibfnamefont{L.~W.} \bibnamefont{Engel}},
  \bibinfo{author}{\bibfnamefont{M.~B.} \bibnamefont{Santos}},
  \bibinfo{author}{\bibfnamefont{M.}~\bibnamefont{Shayegan}}, \bibnamefont{and}
  \bibinfo{author}{\bibfnamefont{D.~C.} \bibnamefont{Tsui}},
  \bibinfo{journal}{Phys. Rev. Lett.} \textbf{\bibinfo{volume}{68}},
  \bibinfo{pages}{1379} (\bibinfo{year}{1992}{\natexlab{a}}).

\bibitem[{\citenamefont{Suen et~al.}(1992{\natexlab{b}})\citenamefont{Suen,
  Santos, and Shayegan}}]{suen-2}
\bibinfo{author}{\bibfnamefont{Y.~W.} \bibnamefont{Suen}},
  \bibinfo{author}{\bibfnamefont{M.~B.} \bibnamefont{Santos}},
  \bibnamefont{and} \bibinfo{author}{\bibfnamefont{M.}~\bibnamefont{Shayegan}},
  \bibinfo{journal}{Phys. Rev. Lett.} \textbf{\bibinfo{volume}{69}},
  \bibinfo{pages}{3551} (\bibinfo{year}{1992}{\natexlab{b}}).

\bibitem[{\citenamefont{Suen et~al.}(1994)\citenamefont{Suen, Manoharan, Ying,
  Santos, and Shayegan}}]{suen-3}
\bibinfo{author}{\bibfnamefont{Y.~W.} \bibnamefont{Suen}},
  \bibinfo{author}{\bibfnamefont{H.~C.} \bibnamefont{Manoharan}},
  \bibinfo{author}{\bibfnamefont{X.}~\bibnamefont{Ying}},
  \bibinfo{author}{\bibfnamefont{M.~B.} \bibnamefont{Santos}},
  \bibnamefont{and} \bibinfo{author}{\bibfnamefont{M.}~\bibnamefont{Shayegan}},
  \bibinfo{journal}{Phys. Rev. Lett.} \textbf{\bibinfo{volume}{72}},
  \bibinfo{pages}{3405} (\bibinfo{year}{1994}).

\bibitem[{\citenamefont{Eisenstein et~al.}(1992)\citenamefont{Eisenstein,
  Boebinger, Pfeiffer, West, and He}}]{eisenstein-bilayer}
\bibinfo{author}{\bibfnamefont{J.~P.} \bibnamefont{Eisenstein}},
  \bibinfo{author}{\bibfnamefont{G.~S.} \bibnamefont{Boebinger}},
  \bibinfo{author}{\bibfnamefont{L.~N.} \bibnamefont{Pfeiffer}},
  \bibinfo{author}{\bibfnamefont{K.~W.} \bibnamefont{West}}, \bibnamefont{and}
  \bibinfo{author}{\bibfnamefont{S.}~\bibnamefont{He}}, \bibinfo{journal}{Phys.
  Rev. Lett.} \textbf{\bibinfo{volume}{68}}, \bibinfo{pages}{1383}
  (\bibinfo{year}{1992}).

\bibitem[{\citenamefont{Halperin}(1983)}]{halperin-331}
\bibinfo{author}{\bibfnamefont{B.~I.} \bibnamefont{Halperin}},
  \bibinfo{journal}{Helv. Phys. Acta} \textbf{\bibinfo{volume}{56}},
  \bibinfo{pages}{783} (\bibinfo{year}{1983}).

\bibitem[{\citenamefont{Luhman et~al.}(2008)\citenamefont{Luhman, Pan, Tsui,
  Pfeiffer, Baldwin, and West}}]{luhman}
\bibinfo{author}{\bibfnamefont{D.~R.} \bibnamefont{Luhman}},
  \bibinfo{author}{\bibfnamefont{W.}~\bibnamefont{Pan}},
  \bibinfo{author}{\bibfnamefont{D.~C.} \bibnamefont{Tsui}},
  \bibinfo{author}{\bibfnamefont{L.~N.} \bibnamefont{Pfeiffer}},
  \bibinfo{author}{\bibfnamefont{K.~W.} \bibnamefont{Baldwin}},
  \bibnamefont{and} \bibinfo{author}{\bibfnamefont{K.~W.} \bibnamefont{West}},
  \bibinfo{journal}{Phys. Rev. Lett.} \textbf{\bibinfo{volume}{101}},
  \bibinfo{pages}{266804} (\bibinfo{year}{2008}).

\bibitem[{\citenamefont{Scarola and Jain}(2001)}]{scarola-jain}
\bibinfo{author}{\bibfnamefont{V.~W.} \bibnamefont{Scarola}} \bibnamefont{and}
  \bibinfo{author}{\bibfnamefont{J.~K.} \bibnamefont{Jain}},
  \bibinfo{journal}{Phys. Rev. B} \textbf{\bibinfo{volume}{64}},
  \bibinfo{pages}{085313} (\bibinfo{year}{2001}).

\bibitem[{vit()}]{vito}
\bibinfo{note}{V.~W. Scarola, C. May, M.~R. Peterson, and M. Troyer,
  arXiv:1004.1636 (unpublished).}

\bibitem[{\citenamefont{Zhang and Das~Sarma}(1986)}]{zds}
\bibinfo{author}{\bibfnamefont{F.~C.} \bibnamefont{Zhang}} \bibnamefont{and}
  \bibinfo{author}{\bibfnamefont{S.}~\bibnamefont{Das~Sarma}},
  \bibinfo{journal}{Phys. Rev. B} \textbf{\bibinfo{volume}{33}},
  \bibinfo{pages}{2903} (\bibinfo{year}{1986}).

\bibitem[{\citenamefont{MacDonald et~al.}(1990)\citenamefont{MacDonald,
  Platzman, and Boebinger}}]{macdonald-1}
\bibinfo{author}{\bibfnamefont{A.~H.} \bibnamefont{MacDonald}},
  \bibinfo{author}{\bibfnamefont{P.~M.} \bibnamefont{Platzman}},
  \bibnamefont{and} \bibinfo{author}{\bibfnamefont{G.~S.}
  \bibnamefont{Boebinger}}, \bibinfo{journal}{Phys. Rev. Lett.}
  \textbf{\bibinfo{volume}{65}}, \bibinfo{pages}{775} (\bibinfo{year}{1990}).

\bibitem[{\citenamefont{Schliemann et~al.}(2001)\citenamefont{Schliemann,
  Girvin, and MacDonald}}]{schliemann}
\bibinfo{author}{\bibfnamefont{J.}~\bibnamefont{Schliemann}},
  \bibinfo{author}{\bibfnamefont{S.~M.} \bibnamefont{Girvin}},
  \bibnamefont{and} \bibinfo{author}{\bibfnamefont{A.~H.}
  \bibnamefont{MacDonald}}, \bibinfo{journal}{Phys. Rev. Lett.}
  \textbf{\bibinfo{volume}{86}}, \bibinfo{pages}{1849} (\bibinfo{year}{2001}).

\bibitem[{\citenamefont{Nomura and Yoshioka}(2002)}]{nomura-1}
\bibinfo{author}{\bibfnamefont{K.}~\bibnamefont{Nomura}} \bibnamefont{and}
  \bibinfo{author}{\bibfnamefont{D.}~\bibnamefont{Yoshioka}},
  \bibinfo{journal}{Phys. Rev. B} \textbf{\bibinfo{volume}{66}},
  \bibinfo{pages}{153310} (\bibinfo{year}{2002}).

\bibitem[{\citenamefont{Park}(2004)}]{park}
\bibinfo{author}{\bibfnamefont{K.}~\bibnamefont{Park}}, \bibinfo{journal}{Phys.
  Rev. B} \textbf{\bibinfo{volume}{69}}, \bibinfo{pages}{045319}
  (\bibinfo{year}{2004}).

\bibitem[{\citenamefont{Simon et~al.}(2003)\citenamefont{Simon, Rezayi, and
  Milovanovic}}]{simon}
\bibinfo{author}{\bibfnamefont{S.~H.} \bibnamefont{Simon}},
  \bibinfo{author}{\bibfnamefont{E.~H.} \bibnamefont{Rezayi}},
  \bibnamefont{and} \bibinfo{author}{\bibfnamefont{M.~V.}
  \bibnamefont{Milovanovic}}, \bibinfo{journal}{Phys. Rev. Lett.}
  \textbf{\bibinfo{volume}{91}}, \bibinfo{pages}{046803}
  (\bibinfo{year}{2003}).

\bibitem[{\citenamefont{M\"oller et~al.}(2008)\citenamefont{M\"oller, Simon,
  and Rezayi}}]{moller-1}
\bibinfo{author}{\bibfnamefont{G.}~\bibnamefont{M\"oller}},
  \bibinfo{author}{\bibfnamefont{S.~H.} \bibnamefont{Simon}}, \bibnamefont{and}
  \bibinfo{author}{\bibfnamefont{E.~H.} \bibnamefont{Rezayi}},
  \bibinfo{journal}{Phys. Rev. Lett.} \textbf{\bibinfo{volume}{101}},
  \bibinfo{pages}{176803} (\bibinfo{year}{2008}).

\bibitem[{\citenamefont{M\"oller et~al.}(2009)\citenamefont{M\"oller, Simon,
  and Rezayi}}]{moller-2}
\bibinfo{author}{\bibfnamefont{G.}~\bibnamefont{M\"oller}},
  \bibinfo{author}{\bibfnamefont{S.~H.} \bibnamefont{Simon}}, \bibnamefont{and}
  \bibinfo{author}{\bibfnamefont{E.~H.} \bibnamefont{Rezayi}},
  \bibinfo{journal}{Phys. Rev. B} \textbf{\bibinfo{volume}{79}},
  \bibinfo{pages}{125106} (\bibinfo{year}{2009}).

\bibitem[{foo()}]{footnote1}
\bibinfo{note}{See J.~R. Schrieffer, \emph{Theory of Superconductivity}
  (Addison-Wesly, Reading, MA, 1964) and Ref.~\onlinecite{read-green} for a
  discussion of the Pfaffian form of the real space BCS pairing wavefunction.}

\bibitem[{\citenamefont{Milovanovi\ifmmode~\acute{c}\else \'{c}\fi{} and
  Read}(1996)}]{mvm-nr}
\bibinfo{author}{\bibfnamefont{M.}~\bibnamefont{Milovanovi\ifmmode~\acute{c}\e%
lse \'{c}\fi{}}} \bibnamefont{and}
  \bibinfo{author}{\bibfnamefont{N.}~\bibnamefont{Read}},
  \bibinfo{journal}{Phys. Rev. B} \textbf{\bibinfo{volume}{53}},
  \bibinfo{pages}{13559} (\bibinfo{year}{1996}).

\bibitem[{\citenamefont{Rezayi and Haldane}(2000)}]{rezayi-haldane}
\bibinfo{author}{\bibfnamefont{E.~H.} \bibnamefont{Rezayi}} \bibnamefont{and}
  \bibinfo{author}{\bibfnamefont{F.~D.~M.} \bibnamefont{Haldane}},
  \bibinfo{journal}{Phys. Rev. Lett.} \textbf{\bibinfo{volume}{84}},
  \bibinfo{pages}{4685} (\bibinfo{year}{2000}).

\bibitem[{\citenamefont{Haldane}(1983)}]{haldane}
\bibinfo{author}{\bibfnamefont{F.~D.~M.} \bibnamefont{Haldane}},
  \bibinfo{journal}{Phys. Rev. Lett.} \textbf{\bibinfo{volume}{51}},
  \bibinfo{pages}{605} (\bibinfo{year}{1983}).

\bibitem[{\citenamefont{Yang and Yang}(1976)}]{wu-yang}
\bibinfo{author}{\bibfnamefont{T.~T.} \bibnamefont{Yang}} \bibnamefont{and}
  \bibinfo{author}{\bibfnamefont{C.~N.} \bibnamefont{Yang}},
  \bibinfo{journal}{Nucl. Phys. B} \textbf{\bibinfo{volume}{107}},
  \bibinfo{pages}{365} (\bibinfo{year}{1976}).

\bibitem[{pap()}]{papic-3}
\bibinfo{note}{Z. Papi\'{c}, M. O. Goerbig, N. Regnault and M. V.
  Milovanovi\'{c}, arXiv:0912.3103 (unpublished).}

\bibitem[{\citenamefont{Yoshioka et~al.}(1983)\citenamefont{Yoshioka, Halperin,
  and Lee}}]{yhl}
\bibinfo{author}{\bibfnamefont{D.}~\bibnamefont{Yoshioka}},
  \bibinfo{author}{\bibfnamefont{B.~I.} \bibnamefont{Halperin}},
  \bibnamefont{and} \bibinfo{author}{\bibfnamefont{P.~A.} \bibnamefont{Lee}},
  \bibinfo{journal}{Phys. Rev. Lett.} \textbf{\bibinfo{volume}{50}},
  \bibinfo{pages}{1219} (\bibinfo{year}{1983}).

\bibitem[{\citenamefont{Yoshioka}(1984)}]{yoshioka-1}
\bibinfo{author}{\bibfnamefont{D.}~\bibnamefont{Yoshioka}},
  \bibinfo{journal}{Phys. Rev. B} \textbf{\bibinfo{volume}{29}},
  \bibinfo{pages}{6833} (\bibinfo{year}{1984}).

\bibitem[{\citenamefont{Haldane}(1985)}]{haldane-torus}
\bibinfo{author}{\bibfnamefont{F.~D.~M.} \bibnamefont{Haldane}},
  \bibinfo{journal}{Phys. Rev. Lett.} \textbf{\bibinfo{volume}{55}},
  \bibinfo{pages}{2095} (\bibinfo{year}{1985}).

\bibitem[{\citenamefont{Wen and Zee}(1998)}]{wz}
\bibinfo{author}{\bibfnamefont{X.-G.} \bibnamefont{Wen}} \bibnamefont{and}
  \bibinfo{author}{\bibfnamefont{A.}~\bibnamefont{Zee}},
  \bibinfo{journal}{Phys. Rev. B} \textbf{\bibinfo{volume}{58}},
  \bibinfo{pages}{15717} (\bibinfo{year}{1998}).

\bibitem[{\citenamefont{Read and Rezayi}(1996)}]{read-rezayi}
\bibinfo{author}{\bibfnamefont{N.}~\bibnamefont{Read}} \bibnamefont{and}
  \bibinfo{author}{\bibfnamefont{E.}~\bibnamefont{Rezayi}},
  \bibinfo{journal}{Phys. Rev. B} \textbf{\bibinfo{volume}{54}},
  \bibinfo{pages}{16864} (\bibinfo{year}{1996}).

\bibitem[{\citenamefont{Shi et~al.}(2008)\citenamefont{Shi, Jolad, Regnault,
  and Jain}}]{shi}
\bibinfo{author}{\bibfnamefont{C.}~\bibnamefont{Shi}},
  \bibinfo{author}{\bibfnamefont{S.}~\bibnamefont{Jolad}},
  \bibinfo{author}{\bibfnamefont{N.}~\bibnamefont{Regnault}}, \bibnamefont{and}
  \bibinfo{author}{\bibfnamefont{J.~K.} \bibnamefont{Jain}},
  \bibinfo{journal}{Phys. Rev. B} \textbf{\bibinfo{volume}{77}},
  \bibinfo{pages}{155127} (\bibinfo{year}{2008}).

\bibitem[{\citenamefont{M\"oller and Simon}(2008)}]{moller}
\bibinfo{author}{\bibfnamefont{G.}~\bibnamefont{M\"oller}} \bibnamefont{and}
  \bibinfo{author}{\bibfnamefont{S.~H.} \bibnamefont{Simon}},
  \bibinfo{journal}{Phys. Rev. B} \textbf{\bibinfo{volume}{77}},
  \bibinfo{pages}{075319} (\bibinfo{year}{2008}).

\end{thebibliography}
\end{document}